\documentclass[12pt, letterpaper]{article}

\usepackage{arxiv}

\usepackage{times}
\usepackage[colorlinks=true, linkcolor=teal, breaklinks=true]{hyperref}
\usepackage{helvet}
\usepackage{enumerate}
\usepackage{pgfplots}
\usepackage{mathtools,relsize}
\usepackage{float}
\usepackage{latexsym}
\usepackage{setspace}
\usepackage{cite}
\usepackage{amsmath,amssymb,amsfonts,amsthm}
\usepackage{algorithm,algorithmic}
\usepackage{graphicx}
\usepackage{textcomp}
\usepackage{xcolor}
\usepackage{multirow}
\usepackage{adjustbox}
\usepackage{nicematrix}
\usepackage{url,xurl}
\usepackage{booktabs}
\usepackage{caption,subcaption}
\usepackage[numbers]{natbib}

\usepackage[scr=boondox]{mathalpha}

\DeclareFontFamily{OT1}{pzc}{}
\DeclareFontShape{OT1}{pzc}{m}{it}{<-> s * [1.10] pzcmi7t}{}
\DeclareMathAlphabet{\mathpzc}{OT1}{pzc}{m}{it}

\usepackage{hyperref}
\hypersetup{
     colorlinks   = true,
     linkcolor    = purple,
     citecolor    = teal
}

\newtheorem{definition}{Definition}

\definecolor{yaleblue}{rgb}{0.06, 0.3, 0.57}
\definecolor{darklavender}{rgb}{0.45, 0.31, 0.59}
\definecolor{upsdellred}{rgb}{0.68, 0.09, 0.13}

\usepackage{tkz-euclide,tikz,picture,xcolor,color,tikz-cd,pst-plot,pgfplots}      
    \usetikzlibrary{arrows,chains}
    \usetikzlibrary{automata,positioning,shapes,calc,patterns,intersections}
    \usetikzlibrary{matrix,trees}
    \usetikzlibrary{shapes.geometric}
    \usetikzlibrary{arrows, arrows.meta,graphs,quotes,shadows}
    \usetikzlibrary{decorations.pathmorphing,backgrounds,fit,shapes.symbols,chains}
    \usetikzlibrary{decorations,decorations.markings,decorations.text}
    \usetikzlibrary{3d, calc}  
    \usetikzlibrary{decorations.pathmorphing,backgrounds,petri}
    \usetikzlibrary{decorations.pathreplacing}
    \usetikzlibrary{decorations.shapes}
    \usetikzlibrary{quotes,angles}
    \usetikzlibrary{hobby}

\tikzset{-dot-/.style={decoration={
  markings,
  mark=at position #1 with {\fill circle (2pt);}},postaction={decorate}}}

\title{A Game‑Theoretic Framework for Incentive‑Compatible AI training Under Renewable‑Energy Constraints}

\usepackage{fancyhdr}
\author{
  \shortstack{${}^1$Konstantinos \hspace{0.2pt} ${}$ \\ Varsos}
  \quad 
  \shortstack{${}^2$Ramin \hspace{0.2pt} ${}$ \\ Khalili}
  \quad 
  \shortstack{${}^1$Adamantia \hspace{0.2pt} ${}$ \\ Stamou}
  \quad 
  \shortstack{${}^1$George D. \hspace{0.2pt} ${}$ \\ Stamoulis}
  \quad 
  \shortstack{${}^1$Vasillios A. \hspace{0.2pt} ${}$ \\ Siris}\\
  ${}$\vspace{1pt}\\
\textit{${}^1$Athens University of Economics and Business, Greece}\\
\textit{\normalsize ${}^2$Independent researcher, Munich Germany}\\
${}$\vspace{1pt}\\
\{kvarsos, stamouad, gstamoul, vsiris\}@aueb.gr, \; raminster@gmail.com}

\date{}

\begin{document}

\maketitle

\begin{abstract}
As artificial intelligence systems increasingly rely on distributed and collaborative training, the energy footprint of these processes becomes a shared responsibility. Modern AI training often unfolds across heterogeneous compute nodes-ranging from cloud clusters to edge devices-whose energy availability is spatially and temporally variable. At the same time, renewable energy grids experience growing levels of excess generation, creating opportunities to align computational workloads with low-carbon energy supply. In this work, we develop a game-theoretic model of carbon-aware AI training in which autonomous agents strategically choose whether to participate and how intensively to train under limited renewable energy availability. Each agent balances diminishing learning returns, rewards for remaining within green-energy budgets, and penalties for grid consumption. While our framework applies broadly to distributed AI training, we examine Federated Learning as a representative case study due to its decentralized structure and flexible scheduling. We analyze equilibrium existence, efficiency, and adaptive dynamics, and provide simulation evidence that appropriately designed incentives can eliminate grid-based energy usage while preserving model performance. Our findings demonstrate how incentive‑compatible training mechanisms can enhance energy efficiency and sharply reduce carbon emissions under renewable‑energy constraints.
\end{abstract}


\section{Introduction}\label{sec:intro}

Artificial intelligence (AI) training has become a significant and growing source of energy consumption. As models increase in scale and complexity, their computational demands translate directly into substantial electricity usage and carbon emissions \cite{Strubell2019EnergyAP,Patterson2021CarbonEA}. Addressing these sustainability challenges requires not only technical innovation but also systemic redesign of digital services. European research efforts, and in particular EXIGENCE \cite{EXIGENCESbD}, exemplify this broader transition toward user-centric incentive models and energy-aware service design, i.e., see in \cite{Varsos2026UserAM,Varsos2026OptimalES}.

Many modern AI systems are trained in distributed environments, where data are generated at the network edge and cannot be centrally aggregated due to privacy, regulatory, or bandwidth constraints \cite{garcia2019estimation}. A prominent architectural paradigm in this setting is Federated Learning (FL) \cite{McMahan2016CommunicationEfficientLO,savazzi2020federated,Wiesner2023FedZeroLR}, in which multiple agents collaboratively train a shared model without transmitting raw data. Instead, agents perform local computation and periodically send model updates to a coordinating service provider.

In this work, we adopt the Federated Learning structure as a canonical model of distributed AI training. The round-based nature of FL, the separation between local computation and aggregation, and the presence of heterogeneous edge participants make it a useful abstraction for studying energy consumption in distributed AI systems. Importantly, the energy challenges we analyze are not specific to FL per se, but arise more broadly in decentralized and edge-based training architectures.

From an energy perspective, distributed training introduces additional inefficiencies compared to centralized training, including repeated communication rounds and execution on heterogeneous, often less energy-efficient devices \cite{Qiu2020AFL}. At the same time, renewable energy grids increasingly experience periods of excess generation, during which energy is curtailed because supply exceeds demand or transmission capacity \cite{Bird2014WindAS}. This stranded renewable energy presents an opportunity: if AI training workloads can be temporally and spatially aligned with renewable availability, their carbon footprint can be significantly reduced \cite{Patterson2021CarbonEA,Lacoste2019QuantifyingTC}.

Carbon-aware computing has therefore emerged as a promising direction for adapting computational workloads to variations in renewable energy supply \cite{Radovanovic2021CarbonAwareCF}. Distributed training architectures such as Federated Learning are particularly suitable for such adaptation because training proceeds in rounds and often lacks strict real-time constraints \cite{Bonawitz2019TowardsFL}. However, most existing work models participating agents as passive resources that execute tasks when scheduled. In practice, participants are autonomous entities with their own objectives, resource constraints, and operational considerations \cite{Kang2019IncentiveDF,Zhan2020ALI}. When renewable availability fluctuates, agents must decide not only whether to participate in a given training round, but also how intensively to train \cite{Lim2022DecentralizedEI}.

This observation motivates a strategic perspective on carbon-aware Federated Learning. We consider a setting in which each agent is modeled as rational and self-interested. In every training round, agents decide whether to participate and, conditional on participation, how many local samples to process. These decisions are constrained by the availability of renewable excess energy in their local environment \cite{Thakur2024GreenFL}. Agents aim to balance the benefits of contributing to the global model--such as monetary rewards, reputation gains, or improved local performance--against the cost of energy consumption and potential penalties for exceeding green energy availability \cite{Sarikaya2019MotivatingWI}.

The central problem we study is twofold. First, \emph{how can participation be incentivized when renewable energy availability is volatile and heterogeneous across agents?} Second, \emph{how can training intensity be endogenously determined so that agents fully utilize available green energy without exceeding it, thereby avoiding carbon-intensive fallback energy sources?} Naively ignoring strategic behavior may lead to inefficient outcomes: some agents may free-ride by under-participating, while others may over-consume resources when incentives are misaligned \cite{Meng2024FederatedLA}. Moreover, uncoordinated participation decisions can increase training time, introduce bias, or create instability in the learning process \cite{Bonawitz2019TowardsFL}.

To address this challenge, we propose modeling carbon-aware Federated Learning as a repeated strategic interaction. Within this framework, agents choose participation and training intensity subject to renewable energy constraints, while a special entity, called AI-service provider, determines penalties, rewards, and aggregation. The objective is to design incentive-compatible mechanisms that (i) encourage sufficient participation, and (ii) align local training effort with locally available green energy. By integrating strategic decision-making with carbon-aware scheduling, this approach aims to achieve sustainable federated training with near-zero operational emissions while preserving efficiency and robustness. Towards this direction, we consider the AI-service provider as a trustworthy mediator that \emph{publicly} signals information to the agents. First, we examine the dynamics generated by decentralized learning rules--specifically, a \emph{fictitious-play heuristic} \cite{brown:fp1951}--to determine whether agents’ adaptive behavior converges to stable outcomes, that is the \emph{Nash equilibria} of the interaction, and how quickly such convergence occurs.

Besides the decentralized heuristic phase, we propose a second approach in which the AI-service provider makes \emph{private} recommendations to each agent on training-intensity plans, in parallel to the information that \emph{publicly} signals. If agents condition their strategies on a common public signal, their decisions may become correlated, leading to outcomes that can be characterized as \emph{correlated equilibria} \cite{Aumann1974SubjectivityAC}. Under this signaling-based mechanism, the AI-service provider can incorporate forecasts of renewable availability and cross-agent heterogeneity into its recommendations. Provided that adherence to the recommendation is incentive-compatible, agents optimally follow the signal. This coordinated approach can outperform purely decentralized adaptation by leveraging heterogeneity in renewable‑energy availability. As a direct consequence, our incentive‑compatible training mechanisms enhance overall energy efficiency leading to zero carbon emissions under renewable‑energy constraints.

\paragraph*{Outline.}
Section \ref{sec:related work} summarizes the key related literature. Section \ref{sec:preliminaries} introduces the necessary preliminaries, including game‑theoretic foundations, training‑accuracy considerations, and the associated energy‑consumption formula. Section \ref{sec:model} presents our framework, followed by two complementary solution concepts: (i) decentralized adaptive dynamics based on fictitious play, and (ii) a signaling‑based mechanism‑design approach that induces correlated strategies through a trusted coordinating intermediary. Section \ref{sec:evaluation} reports our evaluation results, and Section \ref{sec:conclusions} concludes the paper.

\begin{figure}[!t]
    \centering
    \includegraphics[width=0.87\linewidth]{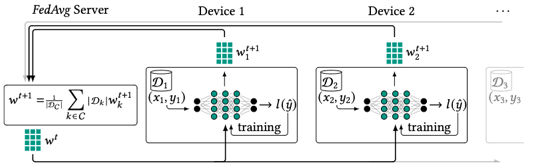}
        \caption{Schematic of a Federated Learning system where, the knowledge of devices  1, $\ldots$, is shared through averaging the NNs’ weights. Each device trains on its disjoint local data $\mathcal{D}_1, \ldots$, producing newly trained weights $w^{t+1}_1, \ldots$. Every round, the new averaged weights $\mathbf{w}^{t+1}$ are distributed to all devices. see in \cite{Pfeiffer2023FederatedLF}.}
    \label{fig:FedZero}
\end{figure}

\section{Related work}\label{sec:related work}

Game theory becomes increasingly important to mitigate the issues and challenges arise from the integration of Federated Learning into into energy-constrained environments \cite{Kang2019IncentiveDF}. Two common approaches are: (i) contract theory, and (ii) auction-based mechanisms. In contract theory models, the AI-service provider offers a list of contracts so that agents self-select according to their private types (e.g., energy budget, data quality), which yields incentive-compatible participation and improves overall learning performance, see in \cite{Kang2019IncentiveDF}. Reverse auctions let agents bid for participation; the aggregator selects participants based on bids to maximize utility while minimizing energy drain on weaker devices, see in \cite{Han2025AFL}.

A second stream of works, utilizes hierarchical leader–follower models (Stackelberg games) are widely used to capture the interaction between a central aggregator (leader), and distributed agents (followers) \cite{Fang2024LossConvergenceDF,Zhan2020ALI}. In such models the leader (AI-service provider) typically announces rewards, precision targets, or pricing, and the followers respond by choosing local CPU frequencies, numbers of local iterations, or transmission power to trade off energy consumption against the server’s accuracy/latency goals, \cite{Khan2019FederatedLF,Sarikaya2019MotivatingWI}. 

A related line of work \cite{Sim2020CollaborativeML} shows how a central coordinator can disclose truthful signals to guide agents toward more efficient participation and training decisions. Building on this idea, we introduce two policies for the AI‑service provider: one that sets only the non‑participation penalty parameter $\gamma$, and another that additionally provides truthful information to promote both agent participation and training efficiency.

Several works formalize the trade-off between number of Federated Learning rounds, such as convergence speed and accuracy, and total energy consumption \cite{Nguyen2022TowardEH}; they show existence (and sometimes uniqueness) of equilibria where the global model converges within bounded rounds while device energy budgets are respected \cite{Le2020AnIM}. These analyses are frequently embedded in joint optimization or dynamic-resource allocation frameworks. However, computing such equilibria is often intractable in practice. To address this limitation, we introduce two practical alternatives: (i) decentralized learning through fictitious play and (ii) coordination via correlated strategies. Additionally, while prior work mitigates inefficiencies by selectively excluding devices from each round using cooperative or evolutionary game‑theoretic tools, our framework keeps all agents eligible to participate in every round.

In \cite{Yin2023AGA}, Federated Learning is modeled as an extensive‑form game that balances energy consumption, privacy preservation, and model accuracy. In contrast, our setting assumes that agents make their decisions simultaneously rather than sequentially. Work in \cite{Zou2019MobileDT} analyzes the trade‑off between energy consumption and Quality of Service (QoS) in FL and derives an evolutionary stable strategy that identifies the optimal behavior for all players; however, this line of work does not address privacy leakage. Complementarily, \cite{Liu2021AnIM} studies the interaction between uploaded sensing data and task performance using a Stackelberg game to derive optimal user‑privacy strategies, but focuses solely on the privacy–accuracy trade‑off and does not incorporate energy consumption during training.
Our framework extends these prior efforts by jointly modeling energy usage, training performance, and participation incentives within a unified game‑theoretic setting. Unlike previous works, we capture simultaneous decision‑making, explicitly integrate renewable‑energy constraints, and design incentive mechanisms that promote both efficient training and sustainable energy use.

\section{Preliminaries}\label{sec:preliminaries}

In this section, we review some elementary notions and notation used in subsequent sections. For any natural number $d \in \mathbb{N}$, $[d]$ denotes the set $\{1, 2, \dots, d\}$. Vectors in $\mathbb{R}^d$ are denoted by bold, e.g., $\boldsymbol{x}$, with coordinates $x_i$, $i \in [d]$, while scalar values by light, e.g., $x$. Further, $\mathbf{1}\{\cdot\}$ denotes the indicator function.

We study the interaction between rational and self-interested agents through the lens of game theory \cite{Osborne1995ACI}. Given a time horizon $T$, a $\mathcal{N}$-player finite normal-form game at time $t \in [T]$, denoted as $G^t$, has a finite number of agents, $\mathcal{N}^t \in \mathbb{N}$, and for each agent $i \in [\mathcal{N}^t]$ a finite set $S^t_i$ of actions. We assume that the number of agents and the number of actions per agent remain constant for all $t \in [T]$, therefore, we simplify the notation to $\mathcal{N}$ and $S_i$, respectively.

The set of pure strategy profiles is the Cartesian product of $S_i$'s, and is denoted by $S$. We denote the set agents other than $i$ by $-i$. Finally, for each agent $i$ and $\mathbf{s} \in S$, $i$ receives a real value, called payoff, and denoted as $u_i(\mathbf{s}, t)$ at $t$. We refer to the set $\{u_i(\mathbf{s}, t)\}_{\mathbf{s} \in S}$ as the payoff table of agent $i$, agglomerating all the payoff tables for all $i \in [\mathcal{N}]$, we take the payoff table of the game, denoted as $\{\mathbf{u}^t_i\}_{i \in \mathcal{N}}$. Succinctly, a $\mathcal{N}$-player finite normal-form game $G^t$ can be written as the tuple $\langle [\mathcal{N}], S, \{\mathbf{u}^t_i\}_{i \in \mathcal{N}}\rangle$ for $t \in [T]$.

For a finite set $A$, let $\Delta(A)$ denote the set of probability distributions over $A$. For $i \in [\mathcal{N}]$, let $\Delta_i := \Delta(S_i)$ denote the set of \emph{mixed strategies} available to agent $i$. That is, the agent $i$ randomizes over its actions $S_i$. Formally,  a mixed strategy for agent $i$ is a distribution on $S_i$, that is, real numbers $x_{i, j} \geq 0$ for each action $j \in S_i$  s.t. $\sum_{j \in S_i} x_{i, j} = 1$. Let $\Delta := \prod_{i \in [\mathcal{N}]} \Delta_i$ denote the set of joint mixed strategies. A mixed strategy profile at $t$ is $\boldsymbol{\sigma}^t = (\boldsymbol{\sigma}^t_1, \boldsymbol{\sigma}^t_2, \ldots, \boldsymbol{\sigma}^t_{\mathcal{N}}) \in \Delta$, where $\boldsymbol{\sigma}^t_{i}$ is the mixed strategy of agent $i$. Similarly, the pure strategy of agent $i$ at $t$ is denoted as $s^t_i$ and $\boldsymbol{s}^t$ is the pure strategy profile at $t$.

\noindent Given a mixed strategy profile $\boldsymbol{\sigma}^t \in \Delta$, the expected payoff of agent $i$ is given by
\[
    u_i(\boldsymbol{\sigma}^t, t) = \sum_{\boldsymbol{s}^t \in S} u_i(\boldsymbol{s}, t) \prod_{i \in [\mathcal{N}]} \boldsymbol{\sigma}^t_i(s_i).
\]
For a mixed strategy profile $\boldsymbol{\sigma} \in \Delta$, the best response of agent $i$ is given by the set-valued function $BR_i : \Delta_{-i} \times [0, T] \to \Delta_i$,
\[
    BR_i(\boldsymbol{\sigma}_{-i}, t) := \text{arg}\max_{\boldsymbol{x} \in \Delta_i} \{u_i((\boldsymbol{x}, \boldsymbol{\sigma}_{-i}), t)\}.
\]
For a mixed strategy profile $\boldsymbol{\sigma}^t \in \Delta$, the joint best response is given by the set-valued function $BR : \Delta \times [0, T] \to \Delta$,
\[
    BR(\boldsymbol{\sigma}, t) := BR_1(\boldsymbol{\sigma}_{-1}, t) \times BR_2(\boldsymbol{\sigma}_{-2}, t) \times \ldots \times BR_{|\mathcal{N}|}(\boldsymbol{\sigma}_{-{|\mathcal{N}}}|, t).
\]
The fundamental solution concept in game theory is that of Nash equilibrium, stating that no agent has an incentive to deviate to another strategy.
\begin{definition}[Nash equilibrium]\label{def:Nash}
Let $G^t = \langle [\mathcal{N}], S, \{\mathbf{u}^t_i\}_{i \in \mathcal{N}}\rangle$ be a normal-form game at time $t$. A Nash equilibrium of $G^t$ is a strategy profile $\boldsymbol{\sigma}^* \in \Delta(S)$ such that, for each $i \in \mathcal{N}$ and each $\boldsymbol{\sigma}^t_i \in \Delta(S_i)$,
\[
    u_i(\boldsymbol{\sigma}^*, t) \geq u_i((\boldsymbol{\sigma}^t_i; \boldsymbol{\sigma}^*_{-i}), t).
\]
\end{definition}
\noindent Nash's theorem \cite{Nash1951NONCOOPERATIVEG} asserts that every finite normal-form game exhibits a Nash equilibrium.

\subsection{Fictitious play}\label{subsec:fictitious play}

In this section we introduce the basic definition of discrete-time fictitious play. Fictitious play refers to a dynamic process where at each stage, agents play a pure best response to the empirical distribution of their opponent’s play, see in \cite{brown:fp1951} and \cite{Robinson1951ANIM}.

Let $s_i^t$ denote the action played by agent $i$ at time $t$. The empirical frequency of agent $i$'s play up to time $t$ is defined as
\begin{equation}
    m_i^t(s_i) = \sum_{\tau = 0}^{t-1} \mathbf{1}\{s_i^\tau = s_i\},
\end{equation}
Thus, $\boldsymbol{m}_i^t$ is an $|S_i|$-dimensional vector whose components count the number of times agent $i$ has played each action. The empirical distribution of agent $i$'s play up to time $t$ is given by
\begin{equation}
    \mu_i^t(s_j) = \frac{m_i^t(s_j)}{t}.
\end{equation}
Finally, let $\boldsymbol{\mu}^t$ denote the joint distribution on $\prod_{i \in \mathcal{N}} S_i$ obtained as the independent product of the individual empirical distributions $\boldsymbol{\mu}_i^t$.

In discrete-time fictitious play, each agent selects an arbitrary action at time $t=0$. For every $t > 0$, agent $i$ plays a pure best response to the product of the marginal empirical distributions of its opponents. Formally, for all $t > 0$ and for each agent $i$, $s_i^t \in \mathrm{BR}_i\!\left(\boldsymbol{\mu}_{-i}^t, t\right)$, where $\boldsymbol{\mu}_{-i}^t$ denotes the product of the empirical distributions of all agents other than $i$, and $\mathrm{BR}_i(\cdot)$ denotes the set of pure best responses of agent $i$.

\begin{algorithm}[h]
\caption{Fictitious play algorithm}
\label{alg:fictitious play algorithm}
\begin{algorithmic}[1]
\REQUIRE A finite normal-form game $\langle [\mathcal{N}], S, \{\mathbf{u}^t_i\}_{i \in [\mathcal{N}]}\rangle$, an initial mixed strategy profile $\boldsymbol{\sigma}$, a finite time $\mathcal{T}$.
\STATE $\boldsymbol{\mu}^0 \gets \boldsymbol{\sigma}$.
\WHILE{$1 \leq \tau \leq \mathcal{T}$}
    \STATE Compute best responses, $s_i^t \in \mathrm{BR}_i\!\left(\boldsymbol{\mu}_{-i}^\tau, t\right)$.
    \STATE Compute empirical frequencies for each action, $m_i^t(s_i) = \sum_{\ell = 0}^{\tau-1} \mathbf{1}\{s_i^\ell = s_i\}$.
    \STATE Update empirical distributions, $\mu_i^t(s_i) = \frac{m_i^t(s_i)}{\tau}$.
\ENDWHILE
\STATE $\boldsymbol{\sigma}^{t+1} \gets \boldsymbol{\mu}^\mathcal{T}$.
\RETURN A mixed strategy profile $\boldsymbol{\sigma}^{t+1}$.
\end{algorithmic}
\end{algorithm}
Implementing the fictitious play algorithm we consider two assumptions, first, we assume that agents move \emph{simultaneously}, and second, the agents put equal weight on every play in the past.

\subsection{Correlated equilibria}\label{subsec: corrlelated equilibria}
Now we twist the previous setting introducing a correlation between the strategies of the agents.  To that end, a trusted mediator signals information and makes recommendations but not enforce behavior, acting as a \emph{correlation device}. The distribution $\chi$ from which the correlation device samples recommendations is \emph{public knowledge}, and in addition the agents only get \emph{private} recommendations from the correlation device. This lead to \emph{correlated equilibrium} solution concept, first introduced in \cite{Aumann1974SubjectivityAC}. Formally, a correlated equilibrium is defined as follows
\begin{definition}[Correlated equilibrium]\label{def: corrlelated equilibria}
Given a finite normal-form game $G^t = \langle [\mathcal{N}], S, \{\mathbf{u}^t_i\}_{i \in [\mathcal{N}]}\rangle$ at time $t$, correlated equilibrium is a distribution $\chi$ over $S$ such that for all agents $i \in [\mathcal{N}]$ and actions $a$, $a' \in S_i$
    \[
        \mathbb{E}_{s \sim \chi} [u_i((a, \mathbf{s}_{-i}), t) \mid s_i = a] \geq \mathbb{E}_{s \sim \chi} [u_i((a', \mathbf{s}_{-i}), t) \mid s_i = a].
    \]
\end{definition}
Nash equilibria are also correlated equilibria, therefore every finite normal-form game has a correlated equilibrium. From the perspective of correlated equilibria, a Nash equilibrium is the special case in which each agent’s actions are drawn from an independent distribution, and hence conditioning on $s$ provides no additional information about $\mathbf{s}_{-i}$.  \\

Importantly, we treat $\langle [\mathcal{N}], S, \{\mathbf{u}^t_i\}_{i \in \mathcal{N}}\rangle$ as a sequence of independent stage games. Our analysis focuses on the equilibrium outcome in each round $t$ separately rather than on a dynamic solution across the horizon $T$. Consequently, we do not consider equilibrium refinements for sequential games.

\subsection{Federated Learning \& Training Component}\label{subsec: local training}

Consider that a set of $\mathcal{N}$ agents interconnected with a power domain, each characterized by energy-related and learning-related attributes. In what follows, the proposed framework in Section \ref{sec:model} is agnostic to the underlying machine-learning methodology and does not depend on a specific optimizer, model architecture, or training algorithm. For analytical simplicity, we assume that the outcome of a local training step depends on the amount and utility of the local training data, as well as on the current state of the global model that is being optimized. Other factors, including the employed learning algorithm, optimizer configuration, learning rate, memory constraints, and computational resources, can be incorporated orthogonally via additional parameters and/or properly defined functions. We consider a finite training horizon $T$, divided into discrete time slots indexed by $t \in \{0, \ldots, T\}$.

Furthermore, we assume that the AI-service provider specifies a target global accuracy threshold, denoted as $\mathscr{a}$, that serves as the \emph{stopping criterion} for the training process. Unlike a centralized setting, where the provider can explicitly terminate training once the desired accuracy is achieved, the proposed decenatralized framework does not allow the provider to enforce a strict training decision. Instead, the accuracy threshold represents a target that is attained through the collective actions and decisions of the participating users. Once the global model reaches this target, no further training is required, thereby avoiding unnecessary computation and energy consumption while maintaining the desired predictive performance. Limiting training beyond the required accuracy can also help mitigate overfitting and improve the model's generalization capability. In privacy-preserving learning settings, it may additionally contribute to balancing the trade-off between model utility and privacy preservation, as discussed in \cite{Chai2023ASF}.
 
We consider a setting where a central AI-service provider assigns a learning model to the $\mathcal{N}$ agents and sets the training accuracy threshold $\mathscr{a}$. At the beginning of each round $t$, the AI-service provider broadcasts model $\mathbf{w}^t$, characterized by predictive accuracy $\Bar{\alpha}^t \in (0, 1)$, to all agents simultaneously. Initially, the training process starts from a baseline model with predictive accuracy $\Bar{\alpha}^{0}$. Upon receiving the global model, each agent $i$ performs local training using its private data, and then returns the updated model to the AI-service provider in a synchronized manner. The AI-service provider then aggregates the received updates to construct a refined global model. This iterative procedure continues until the model reaches a predetermined threshold, e.g., a prescribed predictive accuracy or until exhausting the training horizon.

Each agent $i \in \mathcal{N}$ possesses a private bundle of data $\mathcal{D}_i$. During round $t$ it chooses a subset of samples $d^t_i \in \mathcal{D}_i$ for local training. The quality and usefulness of the selected training data are represented through a time-dependent quality function $\theta_i: \mathcal{D}_i \times [0, T] \to [0, 1]$. Intuitively, $\theta_i(d^t_i, t)$ captures attributes such as data quality, informativeness, diversity, and relevance of the selected samples to the current model state at time $t$.

Starting from the global predictive accuracy $\Bar{\alpha}^t$, each agent performs local training and improves the received model according to its selected training effort and data quality. Specifically, the local predictive accuracy achieved by agent $i$ after local training at round $t+1$ is modeled as $\alpha_i(d^t_i, \theta_i(d^t_i, t+1), t+1) = \Bar{\alpha}^t + \xi_i(d^{t+1}_i, \theta_i(d^{t+1}_i, t+1), t+1)$, where $\xi_i(\cdot) \in [0, 1]$ represents the training gain obtained by agent $i$ at round $t + 1$. The gain function $\xi(\cdot)$ depends on the amount of local samples used, their quality, and potentially other training-related factors, such as the number of local iterations. Since $\xi_i(\cdot) \geq 0$, the local accuracy is non-decreasing over time and satisfies $\alpha_i(\cdot, \theta_i(\cdot, t+1), t+1) \geq \Bar{\alpha}^t$, which is the average accuracy across agents at round $t$.

After local training, agent $i$ submits its updated model $\mathbf{w}_i^{t+1}$ to the AI-service provider. The provider aggregates the received local models to construct the global model $\mathbf{w}^{t+1}$ for the next round. Since agents may contribute different amounts of training data, the aggregation is weighted according to the number of samples used during local training. 

We define the nominal local accuracy achieved by agent $i$ after participating in training at round $t$ as $\alpha_i^{(+)}(t) = \alpha_i(d^t_i, \theta_i(d^t_i, t), t)$. Then the aggregated accuracy of the FL process is defined as
\begin{equation}\label{eq:aggregated training accuracy}
    \Bar{\alpha}^t = \sum_{i \in \mathcal{N}} \left( \frac{d^t_i}{\sum_{j \in \mathcal{N}} d^t_j}\right) \alpha^{(+)}_i(t),    
\end{equation}
where $\frac{d^t_i}{\sum_{j \in \mathcal{N}} d^t_j}$ denotes the normalized contribution of user $i$, proportional to the amount of data used in its local training. The FL training is considered complete once $\alpha^{(+)}_i > \mathscr{a}$ for any $i \in \mathcal{N}$ at some finite iteration $t \leq T$. If $\lim_{t \to \infty} \Bar{\alpha}^t \to \mathscr{a}$, then we say that the FL process converges asymptotically to the target accuracy. Importantly, this criterion can be integrated into the incentivization scheme, see Section \ref{sec:model}, by allowing the AI-service provider to reduce or discontinue rewards once the desired accuracy level has been achieved. 

Note that in a decentralized FL environment, agents may not participate in every training round. When agent $i$ does not train the model for several consecutive rounds, the local model performance may deteriorate due to changes in the underlying data distribution.

Concluding, the predictive accuracy evolves recursively across communication rounds: agents receive the global model from the previous round, improve it locally using private data, and the provider forms a new global model through sample-weighted aggregation. This iterative process continues until the training horizon is reached or a target performance threshold is satisfied. A schematic illustration of this process is shown in Figure \ref{fig:FedZero}.

\subsection{Energy Consumption Component}\label{subsec: energy consumption}

In this section, we model the energy consumption of each agent. Specifically, we define the energy consumption function as a mapping $\varepsilon_i: \mathcal{D}_i \times [0, T] \to \mathbb{R}_{\geq 0}$, where $\varepsilon_i(d^t_i,t)$ denotes the energy consumed by agent $i$ when processing $d^t_i \in \mathcal{D}_i$ bundle of data samples at time $t$. 

The total energy consumption consists of two components: the energy required for local computation, denoted by $\varepsilon_i^{g}(\cdot)$, and the energy required for communication with the AI-service provider, denoted by $\varepsilon_i^{tr}(\cdot)$. We assume that the communication between the agents and the AI-service provider is carried out over wired or fiber-optic links. Since the number of exchanged bits in each communication round is fixed, the corresponding communication energy consumption can be reasonably approximated as constant, say $c^{tr}$. Therefore, for simplicity, we assume that the communication energy between each agent and the AI-service provider satisfies $\varepsilon_i^{tr}(\cdot) = c^{tr}$ for all $i$.

Regarding the energy availability, at round $t$, the system operates under limited green energy availability, denoted by $\mathcal{G}^t$, and by grid energy availability, denoted by $\mathcal{U}^t$. We assume that the AI-service provider first allocates $\mathcal{G}^t$ according to some predefined and exogenous protocol among the users in $\mathcal{N}$. If $\mathcal{G}^t$ is not sufficient to cover the energy required for all tasks, the system can consume grid energy $\mathcal{U}^t$. In cases where both $\mathcal{G}^t$ and $\mathcal{U}^t$ are not sufficient to cover all tasks, then we consider these tasks impractical, and we drop them.

\begin{table*}[t]
\centering
\caption{Summary of notation.}
\label{tab:notation}
\small
\begin{tabular}{ll|ll}
\hline
\textbf{Symbol} & \textbf{Name} &
\textbf{Symbol} & \textbf{Name} \\
\hline
$T$ & Time horizon & $\mathbf{w}_i^t$ & Local model \\

$t\in[T]$ & Time index & $\mathcal{D}_i$ & Local dataset  \\

$G^t$ & Game at time $t$ & $d_i^t$ & Local training samples \\

$\mathcal{N}$ & Number of agents &  $\mathscr{a}$ & Global accuracy threshold  \\ 

$S_i$ & Action set &  $\alpha_i(\cdot)$ & Accuracy function  \\ 

$S$ & Strategy profile space & $\bar{\alpha}^t(\cdot)$ & Global accuracy function \\ 

$s_i^t$ & Pure strategy & $\alpha_i^{(+)}(\cdot)$ & Nominal local accuracy function\\ 

$\mathbf{s}^t$ & Pure strategy profile & $\alpha_i^{(-)}(\cdot)$ & Effective local accuracy function \\

$\boldsymbol{\sigma}_i^t$ & Mixed strategy & $\widetilde{\alpha}_i(\cdot)$ & Previous local accuracy function \\

$\boldsymbol{\sigma}^t$ & Mixed strategy profile & $\alpha_i^{\min}$ & Minimum accuracy \\

$\mathbf{u}_i^t$ & Payoff table & $\theta_i(\cdot)$ & Data quality function\\

$u_i(\cdot)$ & Payoff function & $\xi_i(\cdot)$ & Training gain function\\

$u_i(\cdot)$ & Expected payoff function & $\beta$ & Diminishing returns factor \\

$\mathcal{GW}(\cdot)$ & Green well-being function & $H_i(\cdot)$ & Accuracy drift function \\

$\boldsymbol{\sigma}^*$ & Nash equilibrium & $\mathscr{d}_i$ & Drift rate \\

$\boldsymbol{\chi}$ & Correlated equilibrium  & $\mathscr{t}_i(\cdot)$ & Idle time function\\

\cline{1-2} 

$\mathcal{G}^t$ & Green energy & $h$ & Recovery factor \\

\cline{3-4} 

$\mathcal{U}^t$ & Grid energy & $\gamma$ & Non-participation penalty \\

$\varepsilon_i(\cdot)$ & Energy consumption function & $\eta$ & Relative intensity parameter \\

$\varepsilon^{g}(\cdot)$ & Computation energy function & $\pi$ & Accuracy indicator \\

$\varepsilon^{tr}(\cdot)$ & Communication energy function & $c$ & Accuracy intensity parameter \\

$C_i$ & CPU cycles & $\psi_i$ & Strategy-sample mapping \\

$\kappa$ & Switched capacitance & $P_i(\cdot)$ & Profit function \\

$I_{i,t}$ & Local iterations & $C_i(\cdot)$ & Cost function \\

$f_i$ & Computational capacity & $R_i(\cdot)$ & Reward function \\

\cline{1-2}

$\mathbf{w}^t$ & Global model & $g_i(\cdot)$ & Allocated energy function \\
\hline
\end{tabular}
\end{table*}

\section{Carbon- and Energy-Aware Training Model}\label{sec:model}

We model a synchronous carbon-aware Federated Learning situation as a finite normal-form game played at a finite horizon $T$, in which users strategically choose their local training intensity under limited renewable energy availability. In particular, each user chooses the subset of local data to employ for training, thereby determining the corresponding learning accuracy and energy expenditure. Clearly, the number of local iterations affects both the learning accuracy and energy expenditure. Nevertheless, we consider it as constant throughout the process. The model is designed to capture three fundamental aspects: (i) competition over shared green energy, (ii) aligning with environmental considerations, and (iii) learning returns.

Before formalizing the strategic interaction, we discuss the individual knowledge that each participant possesses. Every participant at each round of the interaction \emph{knows} the model $\mathbf{w}^t$ and the bundle of samples that each agent has. The AI-service provider \emph{knows} the total green energy availability $\mathcal{G}^t$, at each round. Importantly, at each round $t$, the AI-service provider \emph{knows} the available bundle of samples for each agent, but \emph{does not know} their quality $\boldsymbol{\theta}(\mathbf{d}^t, t)$. Since the training process is federated, agents have access \emph{only} to their own local data. In particular, each agent $i$ knows \emph{privately} the quality of its dataset, that is $\theta_i(d^t_i, t)$, as well as the amount of green energy it can utilize, formally defined in eq. \eqref{eq:split energy}. Additional assumptions regarding the information structure are introduced below.

For the strategic interaction, consider the case where $|\mathcal{N}|$ agents participate in a Federated Learning environment. Each agent $i \in [\mathcal{N}]$ selects the number of samples $d^t_i$, considering it quality $\theta_i(d^t_i, t)$, it will use during training at round $t$, therefore $d^t_i \in S_i$. Participation is endogenous such that agent $i$ participates in the training iff $s^t_i > 0$. Moreover, we assume that when agent $i$ does not participate, that is $s^t_i = 0$, then the system imposes to it a penalty $\gamma > 0$. Therefore, the set of actions for agent $i$ is $\mathcal{S}_i = \{0\} \cup S_i$ and $\mathcal{S}$ is the cartesian product of $\mathcal{S}_i$ for all $i \in \mathcal{N}$. Observe that, the set $\mathcal{S}_i$ coincides with the set $\mathcal{D}_i$ and there is a bijection correspondence between each bundle of samples $d_i \in \mathcal{D}_i$ and each action $s_i \in \mathcal{S}_i$. We formalize the latter relationship with the function $\psi_i: \mathcal{S}_i \to \mathcal{D}_i$ for each $i$. In case of a mixed strategy profile $\sigma$, then $\psi_i:  \Delta (\mathcal{S}_i) \to \Delta(\mathcal{D}_i)$, with $\psi_i(\boldsymbol{\sigma}) := \sum_{j \in \mathcal{S}_i} \sigma_{i,j} \cdot \psi_i(s_j) $. 

Naturally, agent $i$ receives a profit $P_i$ and a cost $C_i$ according to the actions it takes during the training process. In particular, the payoff of agent $i$ in a single turn of the procedure is given by the following formula,
\begin{equation}\label{eq:utility}
    u_{i}(\mathbf{s}^t, t) = \mathbf{1}\{s^t_i > 0\} \cdot \left(P_i(\mathbf{s}^t, t) - C_i(\mathbf{s}^t, t) \right) - \mathbf{1}\{s^t_i = 0\}\cdot \pi \cdot \gamma.
\end{equation}
Non-participation yields $-\gamma$ payoff, while active agents balance learning gains and renewable rewards against energy costs and grid penalties. The parameter $\pi$ captures the relationship between the global accuracy threshold $\mathscr{a}$ and model's accuracy at the previous round $t$. If $\Bar{\alpha}^t \leq \mathscr{a}$, then $\pi_i = 1$, otherwise, $\pi = 0$. Therefore, once the global accuracy threshold has been reached the AI-service provider has no considerations regarding the non-participation of the agents. In what follows, we construct the profit and the cost functions for each agent.

Initially, we describe how training intensity translates into energy demand and how renewable energy is allocated. Processing $s^t_i$ samples requires energy $\varepsilon_i(\psi_i(s^t_i), t)$, eq. \eqref{eq:energy consumption agent}, where $\psi_i(s^t_i) = d^t_i$. Let $\mathcal{G}^t$ be the total renewable (green) energy available during round $t$. Then if agent $i$ is an active participant, i.e., $s^t_i > 0 $, can consume up to $g_i(s^t_i, t)$ quantity of $\mathcal{G}^t$. The total energy demand is $\mathcal{E}(\mathbf{s}^t, t) := \sum_{i \in [\mathcal{N}]} \varepsilon_i(\psi_i(s^t_i), t)$, and the green energy allocated as follows,
\begin{equation}\label{eq:split energy}
\displaystyle
    g_i(s^t_i, t) = \left\{\begin{array}{ll}
        \varepsilon_i(\psi_i(s^t_i), t) & , \text{ if } \mathcal{E}(\mathbf{s}^t, t) \leq \mathcal{G}^t,\\
        \mathcal{G}^t \cdot \frac{\varepsilon_i(\psi_i(s^t_i), t)}{\mathcal{E}(\mathbf{s}^t, t)} & , \text{otherwise}
    \end{array}
    \right.,
\end{equation}
with $g_i(0, t) = 0$ for each $t$. In the case where $\varepsilon_i(\psi_i(s^t_i), t) > g_i(s^t_i, t)$, we assume that the remaining energy, $\varepsilon_i(\psi_i(s^t_i), t) - g_i(s^t_i, t)$, required for the completeness of agent's $i$ training is drawn from the grid, which has arbitrarily large availability. Importantly, we assume that each agent $i$ at first consumes green energy, if any, and then it uses grid energy, if necessary. Notably, allocation rule \eqref{eq:split energy} induces interdependence across users: increasing one’s training intensity reduces others’ renewable shares.

We next introduce the reward component, which is designed to align incentives with renewable usage. We define the renewable-alignment term at time $t$ for the agent $i$ as follows,
\[
    R_i(\mathbf{s}^t, t) = \left\{\begin{array}{ll}
        0 & , \text{ if } s^t_i = 0 \ \ \text{or} \ \ \varepsilon_i(\psi_i(s^t_i), t) > g_i(s^t_i, t),\\
        \varepsilon_i(\psi_i(s^t_i), t) & , \text{ if } s^t_i > 0 \ \ \text{and} \ \ \varepsilon_i(\psi_i(s^t_i), t) \leq g_i(s^t_i, t). 
    \end{array}
    \right.
\]
This term rewards users who remain within renewable limits and penalizes users who exceed them. Therefore, the profit $P_i(\cdot)$ of agent $i$ at time $t$ is defined as
\[
    P_i(\mathbf{s}^t, t) = \pi_i \cdot R_i(\mathbf{s}^t, t) \cdot \left( 1 + \eta \cdot \frac{\psi_i(s^t_i)}{\max_{j \in \mathcal{N}} \{\psi_j(s^t_j)\}}\right) + c \cdot \alpha_i(\psi_i(s^t_i), \theta_i(\psi_i(s^t_i), t), t)\footnotemark, 
\]
\footnotetext{In Subsection \ref{subsec: local training}, we define the post-training accuracy $\alpha^{(+)}_i(t)=\alpha_i(\psi_i(s_i^t),\theta_i(\psi_i(s_i^t),t),t)$. In the following, we use the expanded notation $\alpha_i(\psi_i(s_i^t),\theta_i(\psi_i(s_i^t),t),t)$ instead of the compact form $\alpha_i^{(+)}(t)$ to explicitly highlight the dependence of the accuracy on the selected training samples, the local training decision, and the training iteration, which are the factors influencing the resulting utility.
}
where $c, \eta$ are positive constants. Here, the parameter $\pi_i$ plays a role complementary to that in eq. \eqref{eq:utility}. Specifically, once the global accuracy threshold has been reached, the AI-service provider stops offering rewards to the users. The multiplicative factor $\eta \cdot \frac{\psi_i(s^t_i)}{\max_{j \in \mathcal{N}} \{\psi_j(s^t_j)\}}$ rewards relative intensity and captures competitive effort incentives, discouraging free-riding. The parameter $\eta$ controls the strength of competitive effort incentives. If $\eta = 0$, then we have a purely compliance reward. As $\eta$ is increased then the multiplicative factor supports a more competitive interaction. The additive accuracy term, eq. \eqref{eq:accuracy}, ensures that accuracy remains intrinsically valuable.

Although the profit structure contains multiple terms, each component serves a distinct conceptual role. The renewable-alignment reward internalizes the environmental externality of shared green energy, the relative effort factor captures competition among participants, and the accuracy term preserves the intrinsic value of learning progress.

On the other hand, the cost function, $C_i(\cdot)$ captures both renewable usage costs and strong convex penalties for grid consumption in each round of the process. The cost incurred to agent $i$ w.r.t. to the joint decision $\mathbf{s}$ is,
\[  
    C_i(\mathbf{s}^t, t) = c_1 \cdot \max\{0, g_i(s^t_i, t) - \varepsilon_i(\psi_i(s^t_i), t)\} + c_2 \cdot \left(\max\{0, \varepsilon_i(\psi_i(s^t_i), t) - g_i(s^t_i, t)\}\right)^2,
\]
where $c_1$ and $c_2$ are positive constants. The linear term represents baseline renewable energy cost. The quadratic sharing term penalizes grid energy consumption, see in \cite{Kaplan2020ScalingLF}. The convex grid penalty makes carbon-intensive energy strongly undesirable, hence, we assume that $c_1 \ll c_2$, ensuring that exceeding renewable limits is substantially more expensive. Observe that, the decisions of the $-i$ agents implicitly influence functions $P_i(\mathbf{s}^t, t)$ and $C_i(\mathbf{s}^t, t)$ via the green energy split in eq. \eqref{eq:split energy}.

The above components define a finite normal-form game $G^t = \langle [\mathcal{N}], \mathcal{S}, \{\mathbf{u}^t_i\}_{i \in \mathcal{N}}\rangle$, which captures the tension between competitive training effort and collective renewable constraints at time $t$. Agents compete for limited green energy while facing diminishing accuracy returns and convex penalties for grid energy usage.

\begin{figure}[!ht]
\hspace*{1cm}
\begin{tikzpicture}[transform shape, scale=0.92]
\node at (-2,0) {\includegraphics[width=0.87\linewidth]{Fed-Zero-paper.png}};
\filldraw[fill=white,draw=white] (-5.4,-2.0725) rectangle (-1,1.3);
\filldraw[fill=white,draw=white] (3.9,-2.235) rectangle (10,2.1);
\filldraw[fill=white,draw=white] (-1,0.2) rectangle (-8.34,1.49);
\filldraw[fill=white,draw=white] (-6,1.8) rectangle (7.6,2.1);

\filldraw[fill=white,draw=white] (1.7,0.6) rectangle (3.6,1.2);

\filldraw[fill=white,draw=white] (-9,1.7) rectangle (-5,2.2);

\node at (-8,2) {\footnotesize AI-service provider};

\node at (-2,2) {\small \textcolor{gray}{$\cdots$}};
\node at (1.5,2) {\footnotesize \textcolor{red!50!black}{$\text{Device}$ $i$}};
\node at (-3,-2.4) {\footnotesize \textcolor{teal}{$\boldsymbol{\sigma}^t$}};
\node at (-3,1.15) {\footnotesize \textcolor{red!50!black}{$\boldsymbol{\sigma}_i^{t+1}$}};
%
\node at (2.2,0.8) {\small \textcolor{red!50!black}{$\mathbf{w}_i^{t+1}$}};
\filldraw[fill=white,draw=white] (-0.5,0.33) rectangle (0.,-.13);
\node at (-0.25,0.06) {\footnotesize \textcolor{red!50!black}{$\mathcal{D}_i$}};
\filldraw[fill=white,draw=white] (-1.2,-0.24) rectangle (0.5,-1.1);
\node at (-0.3,-.5) {\scriptsize \textcolor{red!50!black}{$(x_i, y_i;\boldsymbol{\sigma}_i^{t+1})$}};
%
%

\draw [very thick, -latex] (-1,-0.1) to[bend right=30] node[midway, above] {\small \textcolor{teal}{$\boldsymbol{\sigma}^t$}} (-2.7,-.1) ;

\draw [very thick, -latex] (-2.7,-1) to[bend right=30] node[midway,below] {\small \textcolor{red!50!black}{$\boldsymbol{\sigma}_i^{t+1}$}} (-1,-1) ;


\node at (-3.7,-.5) {\includegraphics[width=0.11\linewidth]{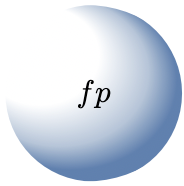}};


\end{tikzpicture}
    \caption{Schematic representation of FL training with the heuristic phase.}\label{fig:fp}
\end{figure}

\paragraph*{Heuristic phase.}
Although Nash equilibrium is a universal and appealing solution concept, it is intractable for large-scale problems, see in \cite{Daskalakis2006TheCO}. Further, Nash equilibrium requires that the agents have full information about the situation, which violates Federated Learning. To mitigate these issues we introduce the following mechanics. Since the training process is federated, agents have access only to their own local data; consequently, for each $t$, game $G^t$ is one of incomplete information. For that, we consider that agents rely on an introspective decision-making process, referred to as the \emph{heuristic phase}. In particular, at time $t$, the AI-service provider assigns the model $\mathbf{w}^t$ to each agent and \emph{publicly} broadcasts the joint strategy profile $\boldsymbol{\sigma}^t$. Agent $i$ then \emph{independently} runs the fictitious play algorithm, Algorithm \ref{alg:fictitious play algorithm}, over a finite horizon $\mathcal{T}$. In other words, at each round agents receive public signal (e.g., observed past profile) and run $\mathcal{T}$ steps of Algorithm \ref{alg:fictitious play algorithm} over the induced one-shot game with estimated opponent mixed strategies $\boldsymbol{\mu}_{-i}$ to output $\boldsymbol{\sigma}^{t+1}_i$. This procedure constitutes the heuristic phase of the training process. The outcome of the heuristic phase \emph{determines} the strategy that agent $i$ applies in the subsequent training round. Finally, agent $i$ returns the updated model $\mathbf{w}^{t+1}$ along with its strategy $\boldsymbol{\sigma}_i^t$, see Figure \ref{fig:fp}.

\begin{figure}[!ht]
\hspace*{1cm}
\begin{tikzpicture}[transform shape, scale=0.92]
\node at (-2,0) {\includegraphics[width=0.87\linewidth]{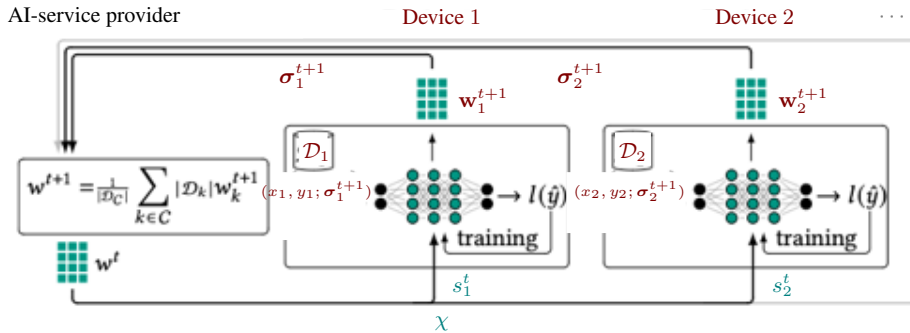}};
\filldraw[fill=white,draw=white] (3.9,-2.235) rectangle (10,2.1);
\filldraw[fill=white,draw=white] (-6,1.8) rectangle (7.6,2.1);

\filldraw[fill=white,draw=white] (1.7,0.6) rectangle (3.6,1.2);

\filldraw[fill=white,draw=white] (-9,1.7) rectangle (-5,2.2);

\node at (-8,2) {\footnotesize AI-service provider};

\node at (3.5,2) {\small \textcolor{gray}{$\cdots$}};
\node at (-3,2) {\footnotesize \textcolor{red!50!black}{$\text{Device}$ $1$}};
\node at (1.5,2) {\footnotesize \textcolor{red!50!black}{$\text{Device}$ $2$}};
\node at (-3,-2.4) {\footnotesize \textcolor{teal}{$\chi$}};
\node at (-2.7,-1.85) {\footnotesize \textcolor{teal}{$s^t_1$}};
\node at (1.9,-1.85) {\footnotesize \textcolor{teal}{$s^t_2$}};
\node at (-1,1.15) {\footnotesize \textcolor{red!50!black}{$\boldsymbol{\sigma}_2^{t+1}$}};
\node at (-5,1.15) {\footnotesize \textcolor{red!50!black}{$\boldsymbol{\sigma}_1^{t+1}$}};
%
\filldraw[fill=white,draw=white] (-2.9,0.6) rectangle (-1.6,1.2);
\node at (-2.4,0.8) {\small \textcolor{red!50!black}{$\mathbf{w}_1^{t+1}$}};
\node at (2.2,0.8) {\small \textcolor{red!50!black}{$\mathbf{w}_2^{t+1}$}};

\filldraw[fill=white,draw=white] (-5.05,0.33) rectangle (-4.65,-.13);
\node at (-4.8,0.06) {\footnotesize \textcolor{red!50!black}{$\mathcal{D}_1$}};
\filldraw[fill=white,draw=white] (-0.5,0.33) rectangle (0.,-.13);
\node at (-0.25,0.06) {\footnotesize \textcolor{red!50!black}{$\mathcal{D}_2$}};
\filldraw[fill=white,draw=white] (-5.4,-0.24) rectangle (-4.1,-1.2);
\node at (-4.8,-.5) {\tiny \textcolor{red!50!black}{$(x_1, y_1;\boldsymbol{\sigma}_1^{t+1})$}};
\filldraw[fill=white,draw=white] (-1.2,-0.24) rectangle (0.5,-1.1);
\node at (-0.3,-.5) {\tiny \textcolor{red!50!black}{$(x_2, y_2;\boldsymbol{\sigma}_2^{t+1})$}};
%
%






\end{tikzpicture}
    \caption{Schematic representation of FL training with the correlated devise.}\label{fig:cd}
\end{figure}

\paragraph*{Correlation device.}
In this approach, given the game $G^t$, we assume that the AI-service provider besides assigning the model $\mathbf{w}^t$ to each agent, it 
\emph{publicly} signals a distribution $\chi$ over the strategy space $\mathcal{S}$, and \emph{privately} recommends actions $s_i \in \mathcal{S}_i$ to each agent $i$, see Figure \ref{fig:cd}. Since the quality of the dataset is private information unknown to the mediator, it suggests distribution $\chi$ using \emph{partial information}. Our objective is to compute a distribution $\chi \in \Delta (\mathcal{S})$ that optimizes a given system-wide green well-being. For that we introduce the function $\mathcal{GW}(\mathbf{s}^t, t) = \sum_{i \in \mathcal{N}} (g_i(s^t_i, t) - \varepsilon_i(\psi_i(s^t_i), t)$, capturing the green energy consumption in the system at round $t$. Clearly, when $\mathcal{GW}(\mathbf{s}^t, t) > 0$ the aggregated energy consumption, w.r.t. $\mathbf{s}^t$, does not overwhelm the amount of aggregated renewable energy at $t$. Contrarily, when $\mathcal{GW}(\mathbf{s}^t, t) < 0$, then agents use grid energy $\mathcal{U}^t$ at some round of the process. So, the AI-service provider wants to find the distribution $\chi$ that maximizes $\mathcal{GW}$. For that, for each $t \in [T]$, we device the following linear program,
\begin{equation}\label{eq:lp}
\resizebox{0.91\hsize}{!}{%
$
\begin{array}{cll}
	\max\limits_{{\chi \sim \Delta(S)}} & \mathcal{GW}(\mathbf{s}^t, t)  &\\
    \text{s.t.} & \sum_{s^t_{-i} \in S_{-i}} (u_i((s^t_i, \mathbf{s}^t_{-i}), t) - u_i((s'_i, \mathbf{s}^t_{-i}), t) \cdot \chi(\mathbf{s}^t_{-i} \mid s_i) \geq 0, \quad \forall i \in [\mathcal{N}] & , s^t_i, s'_i \in \mathcal{S}_i, \\
	& \chi(\mathbf{s}^t) \geq 0 & , \forall \ \mathbf{s}^t \in \mathcal{S} \\
    & \sum_{\mathbf{s}^t \in S} \chi(\mathbf{s}^t) = 1. &
\end{array}
$
}
\end{equation}

In summary, we propose two complementary solution approaches. The first, heuristic phase, is a fully distributed mechanism, where agents independently update their strategies, using locally available information and observations over the choices of the other agents. The second, correlated device, is a semi-centralized mechanism, where a coordinator recommends actions by signaling a distribution over strategies. Together, these two approaches capture the trade-off between decentralization and coordination: the former emphasizes autonomy while the latter leverages centralized guidance to improve global performance.

\section{Evaluation}\label{sec:evaluation}

\paragraph*{Evaluation assumptions.} We begin this section by describing the assumptions used for the energy and accuracy models introduced in Subsections~\ref{subsec: local training} and~\ref{subsec: energy consumption}, respectively. These assumptions are not required by the proposed method in Section~\ref{sec:model}; they are introduced only to facilitate the experimental evaluation. Therefore, they can be replaced by alternative energy and accuracy modeling frameworks without affecting the generality of the proposed approach.

For the accuracy modeling, we adopt a common assumption in learning systems that increasing the amount of training data improves the model accuracy, while the marginal improvement decreases as more data becomes available. To capture this diminishing-return behavior and the iterative nature of the training process, we consider the following gain function,
\begin{equation}\label{eq:accuracy}
    \xi_i(d^t_i, \theta_i(d^t_i, t), t) = 1 - \exp(-\theta_i(d^t_i, t) \cdot (d^t_i)^\beta),
\end{equation}
where the term $(d^t_i)^\beta$ captures the diminishing returns w.r.t. the amount of data. Observe that if $d^t_i = 0$ at round $t$, then $\alpha_i(0, \theta_i(0, t+1), t+1) = \alpha_i(0, \theta_i(0, t), t)$.

Moreover, as stated in Subsection \ref{subsec: local training}, when agent $i$ does not train the model for several consecutive rounds, the local model performance may deteriorate due to changes in the underlying data distribution. Let $\widetilde{\alpha}_i(t)$ denote the last nominal accuracy achieved by agent $i$ before round $t$. The \emph{effective} accuracy before retraining and after drift is modeled as $\alpha_i^{(-)}(t) = \widetilde{\alpha}_i(t) - H_i(t)$, where $H_i(t) = (\widetilde{\alpha}_i(t) - \alpha^{\text{min}}_i) \cdot (1 - e^{- \mathscr{d}_i \mathscr{t}_i(t)})$. Here, $\mathscr{t}_i(t)$ represents the number of consecutive rounds during which agent $i$ does not participate, $\alpha^{\text{min}}_i$ is the minimum achievable accuracy level, and $\mathscr{d}_i$ controls the drift rate. 

When agent $i$ resumes training, its accuracy updated as $\alpha^{(+)}(t) \gets (1 - h) \alpha^{(-)}_i(t) + h \alpha^{(+)}(t) $, where $h \in [0, 1]$ and controls how effectively local training compensates for drift. To be more clear, $\widetilde{\alpha}(t)$ is the pre-drift reference, whereas $\alpha^{(-)}_i(t)$ is the post-drift and pre-trained accuracy. The sequence of quantities regarding accuracy is: $\alpha^{(+)}_i(t-1) \to \widetilde{\alpha}(t) \to \alpha^{(-)}_i(t) \to \alpha^{(+)}_i(t)$.

For the energy consumption, the total energy consumption consists of two components: the energy required for local computation, denoted by $\varepsilon_i^{g}(\cdot)$, and the energy required for communication with the AI-service provider, denoted by $\varepsilon_i^{tr}(\cdot)$. Inspired by \cite{Panagea2026GreenFLagAG}, the energy consumed by agent $i$ for computing the gradients of the local loss is given by
\[
    \varepsilon^{g}_i(d_i, t) = \kappa \cdot C_i \cdot I_{i,t} \cdot d_i \cdot f^2_i(t),
\]
where $\kappa$ is the effective switched capacitance, $C_i$ is the number of CPU cycles required for computing one sample data, and $f_i$ is the computational capacity of agent $i$ at time $t$. The term $I_{i,t}$ is the number of local iteration at agent $i$ at time $t$ and we consider it constant in the work. In general, this quantity depends not only on the selected data subset $d_i$, but also on the current state of the global model $\mathbf{w}^t$ at round $t$. As discussed in Subsection \ref{subsec: energy consumption}, we assume that the communication energy between each agent and the AI-service provider is constant, i.e., $\varepsilon^{tr}_i(\cdot) = c^{tr}$ for all $i$. Therefore, the energy consumption for agent $i$ is
\begin{equation}\label{eq:energy consumption agent}
    \varepsilon_i(d_i, t) = \varepsilon^{g}_i(d_i, t) + c^{tr}.
\end{equation}

Regarding the energy availability at round $t$, the system operates under limited green energy availability, denoted by $\mathcal{G}^t$. Green energy $\mathcal{G}^t$ is \emph{known} to the AI-service provider; users may have only local information. For simplicity, we assume that the AI-service provider allocates this quantity uniformly among the users in $\mathcal{N}$, therefore, the latter can consume up to a specific amount of green energy $\mathcal{G}^t$. If $\mathcal{G}^t$ is not sufficient to cover the energy required for all tasks, the system can consume as much grid energy $\mathcal{U}^t$ as possible.

\paragraph*{Synthetic data.} For our experiments, we evaluate the framework for a population of agents $\mathcal{N} = \{1, 2, 3\}$, where each agent $i \in \mathcal{N}$ has an action set $\mathcal{S}_i$ of size $|\mathcal{S}_i| = 4$. Synthetic datasets are generated to simulate the bundle of samples available to each agent, the quality of these samples, the renewable energy availability, and the penalty parameter $\gamma$. For all the following experiments, the total available renewable energy $\mathcal{G}^t$ in turn $t$ is sampled as an integer uniformly from the interval $[50, 100]$. For each agent, the bundle of samples is generated from the distribution $U[\mathcal{G}^t - \delta_-, \mathcal{G}^t + \delta_+]^4$, where scalars $\delta_-$ and $ \delta_+$ capture the deviation of the agent’s dataset size w.r.t. the total available renewable energy $\mathcal{G}^t$. Here $\mathcal{G}^t$ serves only as a system-wide scaling parameter and does not imply a causal relationship between the ecosystem's green energy availability and agents' energy demand.

Next, the quality of agent's $i$ samples $\theta_i(\mathbf{s}^t_i, t)$ is generated as a strictly increasing sequence w.r.t. $\mathbf{s}^t_i$, with values sampled uniformly from $U[0.01, 1]$ for each $t \in T$. In all experiments, the fictitious play algorithm is executed for $\mathcal{T} = 1000$ iterations. The parameters are set as follows: $I_{i,t} = 10$, $\mathscr{a} = 0.9$, $\Bar{\alpha}^0 = 0.0$, $\alpha^{\min}_i = 0.25$, $\mathscr{d}_i = 0.05$, $c = 10$, $c^{\text{tr}} = 0$, $c_1 = 1$, $c_2 = 1000$, $\eta = 0.52$, $C = 10^2$, $f = 10^3$, $\zeta = 4$ and $\kappa = 10^{-9}$, for any $i$ and $t \leq T$. The penalty parameter $\gamma$ and the parameter $h$ take several non-negative values across experiments. 

For the correlation mechanism, we first compute the distribution $\chi$ and clip any probability value smaller than $10^{-3}$. To analyze the behavior of each player, we compute the marginal distributions induced by $\chi$. Finally, the overall training process runs for $T = 10$ rounds, and in each experiment we simulate each round $t \in T$ for $100$ turns, reporting the mean values.

We focus on three main quantities, for each round: (i) the total energy ($kWh$) consumed during training, $\mathcal{E}(\boldsymbol{\sigma}^\ast, t) := \sum_{i \in [\mathcal{N}]} \varepsilon_i(\psi_i(\boldsymbol{\sigma}^\ast_i), t)$, where $s^\ast_i$ is the solution for agent $i$ provided by either the Algorithm \ref{alg:fictitious play algorithm} or eq. \eqref{eq:lp}, (ii) the training accuracy, eq. \eqref{eq:accuracy}, that each agent achieves, and (iii) the mean accuracy ($\pm$ one standard deviation) achieved. For the case of energy agnostic agents, we compute the correlated equilibrium using the social welfare function $\sum_{i \in [\mathcal{N}]} u^t_i(\boldsymbol{s}^t)$. 

\begin{figure}[h]
    \centering
    \includegraphics[width=.33\textwidth]{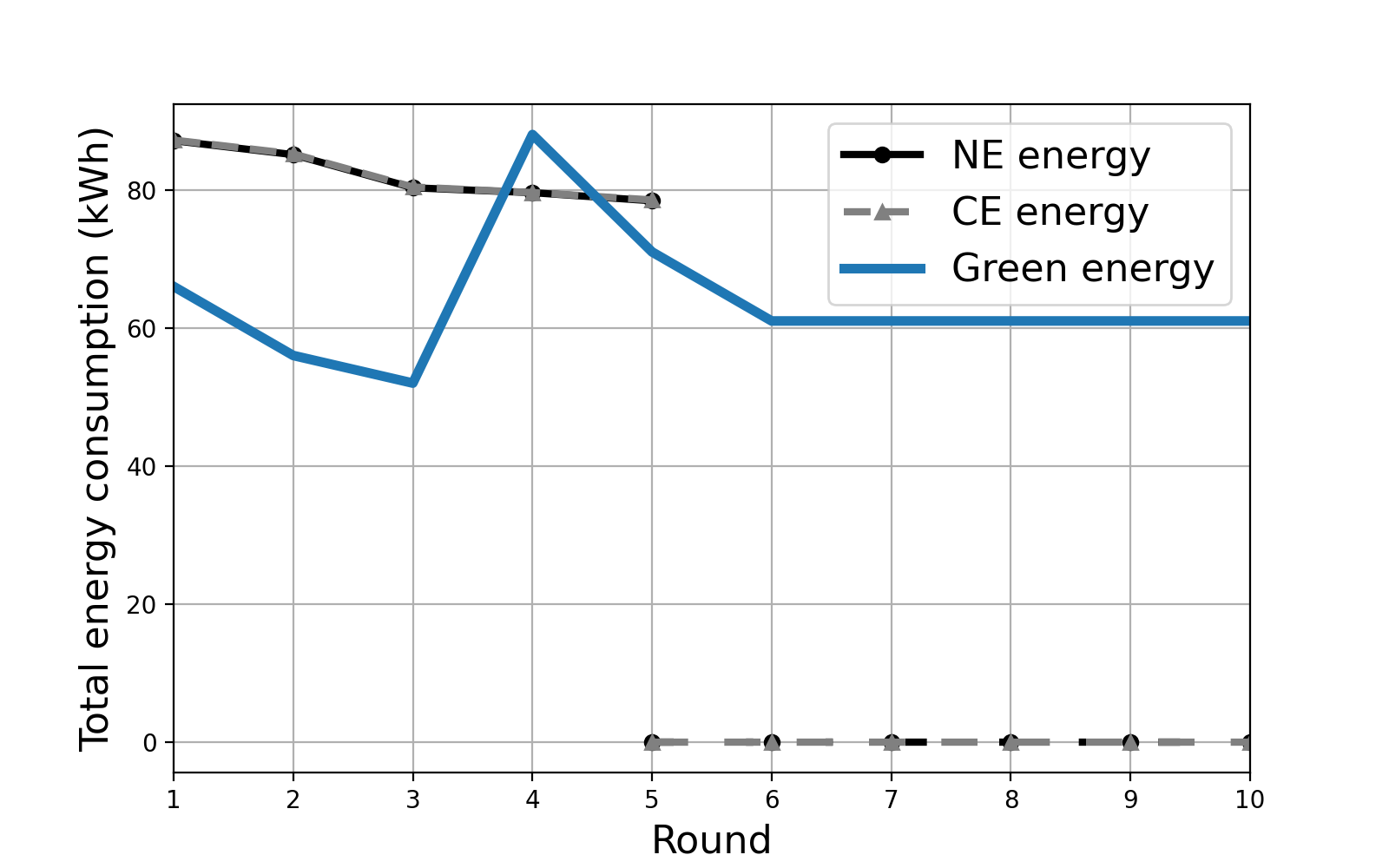}\hfill
    \includegraphics[width=.33\textwidth]{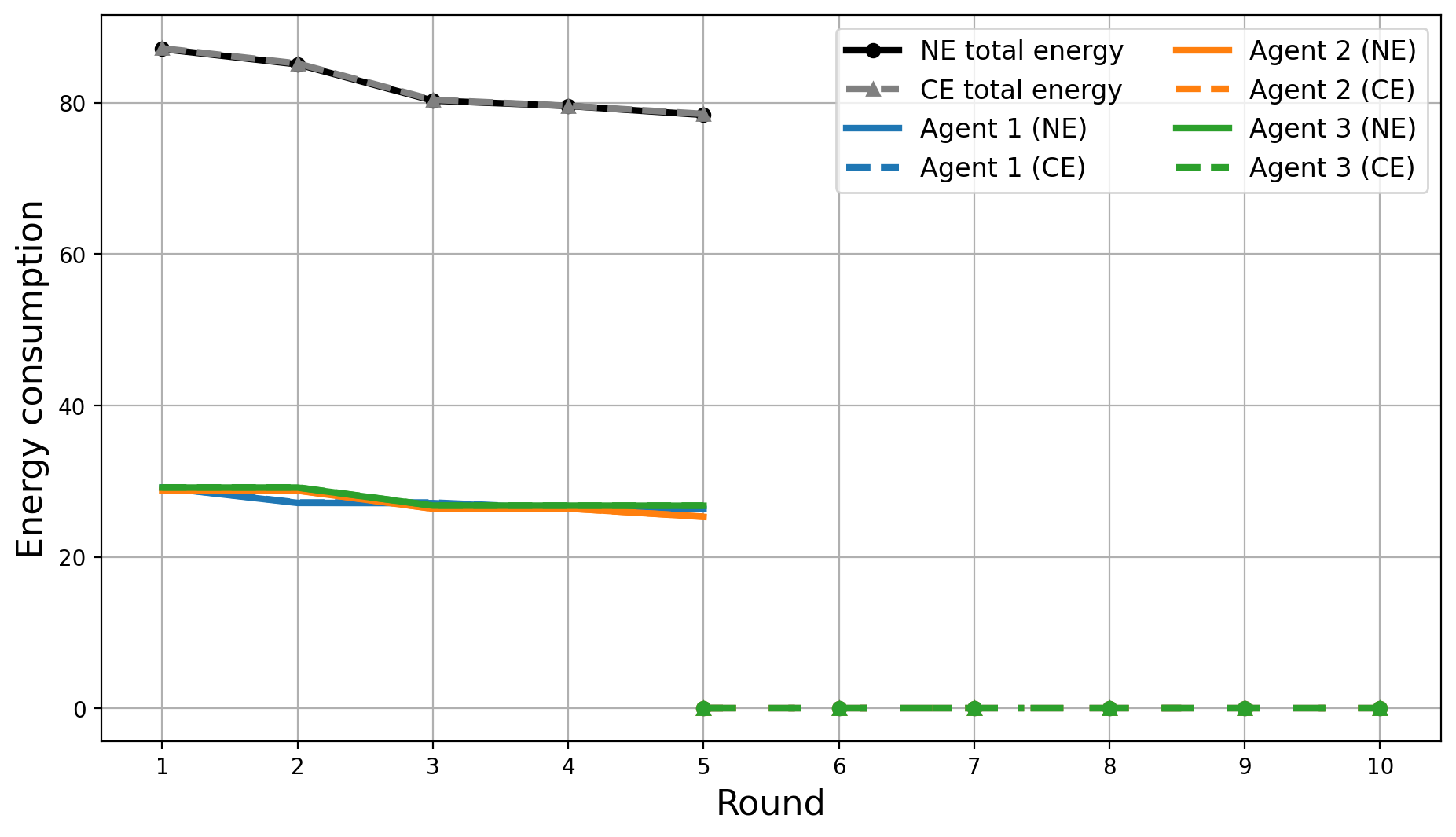}\hfill
    \includegraphics[width=.33\textwidth]{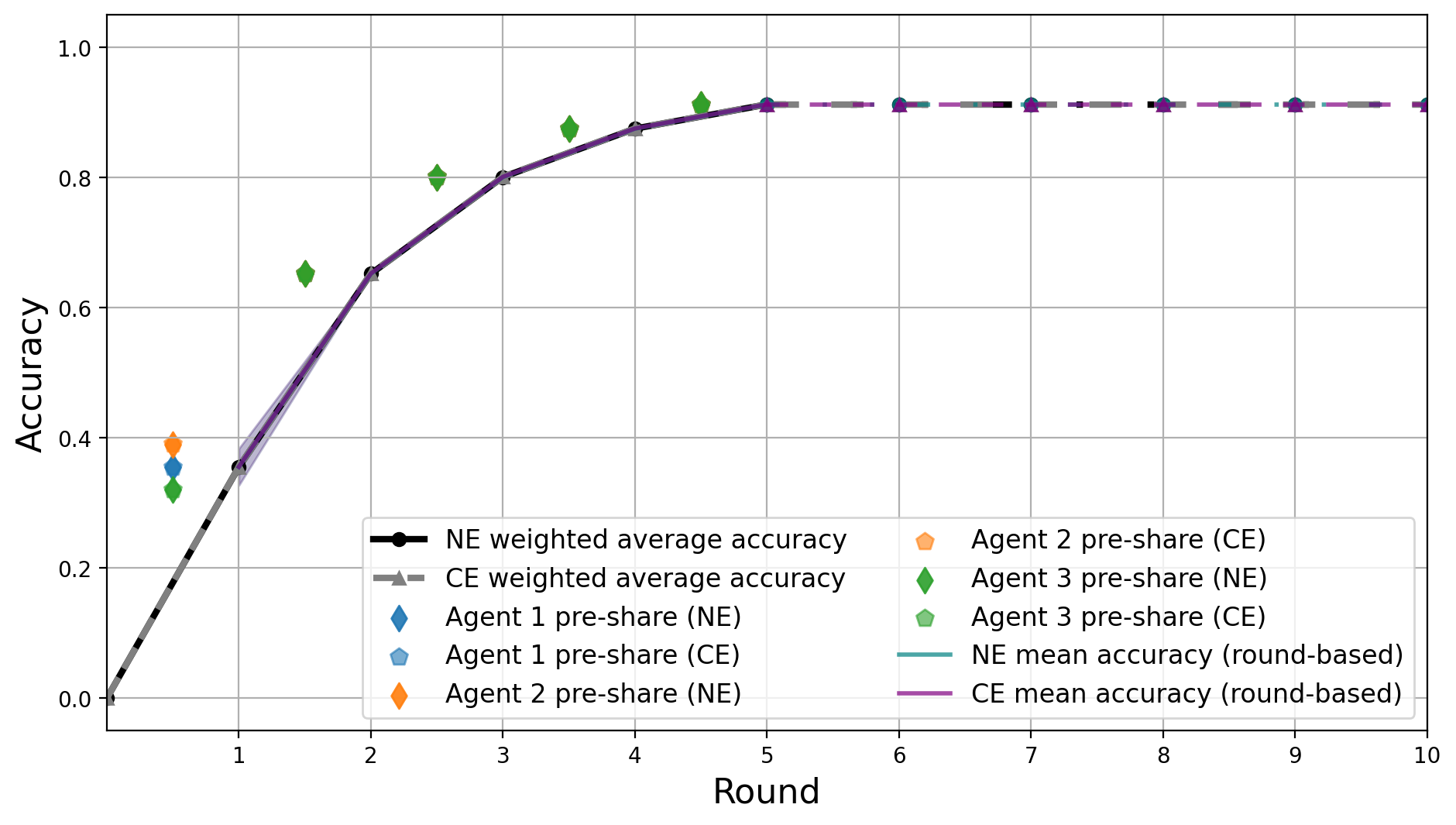}
    \caption{Energy-agnostic training: Total energy consumption (left), energy consumption per agent (middle), total training accuracy and total training accuracy per agent (right), for the case where agents are energy-agnostic with $\delta_- = \delta_+ = 10$.}
\label{fig:energy-agnostic}
\end{figure}

\begin{figure}[h]
    \centering
    \includegraphics[width=.33\textwidth]{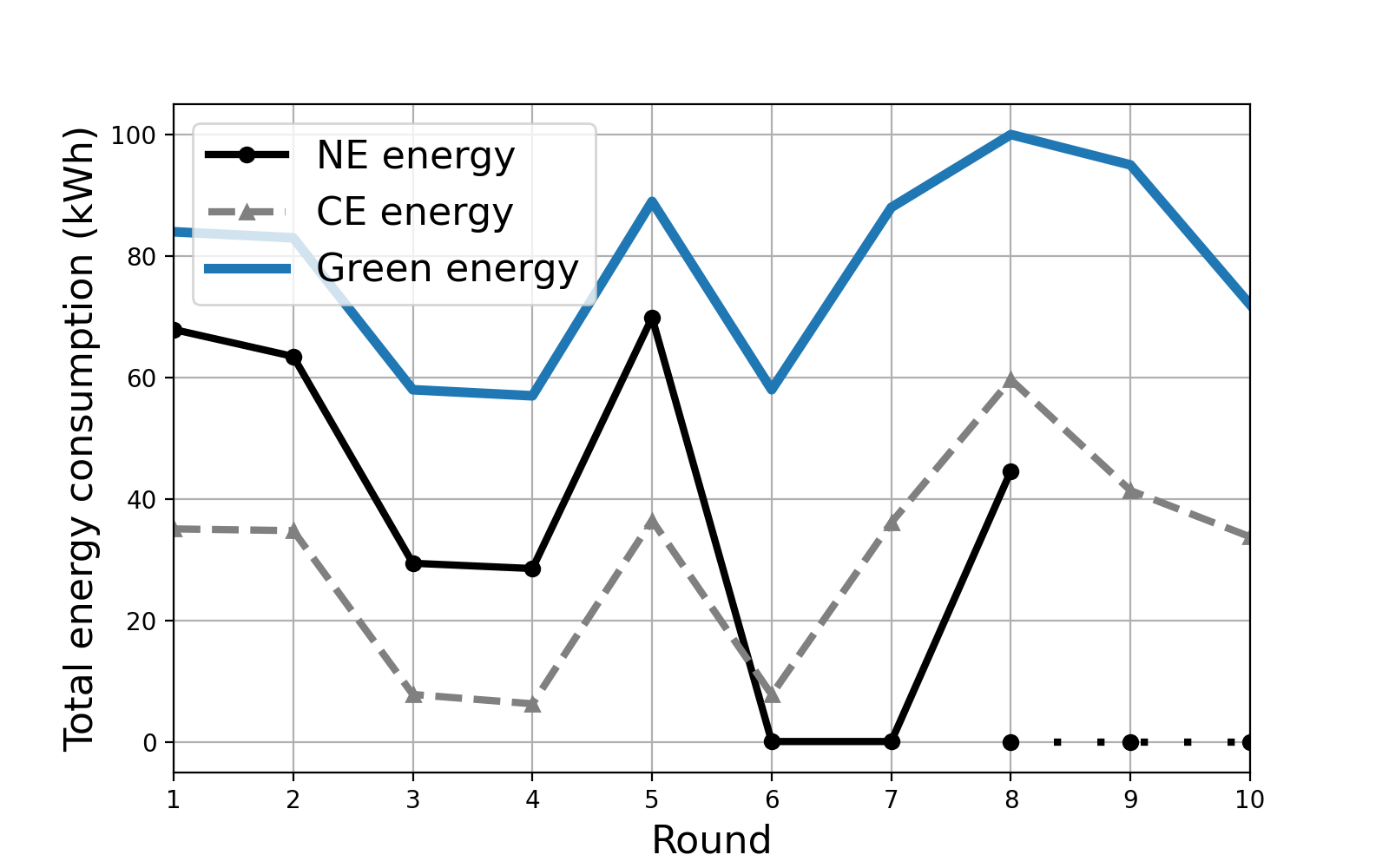}\hfill
    \includegraphics[width=.33\textwidth]{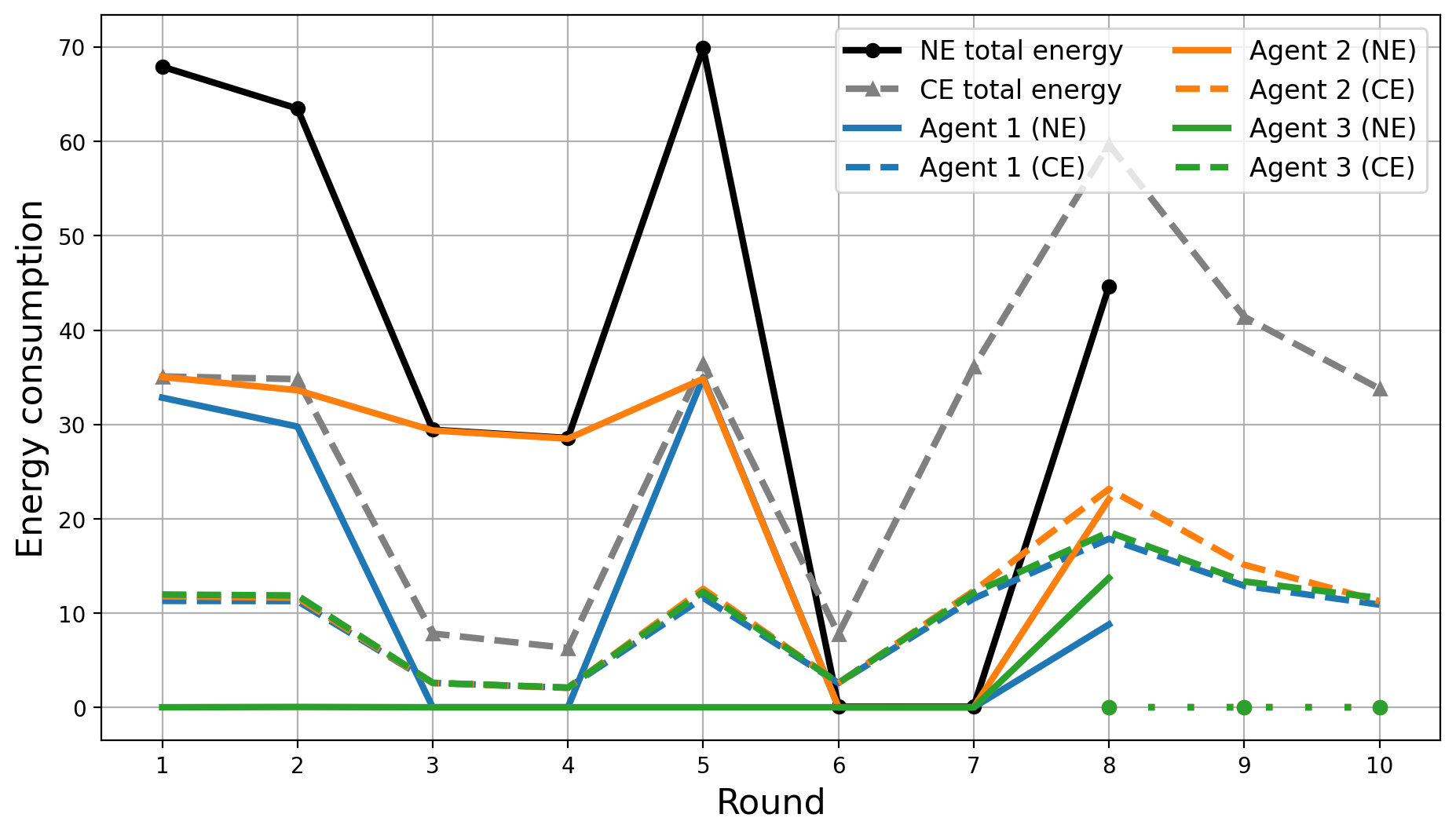}\hfill
    \includegraphics[width=.33\textwidth]{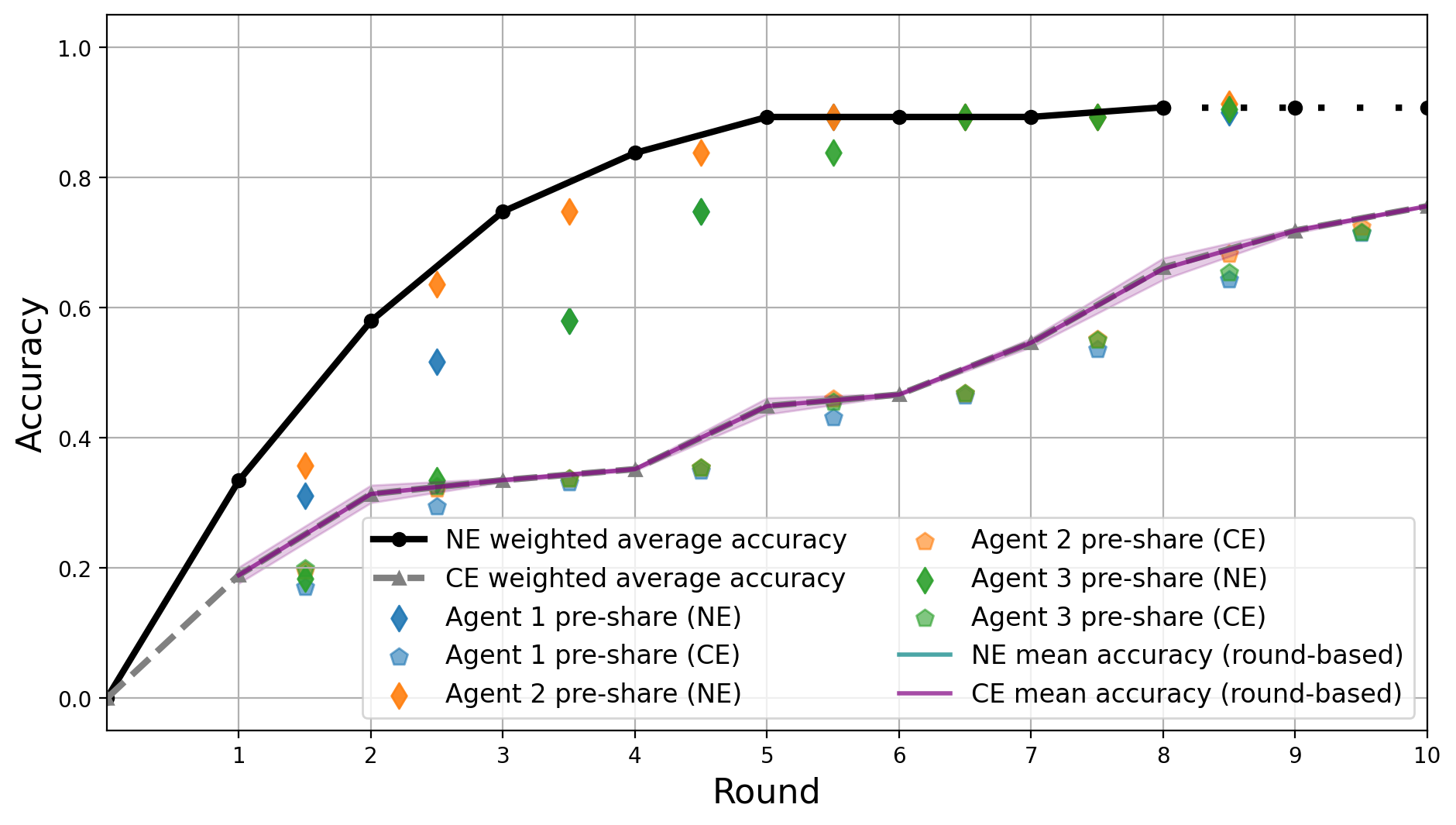}
    \includegraphics[width=.33\textwidth]{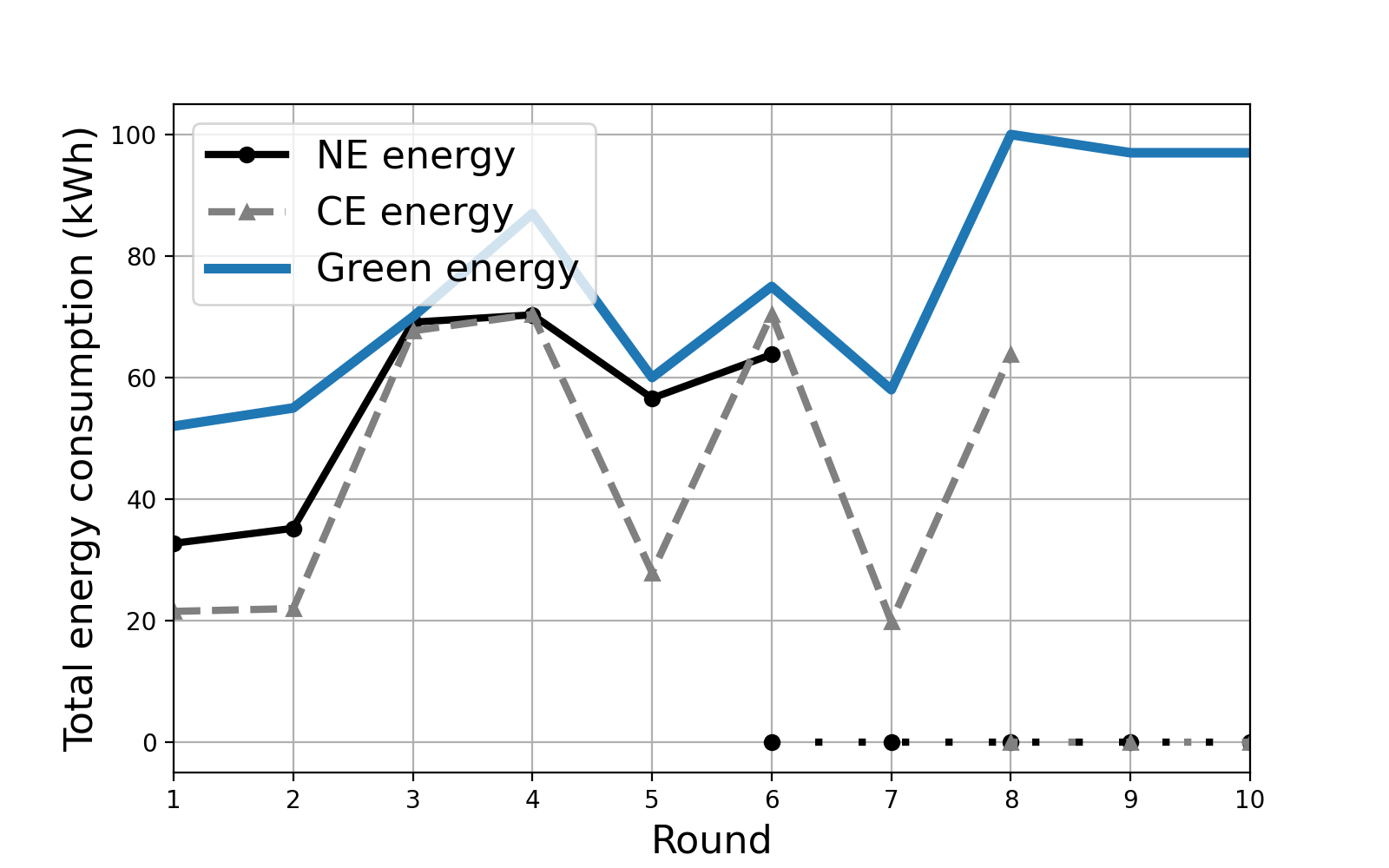}\hfill
    \includegraphics[width=.33\textwidth]{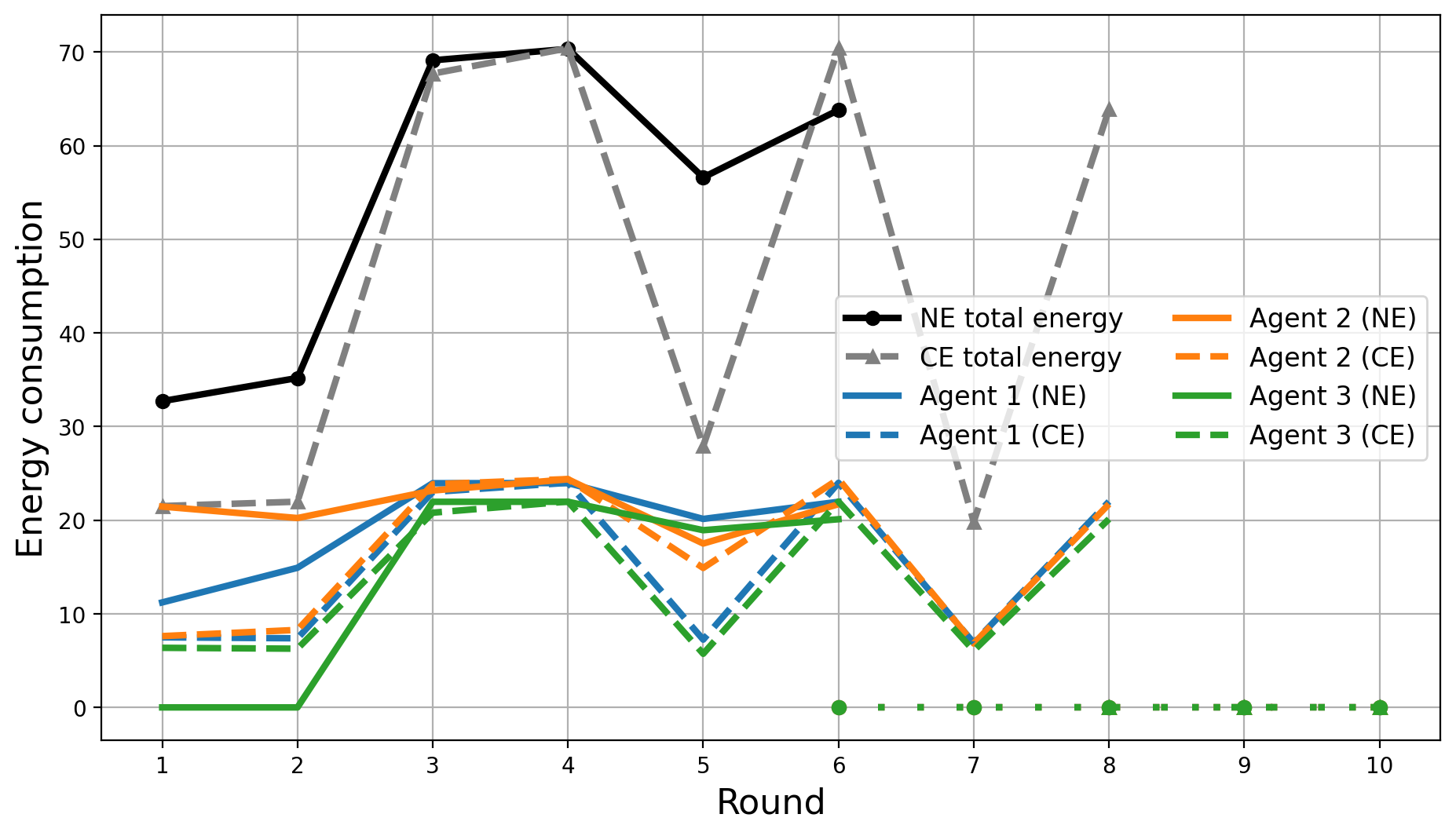}\hfill
    \includegraphics[width=.33\textwidth]{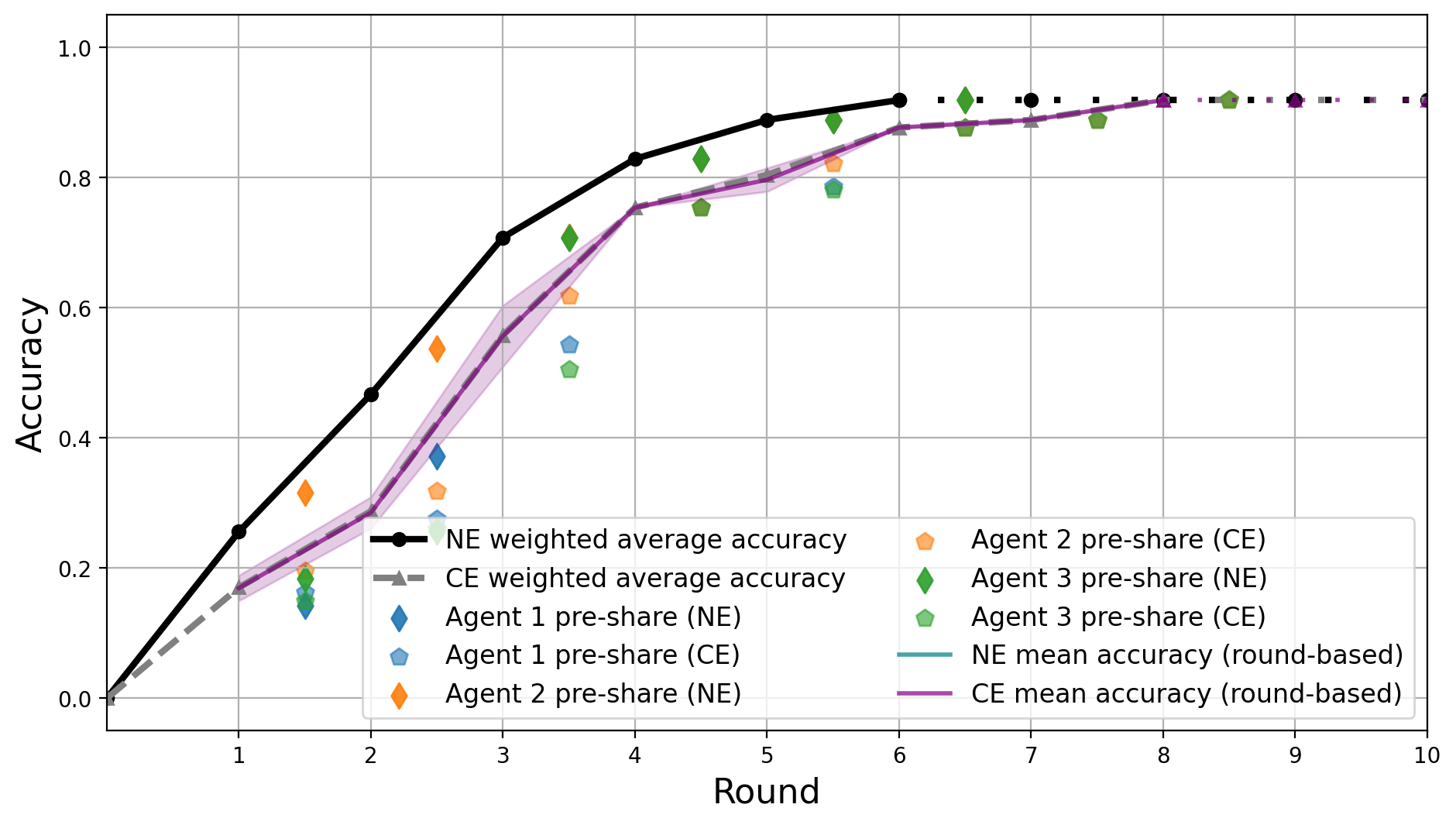}
    \includegraphics[width=.33\textwidth]{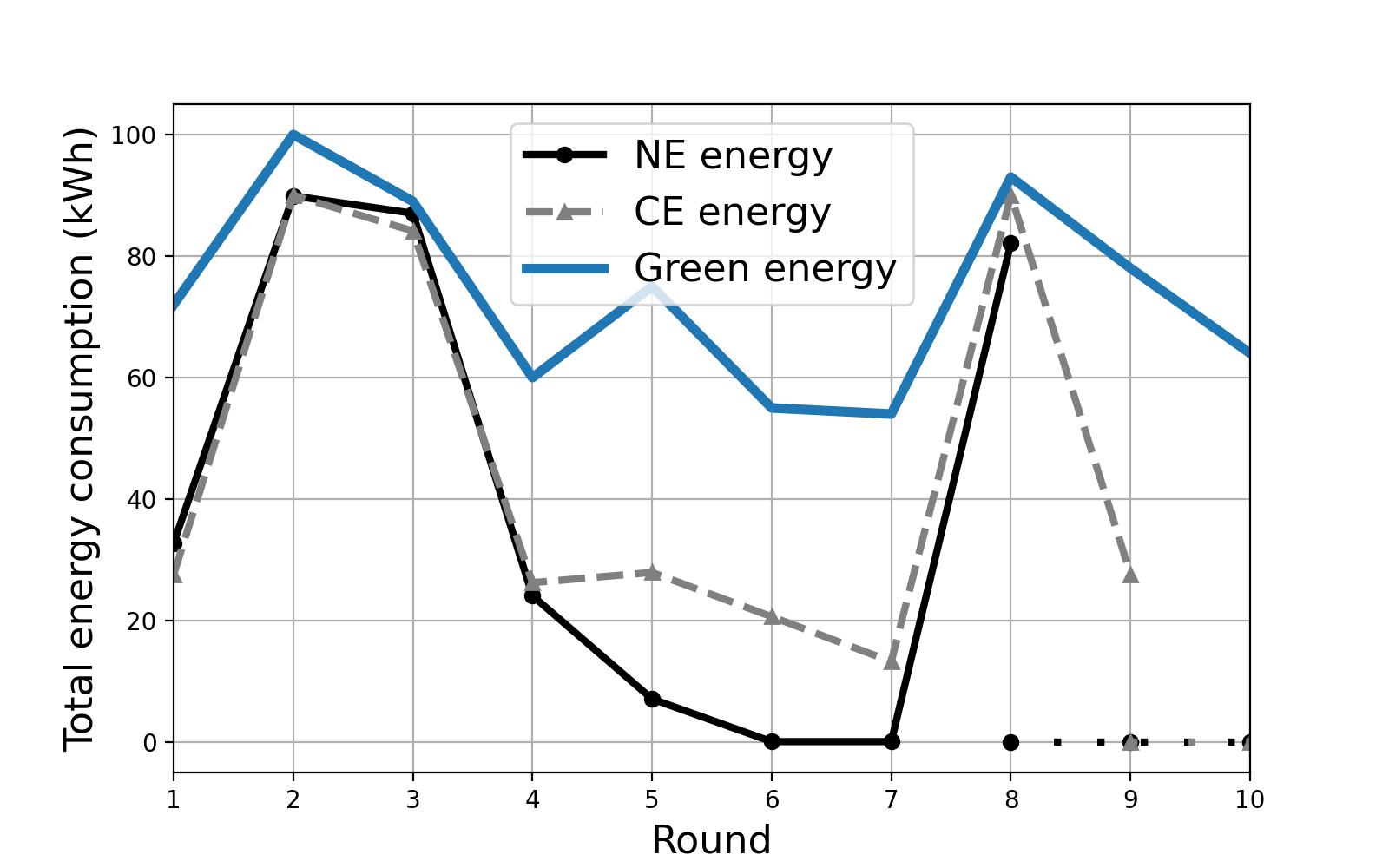}\hfill
    \includegraphics[width=.33\textwidth]{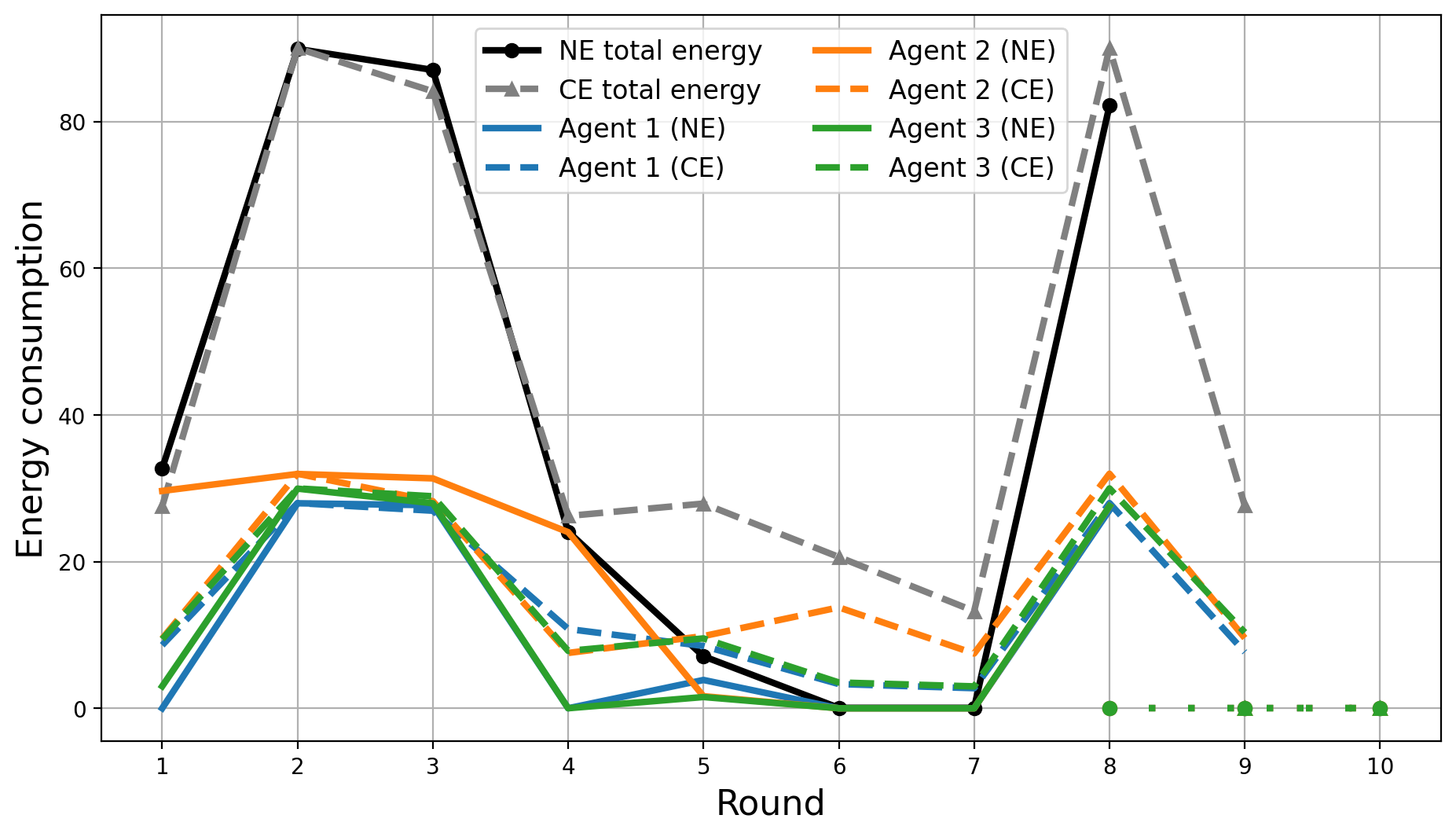}\hfill
    \includegraphics[width=.33\textwidth]{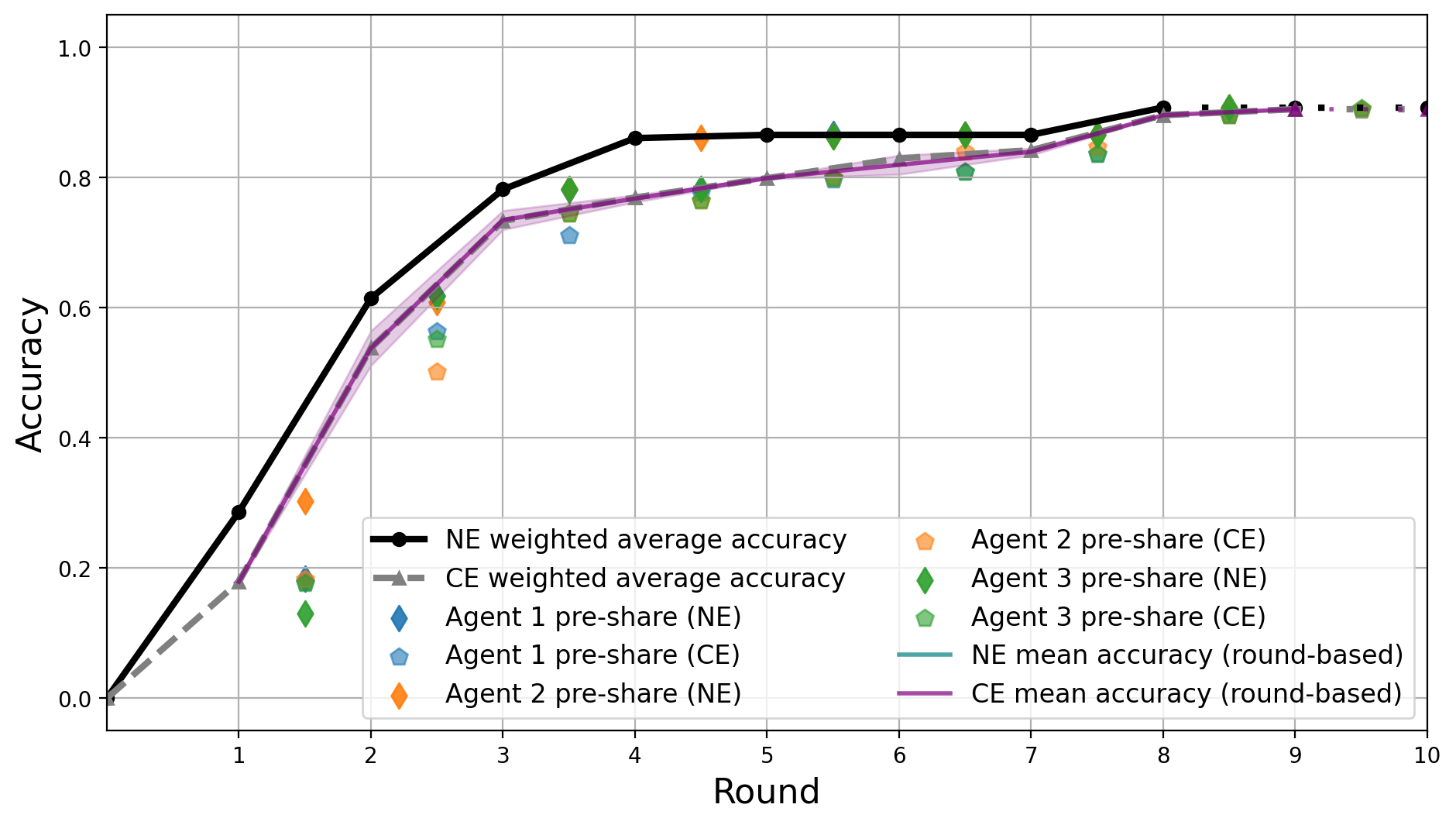}
    \caption{Energy-aware training: Total energy consumption (left column), energy consumption per agent (middle column), training accuracy (right column), for the case where agents are energy-aware, with $\gamma = 1$ (upper row), $\gamma = 10$ (middle row), and $\gamma = 100$ (lower row), with $ h = 0.5$  and $\delta_+ = \delta_- = 10$.}
\label{fig:energy-aware}
\end{figure}

\begin{figure}[h]
    \centering
    \includegraphics[width=.33\textwidth]{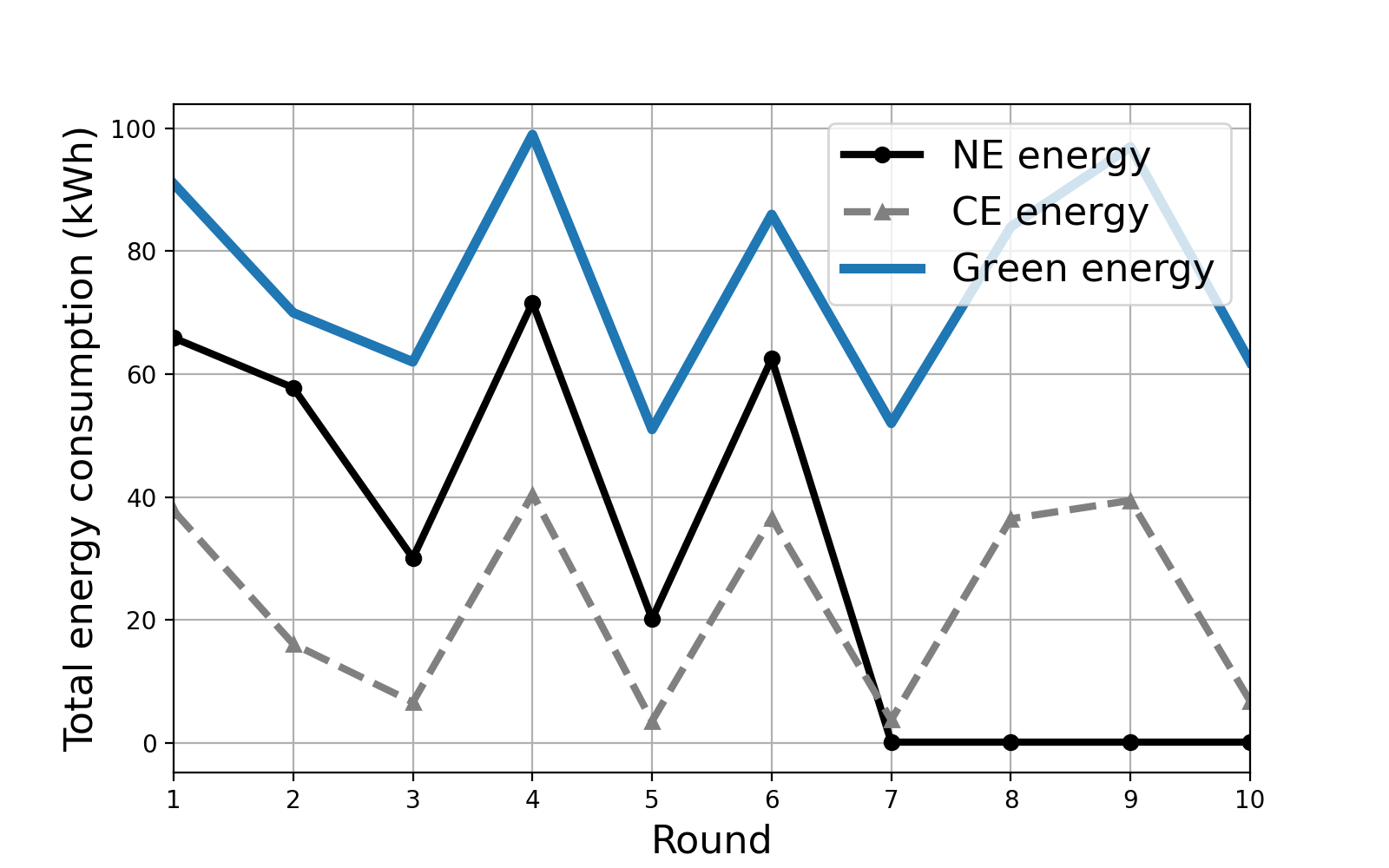}\hfill
    \includegraphics[width=.33\textwidth]{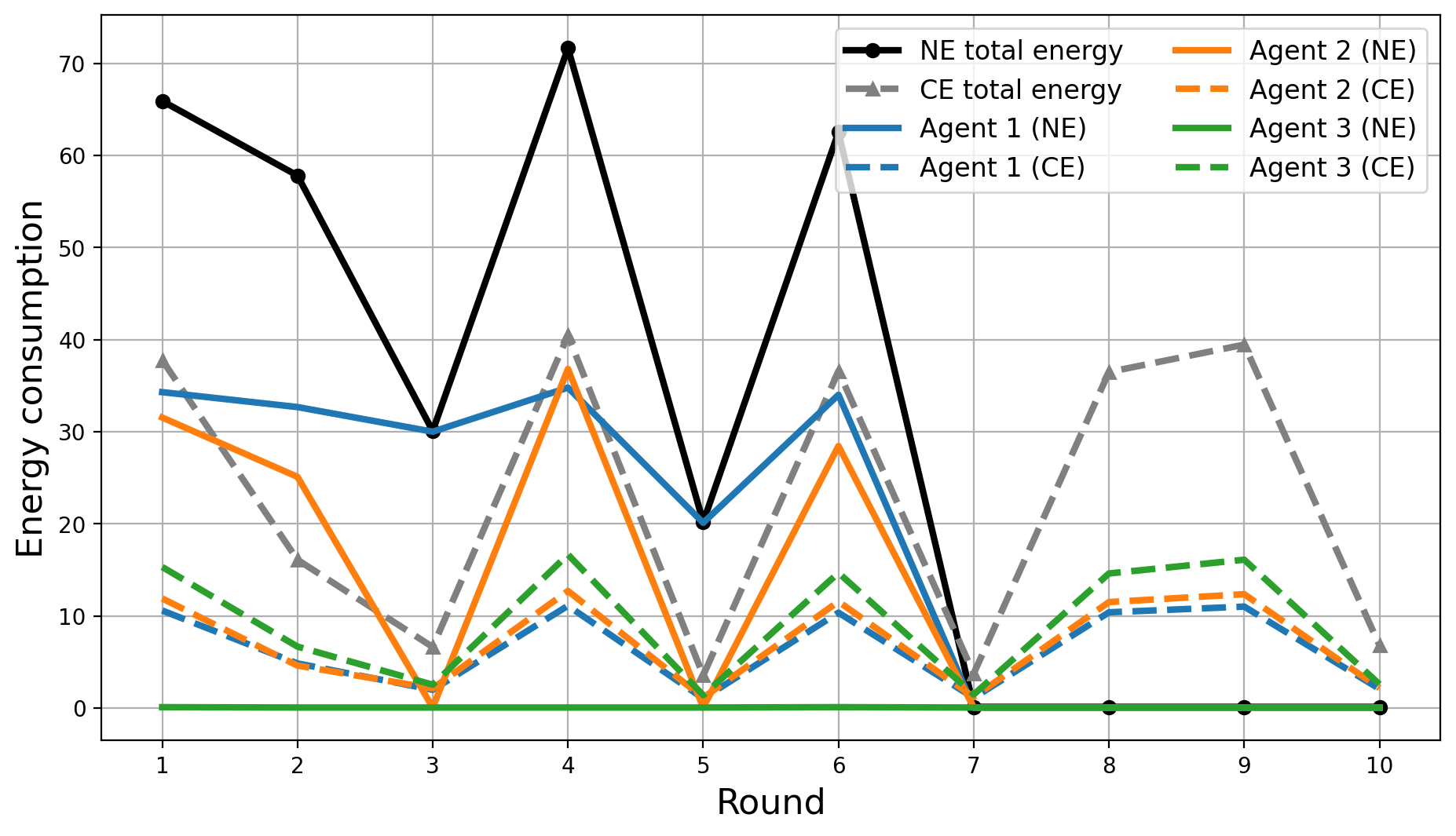}\hfill
    \includegraphics[width=.33\textwidth]{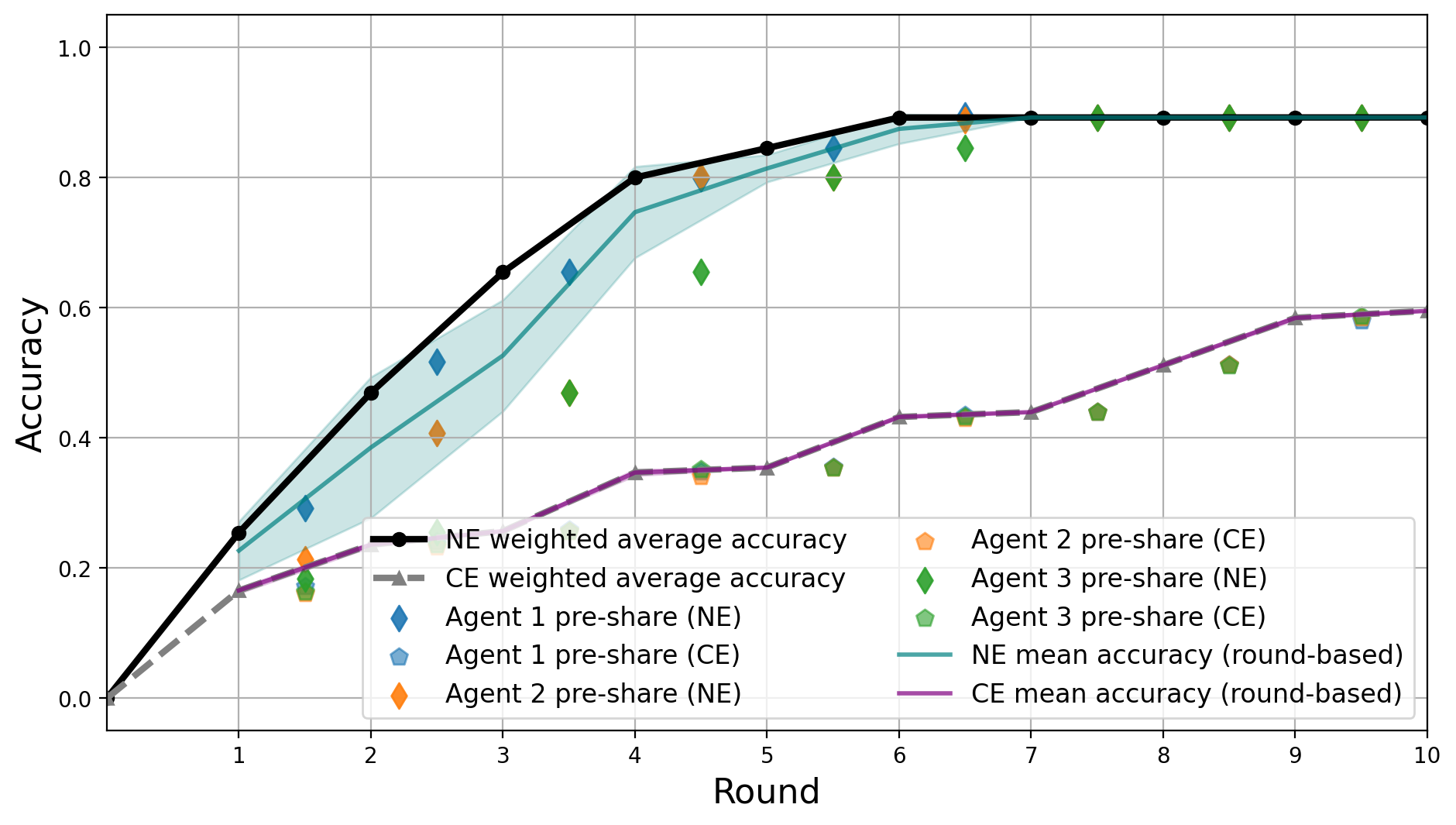}
    \includegraphics[width=.33\textwidth]{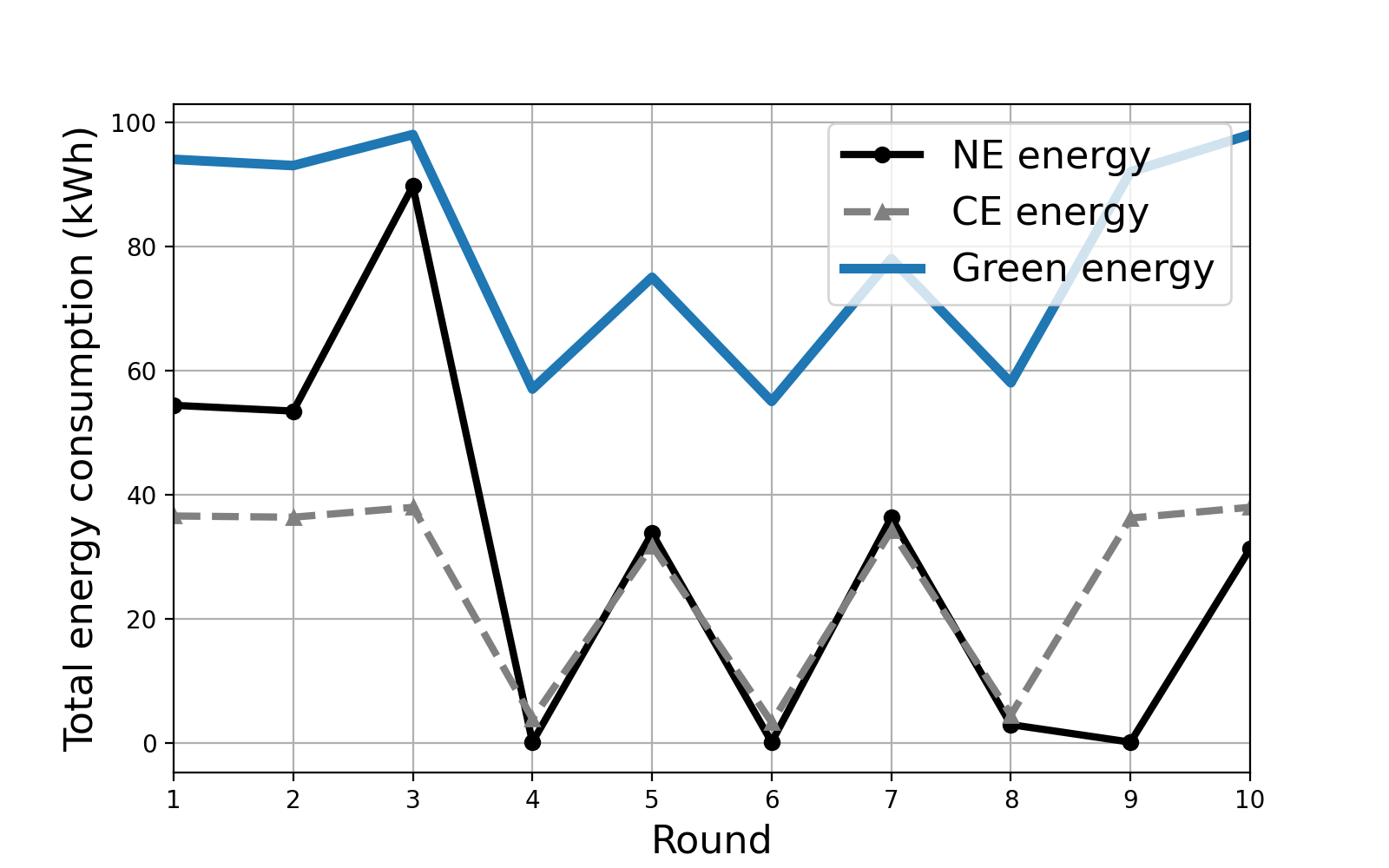}\hfill
    \includegraphics[width=.33\textwidth]{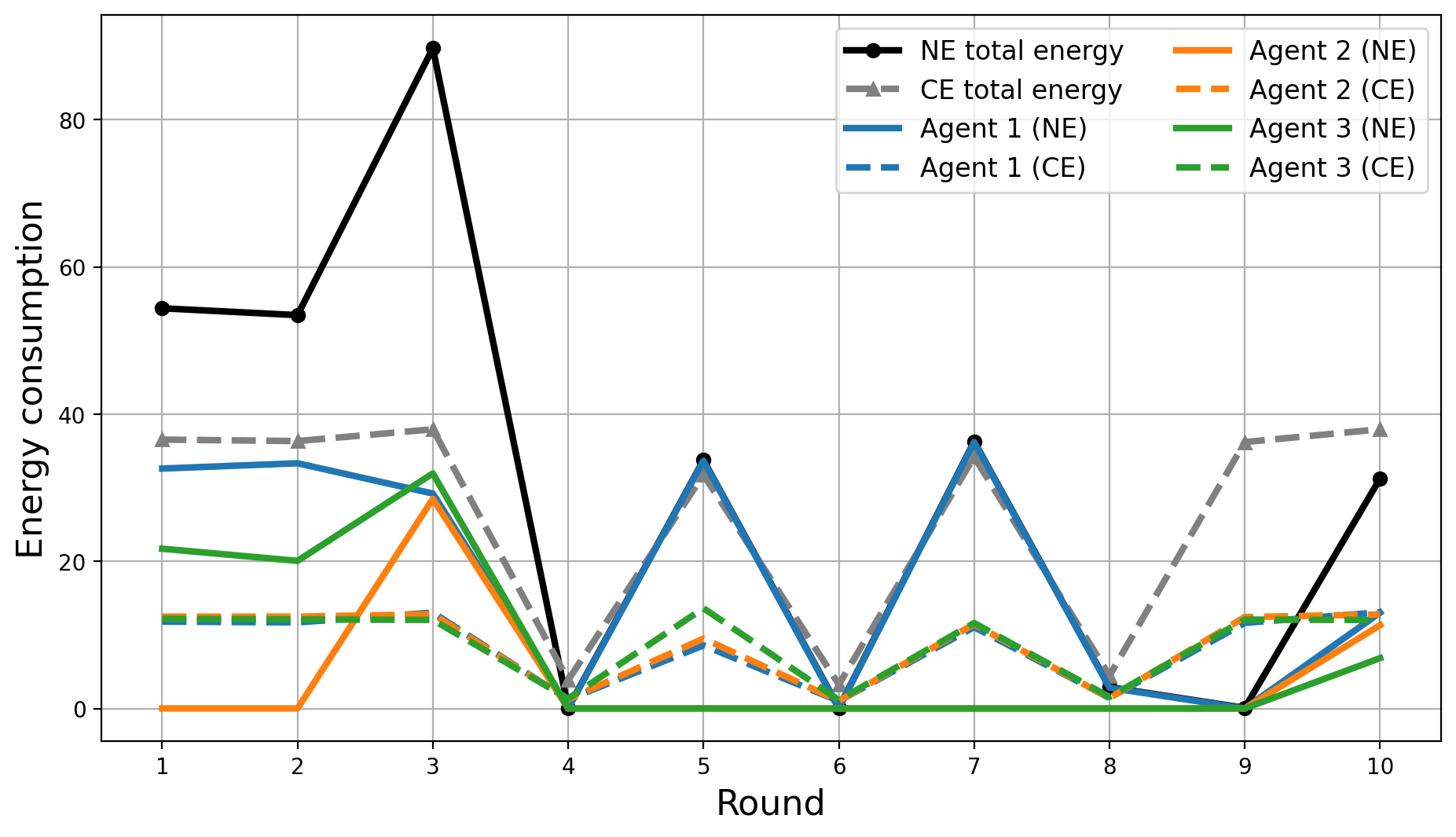}\hfill
    \includegraphics[width=.33\textwidth]{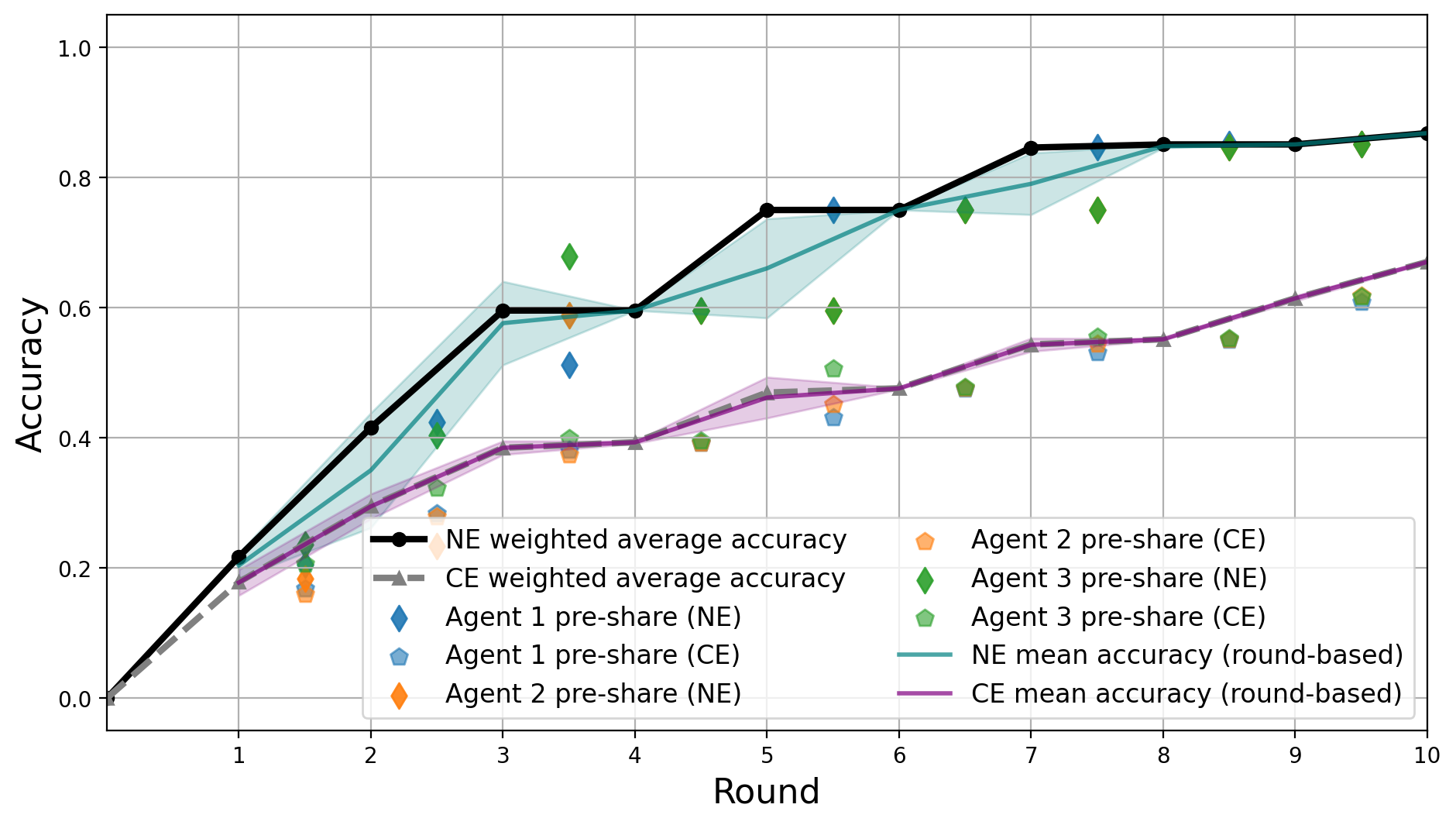}
    \includegraphics[width=.33\textwidth]{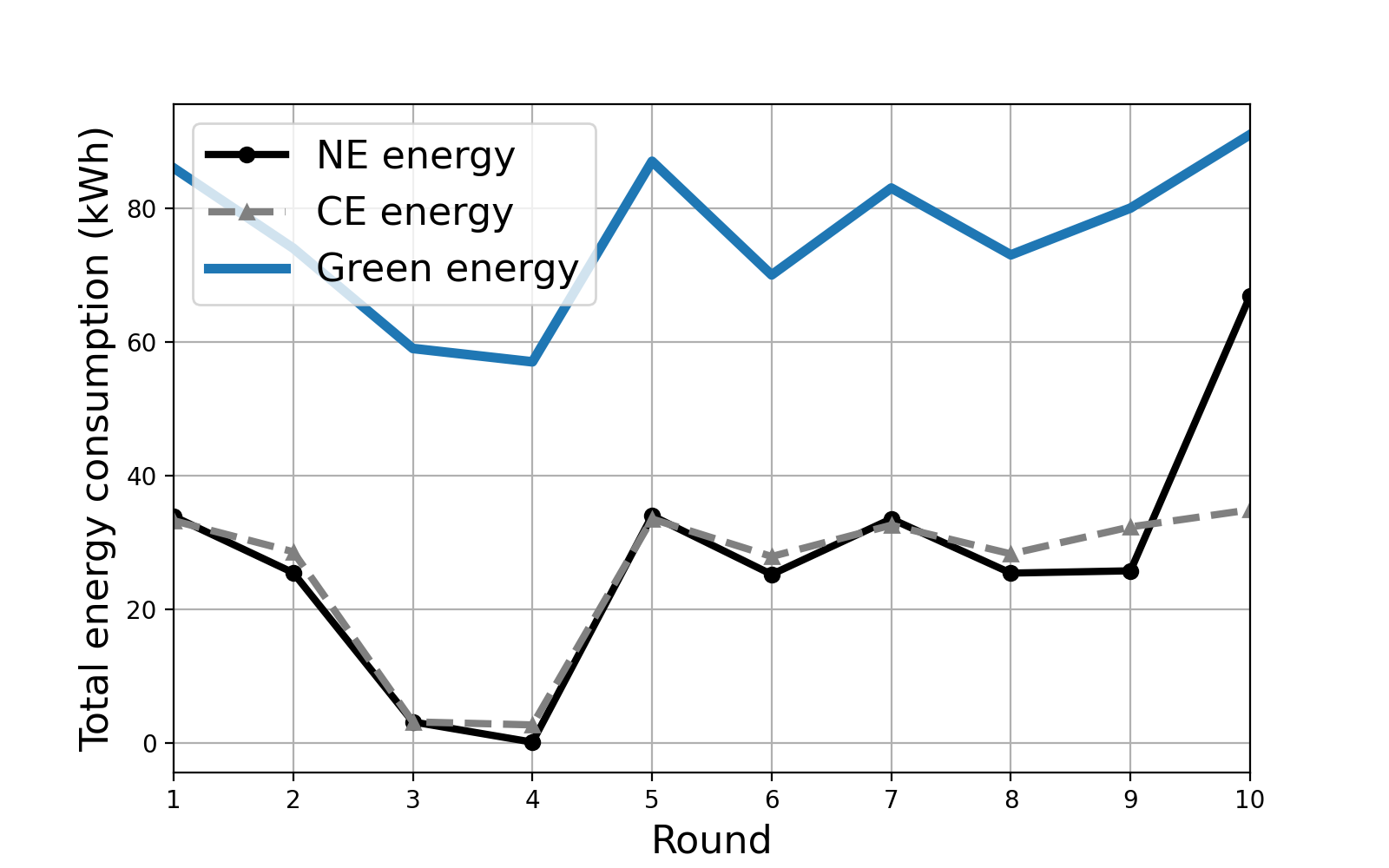}\hfill
    \includegraphics[width=.33\textwidth]{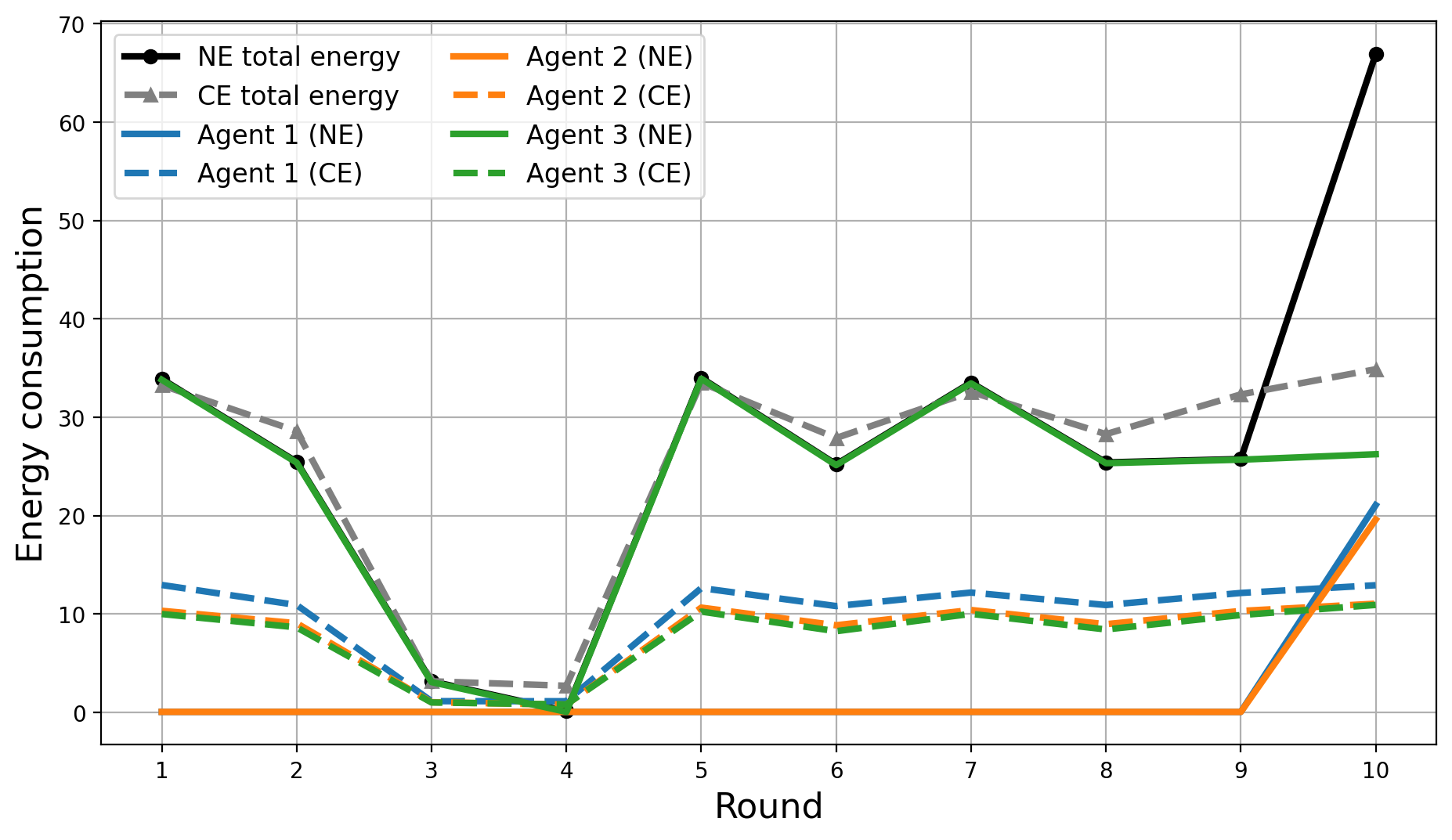}\hfill
    \includegraphics[width=.33\textwidth]{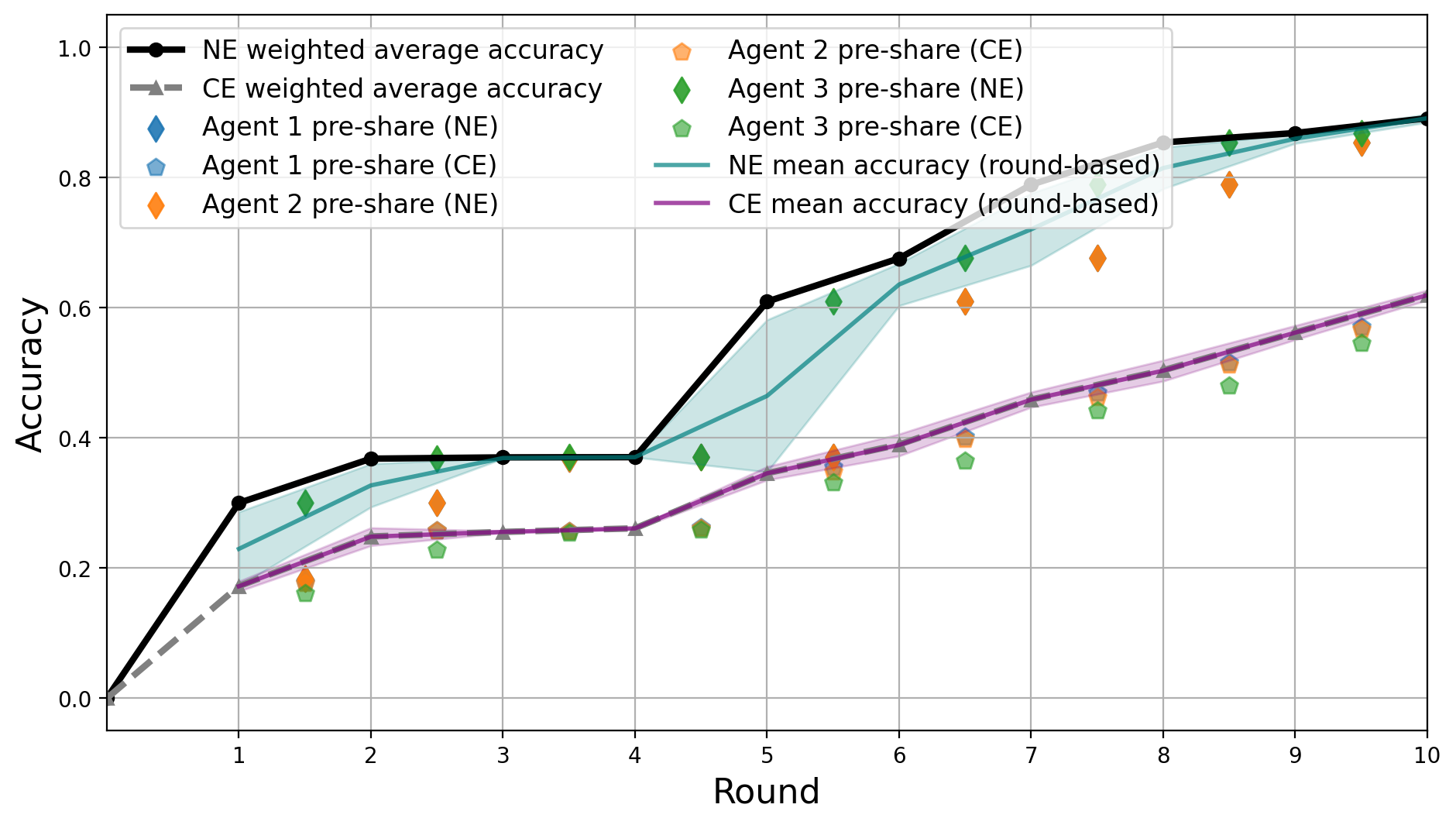}
    \caption{Energy-aware training: Total energy consumption (left column), energy consumption per agent (middle column), training accuracy (right column), for the case where agents are energy-aware, with $\gamma = 1$ (upper row), $\gamma = 10$ (middle row), and $\gamma = 100$ (lower row), with $ h = 0.5$  and $\delta_+ = \delta_- = 20$.}
\label{fig:energy-aware-20}
\end{figure}

\begin{figure}[h]
    \centering
    \includegraphics[width=.33\textwidth]{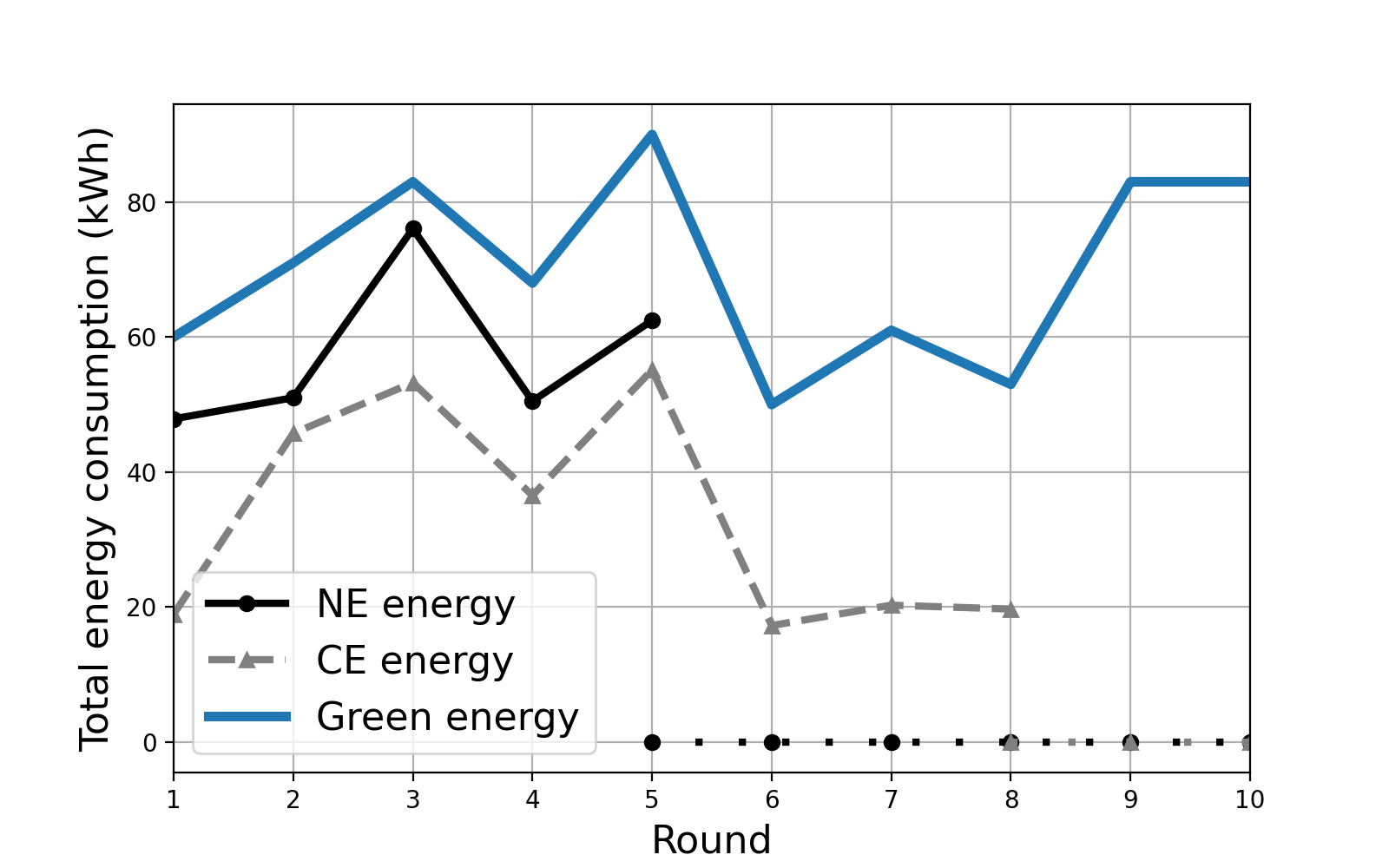}\hfill
    \includegraphics[width=.33\textwidth]{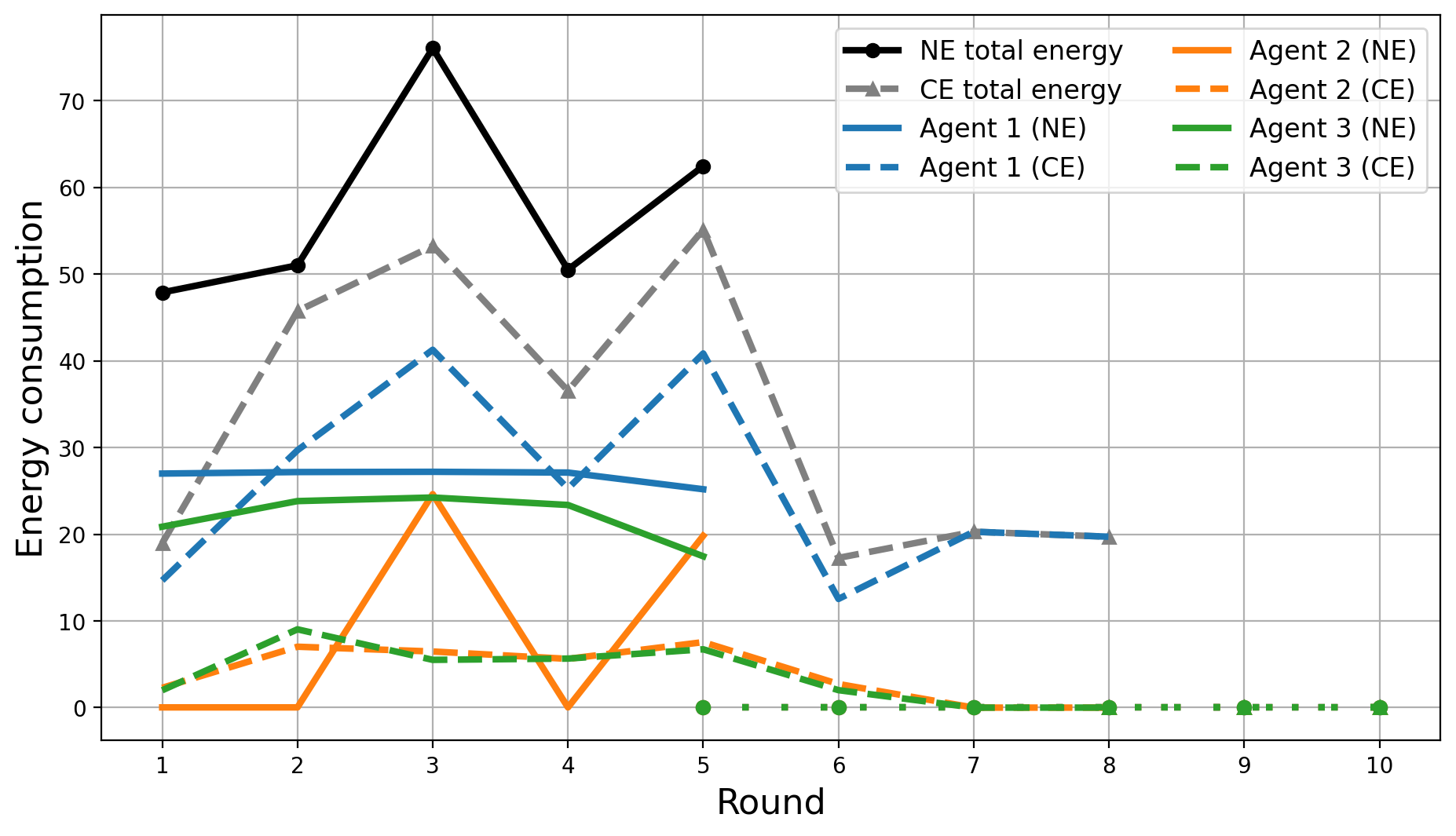}\hfill
    \includegraphics[width=.33\textwidth]{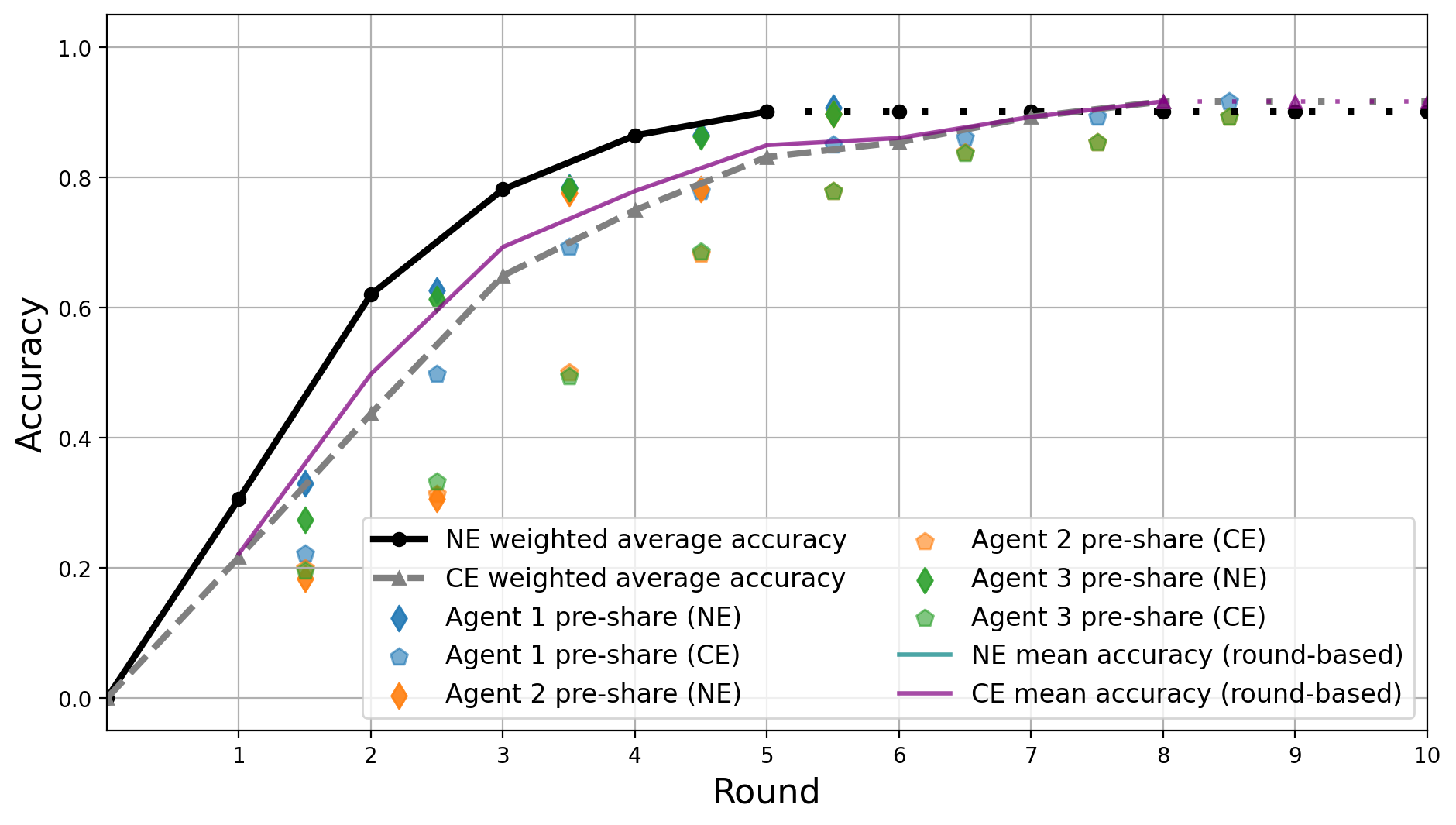}
    \includegraphics[width=.33\textwidth]{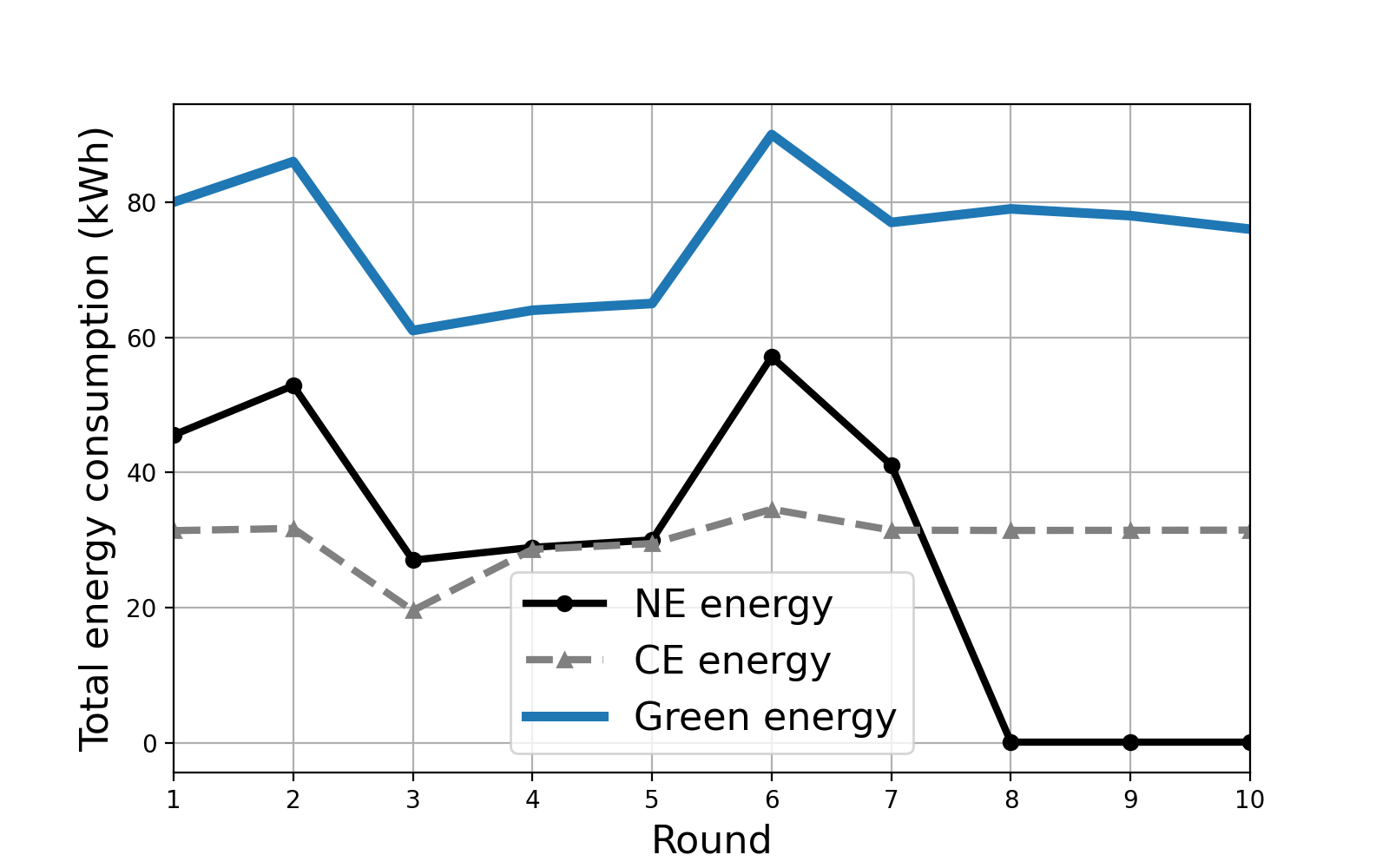}\hfill
    \includegraphics[width=.33\textwidth]{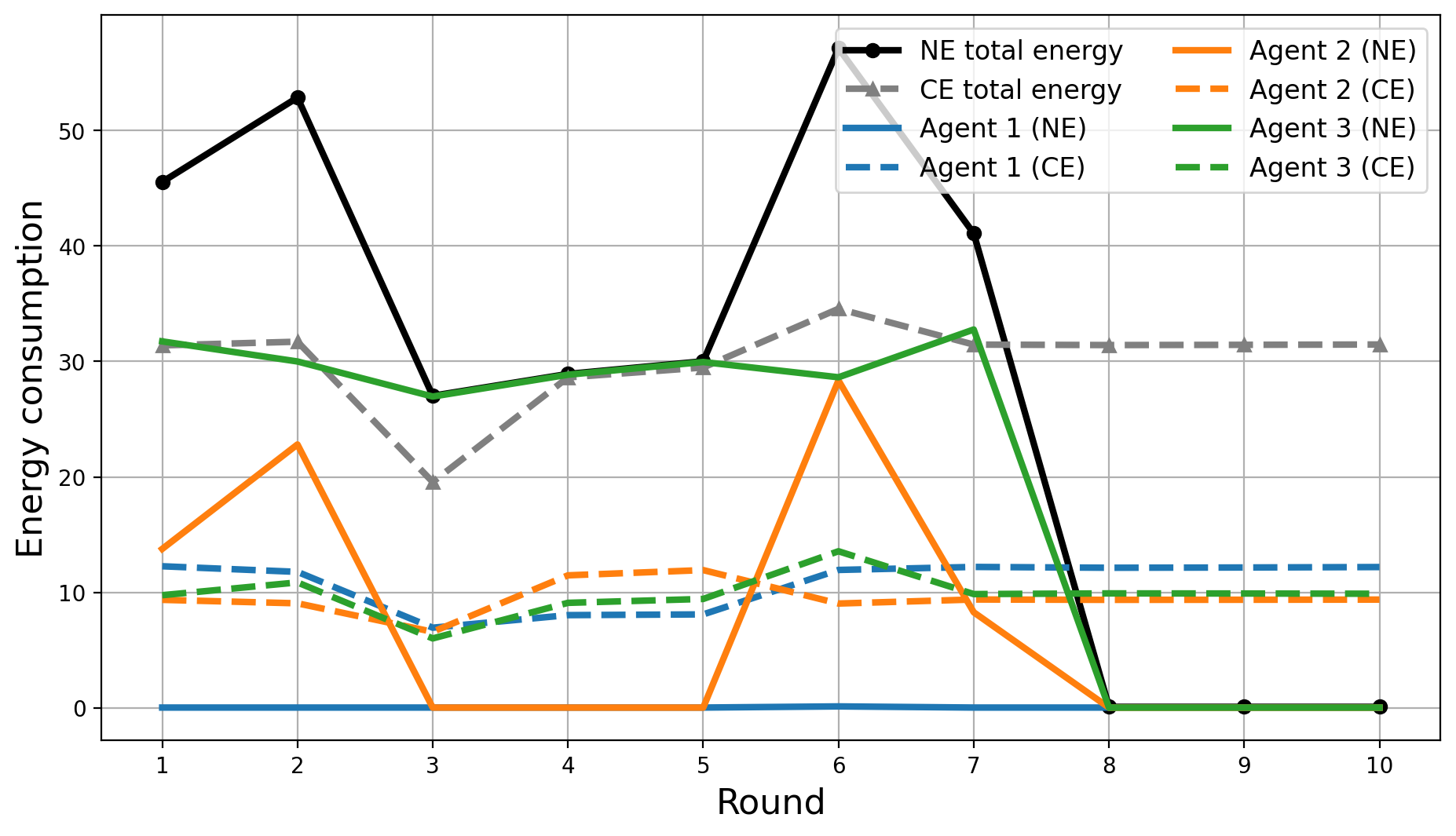}\hfill
    \includegraphics[width=.33\textwidth]{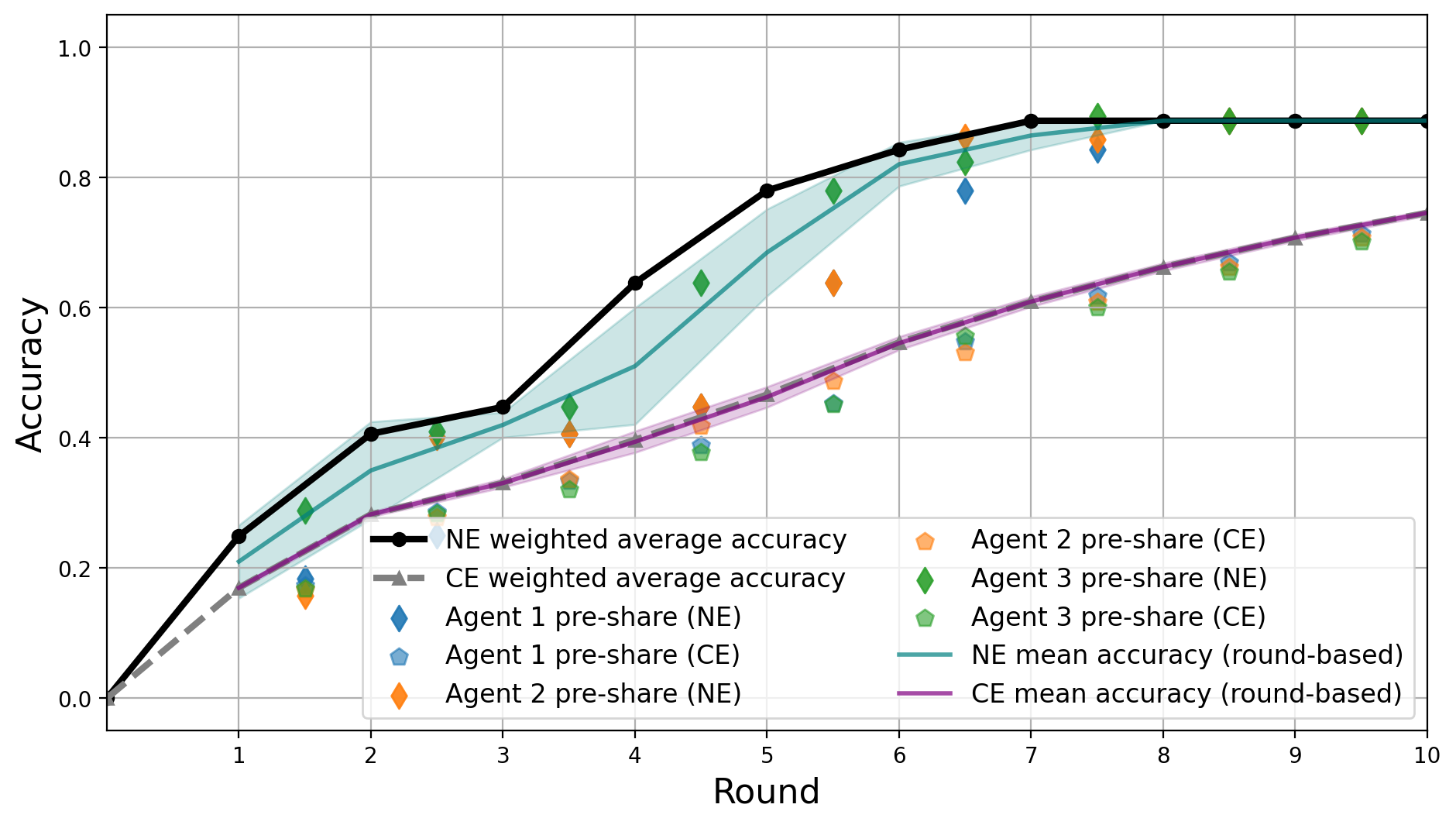}
    \includegraphics[width=.33\textwidth]{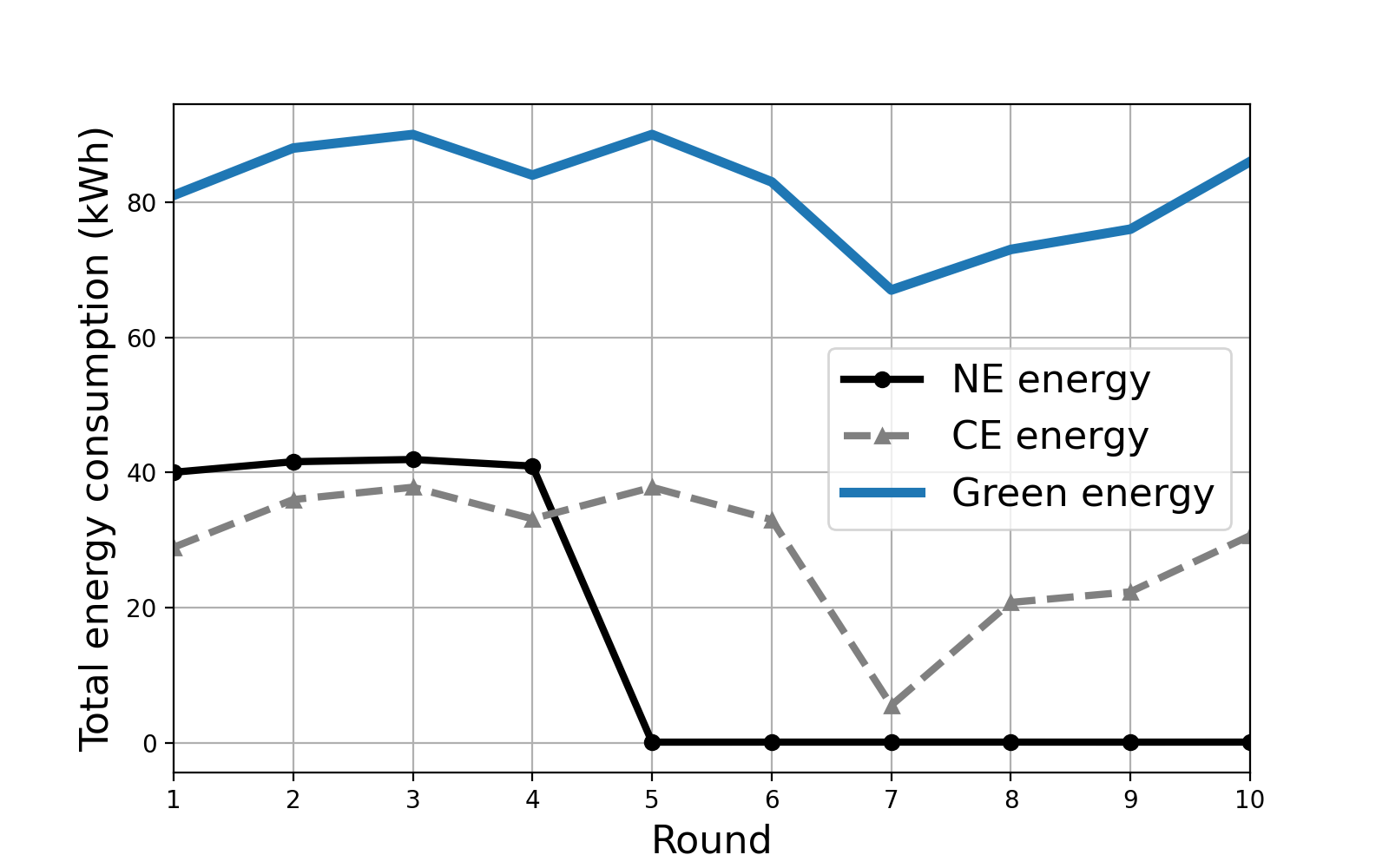}\hfill
    \includegraphics[width=.33\textwidth]{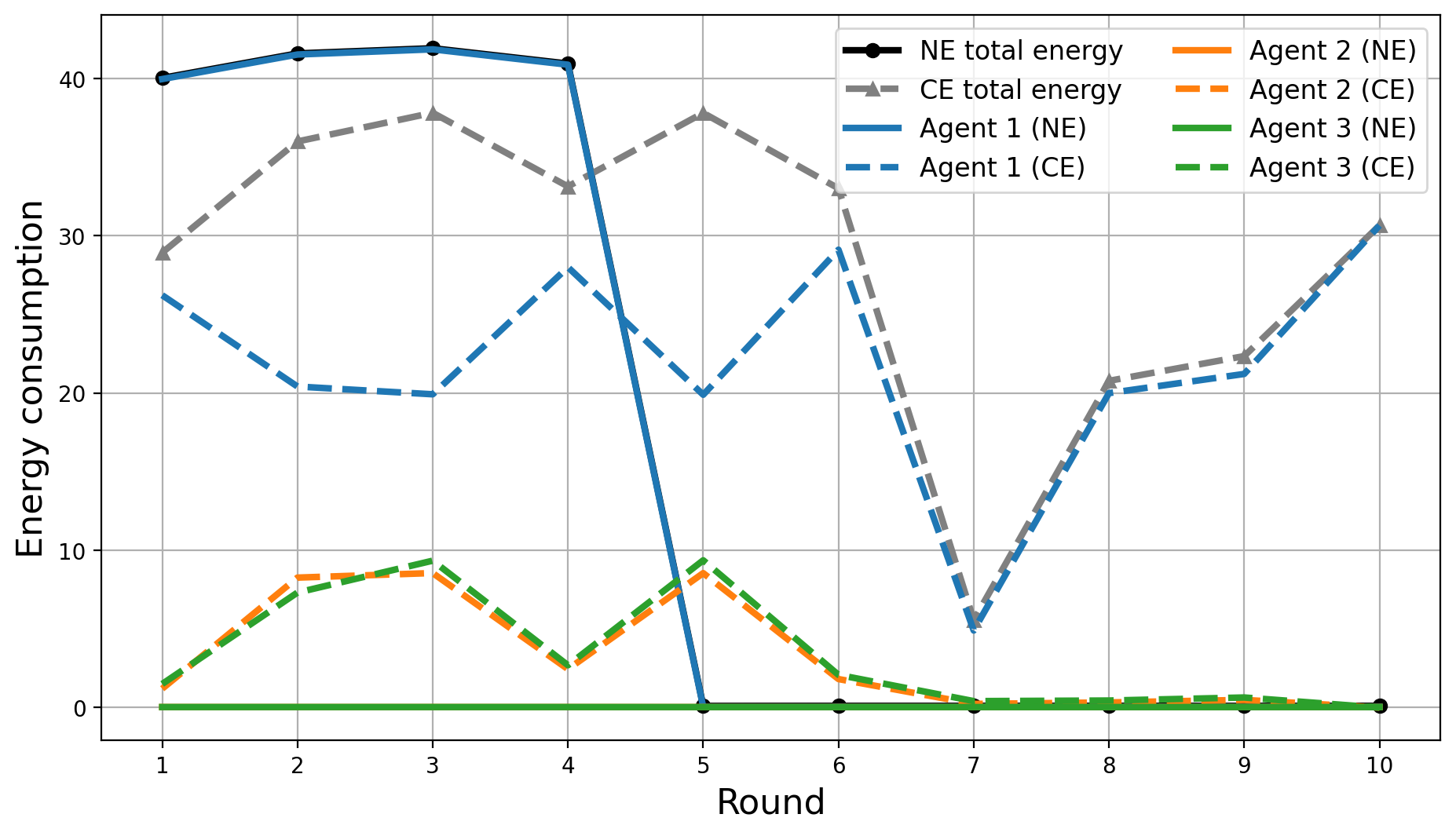}\hfill
    \includegraphics[width=.33\textwidth]{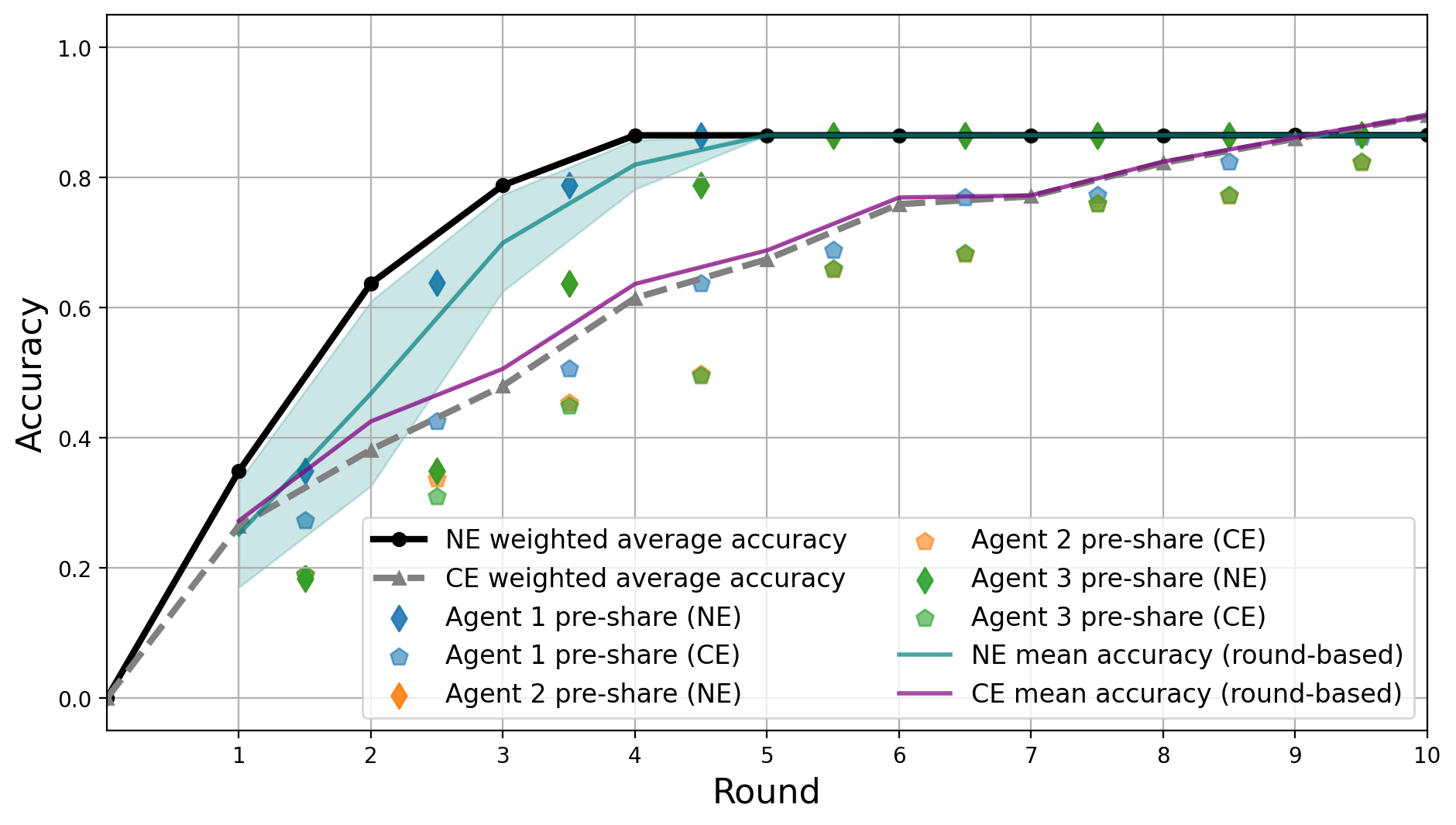}
    \caption{Energy-aware training in a heterogeneous population: Total energy consumption (left column), energy consumption per agent (middle column), training accuracy (right column), for the case where agents are energy-aware, with $\gamma = 1$ (upper row), $\gamma = 10$ (middle row), and $\gamma = 100$ (lower row), and $h = 0.5$.} 
\label{fig:het-agents}
\end{figure}

\begin{figure}[h]
    \centering
    \includegraphics[width=.33\textwidth]{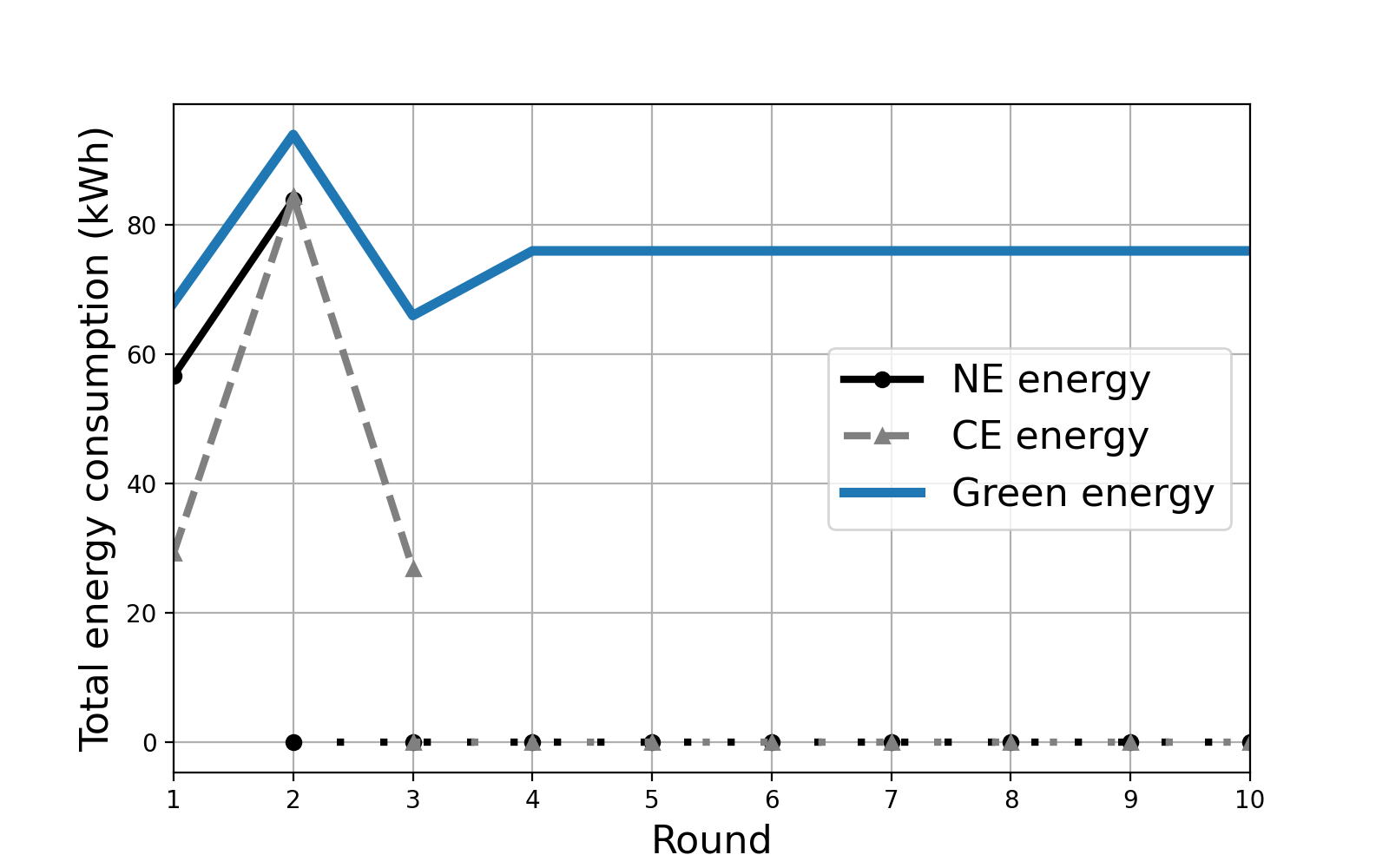}\hfill
    \includegraphics[width=.33\textwidth]{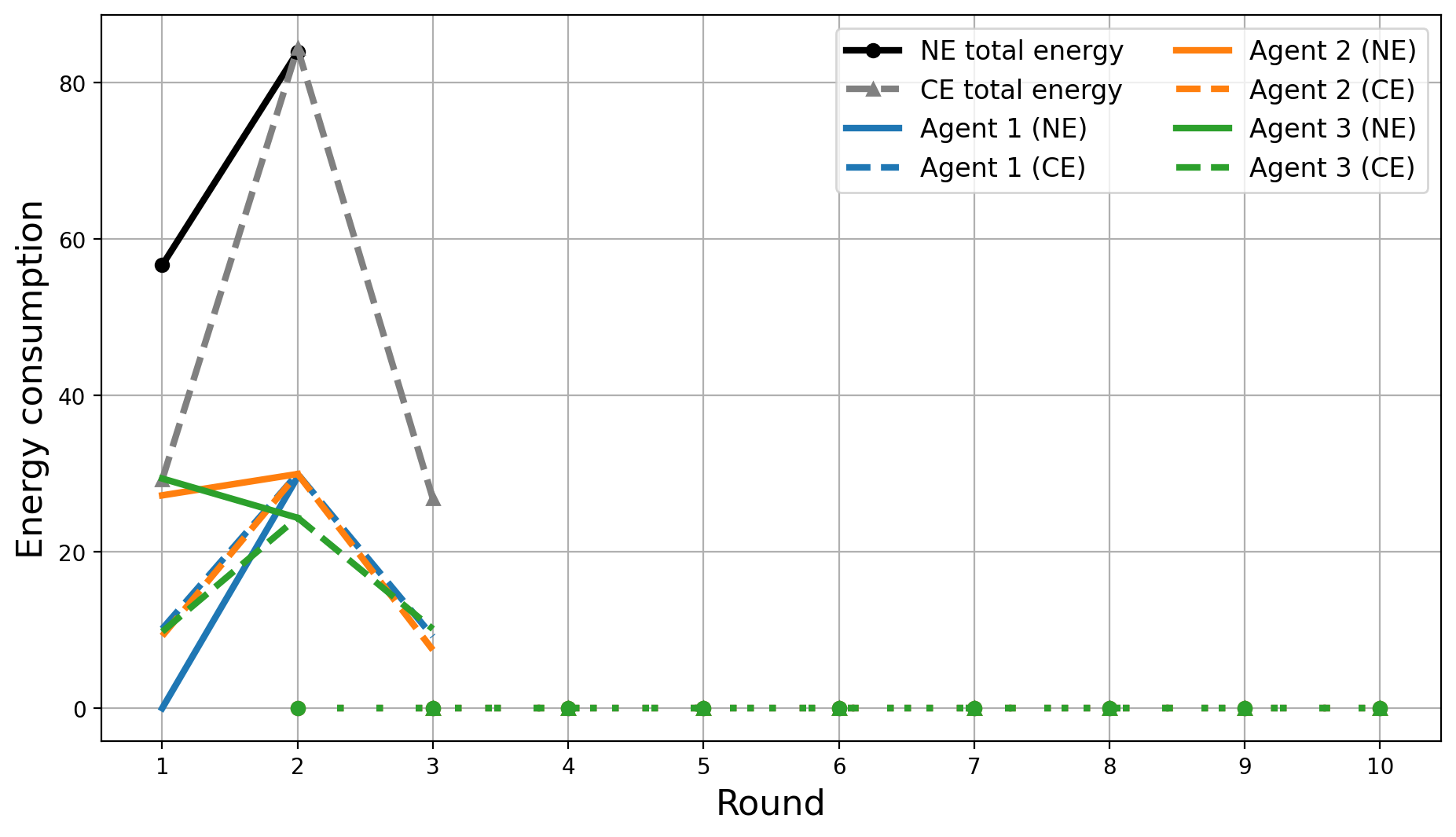}\hfill
    \includegraphics[width=.33\textwidth]{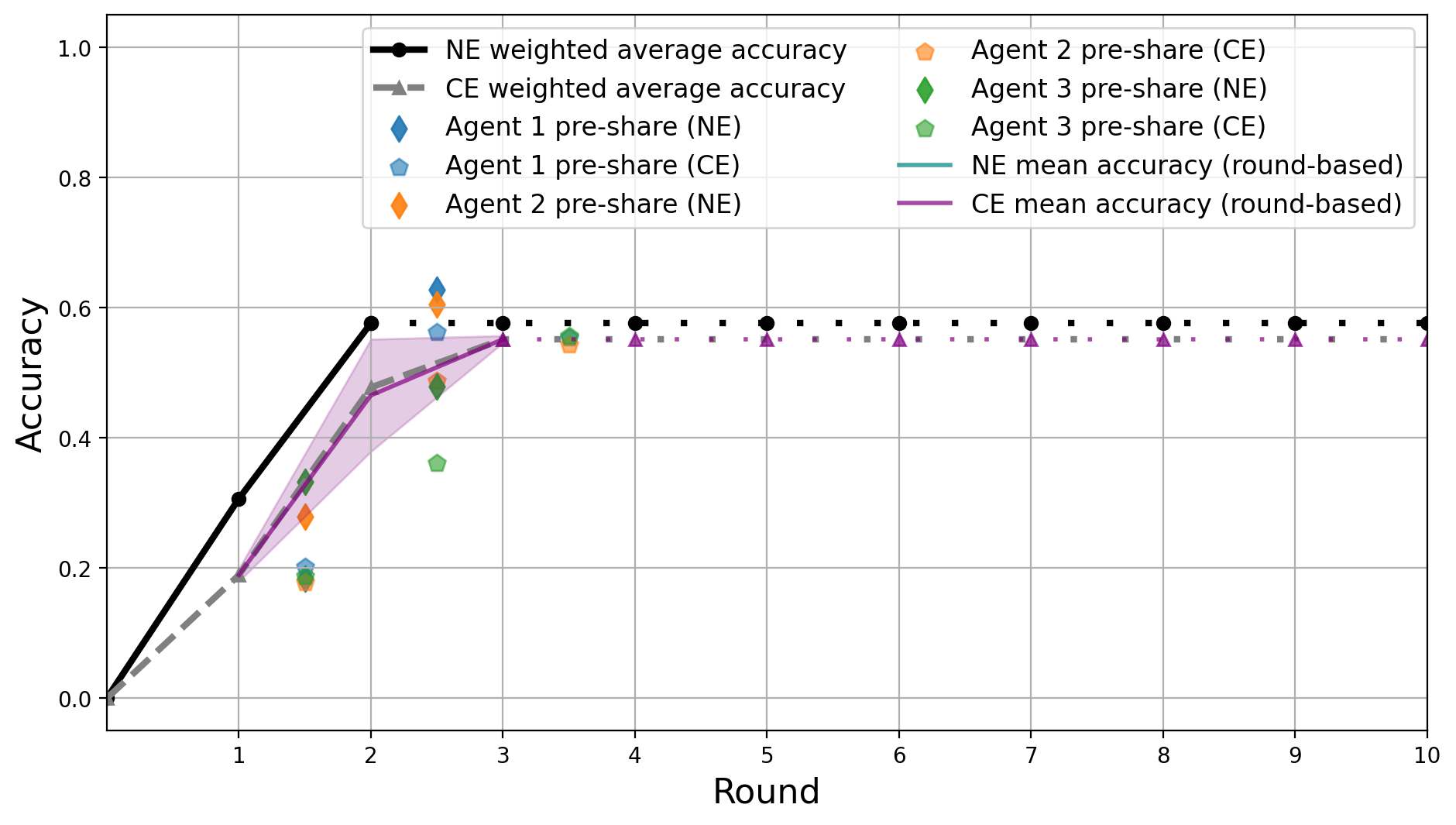}
    \includegraphics[width=.33\textwidth]{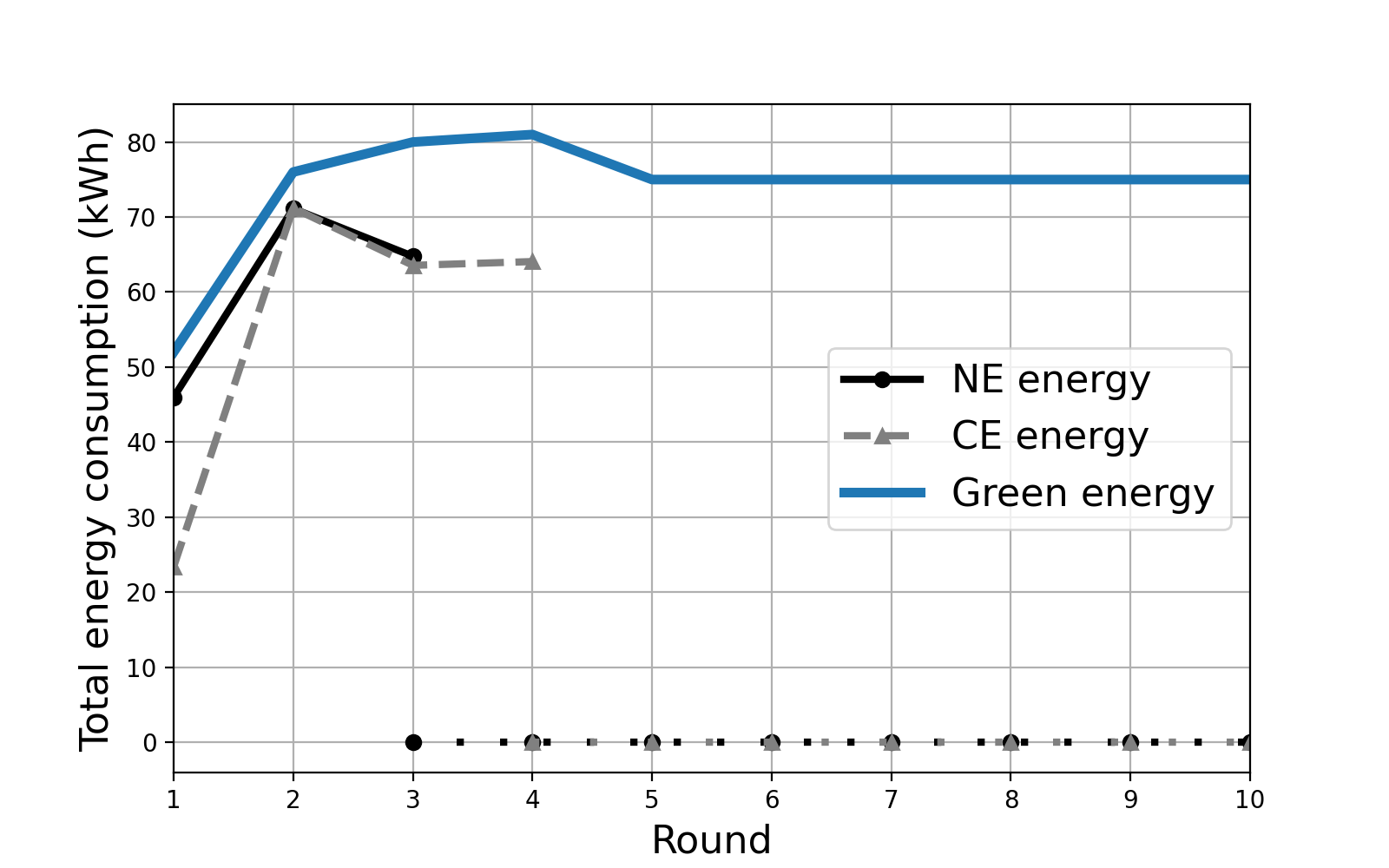}\hfill
    \includegraphics[width=.33\textwidth]{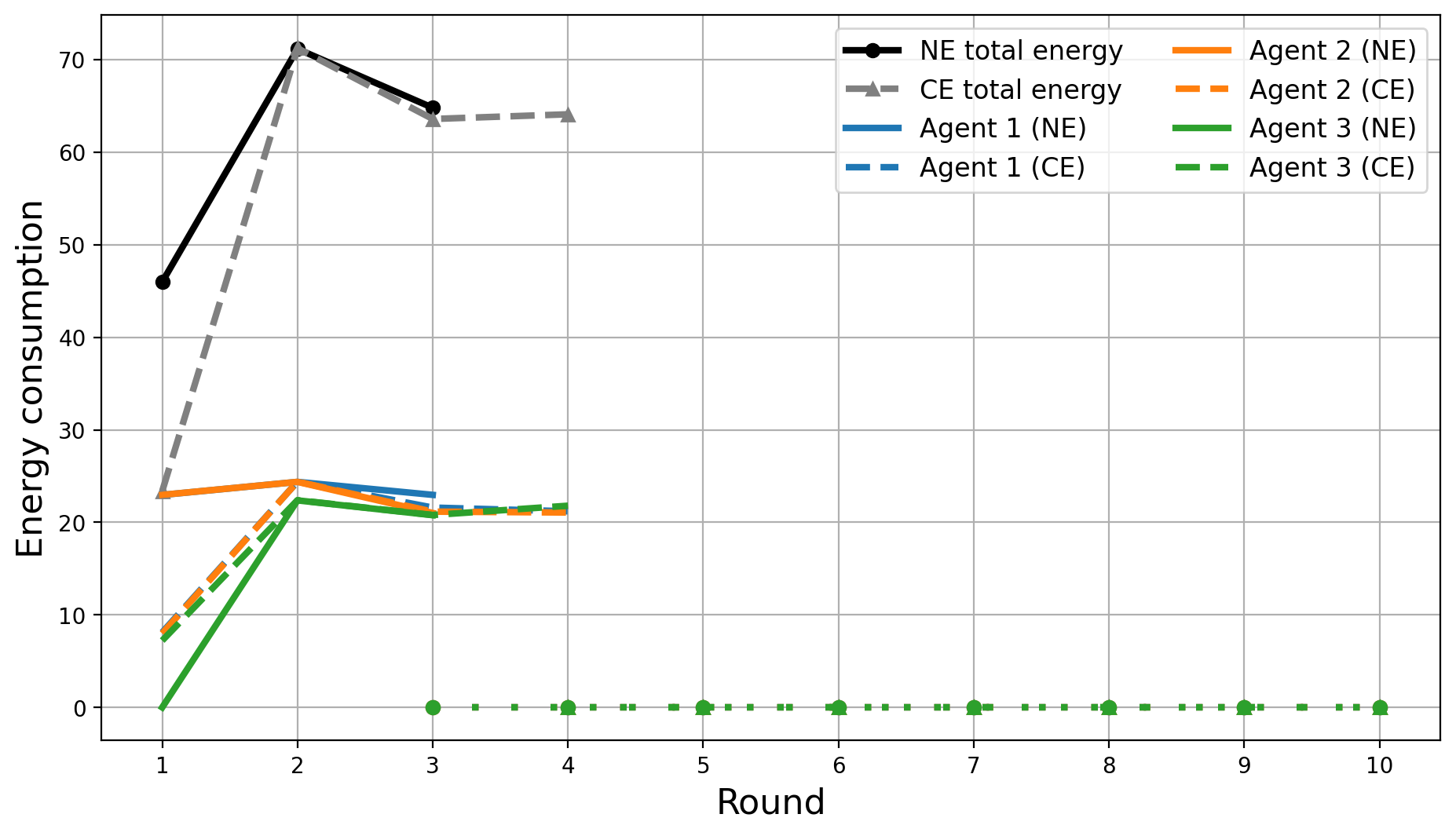}\hfill
    \includegraphics[width=.33\textwidth]{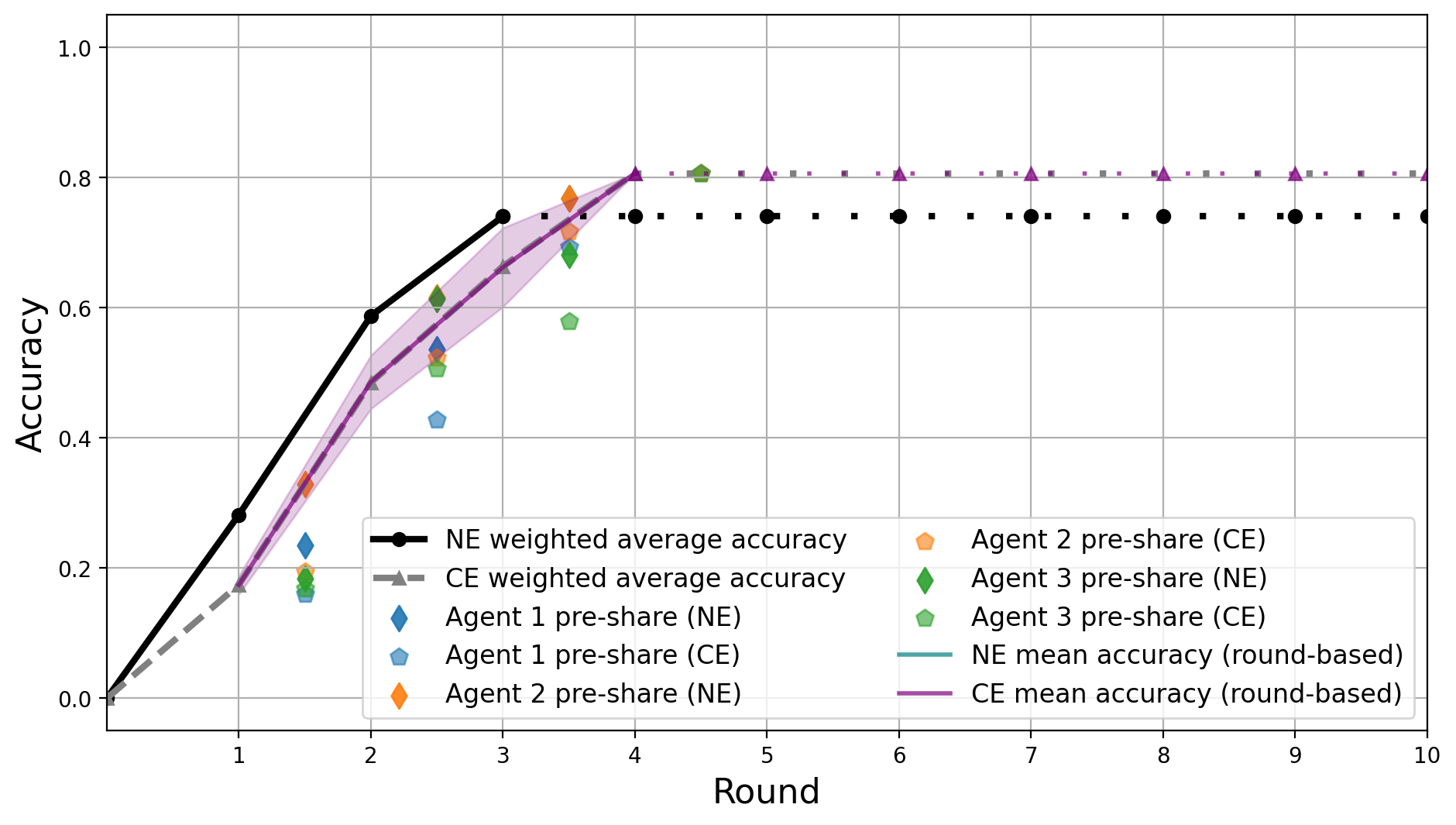}
    \caption{Global accuracy threshold. Total energy reduction (left column), training accuracy per agent (middle column), mean training accuracy (right column), for the cases where: (i) $\mathscr{a} = 0.5$ (upper row), and (ii) $\mathscr{a} = 0.7$ (lower row), with $h = 0.5$, $\gamma = 1$ and $\delta_+ = \delta_- = 10$.} 
\label{fig:global accuracy effect g 1}
\end{figure}

\begin{figure}[h]
    \centering
    \includegraphics[width=.33\textwidth]{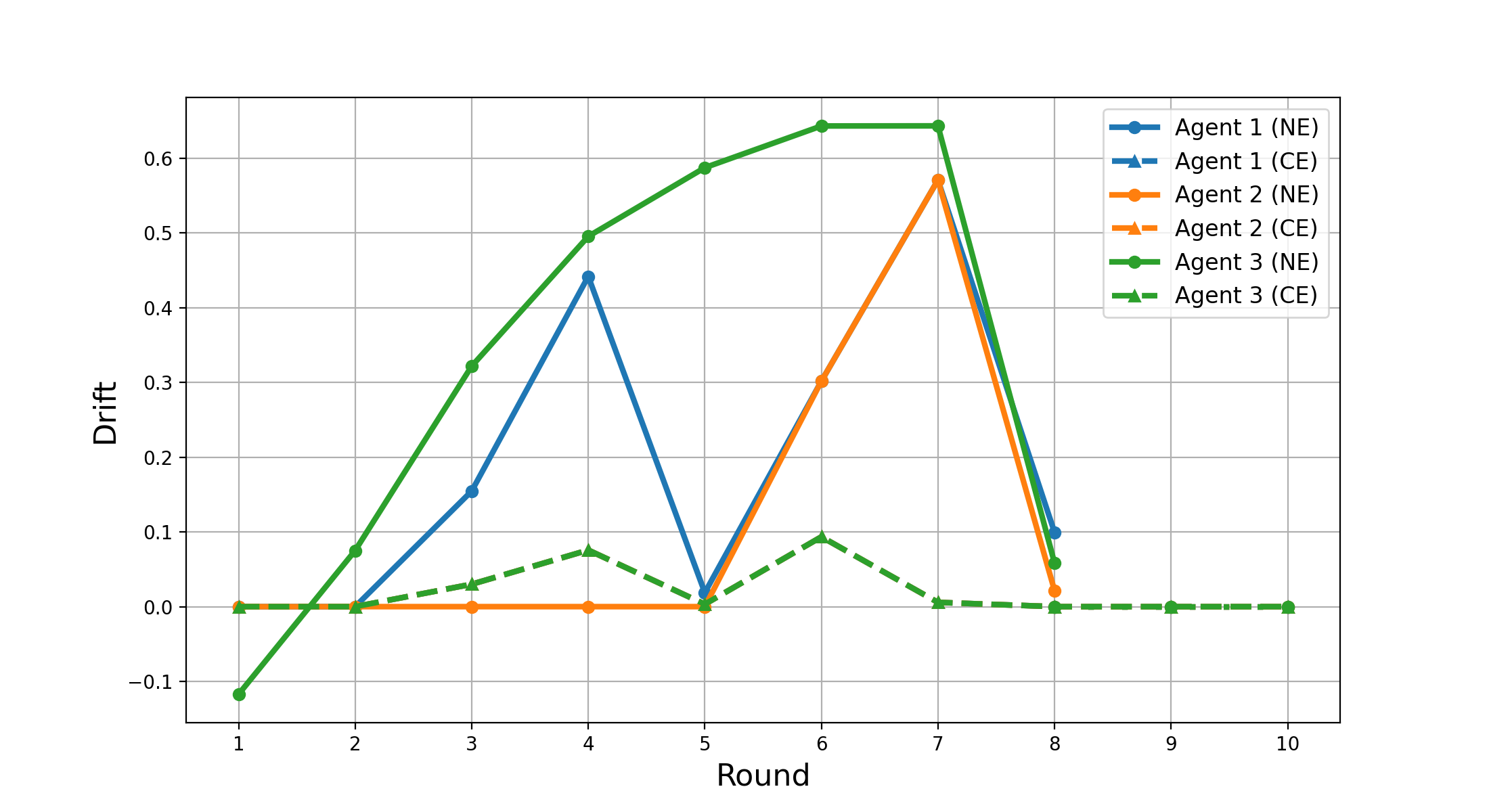}\hfill
    \includegraphics[width=.33\textwidth]{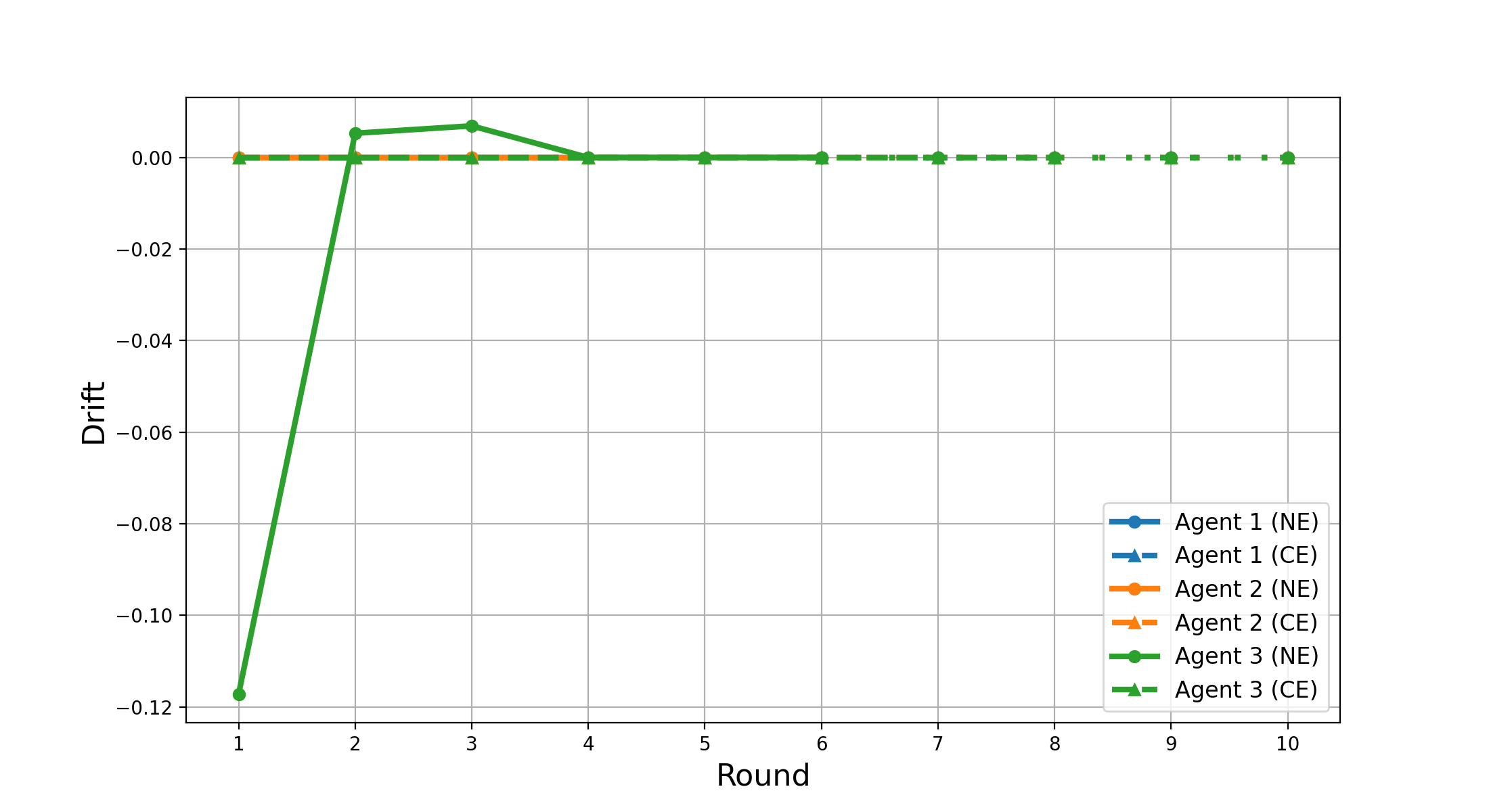}\hfill
    \includegraphics[width=.33\textwidth]{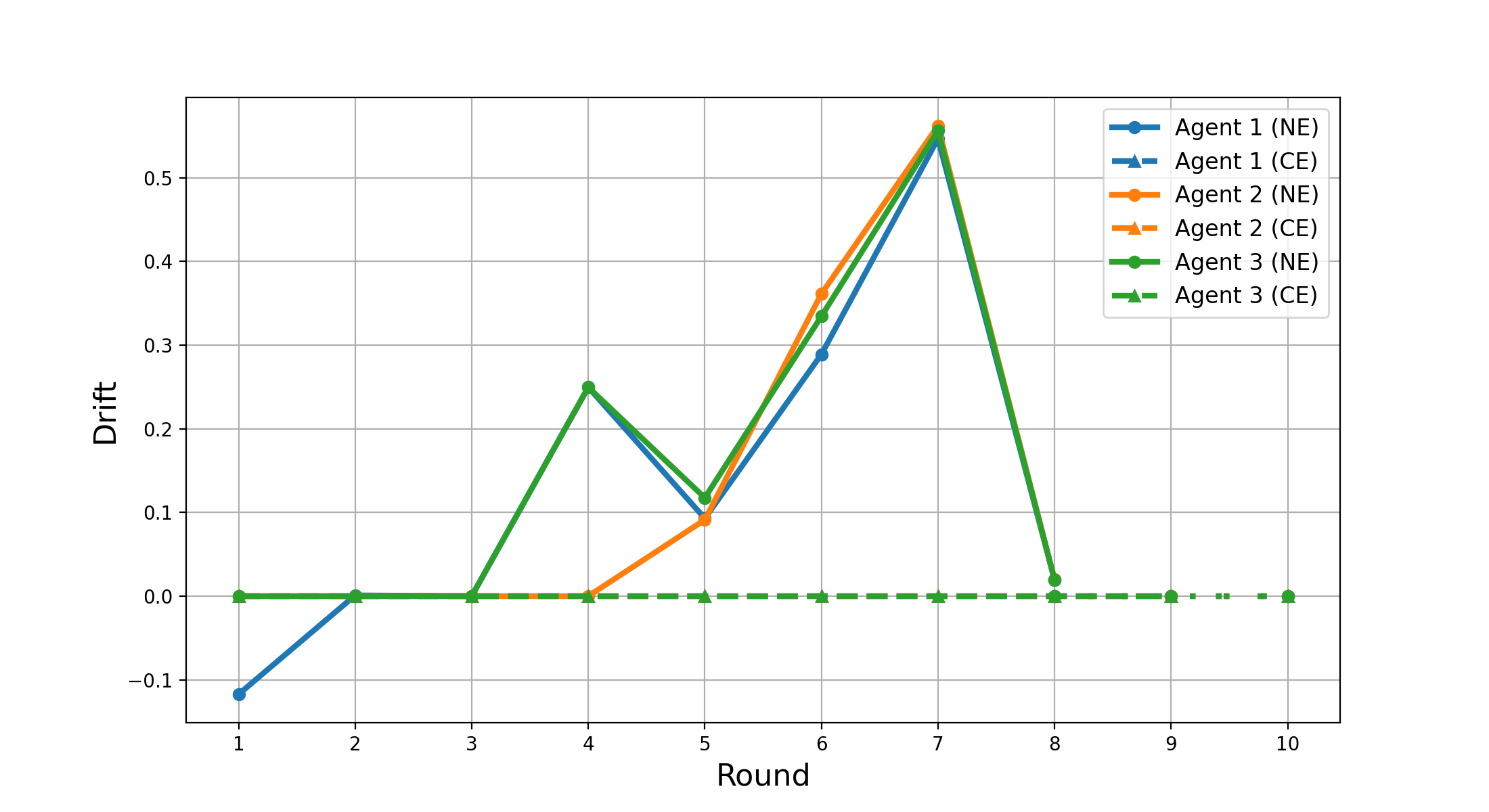}
    \includegraphics[width=.33\textwidth]{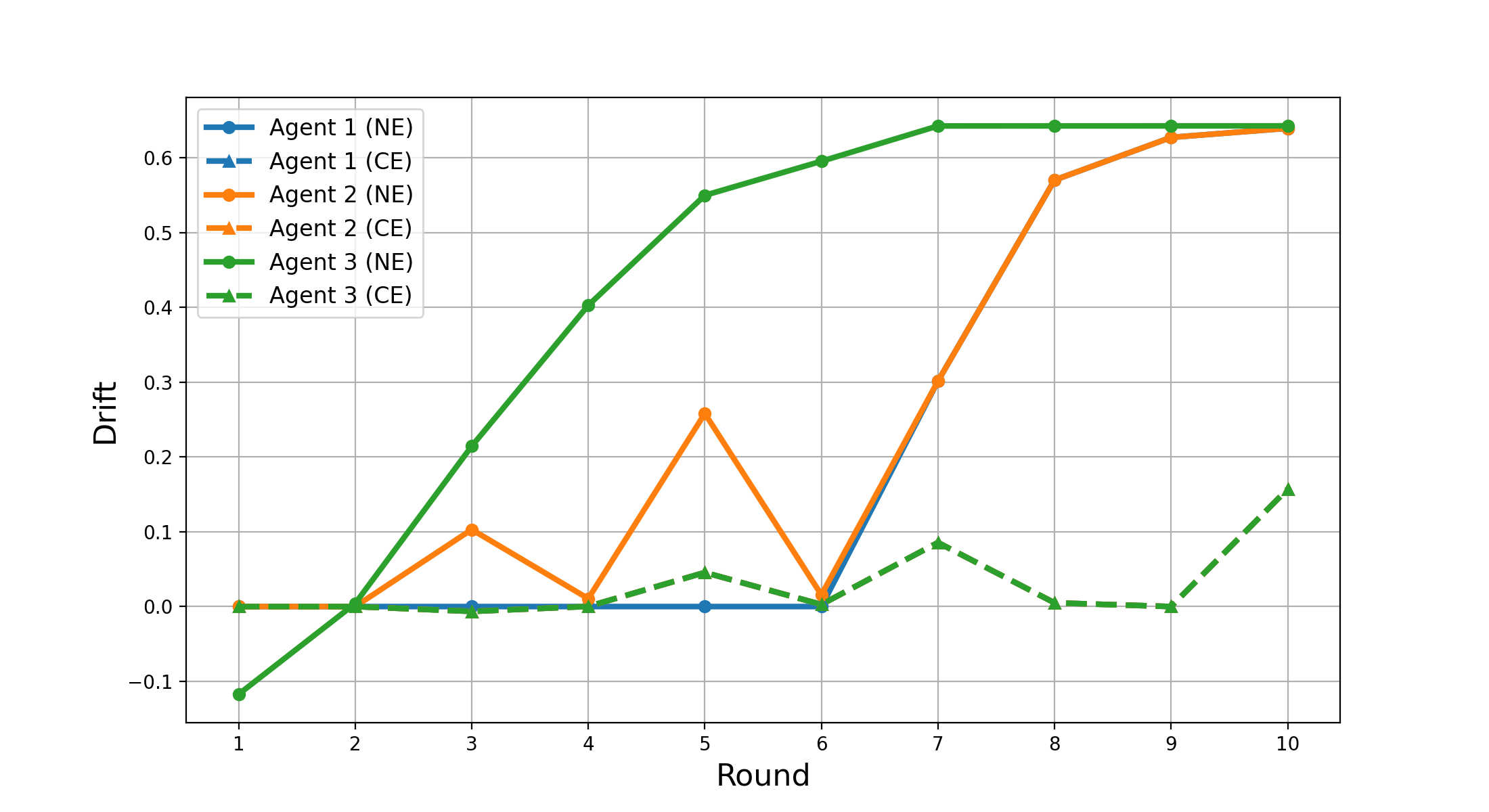}\hfill
    \includegraphics[width=.33\textwidth]{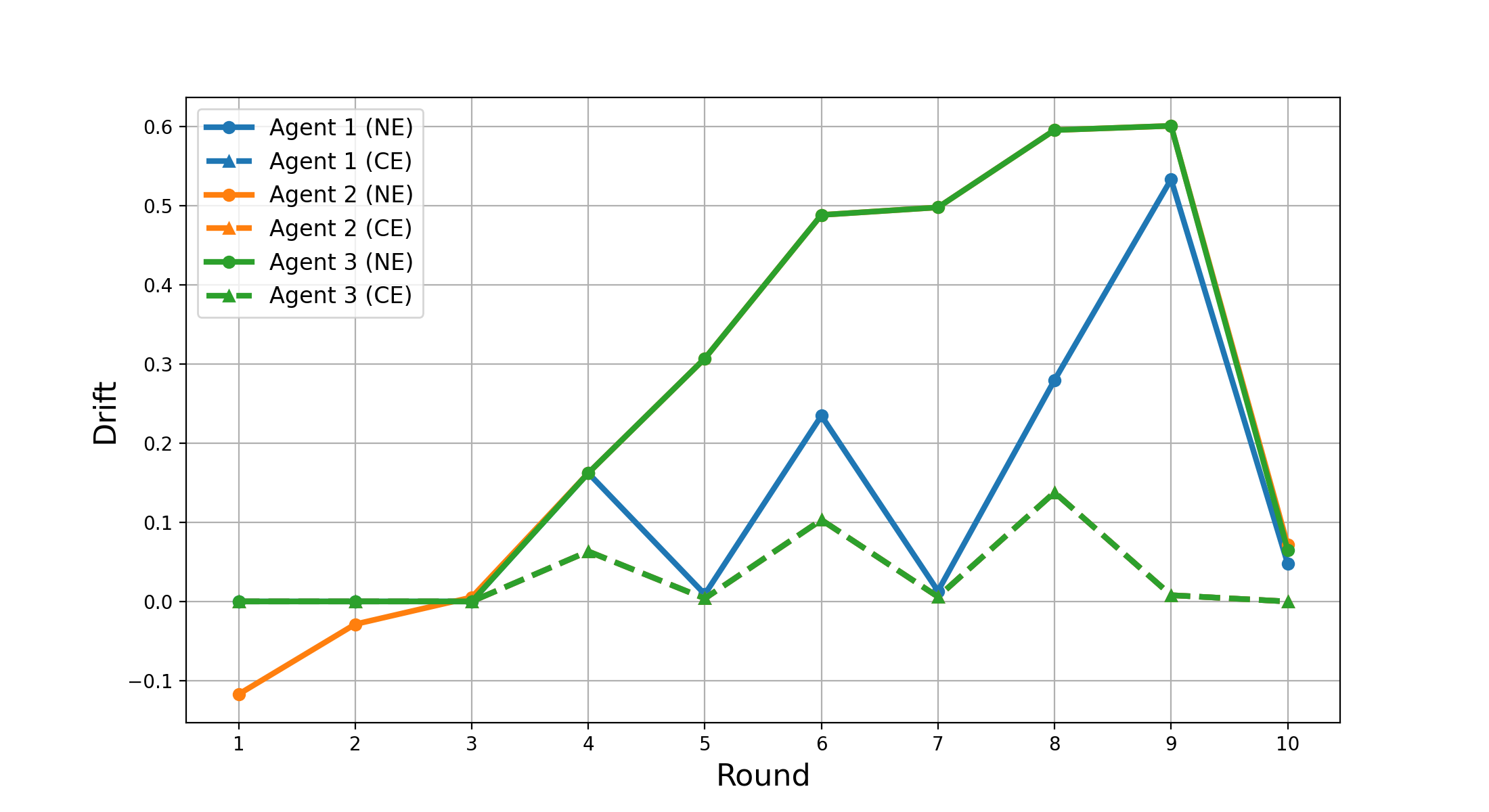}\hfill
    \includegraphics[width=.33\textwidth]{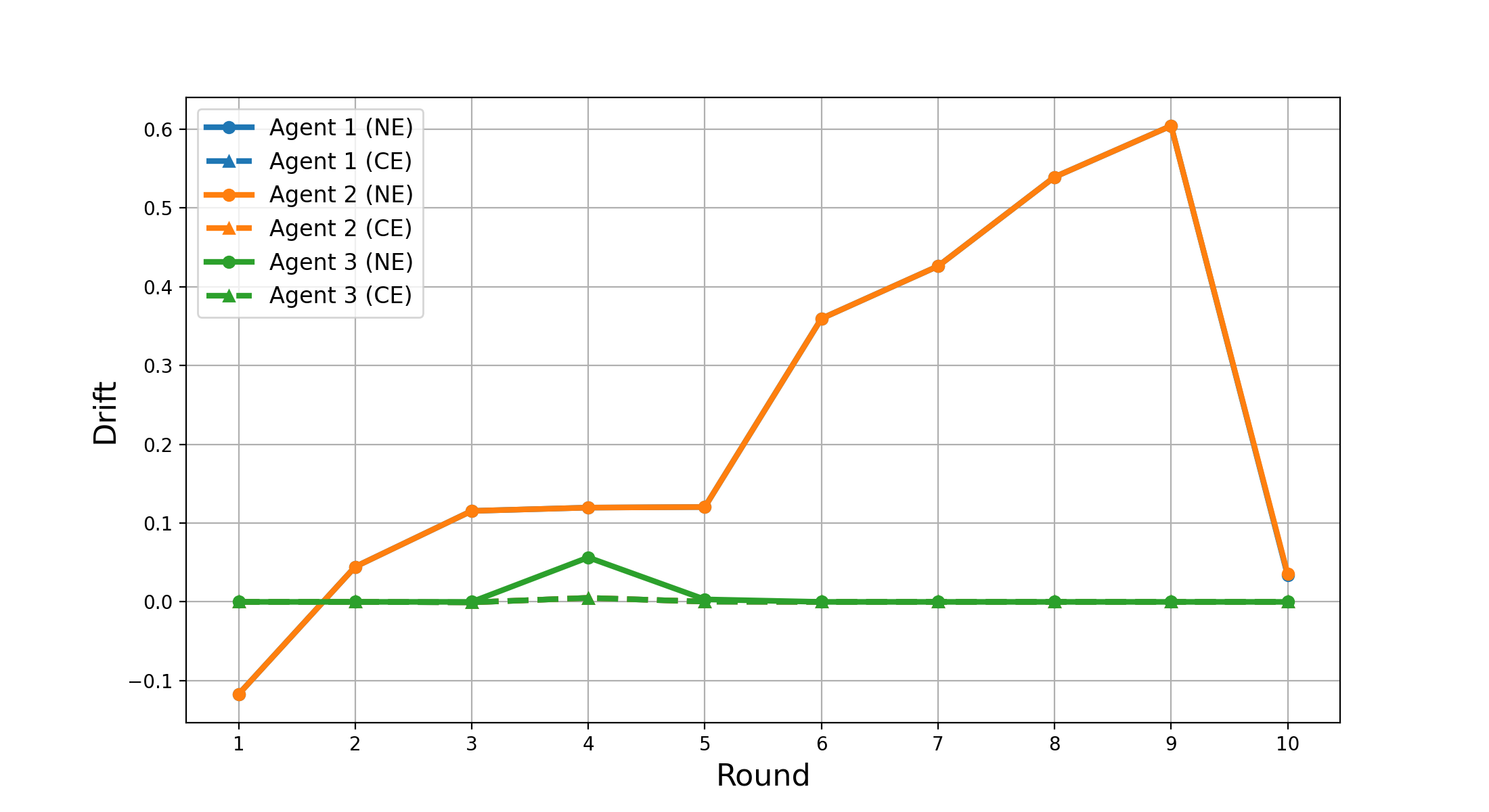}
    \caption{Drifting phenomenon: Here, each agent $i$ takes: (i) $\gamma = 1$ (left column), (ii) $\gamma = 10$ (middle column), and (iii) $\gamma = 100$ (right column), and $\delta_+ = \delta_- = 10$ (upper row), and $\delta_+ = \delta_- = 20$ (lower row), for $h = 0.5$.} 
\label{fig:drifting}
\end{figure}

\begin{figure}[h]
    \centering
    \includegraphics[width=5.5cm, height=3.5cm, keepaspectratio=true]{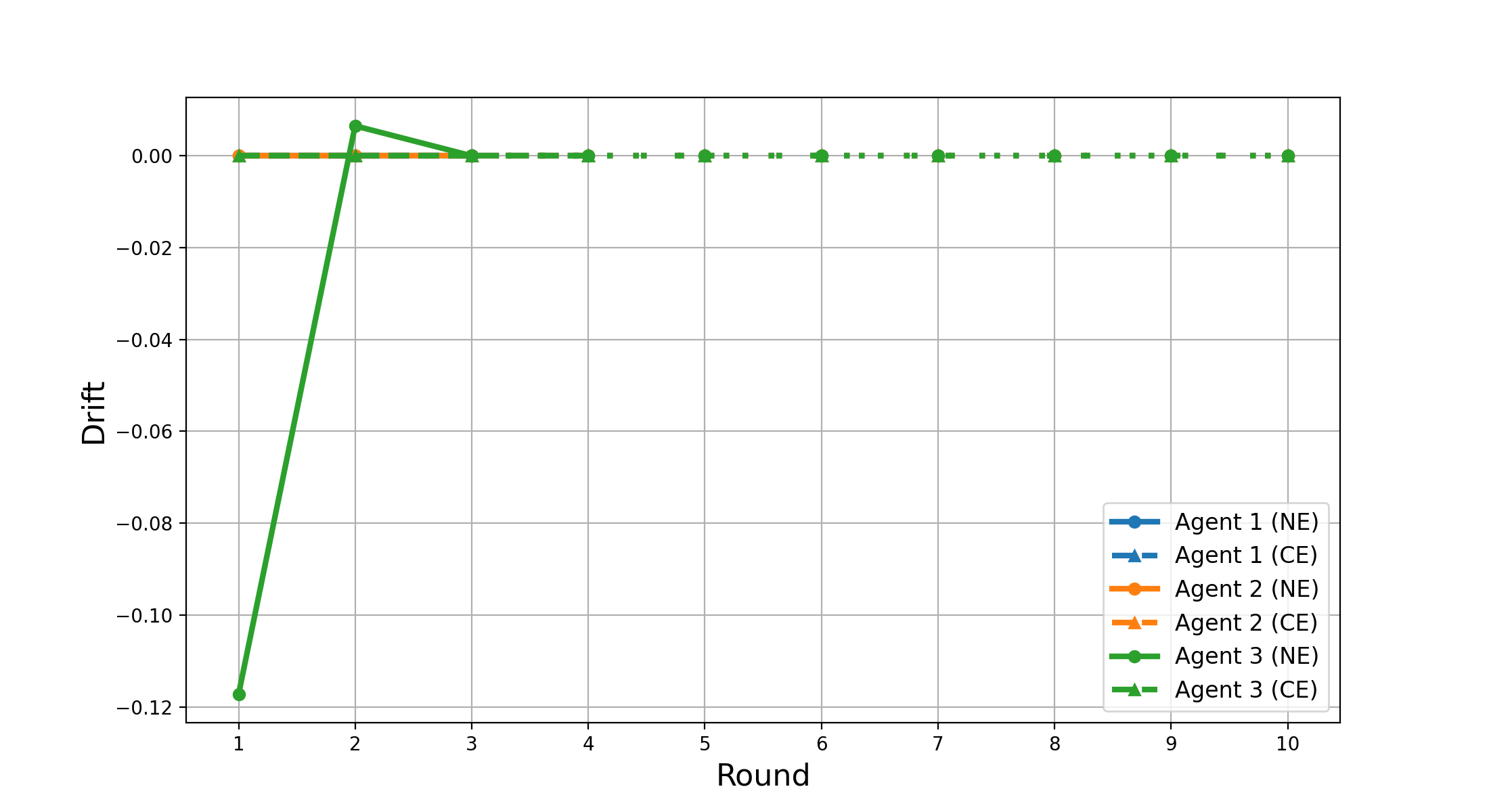}\hfill
    \includegraphics[width=5.5cm, height=2.725cm, keepaspectratio=true]{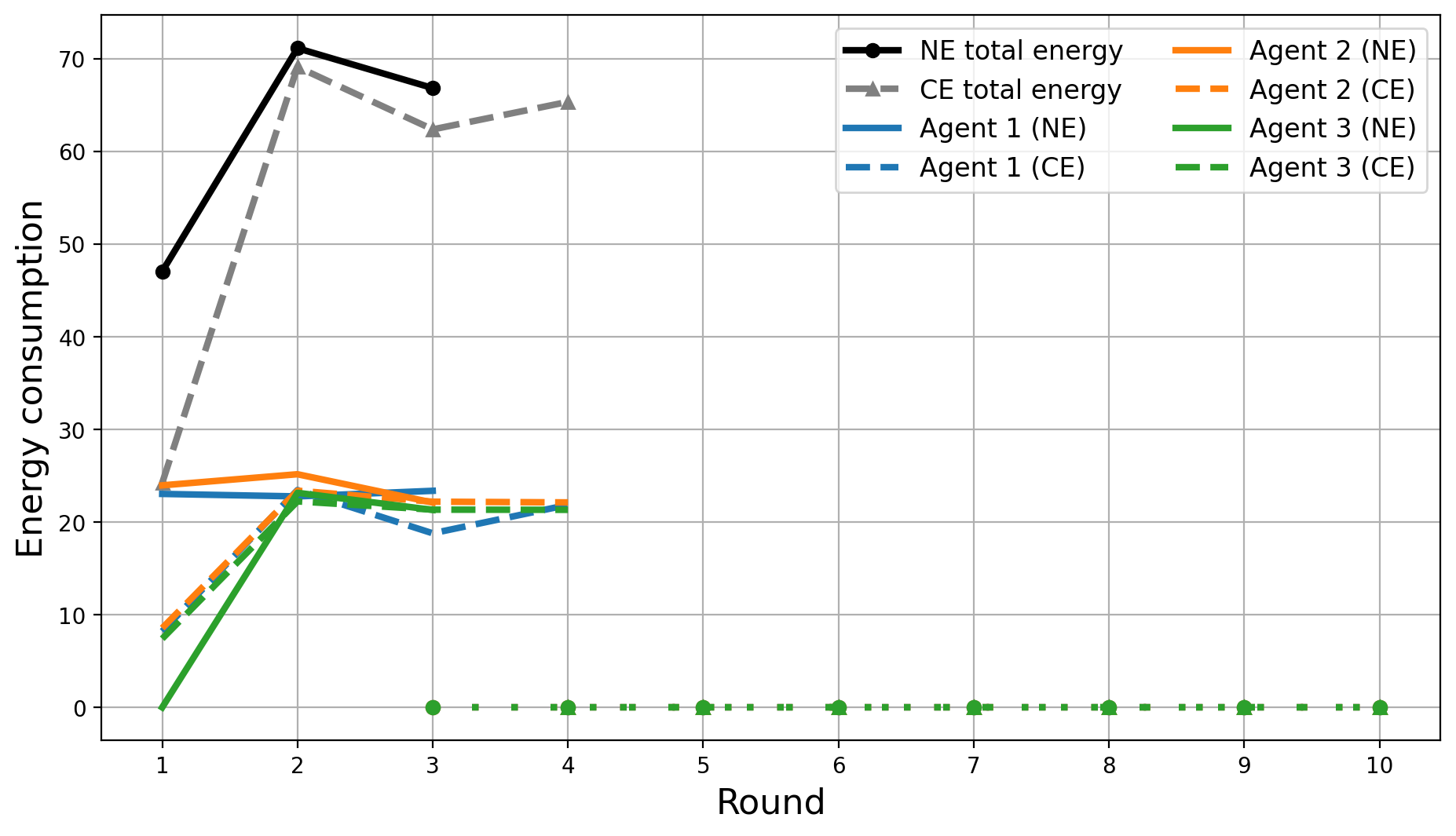}\hfill
    \includegraphics[width=5.5cm, height=2.725cm, keepaspectratio=true]{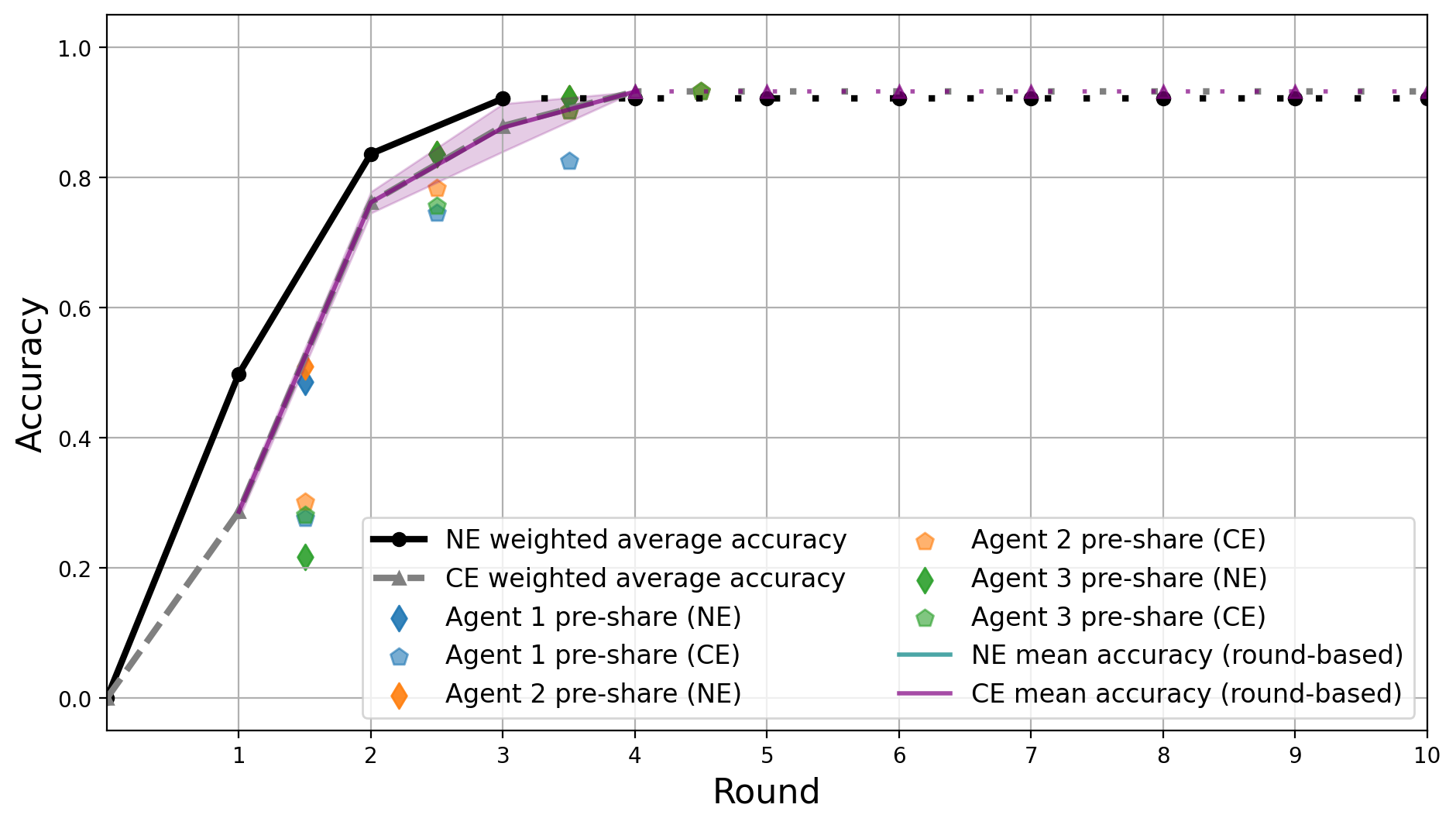}
    \includegraphics[width=5.5cm, height=3.5cm, keepaspectratio=true]{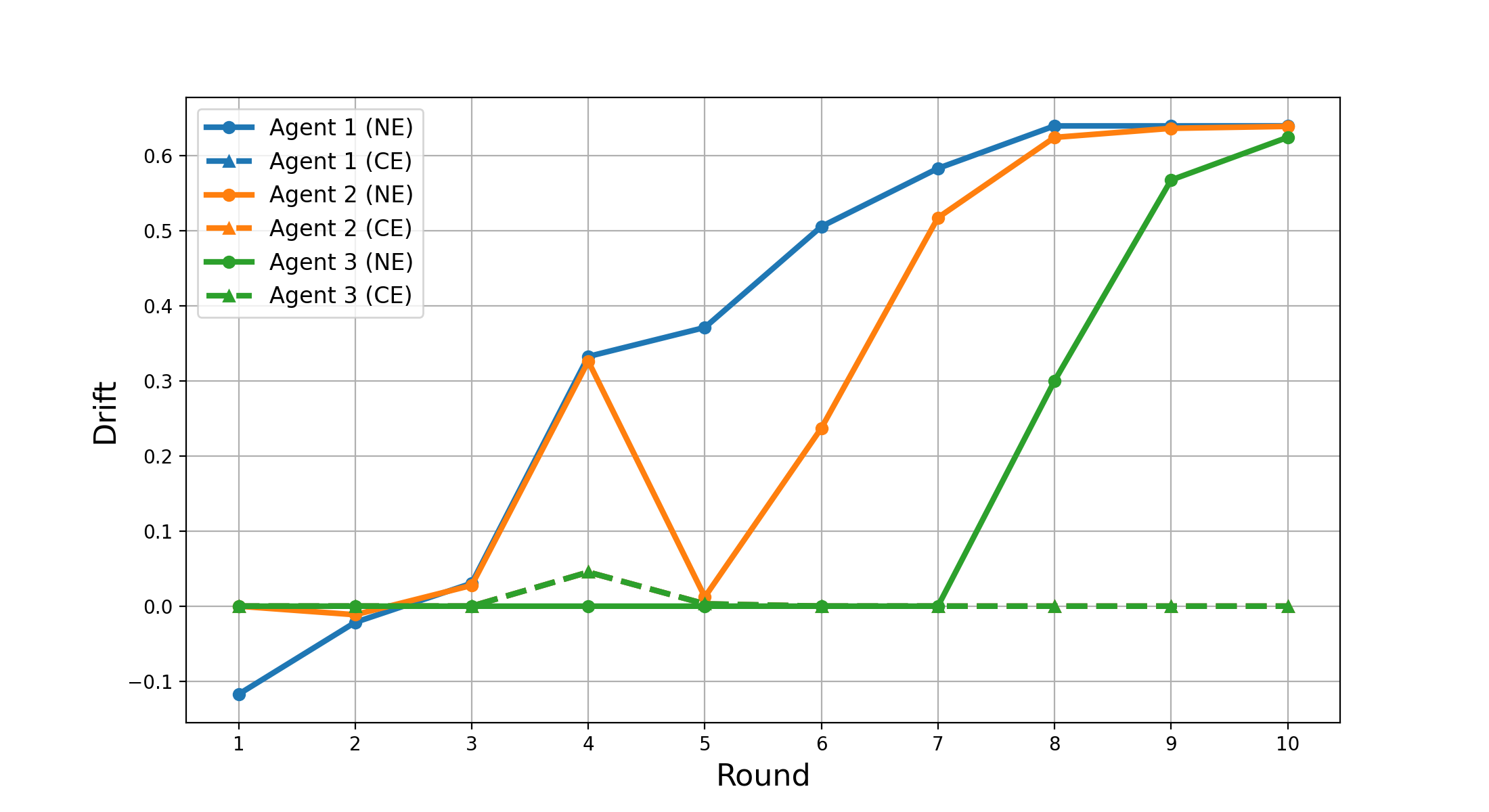}\hfill
    \includegraphics[width=5.5cm, height=2.725cm, keepaspectratio=true]{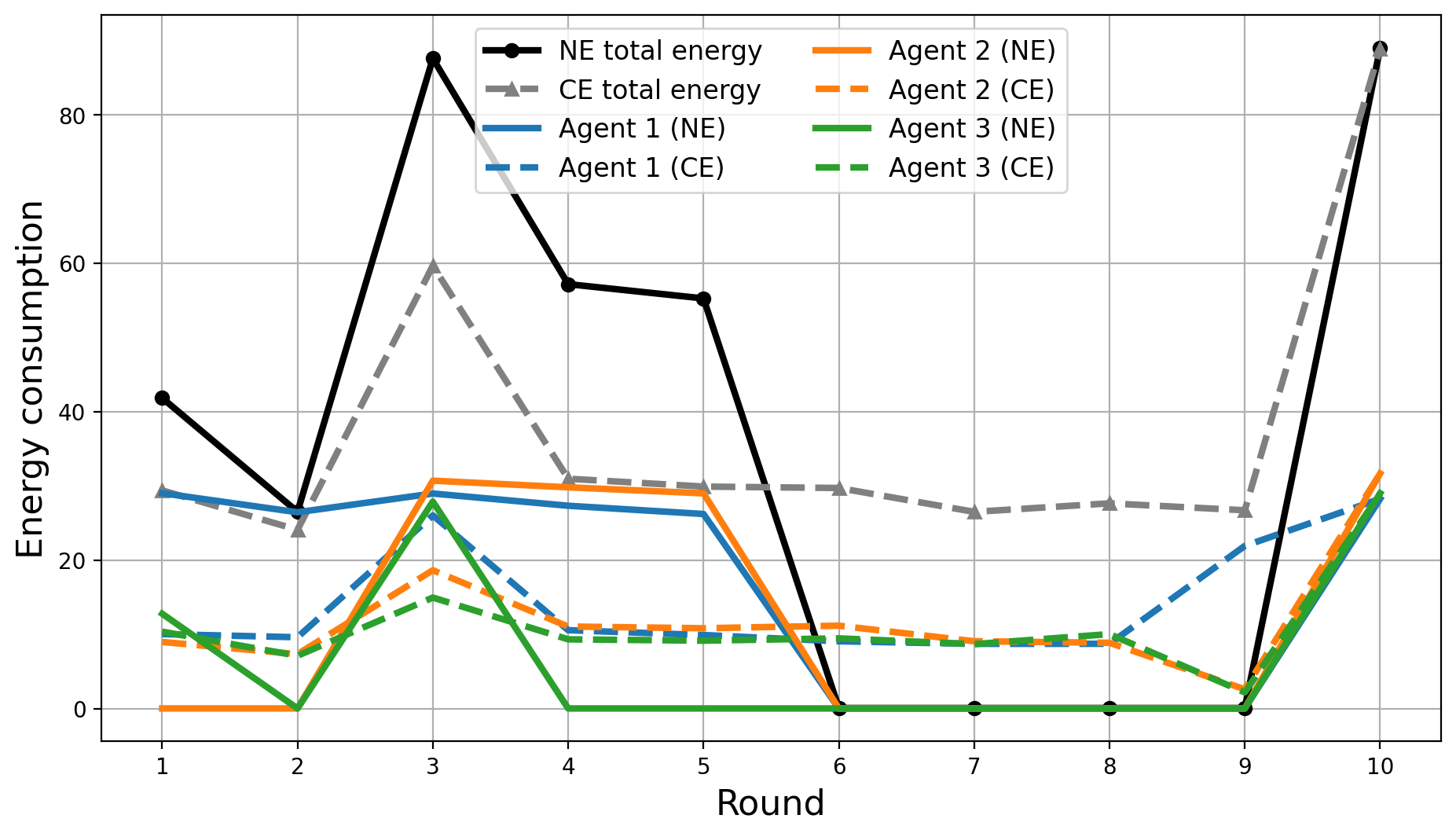}\hfill
    \includegraphics[width=5.5cm, height=2.725cm, keepaspectratio=true]{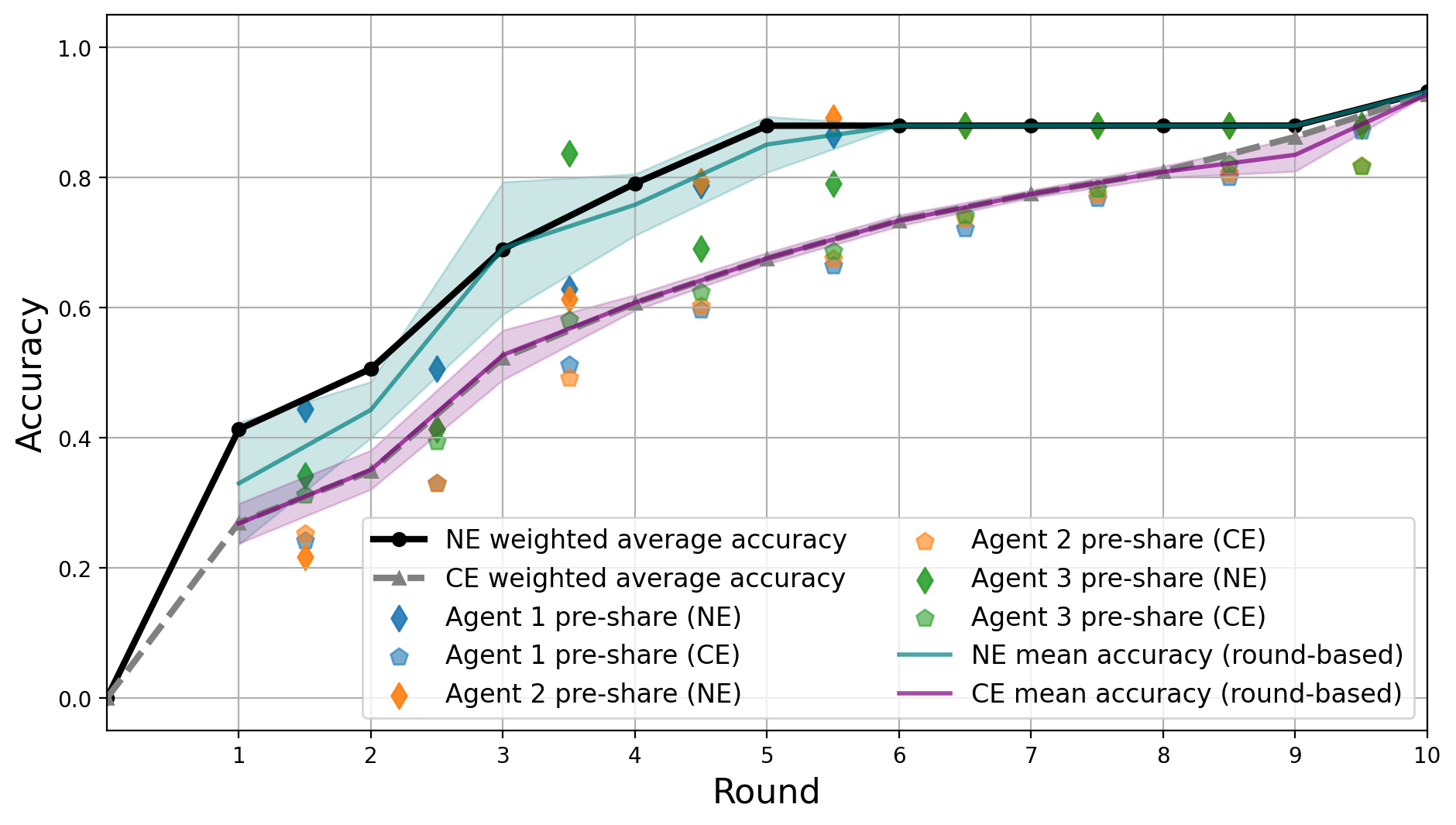}   
    \includegraphics[width=5.5cm, height=3.5cm, keepaspectratio=true]{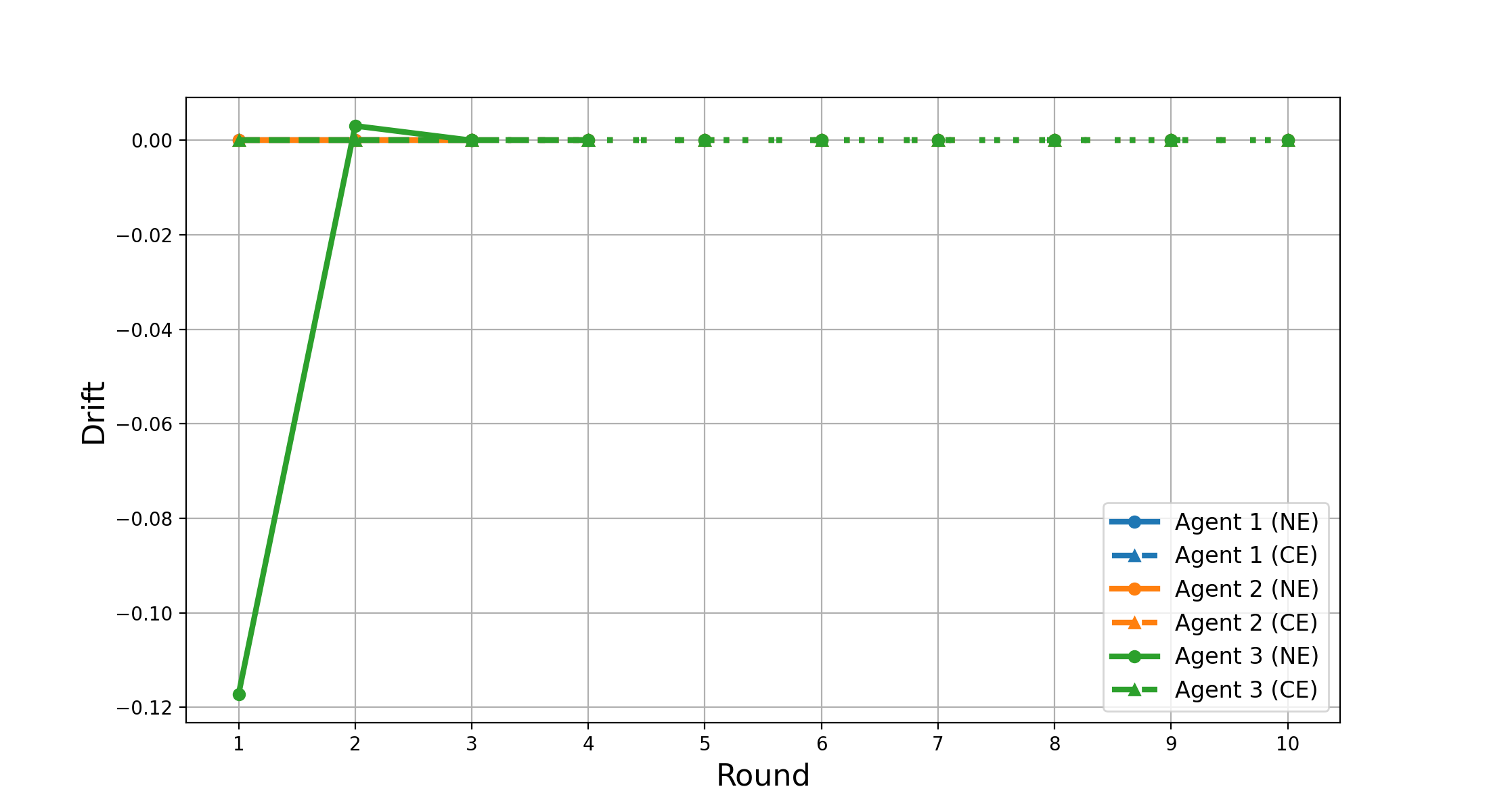}\hfill
    \includegraphics[width=5.5cm, height=2.725cm, keepaspectratio=true]{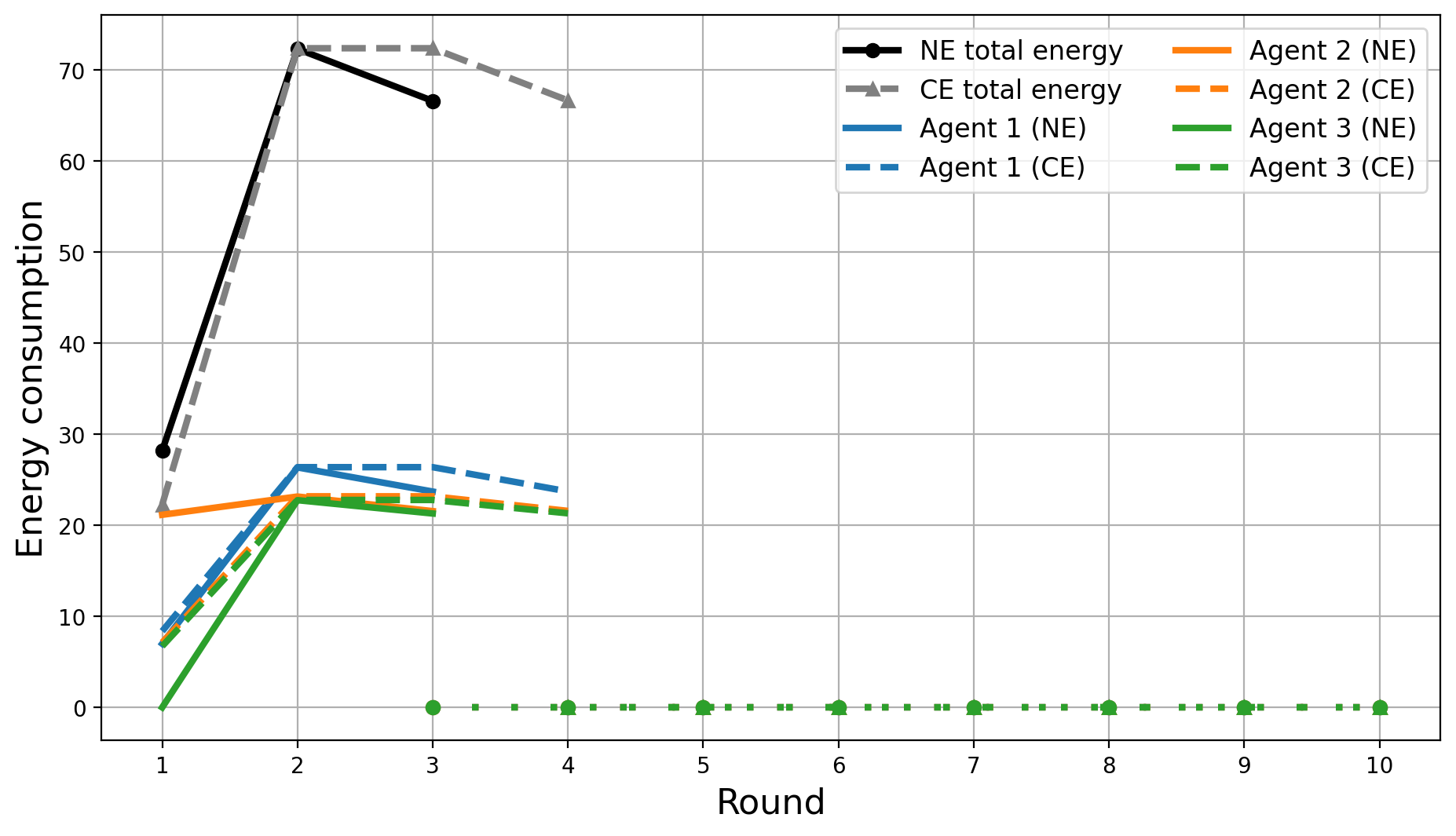}\hfill
    \includegraphics[width=5.5cm, height=2.725cm, keepaspectratio=true]{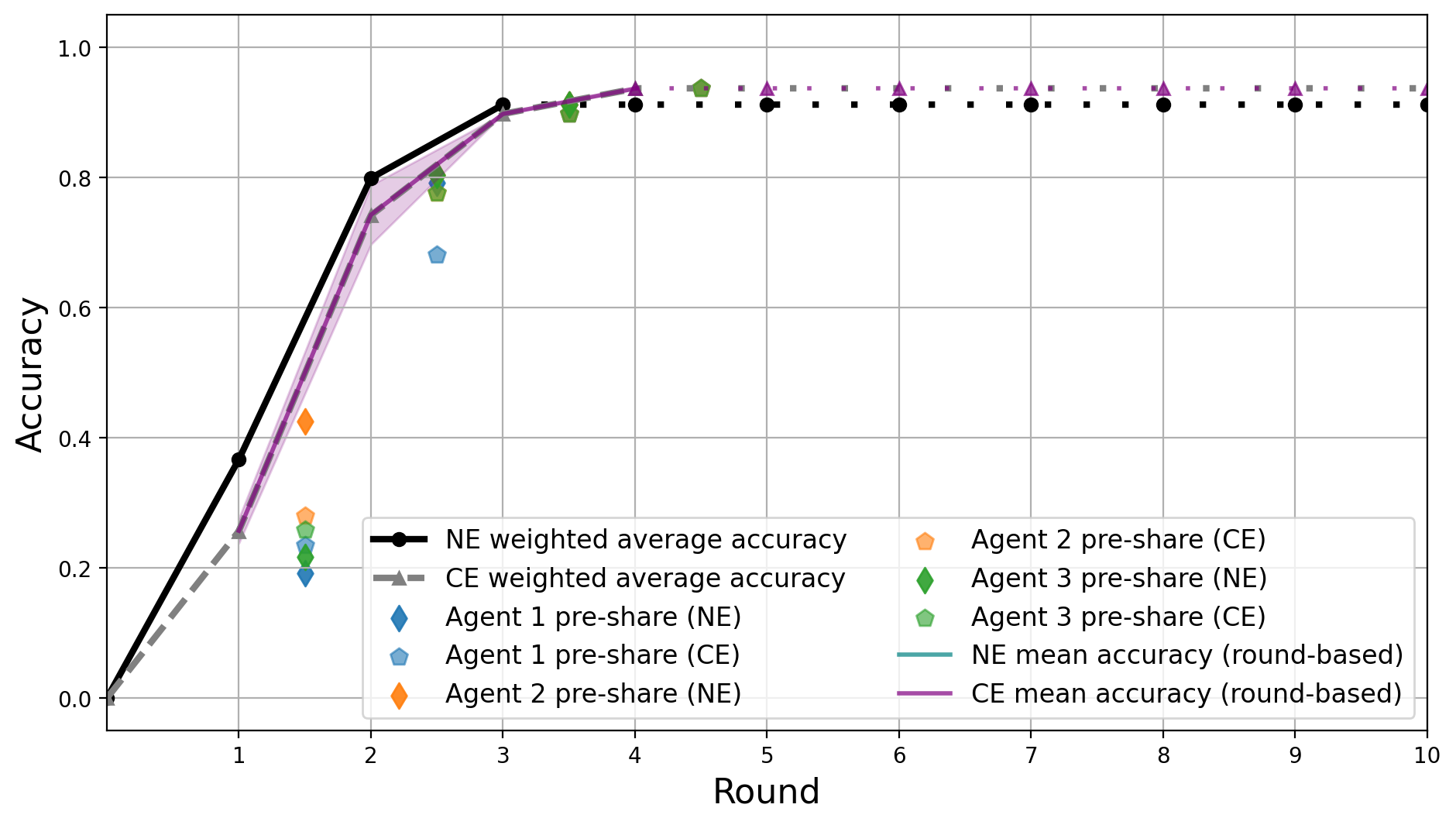} 
    \caption{Blending effect: Drift per user (left column), energy consumption per agent (middle column), training accuracy (right column), for the cases where agent $i$ takes: (i) $\gamma = 1$ (upper row), and (ii) $\gamma = 10$ (middle row), and (iii) $\gamma = 100$ (lower row), with $ h = 0.75$.}\label{fig:drifting 0.75}
\end{figure}

\begin{figure}[h]
    \centering
    \includegraphics[width=5.5cm, height=3.5cm, keepaspectratio=true]{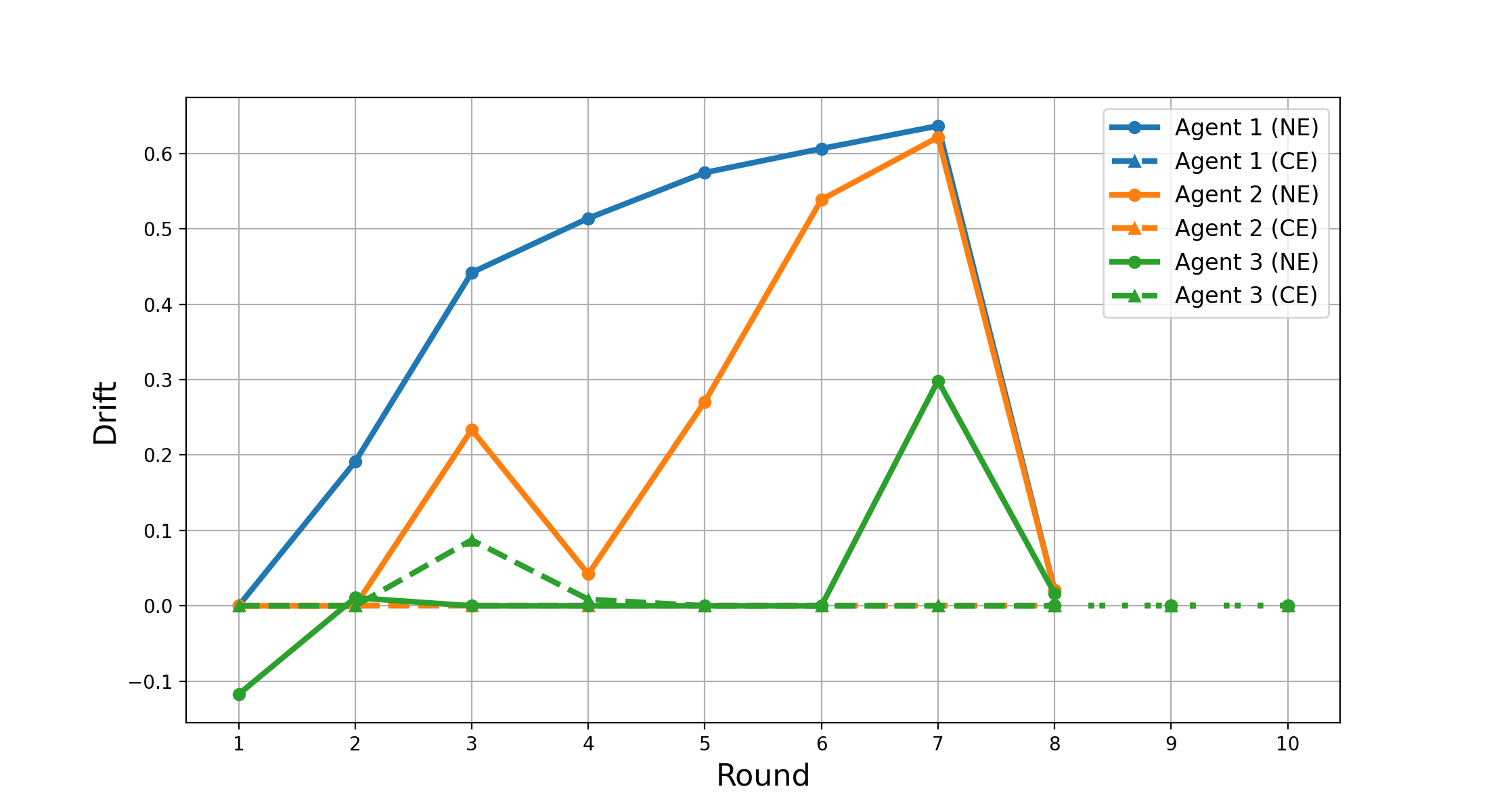}\hfill
    \includegraphics[width=5.5cm, height=2.725cm, keepaspectratio=true]{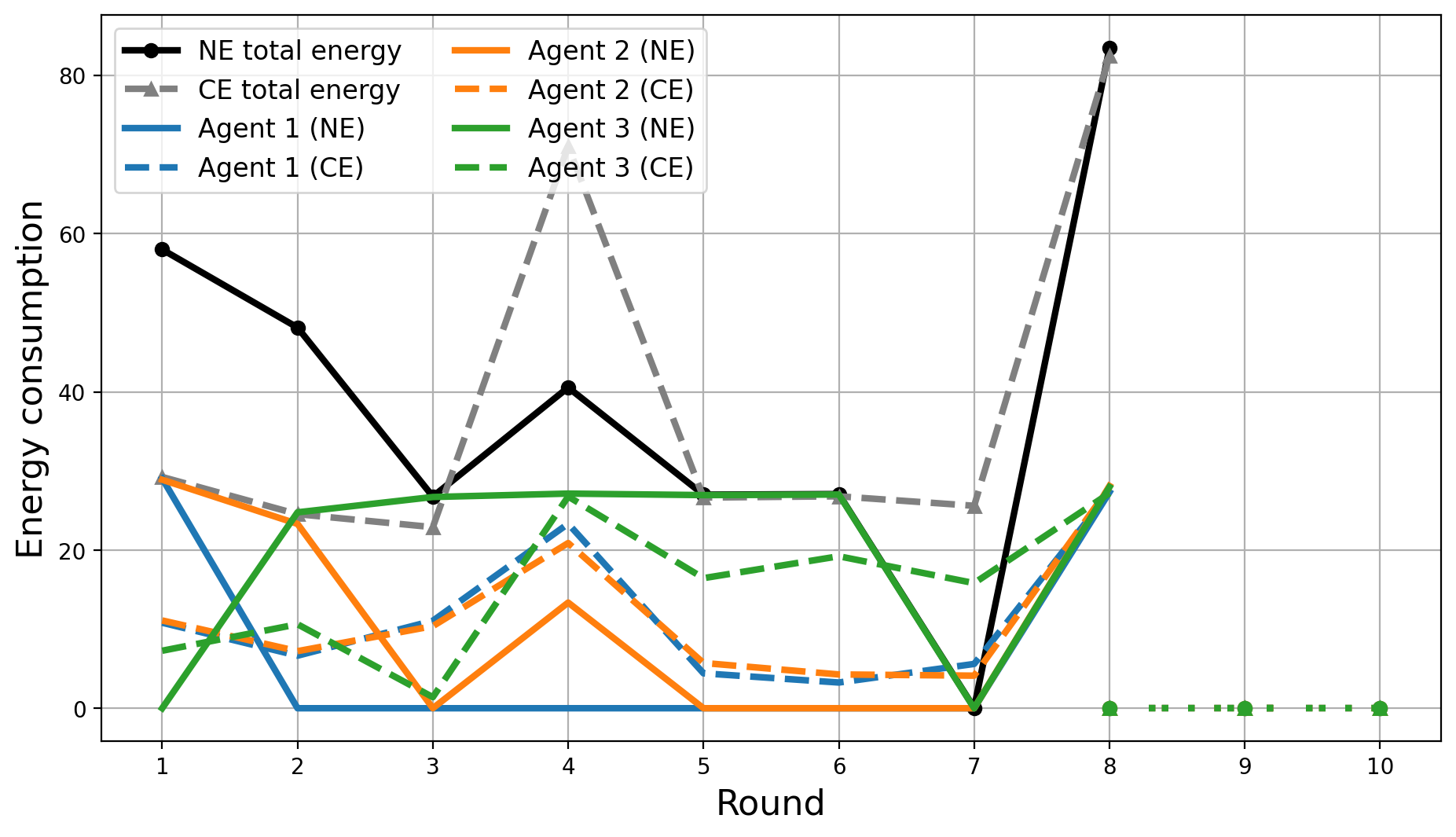}\hfill
    \includegraphics[width=5.5cm, height=2.725cm, keepaspectratio=true]{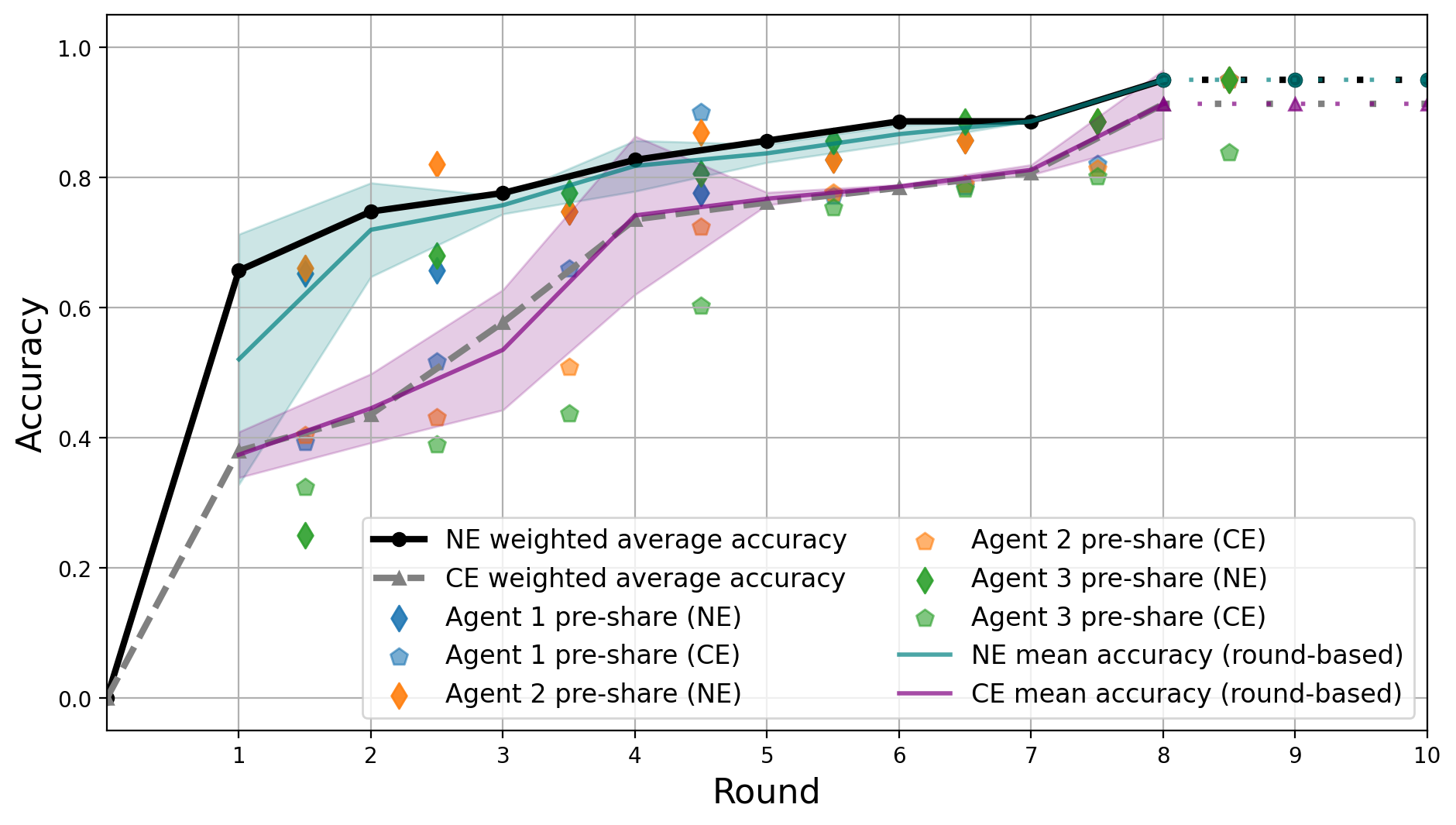}
    \includegraphics[width=5.5cm, height=3.5cm, keepaspectratio=true]{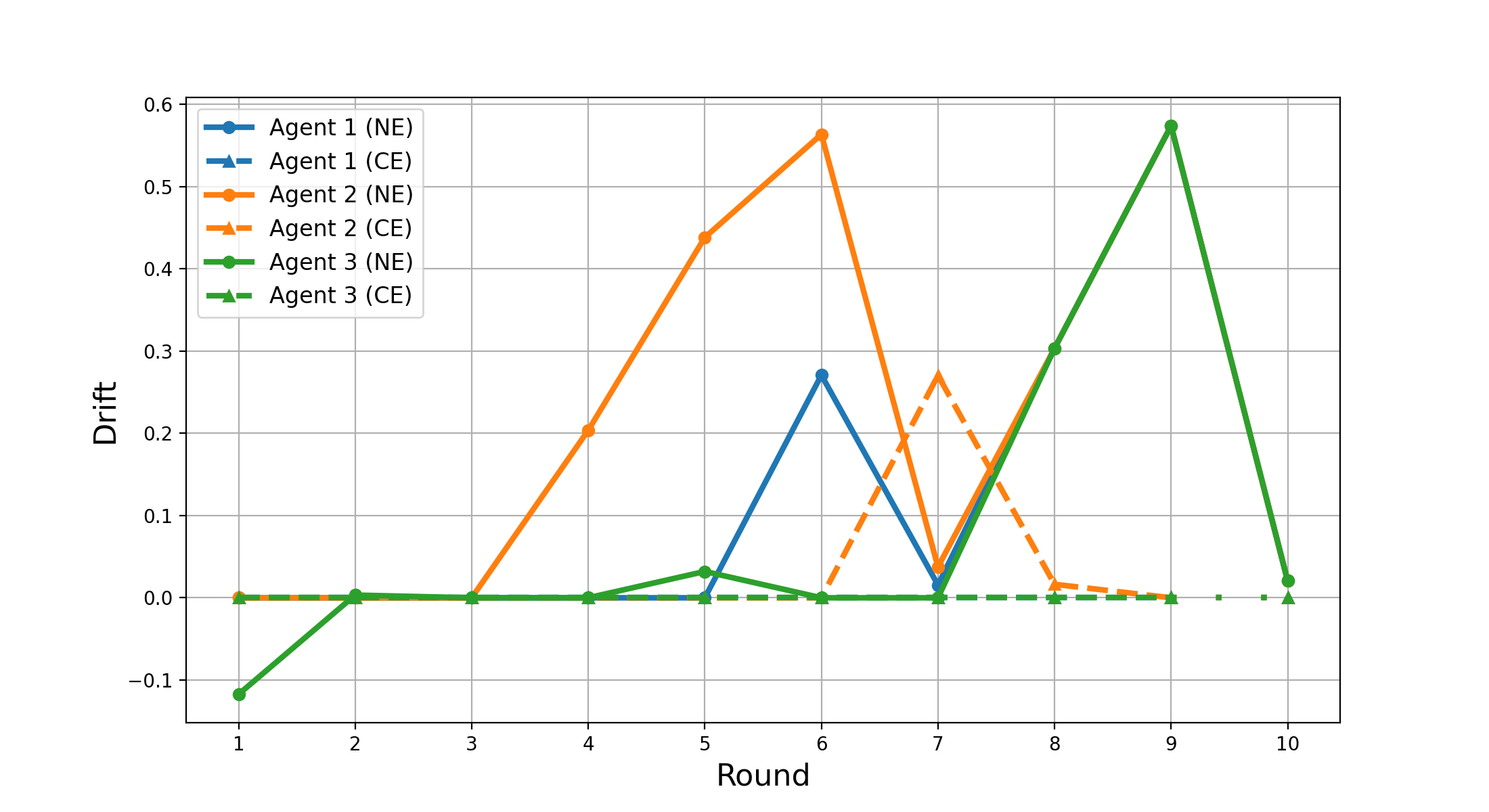}\hfill
    \includegraphics[width=5.5cm, height=2.725cm, keepaspectratio=true]{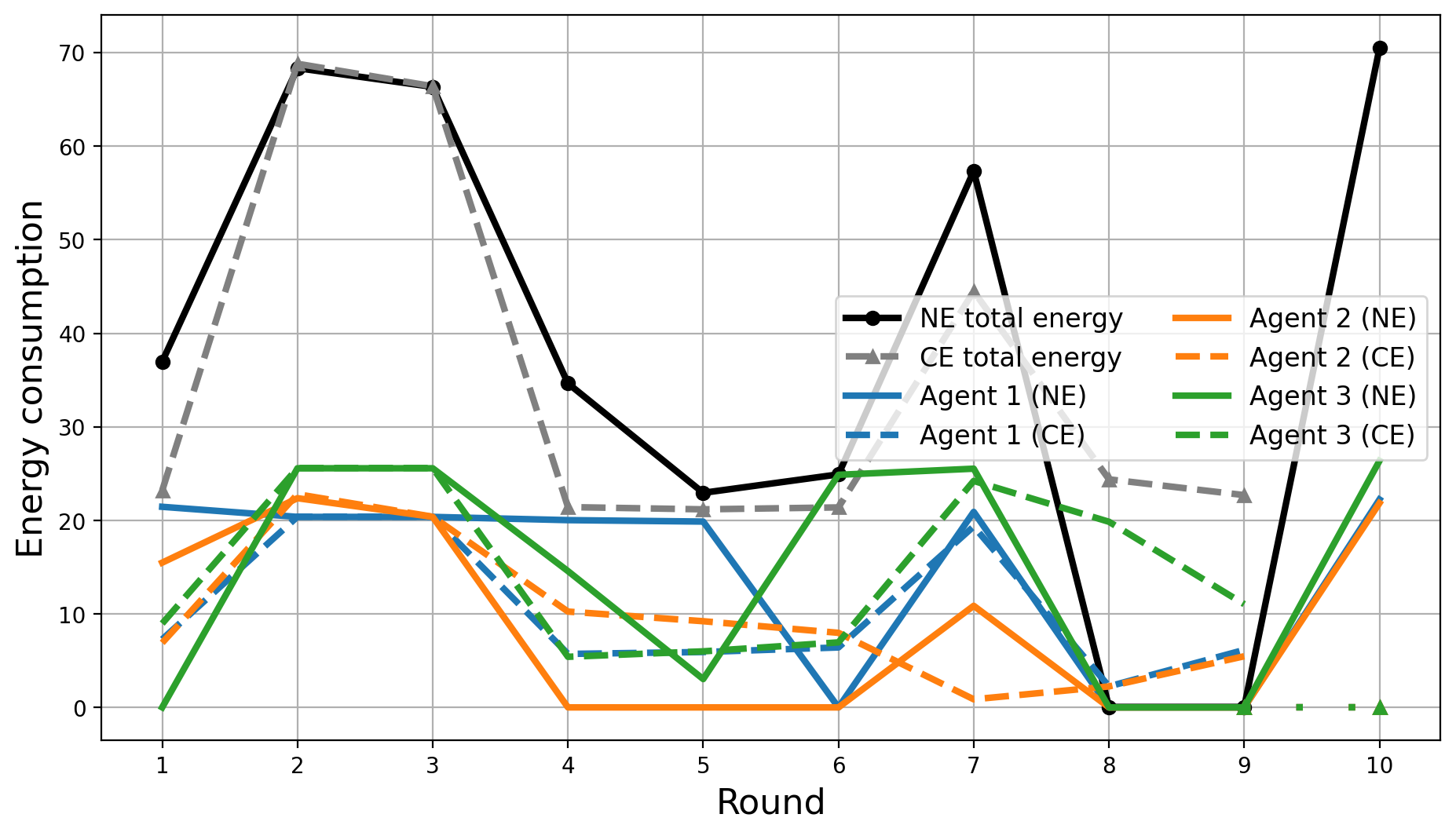}\hfill
    \includegraphics[width=5.5cm, height=2.725cm, keepaspectratio=true]{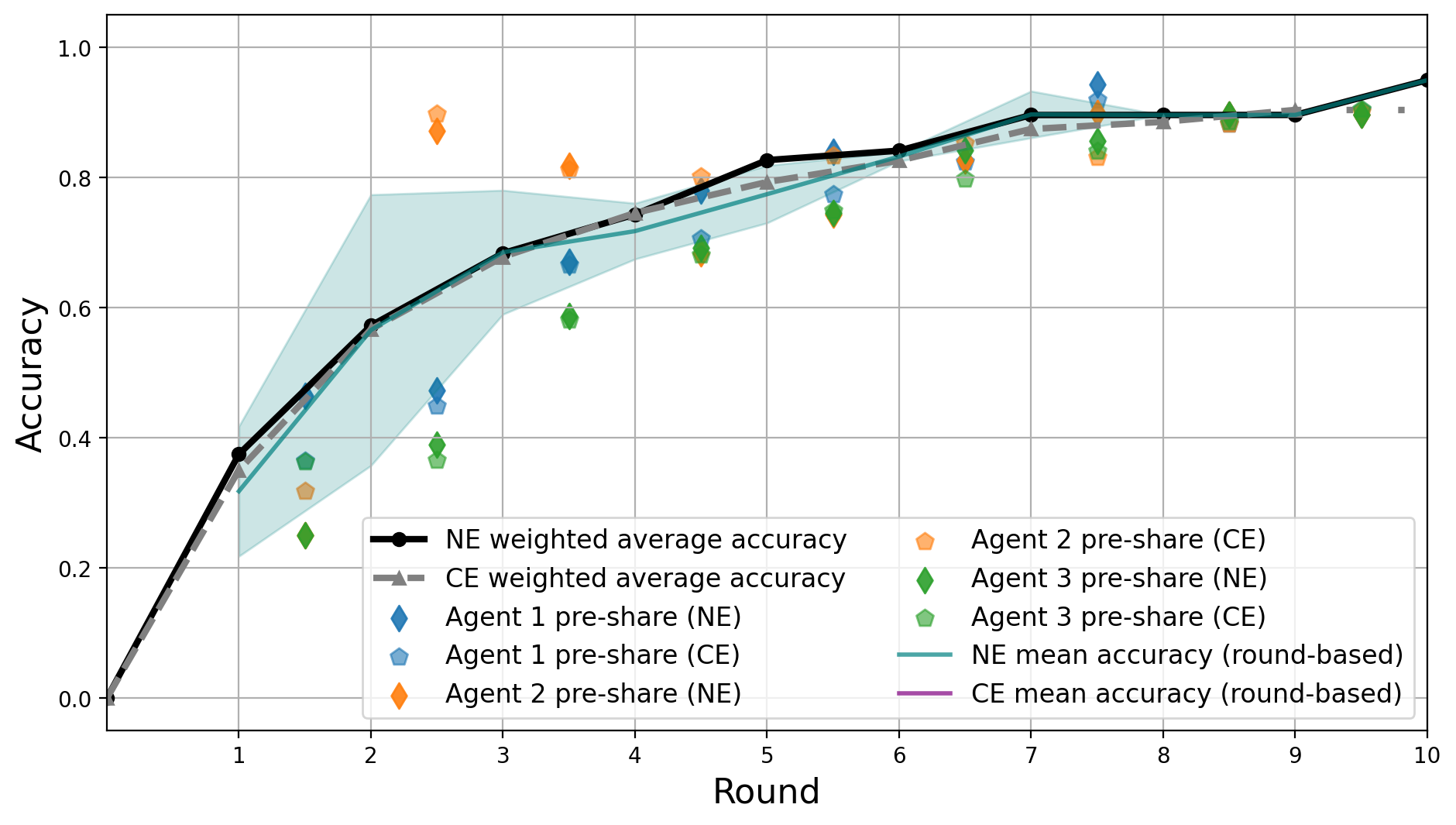}
    \includegraphics[width=5.5cm, height=3.5cm, keepaspectratio=true]{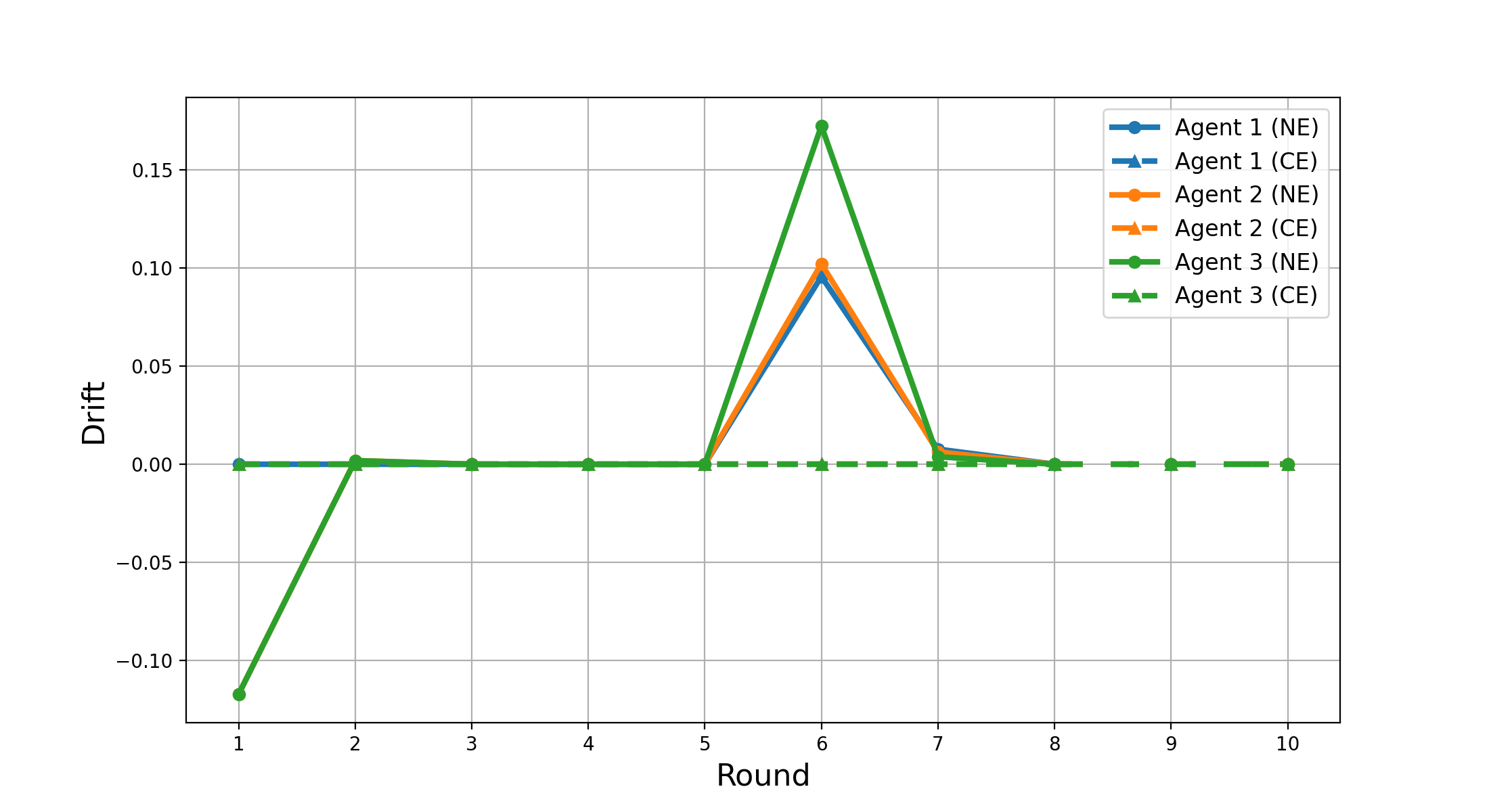}\hfill
    \includegraphics[width=5.5cm, height=2.725cm, keepaspectratio=true]{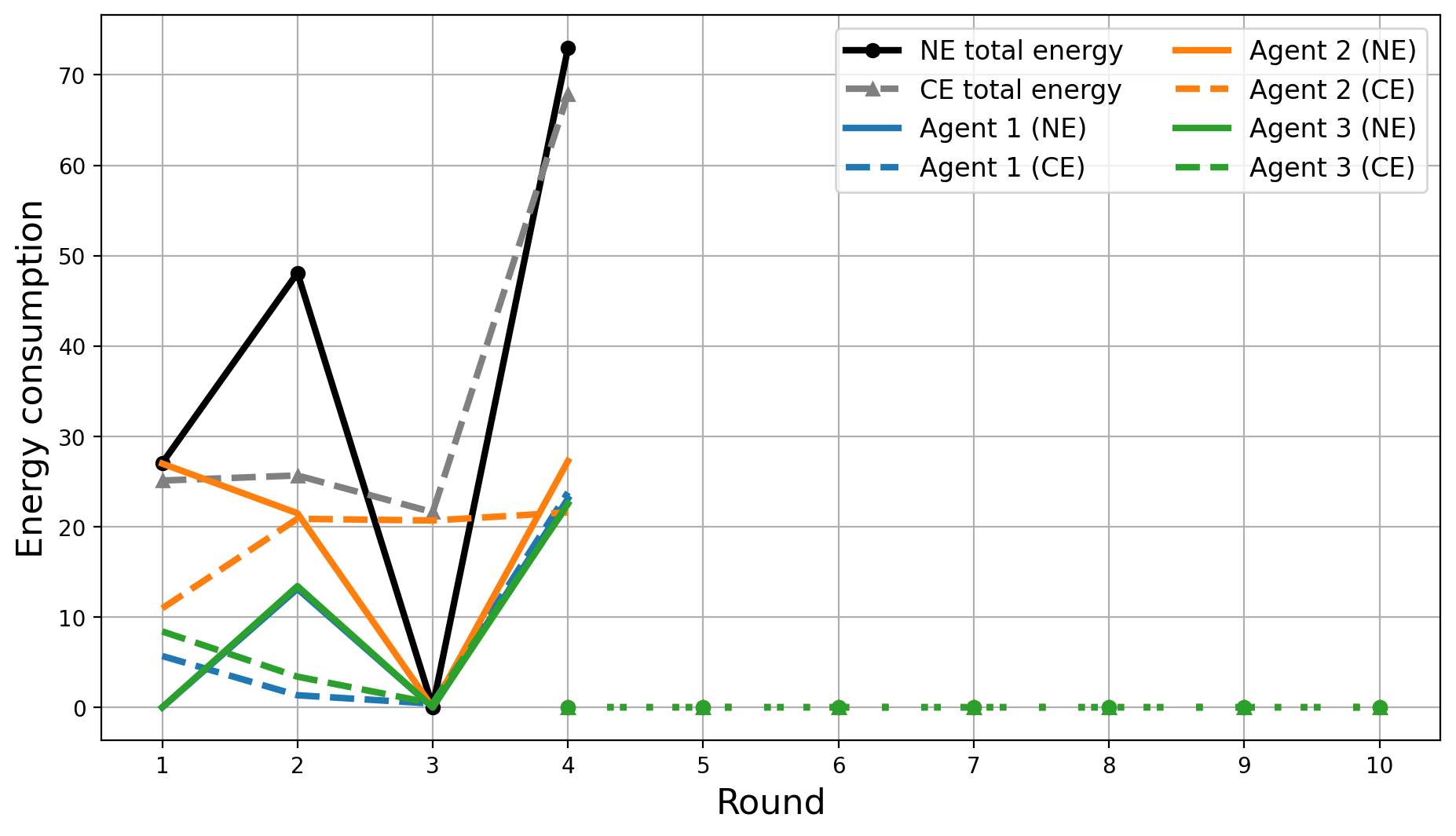}\hfill
    \includegraphics[width=5.5cm, height=2.725cm, keepaspectratio=true]{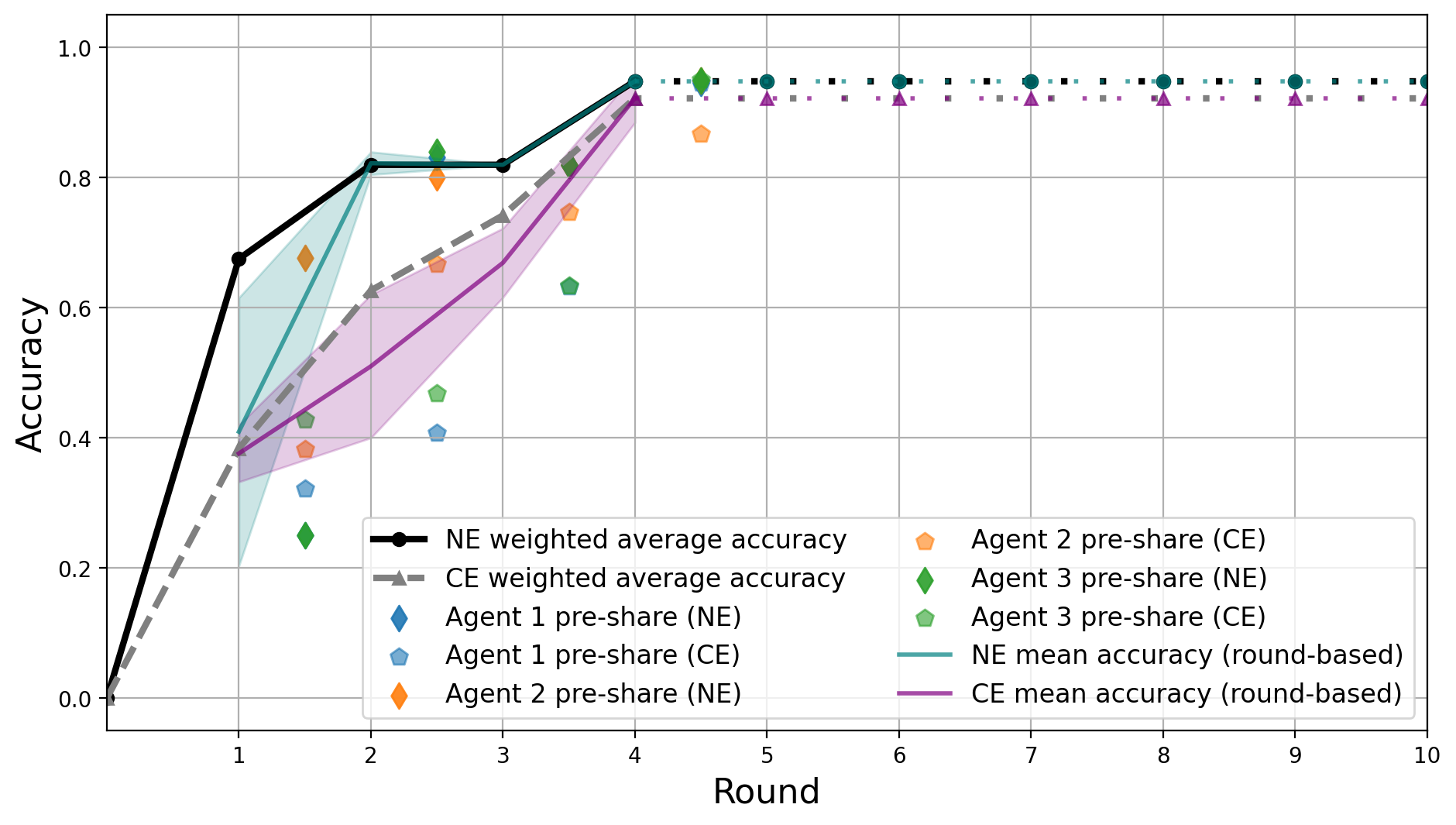}
    \caption{Blending effect. Drift per user (left column), energy consumption per agent (middle column), training accuracy (right column), for the case where agents are energy-aware, with $\gamma = 1$ (upper row), $\gamma = 10$ (middle row), and $\gamma = 100$ (lower row), with $ h = 1$.}\label{fig:drifting 1}
\end{figure}

\begin{figure}[h]
    \centering
    \includegraphics[width=.33\textwidth]{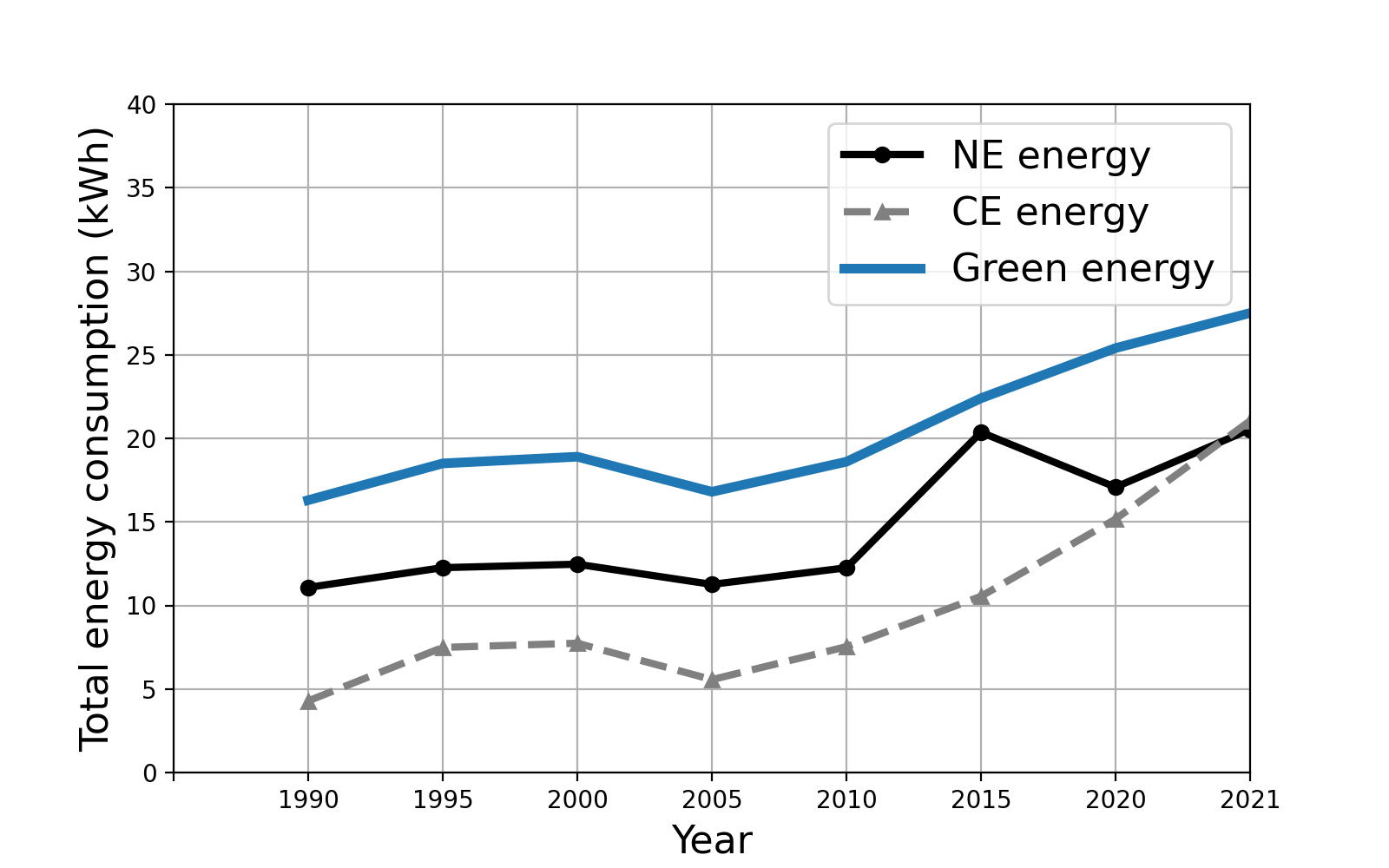}\hfill
    \includegraphics[width=.33\textwidth]{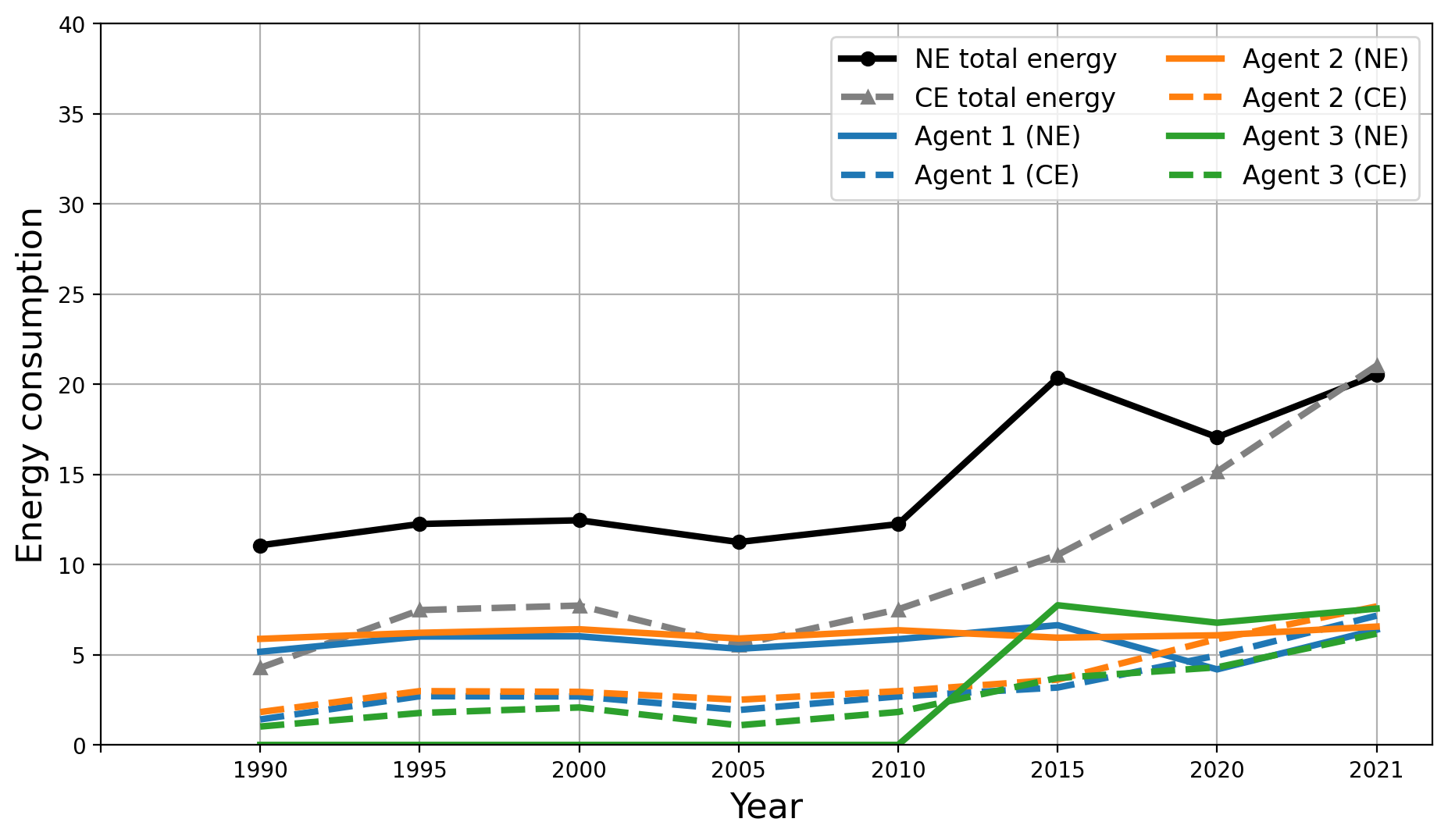}\hfill
    \includegraphics[width=.33\textwidth]{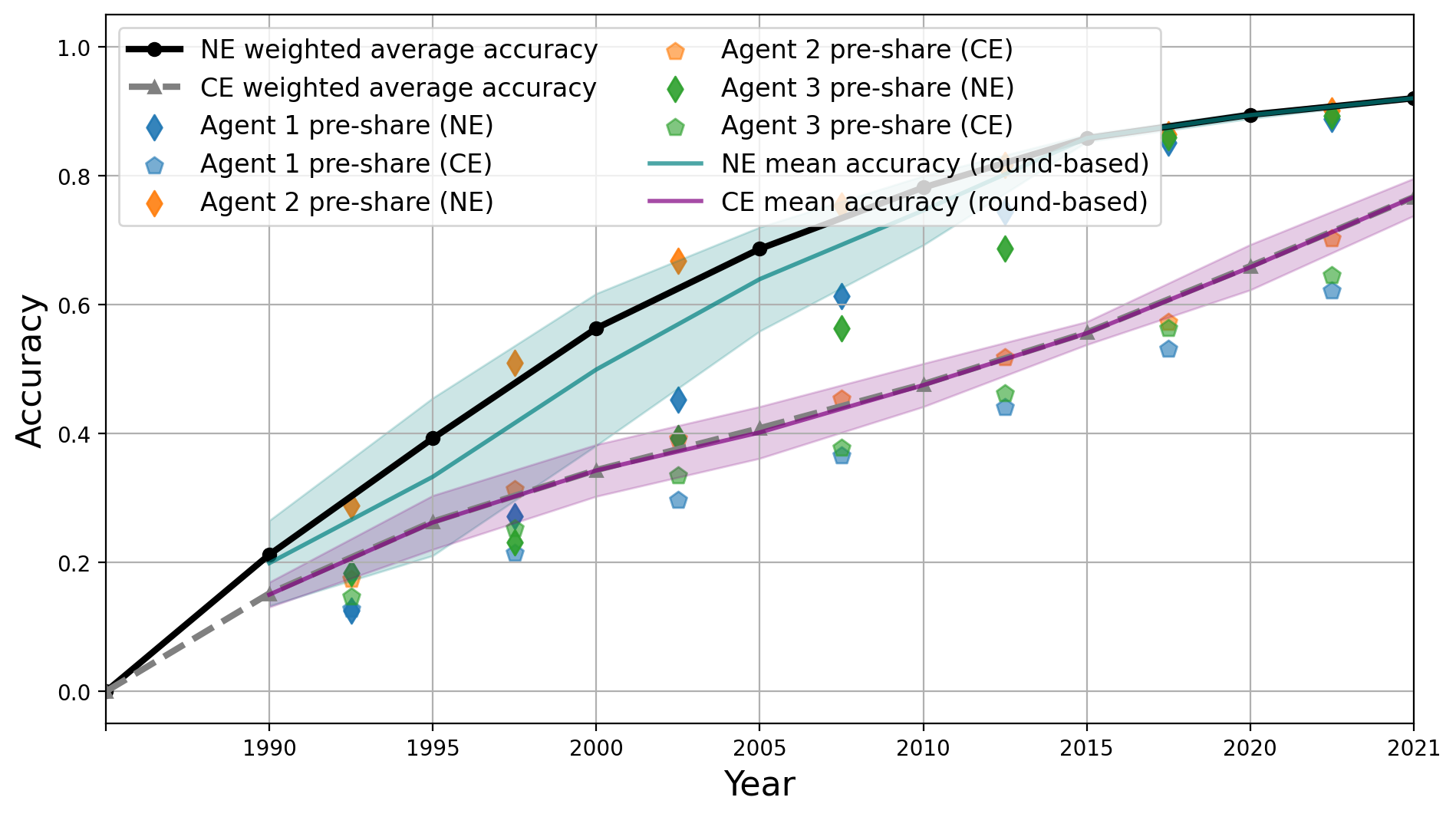}
    \includegraphics[width=.33\textwidth]{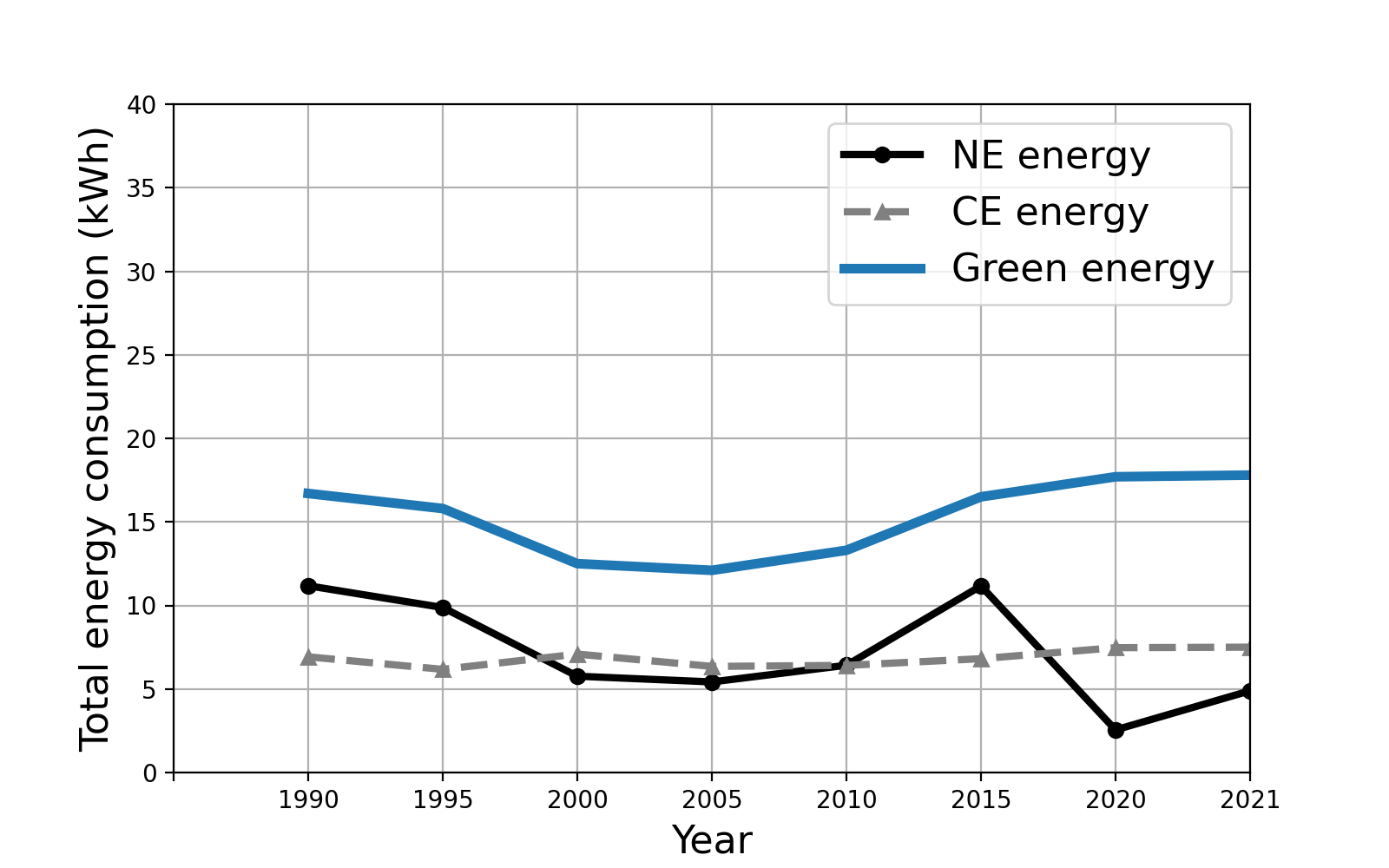}\hfill
    \includegraphics[width=.33\textwidth]{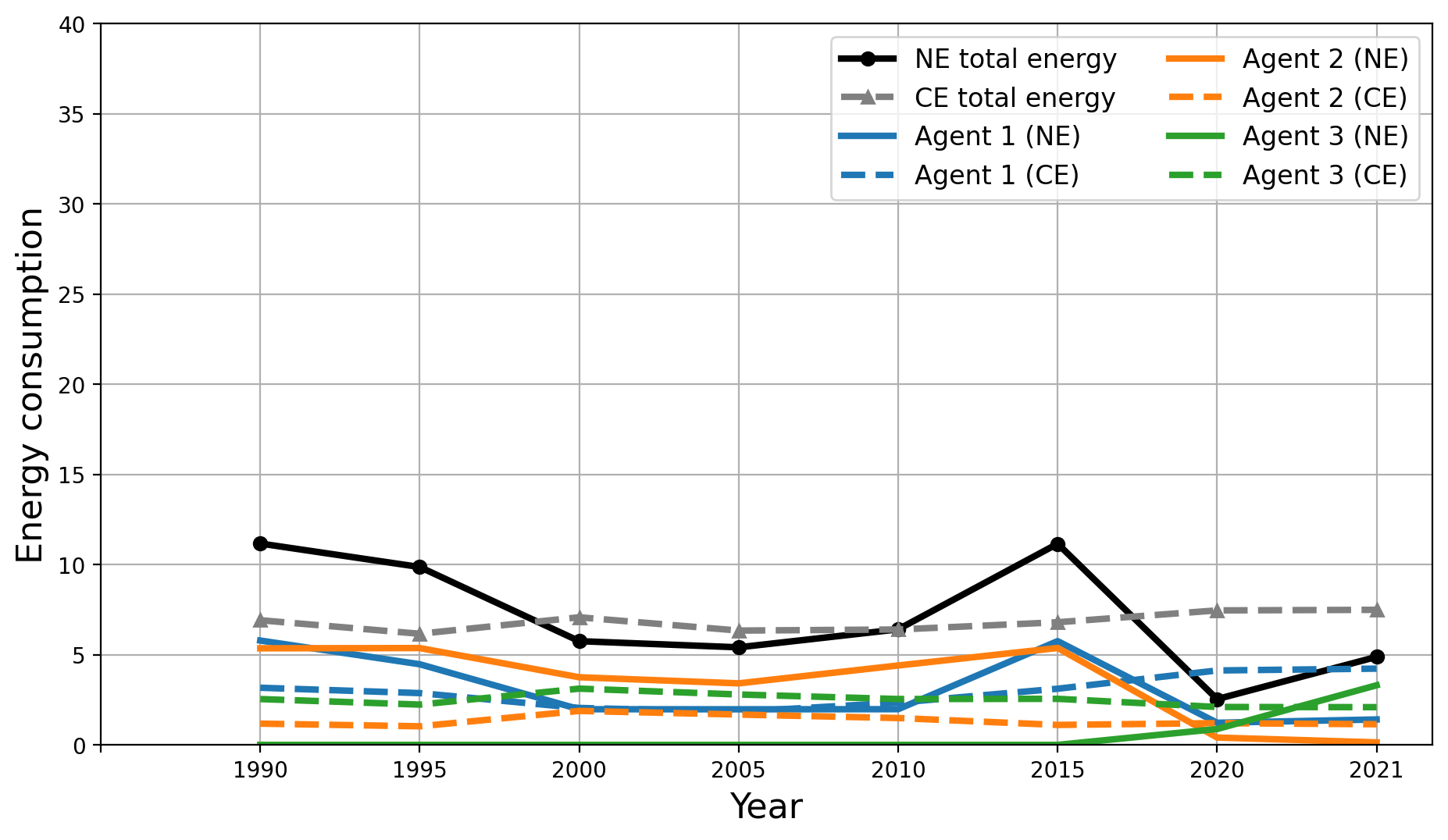}\hfill
    \includegraphics[width=.33\textwidth]{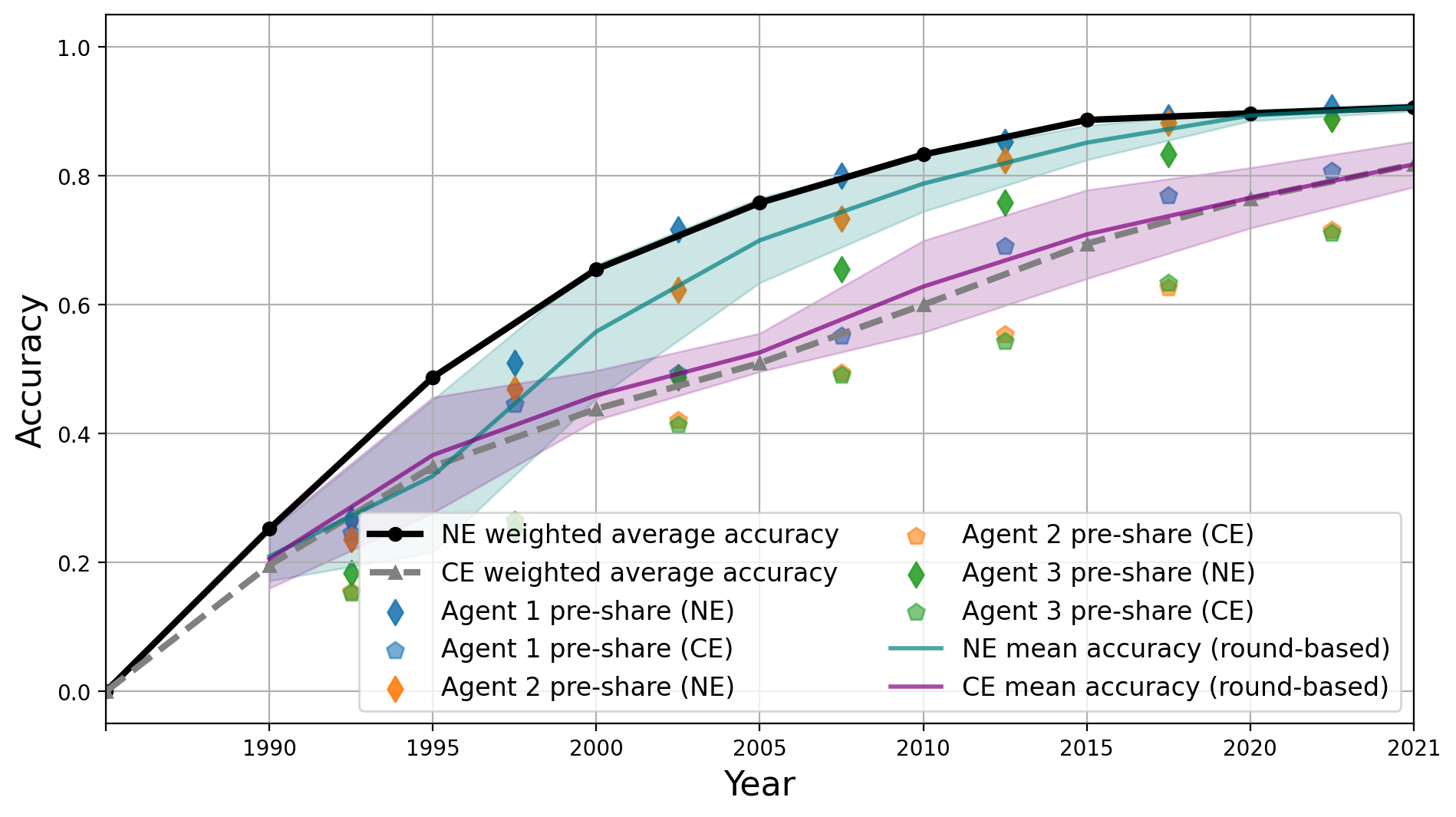}
    \caption{Real-world scenario. Total energy reduction (left column), training accuracy per agent (middle column), mean training accuracy (right column), for the cases Finland (upper row) and Portugal (lower row).} 
\label{fig:real-data}
\end{figure}

In Figures \ref{fig:energy-agnostic}-\ref{fig:real-data}, the y-axis in the left panels is the total energy consumption ($kWh$). The blue dashed curve denotes the total availability of green energy in each round of the procedure. The middle panels in Figures \ref{fig:energy-agnostic}-\ref{fig:real-data} demonstrate the energy consumed by each agent, with a solid curve denoting the energy consumption achieved via the heuristic approach, and the dashed curves denote the energy consumption achieved via the correlation device. The rightmost panels in Figures \ref{fig:energy-agnostic}-\ref{fig:real-data} demonstrate the training accuracy that each agent achieves and the system as a whole. The diamond-shaped points indicate the accuracy that each agent achieved using the heuristic approach, after concluding the local training in each round and before the AI service provider communicates the updated global model for the next round. The black curve denotes the weighted mean accuracy for the heuristic approach. The teal region illustrates the case where the global model's accuracy follows the best(worst) accuracy achieved by agents in each round. Similarly, we use the pentagon-shaped points for the accuracy that each agent achieved using the correlation device. The gray dashed curve denotes the weighted mean accuracy for the correlation device. The magenta region illustrates the case where the global model's accuracy follows the best(worst) accuracy achieved by agents in each round. Further, in all plots, the heuristic phase is denoted as NE and the correlation mechanism as CE. Finally, in case a setting reaches the desired global accuracy threshold before the completion of $T$ rounds, then we depict its performance, e.g., energy consumption, accuracy, etc, with loosely dashed horizontal lines.

\paragraph*{Energy-aware mechanism vs energy-agnostic training.} Figures \ref{fig:energy-agnostic} and \ref{fig:energy-aware} compare two different settings: in Figure \ref{fig:energy-agnostic} agents are energy-agnostic, while in Figure \ref{fig:energy-aware} agents are energy-aware ($\gamma \in \{1, 10, 100\}$), and we set $\delta_- = \delta_+ = 10$ and $h = 0.5$.

In Figure \ref{fig:energy-agnostic}, agents are energy-agnostic, meaning that $g_i(s^t_i, t) = \varepsilon_i(\psi_i(s^t_i), t)$ in eq. \eqref{eq:split energy} and $u_i(\mathbf{s}, t) = \mathbf{1}\{s^t_i > 0\} \cdot P_i(\mathbf{s}^t, t)$ in eq. \eqref{eq:utility} for every $i$, so the parameter $\gamma$ does not affect their decisions. In this setting, the training process is driven exclusively by accuracy considerations. Agents are incentivized only to maximize their predictive performance and therefore consistently select the most informative data samples available. As this behavior remains unchanged over the training rounds, the system rapidly reaches a steady energy consumption level, while the accuracy continues to improve due to the progressive refinement of the global model. The model reaches the desired global accuracy threshold of $\mathscr{a} = 0.9$ in round $5$, and the training process is terminated for both the heuristic and the correlated device approaches.

Moreover, the availability of green energy does not influence the agents' decisions or the resulting system dynamics. Consequently, the system converges to a stable regime in which total energy consumption remains constant across rounds, individual accuracies reach high values, and the mean accuracy exhibits negligible variation, right plot in Figure \ref{fig:energy-agnostic}. This behavior indicates that, without energy-aware incentives, and the agents do not explicitly trade off energy consumption against model performance. These observations apply to both the heuristic approach and the correlation device approach.

In contrast, Figure \ref{fig:energy-aware} illustrates the case where agents are energy-aware. In this setting, total energy consumption exhibits noticeable fluctuations across rounds (see the left column of Figure \ref{fig:energy-aware}), reflecting the continuous adaptation of agents' strategies to balance energy expenditure and learning benefits. Unlike the energy-agnostic case, agents respond to the energy-related incentives and dynamically adjust their decisions rather than following a fixed strategy (see the middle column of Figure \ref{fig:energy-aware}). This adaptive behavior also results in visible variations in the individual strategies.

The impact of this strategic adaptation is further reflected in the training accuracies (see the right column of Figure \ref{fig:energy-aware}). Individual agents achieve different accuracy levels across rounds (represented by the diamond-shaped and pentagon-shaped points); however, all agents contribute to a steady improvement of the global model accuracy (represented by the black solid curve and gray dashed curve). The differences in local accuracy levels, together with the variations in individual energy consumption, highlight the inherent trade-off between energy efficiency and model performance. Agents allocate computational resources when the expected improvement in accuracy justifies the associated energy cost. Comparing the energy efficiency of the two solutions, we observe that both approaches the training process remains within the available renewable-energy budget, in the energy-aware setting, while the heuristic approach better captures the evolution of $\mathcal{G}^t$ (see the left column of Figure \ref{fig:energy-aware}).  

Moving to the learning dynamics, the independent updates produced by Algorithm \ref{alg:fictitious play algorithm} yield higher weighted accuracy but also introduce greater variability across agents (see the right column of Figure \ref{fig:energy-aware}). In contrast, the correlation mechanism reduces variance by steering agents toward more homogeneous, moderate behaviors, at the cost of slightly lower average performance. This difference is further highlighted in the per-agent accuracy points (see right column of Figure \ref{fig:energy-aware}), where fictitious play allows more diverse outcomes, some agents achieve higher accuracy while others limit participation, whereas correlation induces more synchronized behavior among agents with similar characteristics. As a result, fictitious play promotes efficiency through heterogeneity, while correlation emphasizes stability and coordination.

Finally, increasing $\gamma$, which makes the AI service provider’s policy stricter, leads the two approaches to behave more similarly in terms of energy consumption (see left column of Figure \ref{fig:energy-aware}). Starting from $\gamma = 1$, the variance of the performance very quickly narrows for the correlated device. Nevertheless, the black curve (heuristic approach) reaches a plateau\footnote{We provide a more detailed discussion regarding this phenomenon at the end of this Section.}, meaning that small $\gamma$ values do incentivize agents, but this may be insufficient to make them achieve the global accuracy and, instead, remain idle when the global model secures a ``good'' accuracy level. The same trend appears for the heuristic approach, but with a significantly lower pace. 

As $\gamma$ is increased non-participation is discouraged more aggressively, reducing performance variance within and between the heuristic phase and the correlation mechanism, see also Figures \ref{fig:energy-aware-20} and \ref{fig:het-agents}. Moreover, higher $\gamma$ values discourage prolonged non-participation and can therefore help the agents overcome the accuracy plateaus observed for smaller penalty values, as in case where $\gamma = 1$. Interestingly, in all cases displayed in Figure \ref{fig:energy-aware}, the heuristic approach achieves faster the global accuracy threshold $\mathscr{a} = 0.9$, achieving the faster convergence in round $5$ for $\gamma = 10$.

\paragraph*{Energy demands.} Now we alter the experimental setting by examining how the distribution of users' energy demands affect the training dynamics and the resulting energy consumption. Specifically, we consider two scenarios where agents' energy demands are sampled from $U[\mathcal{G}^t - 10, \mathcal{G}^t + 10]^4$, Figure \ref{fig:energy-aware}, and from $U[\mathcal{G}^t - 20, \mathcal{G}^t + 20]^4$, Figure \ref{fig:energy-aware-20}, respectively.

From the system-level perspective, we observe that when narrower energy demand distributions are applied and the $\gamma$ values are relatively small, the heuristic approach produces outcomes that are closer to the actual green energy per active training round. This is mainly because agents experience similar energy conditions and, therefore, their participation decisions remain relatively aligned with the actual energy availability. In contrast, the correlated device approach tends to synchronize agents' behaviors through the provider's signals, resulting in more homogeneous energy consumption trajectories across users. Although this coordination improves consistency, it may reduce the ability of individual agents to adapt to their own local energy conditions.

From the model-level perspective, the reduced participation frequency of some agents directly influences the evolution of local accuracy through the model drift mechanism. Agents that remain inactive for several consecutive rounds experience a gradual degradation of their local model performance, as their accuracy decreases from the last achieved value toward the baseline accuracy level, see the evaluation assumptions for more details. When these agents resume participation, the parameter $h$ determines the extent to which the accumulated drift is compensated. Since $h < 1$, the recovery is only partial, and agents require multiple participation rounds to fully restore their local accuracy. Therefore, under heterogeneous energy conditions, independent decision-making may lead to more irregular participation patterns, resulting in fluctuations in the local and aggregated accuracy trajectories.

However, as show in Figure \ref{fig:energy-aware-20} increasing the energy-demand variability does not necessarily degrade the learning performance. In fact, for wider energy-demand distributions, the two mechanisms exhibit more similar aggregate consumption profiles. This suggests that the larger variability in individual energy demands changes the participation decisions in a way that reduces the differences between the two approaches. This behavior is consistent with agents adjusting their participation more frequently in response to the mismatch between their energy demands and the available renewable energy.

Nevertheless, the heuristic approach generally achieves higher global accuracy than the correlated device approach across the two energy-demand distributions. This suggests that allowing agents to independently adapt their participation decisions enables a more effective selection of training contributions, whereas excessive coordination may constrain agents' ability to exploit favorable individual energy conditions.

In contrast to the case of narrower energy-demand distributions, Figure \ref{fig:energy-aware}, in Figure \ref{fig:energy-aware-20} the global accuracy threshold is never reached. Nevertheless, the heuristic approach maintains higher accuracy than the correlation mechanism throughout the experiment, while the two approaches exhibit more similar energy-consumption patterns than in the narrower-demand setting. Thus, increasing the heterogeneity of energy demands does not simply translate into lower energy consumption or higher accuracy; rather, it changes the participation dynamics and the resulting trade-off between energy consumption and learning performance. Within the considered horizon, the heuristic approach provides a more favorable accuracy outcome, although neither mechanism achieves the prescribed accuracy target.

For $\gamma = 1$, in both energy-demand distributions, the heuristic approach reaches quickly a relatively good global accuracy. However, for $\delta_+ = \delta_- = 20$, the wider range of energy demands leads to less favorable participation patterns under the same energy incentive, and the global accuracy reaches a plateau below $\mathscr{a}$. In contrast, when $\delta_+ = \delta_- = 10$, the energy demands remain closer to the available renewable energy, allowing participation to be sustained and the global accuracy to continue improving until the target is reached. 

For $\gamma = 10$ and wider energy-demand distributions, middle row in Figure \ref{fig:energy-aware-20}, the correlated device exhibits a step-like accuracy evolution due to more synchronized participation, whereas the heuristic approach improves more smoothly through independent participation decisions. In contrast to $\gamma = 1$, heuristic approach appears to overcome the saturation and continue improving toward the target accuracy.

For $\gamma = 100$, the gap between the heuristic and correlated-device approaches further narrows, as in the previous case. The heuristic approach exhibits a more step-like accuracy evolution, while the correlated device produces a more gradual, almost linear improvement. The stronger participation incentive also enables agents to exit the saturation regime earlier, allowing agents to make further progress toward $\mathscr{a}$. This behavior reflects the stronger pressure to participate when the expected training benefit justifies the energy cost, reducing prolonged periods of non-participation and making the two approaches behave more similarly.

Overall, the results highlight a trade-off between coordinated participation and individual adaptability. Wider energy-demand distributions reduce differences in energy consumption across agents, but the learning performance depends on how participation decisions interact with model drift and recovery. While the correlated device approach can create more synchronized participation patterns, the heuristic approach benefits from decentralized adaptation, enabling agents to exploit their individual conditions and achieve higher accuracy. These findings emphasize that coordination mechanisms should balance energy alignment with preserving sufficient flexibility for agents to contribute when their participation is most beneficial.

\paragraph*{Heterogeneous population.} We next consider a heterogeneous population of agents. In particular, agent $1$ has a significantly larger dataset bundle, with $S_1 \sim U[\mathcal{G}^t - \delta_-, \mathcal{G}^t + \delta_+]^4$, where $\delta_- = 20$ and $\delta_+ = 100$. From eq. \eqref{eq:energy consumption agent}, agent $1$ can thus be interpreted as a high energy-demand agent. The remaining agents have moderate dataset bundles, i.e., $S_i \sim U[\mathcal{G}^t - 10, \mathcal{G}^t + 10]^4$ for all $i \neq 1$.  Furthermore, all agents experience penalty $\gamma \in \{1, 10, 100\}$.

In Figure \ref{fig:het-agents}, the energy consumption trajectories (left column) remain comparable across the two approaches, particularly for moderate and high penalty values. For $\gamma\in \{10, 100\}$, both mechanisms adapt their decisions to the available renewable energy and exhibit similar aggregate consumption patterns across training rounds. For small penalty values, the correlated device produces smoother energy trajectories compared with the heuristic approach. This occurs because the provider's signal coordinates agents' decisions, reducing uncertainty in individual participation choices and avoiding abrupt changes in energy usage. In contrast, the heuristic approach relies on decentralized decisions, where each agent independently evaluates its own trade-off between energy consumption and participation benefits, leading to larger fluctuations. As $\gamma$ increases, the two approaches become more aligned since the higher cost of non-participation dominates individual preferences and encourages agents to adjust their behavior more conservatively.

Despite the similarity in aggregate energy consumption, the two approaches lead to different participation patterns. In contrast to the homogeneous case, where different agents may become dominant depending on the realization of energy demands, the heterogeneous setting creates a persistent asymmetry due to the larger energy requirements of agent $1$ (blue curves in the middle column of Figure \ref{fig:het-agents}). For $\gamma = 1$, agent $1$ becomes an important contributor to the training process, but does not completely dominate the learning dynamics. Further for $\gamma = 100$ it does dominate the heuristic approach. The remaining agents either participate intermittently or reduce their contribution depending on their individual energy conditions. Consequently, agent $1$ consumes a moderately larger share of the available green energy budget.

For $\gamma = 10$, however, the medium non-participation penalty significantly alters the agents' decisions. Although abstaining becomes costly, agents do not necessarily increase their contribution uniformly. In particular, agents $2$ and $3$ frequently abstain, while agent $1$ also reduces its participation because its higher energy demand makes participation more costly. In this experimental setting, this reduces the attractiveness of training for the high-demand agent and may ultimately lead to reduced participation despite the penalty. This result illustrates that excessive penalties may create undesirable incentives: \emph{agents may participate less efficiently or reduce their training contribution rather than providing additional useful updates}.

The accuracy trajectories (right column in Figure \ref{fig:het-agents}) further highlight the interaction between heterogeneous energy availability, participation decisions, and model drift. Overall, the heuristic approach achieves higher accuracy than the correlated device approach, although the difference depends on the penalty level. For $\gamma = 1$, almost all agent are constantly engaged in the training and both approaches converge quickly to the $\mathscr{a}$, and the accuracy trajectories of the two approaches are very close. However, as training progresses, the correlated approach exhibits larger variability across agents. The coordinated decisions may cause some agents to become systematically less active, leading to the accumulation of model drift at those agents. Since the recovery parameter satisfies $h < 1$, returning agents only partially recover their lost accuracy after retraining, resulting in persistent differences across local models.

For $\gamma = 10$, both approaches slow down their performance, with the correlated mechanism experience the larger degradation. For $\gamma = 100$, the divergence between agents becomes more pronounced. Although the correlated device maintains a more coordinated participation structure, this coordination does not necessarily translate into a more balanced learning process. Instead, the provider's signal can concentrate training activity among a subset of agents, causing other agents to contribute less frequently and experience larger drift effects. Conversely, in the heuristic approach, participation is determined independently, and although some agents may remain inactive, the learning process can be driven effectively by the most active contributors. In the considered setting, agent $1$ becomes the major contributor, resulting in a global accuracy improvement primarily driven by its local training effort. For $\gamma \in \{10, 100 \}$, the heuristic approach experiences saturation.

This behavior highlights an important distinction between coordination and fairness. The correlated device reduces randomness in participation decisions but may introduce a stronger form of participation bias, as agents respond similarly to the provider's signal and some agents become systematically underrepresented. The heuristic approach, despite producing more heterogeneous individual behaviors, allows agents to exploit their own energy conditions and can avoid excessive concentration of training effort. From the perspective of model drift, this suggests that \emph{coordinated inactivity can be particularly harmful because multiple agents may simultaneously accumulate accuracy degradation, whereas decentralized decisions naturally distribute participation opportunities over time}.

Overall, the heterogeneous experiments reveal a trade-off between coordinated energy management and learning diversity. The heuristic mechanism promotes selective participation based on individual incentives and energy availability, whereas the correlation device provides stronger coordination but may unintentionally amplify participation imbalance. Therefore, while coordination can improve predictability of energy consumption, it must be carefully designed to avoid reducing the representation of specific agents and increasing drift-related accuracy degradation.

\paragraph*{Accuracy vs Green energy efficiency.} In  Figures \ref{fig:energy-aware} - \ref{fig:het-agents}, we observe a trade-off between accuracy and total energy consumption, together with a strong dependence on the penalty parameter $\gamma$. In both approaches, the energy contribution of grid energy remains zero. This indicates that the system consistently adapts to operate within the renewable energy budget rather than exceeding it. For small values of $\gamma$, e.g., $\gamma = 1$, participation is selective and heterogeneous. In the energy-aware homogeneous case, see Figures  \ref{fig:energy-aware} and \ref{fig:energy-aware-20}, agents dynamically adjust their training intensity, leading to fluctuating energy consumption and diverse accuracy outcomes (the diamond- and pentagon-shaped points). In the heterogeneous case, see Figure \ref{fig:het-agents}, this effect is clearer: under the heuristic phase, some agents, e.g., agent $1$, are the only agents not abstaining, and take on most of the workload; under the correlation mechanism, participation is instead concentrated on all agents. In both cases, the outcomes reflect how the limited renewable-energy budget is allocated across agents with different energy demands and training incentives, which leads to variability in both individual and mean accuracy. As $\gamma$ increases, e.g., $\gamma = 100$, abstaining becomes less favorable if the global accuracy threshold has not been achieved, and agents exhibit more consistent, but conservative, participation. This results in more homogeneous behavior across agents and reduces variability in both energy consumption and accuracy. However, rather than significantly increasing performance, the stricter penalty leads agents to operate cautiously within the renewable energy constraint, which keeps accuracy improvements modest. Overall, $\gamma$ acts as a key control parameter that regulates participation incentives under energy constraints. Lower and intermediate values allow flexible, uneven participation that can improve efficiency but introduce variability, while higher values enforce more uniform behavior, reducing variance at the cost of limiting performance gains.

\paragraph{Global accuracy threshold effect.} Here we examine the effect of global accuracy threshold $\mathscr{a}$. In particular, we set $\gamma = 1$, $h = 0.5$, $\delta_{+} = \delta_{-} = 10$, and $\mathscr{a} \in \{0.5, 0.7, 0.9\}$. 

The upper row of Figure \ref{fig:energy-aware} illustrates the case where $\mathscr{a} = 0.9$. Under the heuristic approach, the target accuracy is achieved after five training rounds, at which point the provider discontinues incentives and the training process terminates. In contrast, under the correlated device, the aggregated accuracy reaches only approximately $77\%$ within the available ten rounds. This behavior is explained by the coordinated recommendations, which induce some agents to postpone participation in specific rounds. As a result, model drift accumulates at inactive agents, and because $h=0.5$ only partially compensates for the lost accuracy upon re-entry, the recovery of the aggregated accuracy becomes slower, preventing the system from reaching the desired threshold within the considered time horizon.

For $\mathscr{a} = 0.7$, shown in the lower row of Figure \ref{fig:global accuracy effect g 1}, the heuristic approach reaches the target accuracy after three rounds. Further, the correlated device does not require the entire training horizon to attain the prescribed threshold, but only four rounds. Noteworthy, the aggregated accuracies increase monotonically exhibiting periods of different growth rate, due to the fact that in some rounds some agents may abstain from training. This behavior reflects the dynamic interaction between participation decisions and drift. During rounds with limited participation, the accumulated drift slows the improvement of the global model. When previously inactive agents resume training, the blending $h$ partially restores their local accuracy, producing temporary accelerations in the convergence process. Consequently, the global accuracy evolves in a non-uniform manner before eventually reaching the prescribed threshold.

Clearly, for lower $\mathscr{a}$ values the training process is terminated in earlier rounds, e.g., round $2$ (heuristic phase) and round $3$ (correlated device) when $\mathscr{a} = 0.5$, and round $3$ (heuristic phase) and round $4$ (correlated device) when $\mathscr{a} = 0.7$, in Figure \ref{fig:global accuracy effect g 1}.

For all considered values of $\mathscr{a}$, we observe that agents under the correlated device participate more frequently than under the heuristic approach. Nevertheless, the participation dynamics remain noticeably different for different values of $\mathscr{a}$. While the heuristic approach exhibits relatively stable participation throughout the incentivized period, the correlated device leads several agents to temporarily suspend their participation and subsequently re-enter the training process. Owing to the relatively low target accuracy, these temporary inactivity periods do not significantly affect the stopping time. However, they demonstrate the ability of the correlated recommendations to coordinate participation while exploiting the recovery mechanism. In the considered experiments, this suggests that intermittent participation can reduce training activity without preventing the system from reaching moderate accuracy targets. As the required accuracy increases, however, the accumulated effects of model drift become more pronounced, making sustained participation increasingly important for convergence.

\paragraph{Drifting phenomenon.} To evaluate the impact of intermittent agent participation on model performance, we investigate the emergence of drift under penalty levels $\gamma \in \{1, 10, 100\}$, $\mathscr{a} = 0.9$ and $h = 0.5$, Figure \ref{fig:drifting}. Further, we consider two different regimes regarding the agents' bundle of samples, $\delta_+ = \delta_- = 10$, and $\delta_+ = \delta_- = 20$. Specifically, we analyze how the incentive mechanism influences agents' training decisions and how these decisions affect the evolution of local model accuracy over time. 

For the case where the agents have more balanced data, that is $\delta_+ = \delta_- = 10$, increasing $\gamma$ leads to increasing participation for some agents, see also first row in Figure \ref{fig:energy-aware}. When $\gamma = 1$, for the heuristic approach, agent $1$ is almost never engaged in the training, resulting into high drift. remain engaged until the global accuracy threshold $\mathscr{a}$ is reached. For $\gamma = 10$, agents $3$ abstains from the first round of the training, while all other agents participate, see second row in Figure \ref{fig:energy-aware}, so only agents $3$ experiences drifting. Finally, for $\gamma = 100$, agents oscillate between participation and non-participation, see lower row in Figure \ref{fig:energy-aware}. Hence, the drift follows the same pattern. Finally, for $\gamma = 100$ case,  engagement to training is almost complete, so agents experience small drift. For all $\gamma$ values using the correlated device, all agents are almost fully committed to the training process, so the drifting phenomenon is negligible.

When agents have wider energy demand distributions, that is $\delta_+ = \delta_- = 20$, the participation follows different trend. The main difference is that when $\gamma = 10$, all agents may abstain from training, resulting into significant drift. For higher penalty values, $\gamma \in \{ 10, 100\}$, we observe the same behavior as the case where $\delta_+ = \delta_- = 10$.

These results highlight that the effect of the non-participation penalty is not independent of the local data distribution. When agents have highly uneven data availability, increasing $\gamma$ to moderate values may discourage participation and amplify model drift. However, when data availability is more balanced, the same penalty can reinforce participation and improve model stability. Therefore, the effectiveness of incentive mechanisms depends not only on the penalty design but also on the distribution and relative contribution of local training data across agents.

\paragraph{Blending effect.} Here we examine the effect of blending both from systemic and agent perspective, setting $h \in \{0.5, 0.75, 1\}$, see Figures \ref{fig:energy-aware}, \ref{fig:drifting}, \ref{fig:drifting 0.75}, and \ref{fig:drifting 1}. When $h = 1$ we examine immediate recovery, Figure \ref{fig:drifting 1}. Further, we set $\mathscr{d}_i = 0.05$, $\mathscr{a} = 0.9$, $\gamma \in \{1, 10, 100\}$ and $\delta_+ = \delta_- = 10$.

For $h = 0.5$, the recovery from model drift is partial, meaning that when an agent resumes participation after a period of inactivity, only half of the accumulated degradation is compensated in the current training round. For $\gamma = 1$, the low penalty for non-participation is insufficient to discourage agents from avoiding training, leading to more frequent participation decisions being skipped, see first row in Figure \ref{fig:energy-aware}. So agents that skip several training rounds remain with lower accuracy even after returning, see agent $3$ (green diamond) in Figure \ref{fig:energy-aware}. Increasing the penalty to $\gamma = 10$ encourages more consistent participation, reducing the occurrence of drift and improving the convergence speed, see upper middle panel in Figure \ref{fig:drifting} and middle row in \ref{fig:energy-aware}. For $\gamma = 100$, the large penalty makes agents to choose either low intensity training or short abstain periods. Consequently, model drift accumulates, resulting in lower per-agent accuracy and requiring additional training rounds to reach the saturation level, see lower right panel in Figure \ref{fig:energy-aware}.

In general, since $h$ remains moderate, agents that experience inactivity require multiple rounds to fully recover their accuracy, leading to a longer incentivization period and higher energy consumption compared with larger values of $h$, see upper left and right panels in Figure \ref{fig:drifting}.

For $h = 0.75$, agents recover a larger portion of their lost accuracy after participating in training. Under $\gamma = 1$, although some agents may still choose not to participate due to the lower penalty, the impact of such decisions is less severe compared with $h = 0.5$, as the accumulated drift is compensated more effectively once agents return. This results in improved accuracy stability and fewer corrective training rounds, see upper row in Figure \ref{fig:drifting 0.75}. For $\gamma = 10$, the agents are more indifferent regarding the training process via the heuristic approach. Nevertheless, the high blending value allows them to recovery quickly and the model's accuracy reaches a saturation point. For $\gamma = 100$, participation is already strongly encouraged; increasing $\gamma$ hen further mitigates the accuracy cost of occasional non-participation by accelerating recovery. 

For $h = 1$, local training fully eliminates the impact of accumulated drift after a single participation round. With $\gamma = 1$, agents may still avoid participation due to the low non-participation penalty, but the resulting degradation has a limited impact because agents completely recover when they return to training. Therefore, the accuracy remains more stable compared with $h = 0.5$ and $h = 0.75$, although occasional non-participation may still increase the number of rounds required to reach the target accuracy. For $\gamma \in \{1, 10\}$, we observe that the absence of drifting makes the agents to neglect more often the training process, but the full recovery allows them to terminate the training, see the upper row in \ref{fig:drifting 1}, or to be really close, see the middle row in \ref{fig:drifting 1}. For $\gamma = 100$, agents are strongly encouraged to participate, and the complete drift recovery mechanism ensures that local accuracies quickly return to their optimal values, see lower row in Figure \ref{fig:drifting 1}. Consequently, the aggregated accuracy converges faster, the incentive phase terminates earlier, resulting in the lowest total energy consumption among the three recovery settings.

Interestingly, from the above analysis we observe a tension between $\gamma$ and $h$, e.g, middle row in Figure \ref{fig:drifting 0.75} ($\gamma = 10$ and $h = 0.75$) . The reason is that they affect two different mechanisms, $\gamma$ controls participation choice, while $h$ models the consequences of non-participation. In the second row in Figure \ref{fig:drifting 0.75} we observe that a higher drift recovery $h$ can partially compensate for lower participation incentives $\gamma$, creating a trade-off. Intuitively, when $h$ is low, e.g., $h = 0.5$, missing training rounds has a significant cost because agents cannot fully recover from drift immediately. Therefore, the training process relies more heavily on $\gamma$ to maintain participation. A low $\gamma$ may lead to frequent non-participation, accumulated drift, slower convergence, and higher energy consumption due to additional training rounds. When $h$ is higher, agents can faster recover their accuracy after a single participation round. So, the system becomes more tolerant to intermittent participation. Therefore, the provider may not need to impose a high penalty $\gamma$ because drift effects are quickly corrected. This creates a potential strategic tension: a high $h$ reduces the negative impact of skipping training. Hence, agents may become more willing to avoid participation because they know that their accuracy can be restored later. Therefore, increasing $h$ may unintentionally reduce the effectiveness of $\gamma$ as an incentive mechanism. In a nutshell, the proper penalty $\gamma$ also depends on the drifting value.

Furthermore, the interaction between $h$ and $\gamma$ may introduce representation bias in the FL process. Specifically, larger values of $h$ allow agents to recover quickly from periods of inactivity, potentially reducing their incentive to participate consistently. Consequently, the global model may reach the target accuracy primarily through the contributions of a subset of frequently participating agents, while the data distributions of less active agents remain underrepresented, e.g., agents $1$ and $2$ in Figure \ref{fig:global accuracy effect g 1} for $\gamma = 1$ and $\mathscr{a}$. This highlights a trade-off between learning efficiency and fairness, suggesting that stopping criteria based solely on the aggregated accuracy may not always guarantee balanced learning across all participating agents. 

\paragraph*{Real-world energy-availability scenario.} Figure \ref{fig:real-data} illustrates the evolution of renewable energy shares of total energy production for Finland, and Portugal (Table \ref{tab:real data}), and their interaction with total energy consumption under the two mechanisms, heuristic phase and correlation device in Section \ref{sec:model}. Importantly, the proposed framework operates in iterative rounds; the yearly data points should not be interpreted as the duration of individual training rounds. Instead, each year represents a different observation of the real-world environment, capturing the evolution of data characteristics over time. The experimental setup uses these yearly snapshots to simulate successive states encountered during deployment. At each state, local agents perform training using the available data representation, return model updates, and the central system aggregates these updates to obtain the next model state. Therefore, the evaluation focuses on the ability of the proposed approach to adapt to evolving real-world conditions rather than assuming that one training round corresponds to one calendar year.

For the proposes of our model, given the Table \ref{tab:real data}, we assume the following values for each country: Finland $\mathcal{G}^t_f \sim U[15, 30]$, and $S_i \sim U[\mathcal{G}^t_f - 5, \mathcal{G}^t_f + 10]^4$, and Portugal $\mathcal{G}^t_p \sim U[10, 20]$, and $S_i \sim U[\mathcal{G}^t_p - 10, \mathcal{G}^t_p + 20]^4$, for each $i \in \{1, 2, 3\}$. For simplicity, we normalize the total energy production of each country to one unit. Therefore, the reported renewable energy percentages are directly used as normalized green energy availability values. Further, we set $\gamma = 10$, $h = 0.5$, $\mathscr{d} = 0.05$, and $\mathscr{a} = 0.9$.

\begin{table}[]
    \centering
    \begin{tabular}{lcccccccc}
        \toprule
        & 1990 & 1995 & 2000 & 2005 & 2010 & 2015 & 2020 & 2021 \\
        \midrule
        \text{Finland} & 16.3 & 18.5 & 18.9 & 16.8 & 18.6 & 22.4 & 25.4 & 27.5  \\
        \text{Portugal} & 16.7 & 15.8 & 12.5 & 12.1 & 13.3 & 16.5 & 17.7 & 17.8  \\
        \bottomrule
    \end{tabular}
    \caption{Percentage of renewable energy in total energy calculated on the level of Energy Available for Consumption\protect\footnotemark.}
    \label{tab:real data}
\end{table}
\footnotetext{Share of renewable energy based on Directive 2009/28/EC and Directive (EU) 2018/2001 (SHARES tool): 
\href{https://ec.europa.eu/eurostat/statistics-explained/index.php?title=Calculation_methodologies_for_the_share_of_renewables_in_energy_consumption\#cite_note-1}{Calculation methodologies for the share of renewables in energy consumption}.}

From the perspective of the induced energy consumption, both mechanisms operate within a relatively comfortable margin relative to green supply. However, the gap between NE and CE persists, indicating that coordination continues to yield efficiency gains even when renewable penetration is substantial. Nevertheless, the independent behaviours of the agents are more sensitive to the changes in the availability of green energy.

Overall, the figure highlights three main insights. The results indicate that increasing renewable-energy availability does not by itself eliminate the differences between decentralized and coordinated decision-making in the considered scenarios. There is a trade-off between the policies an AI-service provider can follow: coordination is stable, but best responses are more sensitive. Second, the gains from best responses are particularly valuable in early-stage or low-renewable environments, where system constraints are tighter. Third, as renewable shares increase, the system becomes more robust, but strategic control mechanisms still influence stability and variance of total consumption. \\

\vspace{1pt}

Across several experimental settings, the training process may exhibit a saturation point at an accuracy level close to, but below, $\mathscr{a}$, e.g., middle row in Figure \ref{fig:energy-aware-20} and lower row in Figure \ref{fig:het-agents}. Once the aggregated accuracy reaches this regime, further improvements become increasingly limited because the incentive to participate is determined jointly by the expected training benefit and the energy cost of participation. When the current accuracy is already relatively high, some agents may therefore find continued participation insufficiently attractive and temporarily abstain from training. Their inactivity, in turn, causes their local models to drift toward the baseline accuracy level, see lower row in Figure \ref{fig:drifting}. When they subsequently re-enter the training process, the recovery mechanism only partially compensates for the accumulated drift when $h < 1$, so several additional participation rounds may be required to recover the lost accuracy. Consequently, the gains obtained from newly available training contributions can be largely offset by the degradation accumulated during inactive periods, causing the aggregated accuracy to remain in a narrow region and giving rise to the observed saturation. This effect becomes more pronounced when the target accuracy is high, since the system must maintain participation for longer and even temporary inactivity can have a cumulative impact on the global model. Experimentally, we observed that larger penalty may mitigate this issue, see middle and lower row in Figure \ref{fig:energy-aware-20}. Summarizing, the observed saturation is not necessarily caused by a lack of available learning capacity, but can emerge \emph{endogenously} from the interaction between high current accuracy, participation incentives, model drift, and incomplete recovery.

\section{Conclusions}\label{sec:conclusions}

In this work, we studied carbon-aware Federated Learning under limited renewable energy availability through a game-theoretic perspective. We modeled each training round as a finite normal-form game in which agents strategically select their local training intensity while competing for shared green energy resources. The proposed framework captures the trade-offs between learning accuracy, renewable energy utilization, and carbon-intensive grid consumption.

The payoff structure incorporates diminishing returns in training accuracy, proportional renewable energy allocation, and convex penalties for grid energy usage. This formulation creates strategic interdependence among agents, as changes in one agent's training decision affect the energy resources available to others. To address the decentralized and information-limited nature of Federated Learning, we introduced a heuristic decision mechanism based on discrete-time fictitious play, allowing agents to adapt their strategies using empirical observations. In addition, we considered a provider-assisted correlation device that generates coordinated green-aware recommendations based on a system-level objective.

Our results demonstrate that appropriate incentive mechanisms can enable sustainable Federated Learning with zero-grid energy consumption under the considered renewable energy availability conditions, while maintaining competitive learning performance. The heuristic mechanism generally achieves higher accuracy by allowing agents to adapt their decisions to individual energy conditions, whereas the correlation device provides more coordinated energy consumption patterns. However, increased coordination may also lead to less balanced participation and amplify the effects of model drift when some agents become underrepresented. These results highlight the trade-off between coordinated energy management and preserving diverse contributions to the global model.

Overall, this work connects Federated Learning, renewable energy allocation, and non-cooperative game theory by providing a framework for analyzing sustainable distributed AI training. Future research directions include extending the model to dynamic renewable availability, heterogeneous energy pricing, asynchronous updates, and scenarios with incomplete observability of global energy conditions.

\section*{Acknowledgment}

This work has been partly developed in the scope of the project EXIGENCE, which has received funding from the Smart Networks and Services Joint Undertaking (SNS JU) under the European Union (EU) Horizon Europe research and innovation programme under Grant Agreement No 101139120. Views and opinions expressed are however those of the author(s) only and do not necessarily reflect those of the EU or SNS JU.

\bibliographystyle{plain}
\bibliography{ref-ai}

@article{Aumann1974SubjectivityAC,
  title={Subjectivity and Correlation in Randomized Strategies},
  author={Robert J. Aumann},
  journal={Journal of Mathematical Economics},
  year={1974},
  volume={1},
  pages={67-96}
}

@article{Nash1951NONCOOPERATIVEG,
  title={NON-COOPERATIVE GAMES},
  author={John F. Nash},
  journal={Classics in Game Theory},
  year={1951}
}

@article{Daskalakis2006TheCO,
  title={The complexity of computing a Nash equilibrium},
  author={Constantinos Daskalakis and Paul W. Goldberg and Christos H. Papadimitriou},
  journal={Electron. Colloquium Comput. Complex.},
  year={2006},
  volume={TR05}
}

@inproceedings{McMahan2016CommunicationEfficientLO,
  title={Communication-Efficient Learning of Deep Networks from Decentralized Data},
  author={H. B. McMahan and Eider Moore and Daniel Ramage and Seth Hampson and Blaise Ag{\"u}era y Arcas},
  booktitle={International Conference on Artificial Intelligence and Statistics},
  year={2016},
}

@article{Wiesner2023FedZeroLR,
  title={FedZero: Leveraging Renewable Excess Energy in Federated Learning},
  author={Philipp Wiesner and Ramin Khalili and Dennis Grinwald and Pratik Agrawal and Lauritz Thamsen and Odej Kao},
  journal={Proceedings of the 15th ACM International Conference on Future and Sustainable Energy Systems},
  year={2023}
}

@article{Kang2019IncentiveDF,
  title={Incentive Design for Efficient Federated Learning in Mobile Networks: A Contract Theory Approach},
  author={Jiawen Kang and Zehui Xiong and Dusit Tao Niyato and Han Yu and Ying-Chang Liang and Dong In Kim},
  journal={2019 IEEE VTS Asia Pacific Wireless Communications Symposium (APWCS)},
  year={2019},
  pages={1-5}
}

@incollection{brown:fp1951,
	Address = {New York},
	Author = {George W. Brown},
	Booktitle = {Activity Analysis of Production and Allocation},
	Publisher = {Wiley},
	Title = {Iterative Solution of Games by Fictitious Play},
	Year = {1951}}

@article{Robinson1951ANIM,
  title={AN ITERATIVE METHOD OF SOLVING A GAME},
  author={Julia Jean Robinson},
  journal={Classics in Game Theory},
  year={1951}
}

@article{Patterson2021CarbonEA,
  title={Carbon Emissions and Large Neural Network Training},
  author={David A. Patterson and Joseph Gonzalez and Quoc V. Le and Chen Liang and Llu{\'i}s-Miquel Mungu{\'i}a and Daniel Rothchild and David R. So and Maud Texier and Jeff Dean},
  journal={ArXiv},
  year={2021},
  volume={abs/2104.10350}
}

@article{garcia2019estimation,
  title={Estimation of energy consumption in machine learning},
  author={Garcia-Martin, Eva and Lavesson, Niklas and Grahn, H{\aa}kan and Casalicchio, Emiliano and Boeva, Veselka},
  journal={Journal of Parallel and Distributed Computing},
  volume={134},
  pages={75--88},
  year={2019}
}

@article{savazzi2020federated,
  title={Federated learning with cooperating devices: A consensus approach for wireless IoT networks},
  author={Savazzi, Stefano and Nicoli, Monica and Rampa, Vittorio and Kianush, Sanaz},
  journal={IEEE Transactions on Communications},
  volume={68},
  number={12},
  pages={7644--7659},
  year={2020}
}

@article{Strubell2019EnergyAP,
  title={Energy and Policy Considerations for Deep Learning in NLP},
  author={Emma Strubell and Ananya Ganesh and Andrew McCallum},
  journal={ArXiv},
  year={2019},
  volume={abs/1906.02243}
}

@article{Qiu2020AFL,
  title={A first look into the carbon footprint of federated learning},
  author={Xinchi Qiu and Titouan Parcollet and Daniel J. Beutel and Taner Topal and Akhil Mathur and Nicholas D. Lane},
  journal={ArXiv},
  year={2020},
  volume={abs/2102.07627}
}

@inproceedings{Bird2014WindAS,
  title={Wind and Solar Energy Curtailment: Experience and Practices in the United States},
  author={Lori Bird and Jaquelin M. Cochran and Xi Wang},
  year={2014}
}

@article{Lacoste2019QuantifyingTC,
  title={Quantifying the Carbon Emissions of Machine Learning},
  author={Alexandre Lacoste and Alexandra Sasha Luccioni and Victor Schmidt and Thomas Dandres},
  journal={ArXiv},
  year={2019},
  volume={abs/1910.09700}
}

@article{Radovanovic2021CarbonAwareCF,
  title={Carbon-Aware Computing for Datacenters},
  author={Ana Radovanovic and Ross Koningstein and Ian Schneider and Bokan Chen and Alexandre Nobrega Duarte and Binz Roy and Diyue Xiao and Maya Haridasan and Patrick Hung and Nick Care and Saurav Talukdar and E. Mullen and Kendal Smith and MariEllen Cottman and Walfredo Cirne},
  journal={IEEE Transactions on Power Systems},
  year={2021},
  volume={38},
  pages={1270-1280}
}

@article{Bonawitz2019TowardsFL,
  title={Towards Federated Learning at Scale: System Design},
  author={Keith Bonawitz and Hubert Eichner and Wolfgang Grieskamp and Dzmitry Huba and Alex Ingerman and Vladimir Ivanov and Chlo{\'e} Kiddon and Jakub Konecn{\'y} and Stefano Mazzocchi and H. B. McMahan and Timon Van Overveldt and David Petrou and Daniel Ramage and Jason Roselander},
  journal={ArXiv},
  year={2019},
  volume={abs/1902.01046}
}

@article{Zhan2020ALI,
  title={A Learning-Based Incentive Mechanism for Federated Learning},
  author={Yufeng Zhan and Peng Li and Zhihao Qu and Deze Zeng and Song Guo},
  journal={IEEE Internet of Things Journal},
  year={2020},
  volume={7},
  pages={6360-6368}
}

@article{Lim2022DecentralizedEI,
  title={Decentralized Edge Intelligence: A Dynamic Resource Allocation Framework for Hierarchical Federated Learning},
  author={Wei Yang Bryan Lim and Jer Shyuan Ng and Zehui Xiong and Jiangming Jin and Yang Zhang and Dusist Niyato and Cyril Leung and Chunyan Miao},
  journal={IEEE Transactions on Parallel and Distributed Systems},
  year={2022},
  volume={33},
  pages={536-550}
}

@inproceedings{Osborne1995ACI,
  title={A Course in Game Theory},
  author={Martin J. Osborne and Ariel Rubinstein},
  year={1995}
}

@article{Thakur2024GreenFL,
  title={Green Federated Learning: A New Era of Green Aware AI},
  author={Dipanwita Thakur and Antonella Guzzo and Giancarlo Fortino and Francesco Piccialli},
  journal={ACM Computing Surveys},
  year={2024},
  volume={57},
  pages={1 - 36}
}

@article{Sarikaya2019MotivatingWI,
  title={Motivating Workers in Federated Learning: A Stackelberg Game Perspective},
  author={Yunus Sarikaya and Ozgur Ercetin},
  journal={IEEE Networking Letters},
  year={2019},
  volume={2},
  pages={23-27}
}

@article{Meng2024FederatedLA,
  title={Federated Learning and Free-riding in a Competitive Market},
  author={Jiajun Meng and Jing Chen and Dongfang Zhao and Lin Liu},
  journal={ArXiv},
  year={2024},
  volume={abs/2410.12723}
}

@article{Yin2023AGA,
title = {A game incentive mechanism for energy efficient federated learning in computing power networks},
journal = {Digital Communications and Networks},
volume = {10},
number = {6},
pages = {1741-1747},
year = {2024},
author = {Xiao Lin and Ruolin Wu and Haibo Mei and Kun Yang}
}

@article{Fang2024LossConvergenceDF,
  title={Loss-Convergence- Driven Federated Learning with Energy Optimization: A Stackelberg Game Approach},
  author={Yanbo Fang and Tengfei Cao and Yiming Zhang},
  journal={2024 10th International Conference on Computer and Communications (ICCC)},
  year={2024},
  pages={2189-2193}
}

@article{Khan2019FederatedLF,
  title={Federated Learning for Edge Networks: Resource Optimization and Incentive Mechanism},
  author={Latif Ullah Khan and Nguyen H. Tran and Shashi Raj Pandey and Walid Saad and Zhu Han and Minh N. H. Nguyen and Choong Seon Hong},
  journal={IEEE Communications Magazine},
  year={2019},
  volume={58},
  pages={88-93}
}

@article{Zou2019MobileDT,
  title={Mobile Device Training Strategies in Federated Learning: An Evolutionary Game Approach},
  author={Yuze Zou and Shaohan Feng and Dusit Tao Niyato and Yutao Jiao and Shimin Gong and Wenqing Cheng},
  journal={2019 International Conference on Internet of Things (iThings) and IEEE Green Computing and Communications (GreenCom) and IEEE Cyber, Physical and Social Computing (CPSCom) and IEEE Smart Data (SmartData)},
  year={2019},
  pages={874-879}
}

@article{Liu2021AnIM,
  title={An Incentive Mechanism for Privacy-Preserving Crowdsensing via Deep Reinforcement Learning},
  author={Yang Liu and Hongsheng Wang and Mugen Peng and Jianfeng Guan and Yu Wang},
  journal={IEEE Internet of Things Journal},
  year={2021},
  volume={8},
  pages={8616-8631}
}

@article{Kaplan2020ScalingLF,
  title={Scaling Laws for Neural Language Models},
  author={Jared Kaplan and Sam McCandlish and T. J. Henighan and Tom B. Brown and Benjamin Chess and Rewon Child and Scott Gray and Alec Radford and Jeff Wu and Dario Amodei},
  journal={ArXiv},
  year={2020},
  volume={abs/2001.08361}
}

@article{Han2025AFL,
  title={A federated learning-based selection and incentive system using blockchain technology},
  author={Yang Han and Tasiu Muazu and Samuel Omaji and Shiyu Miao},
  journal={Pervasive Mob. Comput.},
  year={2025},
  volume={112},
  pages={102091}
}

@inproceedings{Sim2020CollaborativeML,
  title={Collaborative Machine Learning with Incentive-Aware Model Rewards},
  author={Rachael Hwee Ling Sim and Yehong Zhang and Mun Choon Chan and Bryan Kian and Hsiang Low},
  booktitle={International Conference on Machine Learning},
  year={2020}
}

@article{Nguyen2022TowardEH,
  title={Toward Efficient Hierarchical Federated Learning Design Over Multi-Hop Wireless Communications Networks},
  author={Tu Viet Nguyen and Nhan Duc Ho and Hieu Thien Hoang and Cuong Danh Do and Kok-Seng Wong},
  journal={IEEE Access},
  year={2022},
  volume={10},
  pages={111910-111922}
}

@article{Le2020AnIM,
  title={An Incentive Mechanism for Federated Learning in Wireless Cellular Networks: An Auction Approach},
  author={Tra Huong Thi Le and Nguyen Hoang Tran and Yan Kyaw Tun and Minh N. H. Nguyen and Shashi Raj Pandey and Zhu Han and Choong Seon Hong},
  journal={IEEE Transactions on Wireless Communications},
  year={2020},
  volume={20},
  pages={4874-4887}
}

@misc{EXIGENCESbD,
    author = "{The EXIGENCE-project}",
    title = "{Sustainable by Default: Driving Energy-Efficient Streaming and AI Services through Smart Incentives}",
    howpublished = "[Online] Available: \url{https://projectexigence.eu/sustainable-by-default-driving-energy-efficient-streaming-and-ai-services-through-smart-incentives/}",
    note = "Accessed: 12 Feb 2026"
}

@article{Varsos2026UserAM,
  title={User Acceptance Model for Smart Incentives in Sustainable Video Streaming towards 6G},
  author={Konstantinos Varsos and Adamantia Stamou and George D. Stamoulis and Vasillios A. Siris},
  journal={ICC 2026 - IEEE International Conference on Communications},
  year={2026},
  pages={1-6}
}

@inproceedings{Varsos2026OptimalES,
  title={Optimal Energy-Aware Service Management in Future Networks with a Gamified Incentives Mechanism},
  author={Konstantinos Varsos and Adamantia Stamou and George D. Stamoulis and Vasillios A. Siris},
  year={2026}
}

@article{Pfeiffer2023FederatedLF,
  title={Federated Learning for Computationally Constrained Heterogeneous Devices: A Survey},
  author={Kilian Pfeiffer and Martin Rapp and Ramin Khalili and J{\"o}rg Henkel},
  journal={ACM Computing Surveys},
  year={2023},
  volume={55},
  pages={1 - 27}
}

@inproceedings{Panagea2026GreenFLagAG,
  title={GreenFLag: A Green Agentic Approach for Energy-Efficient Federated Learning},
  author={Theodora Panagea and Nikolaos Koursioumpas and Lina Magoula and Ramin Khalili},
  year={2026}
}

@article{Chai2023ASF,
  title={A Survey for Federated Learning Evaluations: Goals and Measures},
  author={Di Chai and Leye Wang and Liu Yang and Junxue Zhang and Kai Chen and Qian Yang},
  journal={IEEE Transactions on Knowledge and Data Engineering},
  year={2023},
  volume={36},
  pages={5007-5024}
}

\newpage

\appendix

\section{Additional Experiments}

\subsection*{Heuristic phase vs Nash equilibrium.} 
\begin{figure}[htp!]
    \centering
    \includegraphics[width=.33\textwidth]{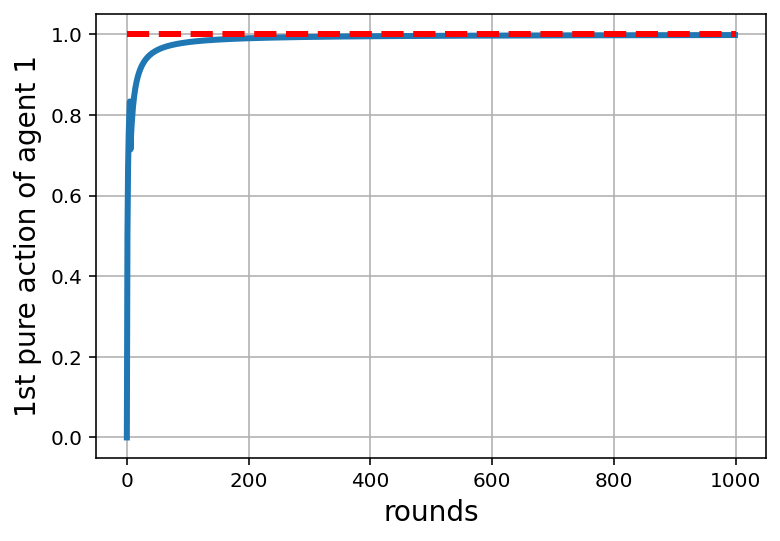}\hfill
    \includegraphics[width=.33\textwidth]{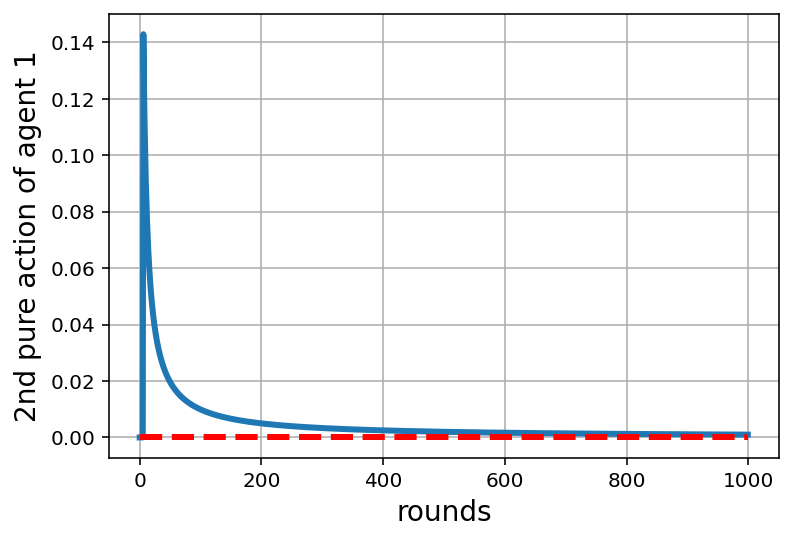}\hfill
    \includegraphics[width=.33\textwidth]{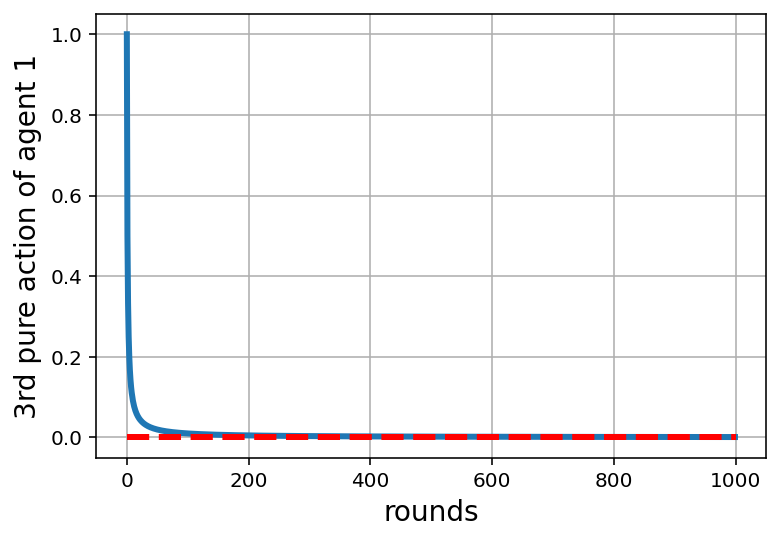}\\
    \includegraphics[width=.33\textwidth]{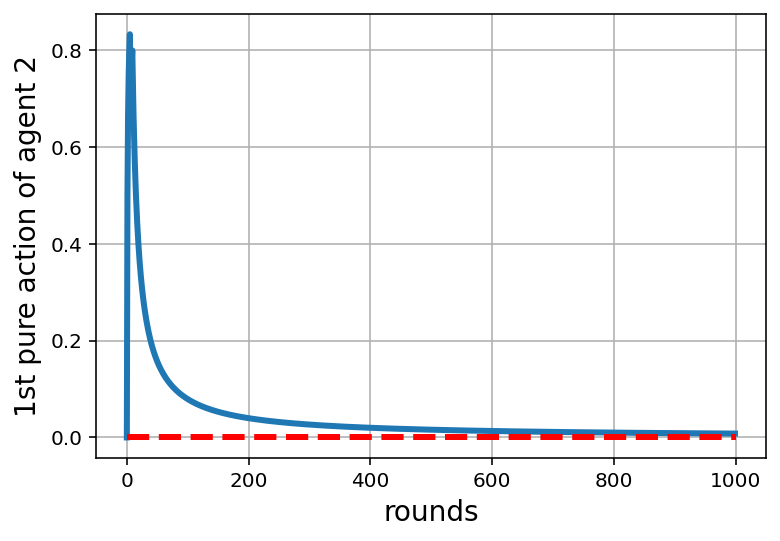}\hfill
    \includegraphics[width=.33\textwidth]{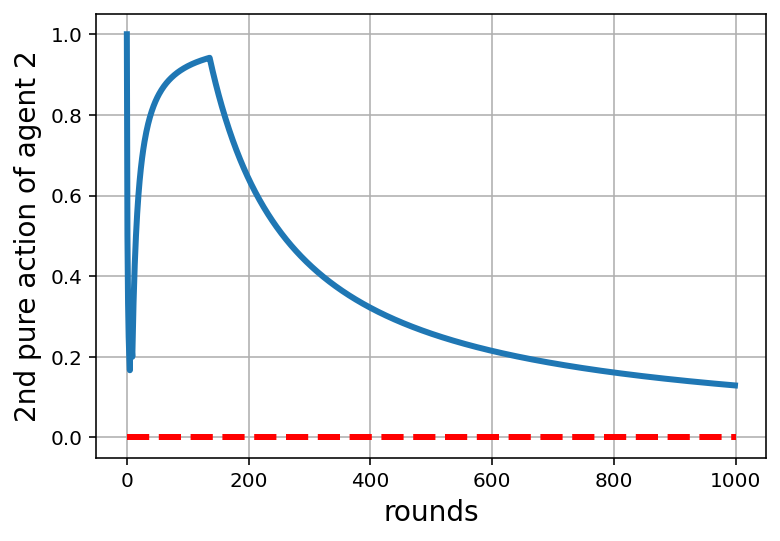}\hfill
    \includegraphics[width=.33\textwidth]{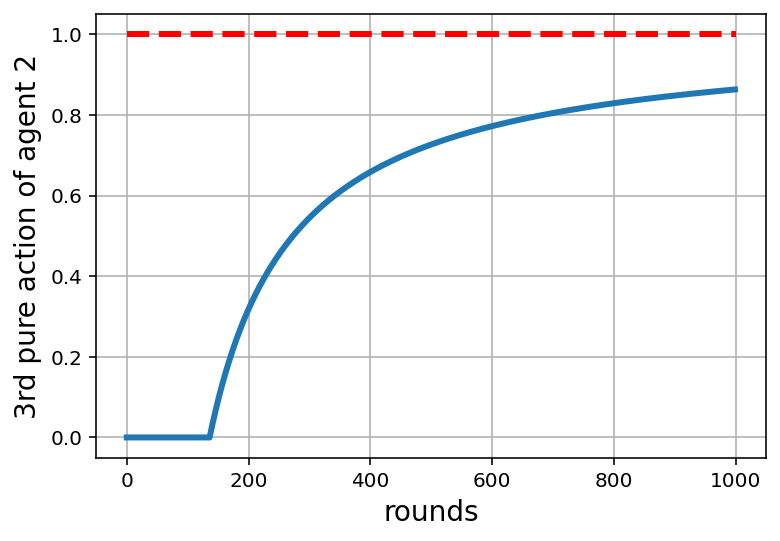}\    
    \caption{Energy flexibility: (i) $\gamma = 0$ (left), (ii) $\gamma = 10$ (middle), and (iii) $\gamma = 1000$ (right).}
\label{fig:fp vs Nash}
\end{figure}

Now we consider the case where two agents are engaged in the game with three actions each, and they follow the heuristic phase, also we set $\gamma = 10$. Therefore we have a $3 \times 3$ bimatrix game. In Figure \ref{fig:fp vs Nash}, we compare the solution provided by the fictitious play algorithm $\ref{alg:fictitious play algorithm}$ with the accurate solution of the game. The game has a Nash equilibrium with strategy profile $\left((1, 0, 0), (0, 0, 1)\right)$. We initialize the algorithm in the pure strategy profile $\left((0, 0, 1), (1, 0, 0)\right)$. We observed that very fast the fictitious play algorithm converges to Nash equilibrium of the game.

\subsection*{Rewards} 

Here, we consider a homogeneous population of agents, $\mathcal{N} = \{1, 2, 3\}$, s.t. $S_i \sim U[\mathcal{G}^t - \delta_-, \mathcal{G}^t + \delta_+]^4$, with $\delta_- = \delta_+ = 10$, $\mathscr{a} = 0.9$, $h = 0.5$, $\theta_i \sim U[0, 1]$, and $\gamma \in \{1, 10, 100\}$.

\begin{figure}[htp!]
    \centering
    \includegraphics[width=.33\textwidth]{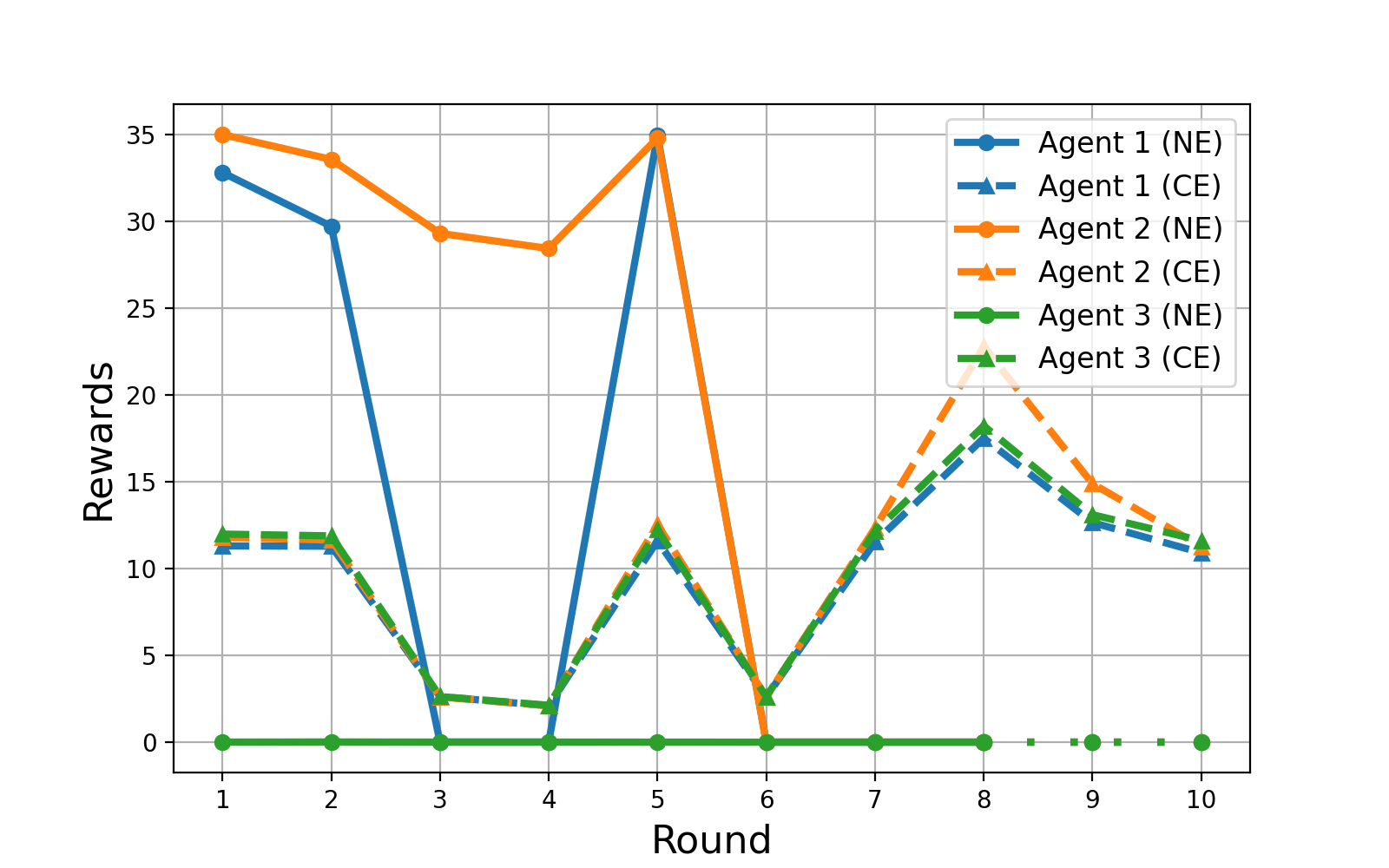}\hfill
    \includegraphics[width=.33\textwidth]{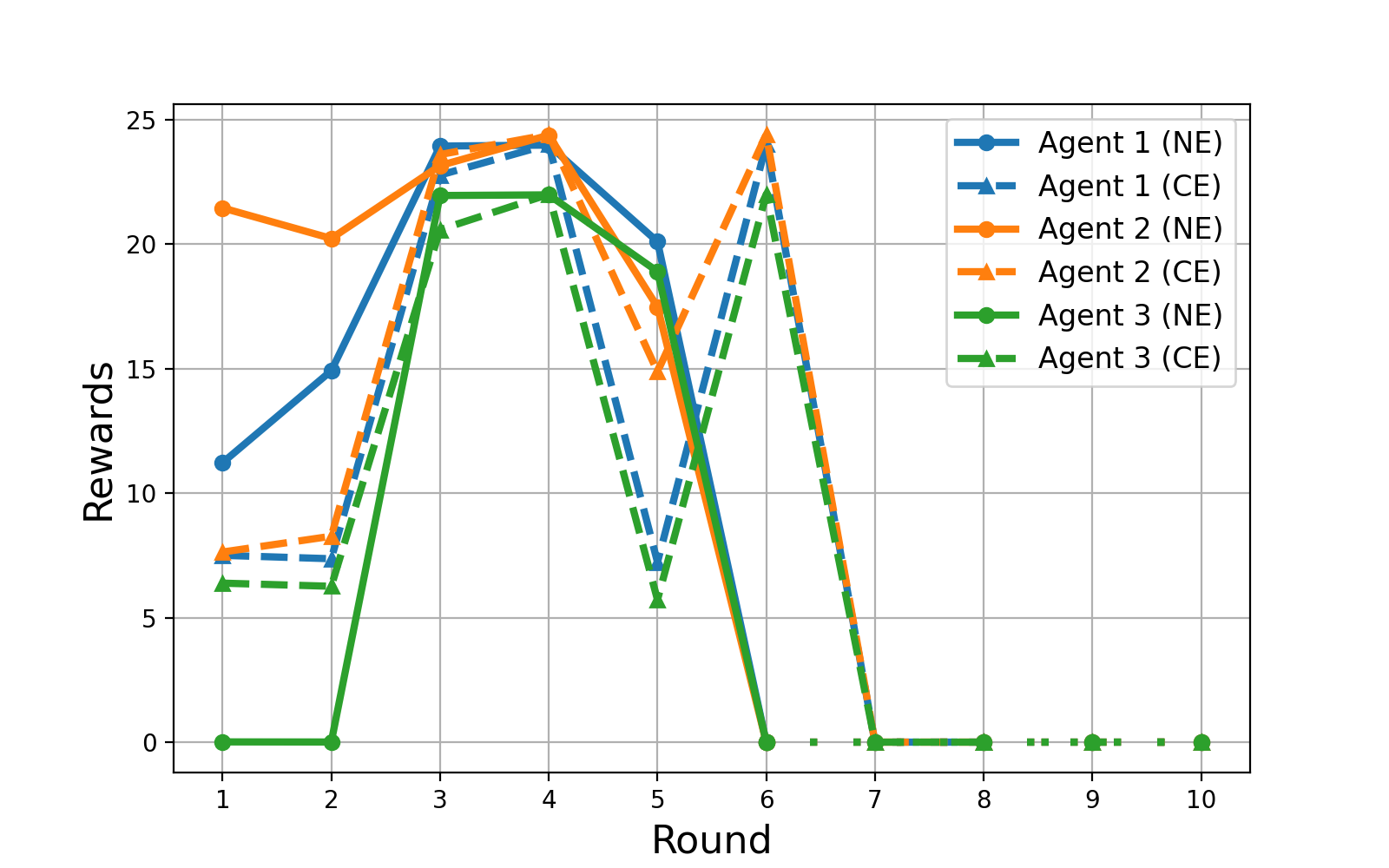}\hfill
    \includegraphics[width=.33\textwidth]{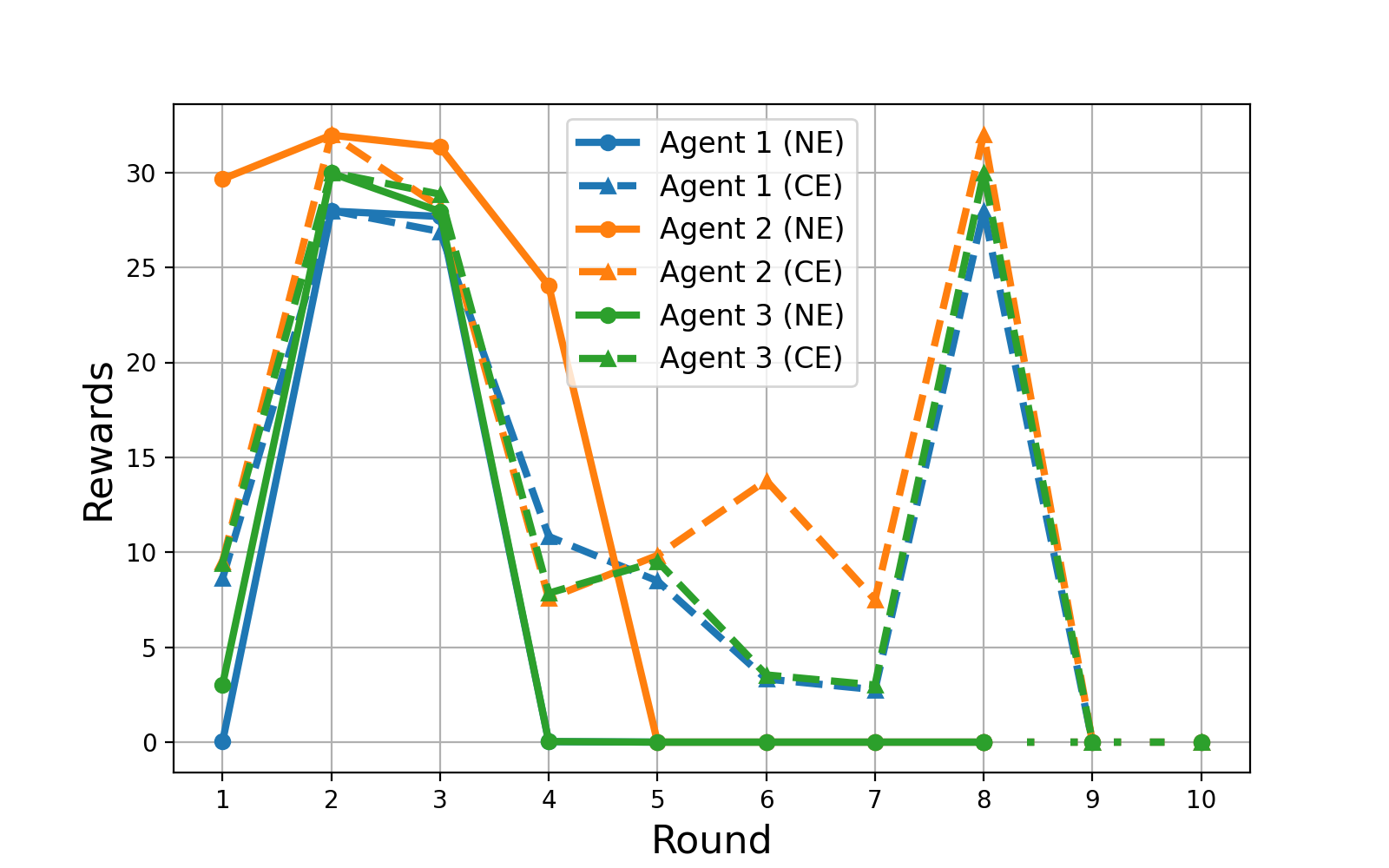}
    \caption{Rewards and accuracy per agent: (i) $\gamma = 1$ (left), (ii) $\gamma = 10$ (middle), and (iii) $\gamma = 100$ (right).}
\label{fig:rewards 1}
\end{figure}

Figure \ref{fig:rewards 1} presents the evolution of rewards across training rounds, with three panels corresponding to $\gamma \in \{1, 10, 100\}$. For $\gamma = 1$, the rewards w.r.t. heuristic algorithm exhibit differences across agents. Agents $1$ secures almost constant rewards, and agent $2$ receives rewards partially. Further, they consume energy in all training rounds, see Figure \ref{fig:energy-aware}. On the other hand, agent $3$ takes zero rewards in the rounds where it does not participates in the training. Under the correlation mechanism, the rewards are generally higher, in total, than under the heuristic mechanism, reflecting the more coordinated participation pattern observed in the corresponding energy trajectories. For all $\gamma \in \{1, 10, 100\}$ we observe that rewards follow the energy patterns in Figure \ref{fig:energy-aware}.

Comparing the two mechanisms, the correlation device generally assigns higher rewards in the considered experiments, while the heuristic mechanism produces more differentiated reward profiles across agents. This difference is consistent with the respective participation structures: coordination tends to synchronize the agents' decisions, whereas the heuristic mechanism allows agents to respond independently to their individual conditions. The reward results therefore complement the energy and accuracy experiments by showing that the two mechanisms do not only differ in their aggregate energy-consumption patterns, but also in how the available incentives are distributed across agents.

\begin{figure}[htp!]
    \centering
    \includegraphics[width=.33\textwidth]{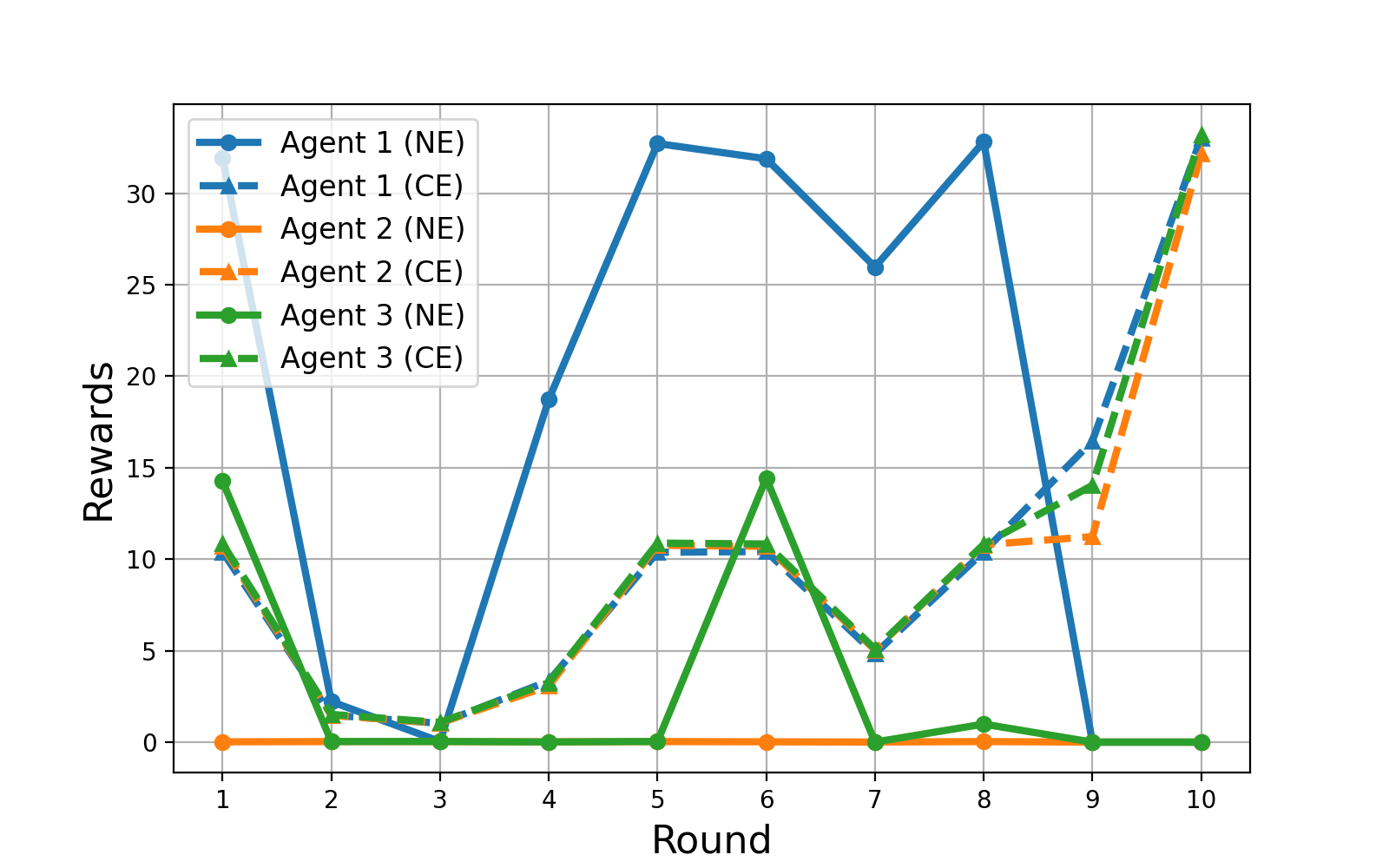}\hfill
    \includegraphics[width=.33\textwidth]{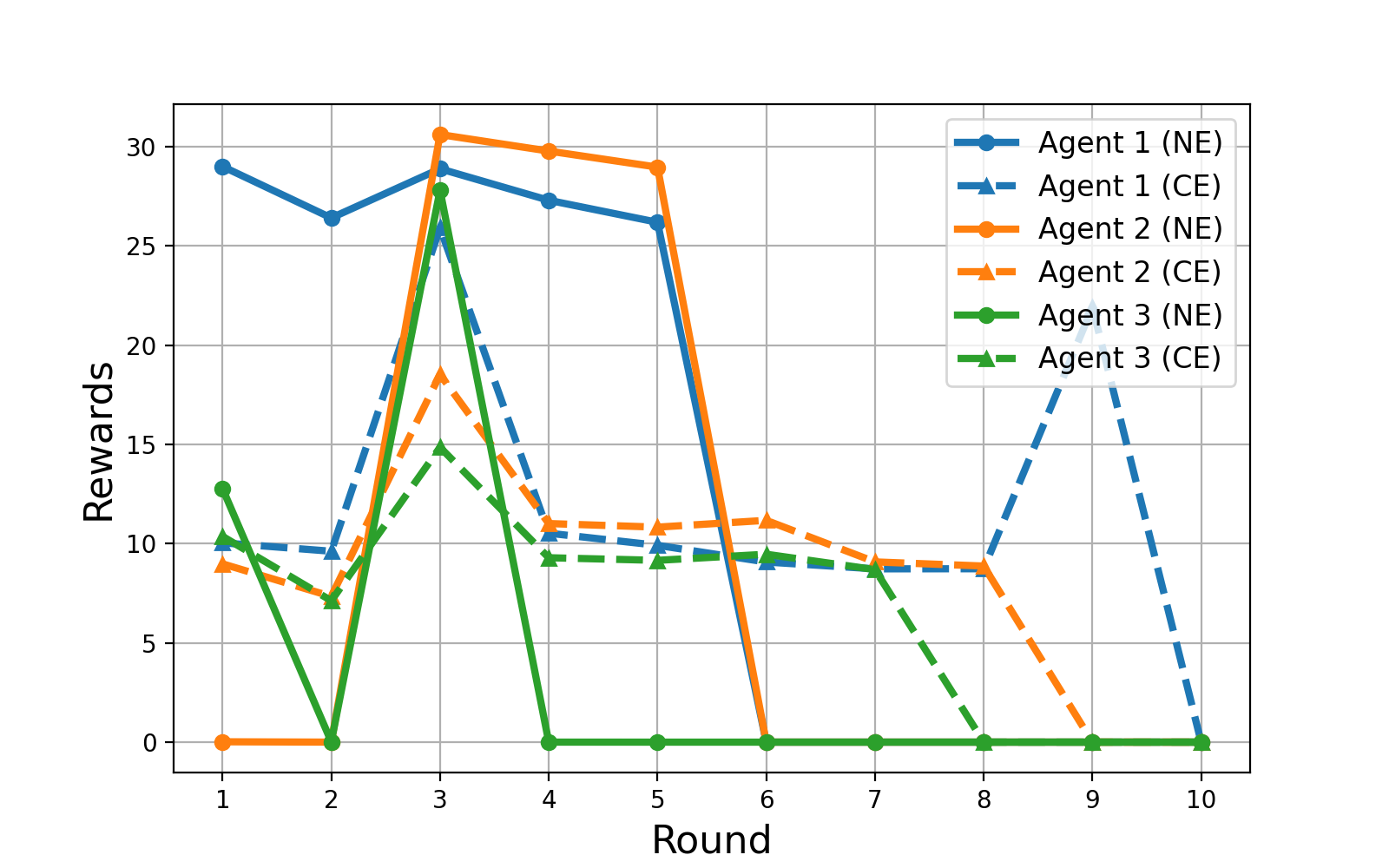}\hfill
    \includegraphics[width=.33\textwidth]{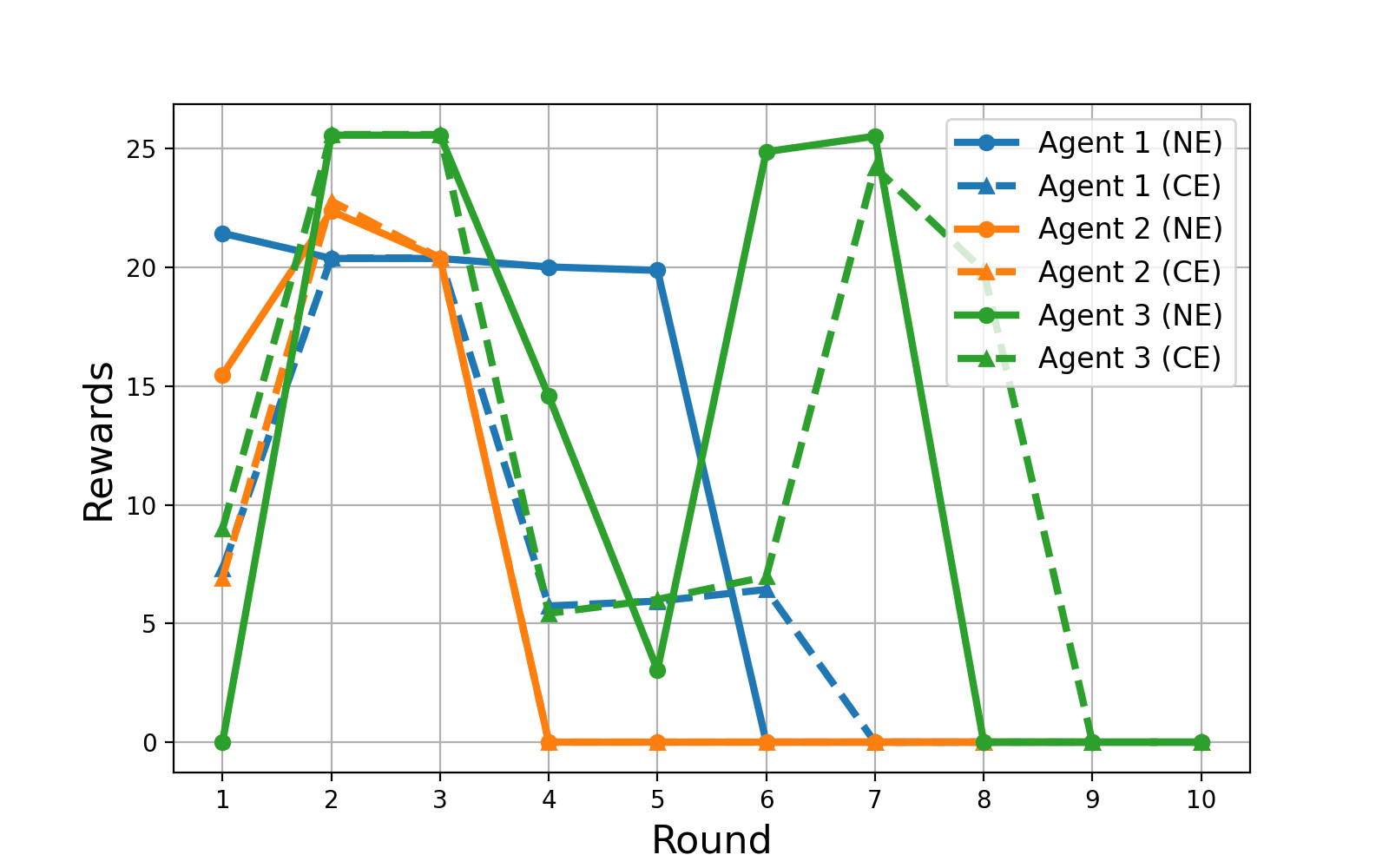}
    \caption{Rewards and accuracy per agent: (i) $h = 0.5$ (left), (ii) $h = 0.75$ (middle), and (iii) $h = 1$ (right), when $\gamma = 10$.}
\label{fig:rewards 2}
\end{figure}

Finally, the reward trajectories should be interpreted jointly with the accuracy curves. In particular, a zero reward corresponds to a round in which an agent does not participate, and therefore does not imply poor learning performance by itself. Its effect on accuracy depends on the subsequent evolution of the local model and on the recovery parameter $h$, see Figure \ref{fig:rewards 2}. With $h = 0.5$, inactive agents only partially recover the accuracy lost through model drift when they return to training, so repeated periods of non-participation can affect the subsequent accuracy trajectory. This distinction is important because the reward experiment demonstrates the relationship between participation and incentives, whereas the learning consequences of those participation decisions are captured by the accuracy and drift experiments.

\subsection*{Quality of data.}
Here, we consider a homogeneous population of agents, $\mathcal{N} = \{1, 2, 3\}$, s.t. $S_i \sim U[\mathcal{G}^t - \delta_-, \mathcal{G}^t + \delta_+]^4$, with with $\delta_- = \delta_+ = 10$, $\mathscr{a} = 0.9$, $h = 0.5$, and $\gamma \in \{1, 10, 100\}$. We discuss three cases: (i) $\theta_i \sim U[0, 0.5]$ (relatively poor training data), (ii) $\theta_i \sim U[0.25, 0.75]$ (moderately heterogeneous data quality), and (iii) $\theta_i \sim U[0.5, 1]$ (relatively good training data).

\begin{figure}[h]
    \centering
    \includegraphics[width=.33\textwidth]{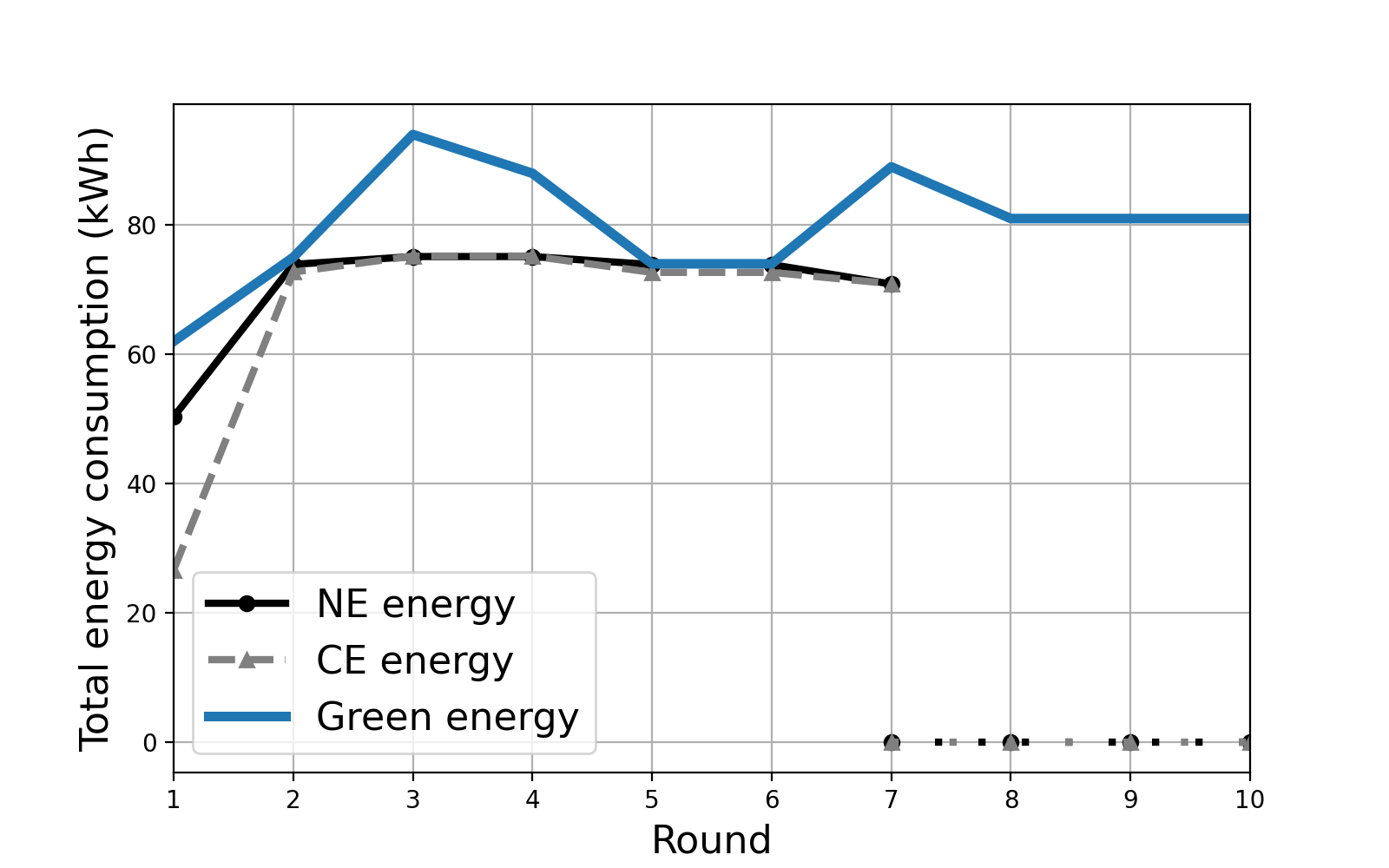}\hfill
    \includegraphics[width=.33\textwidth]{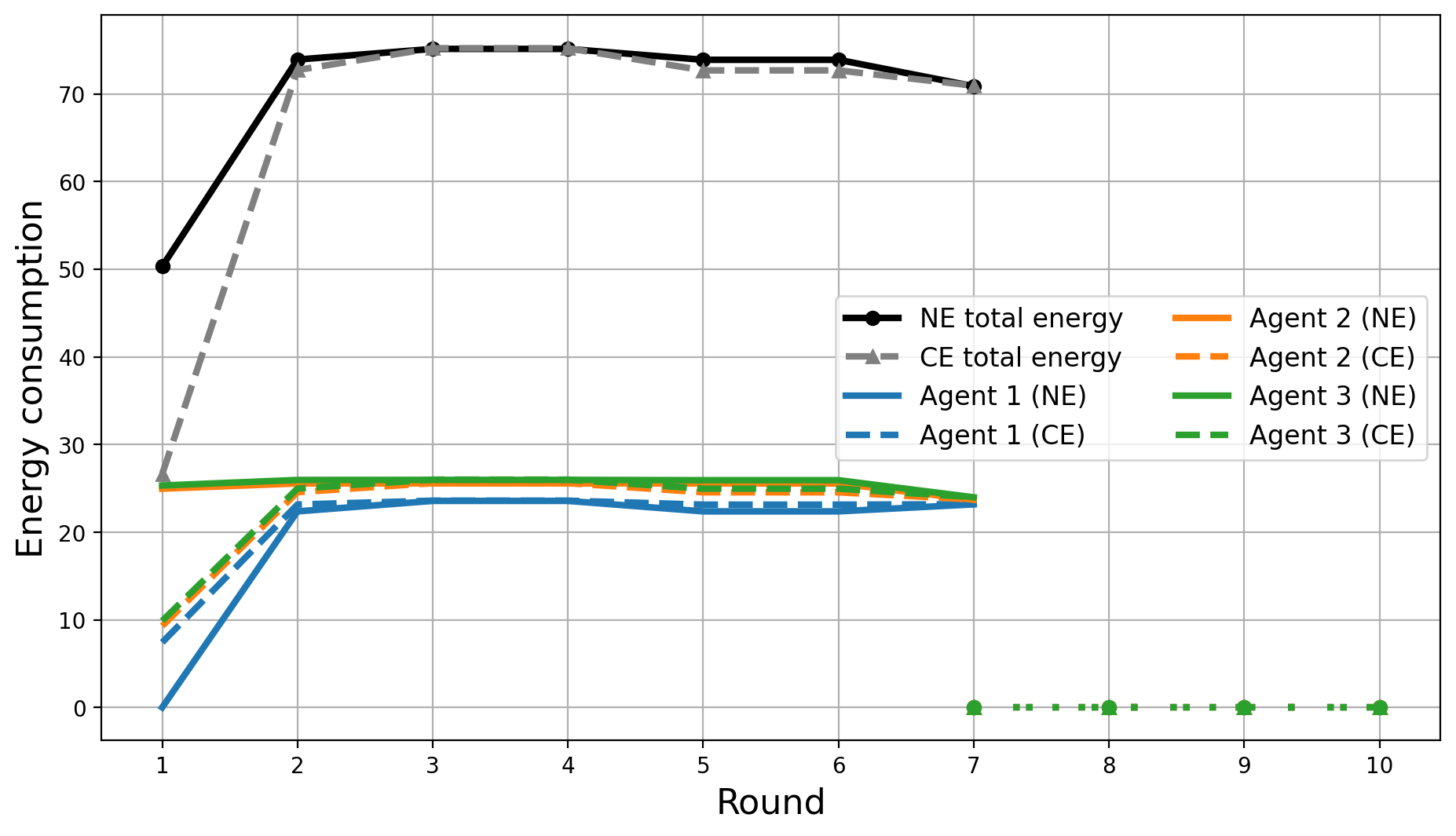}\hfill
    \includegraphics[width=.33\textwidth]{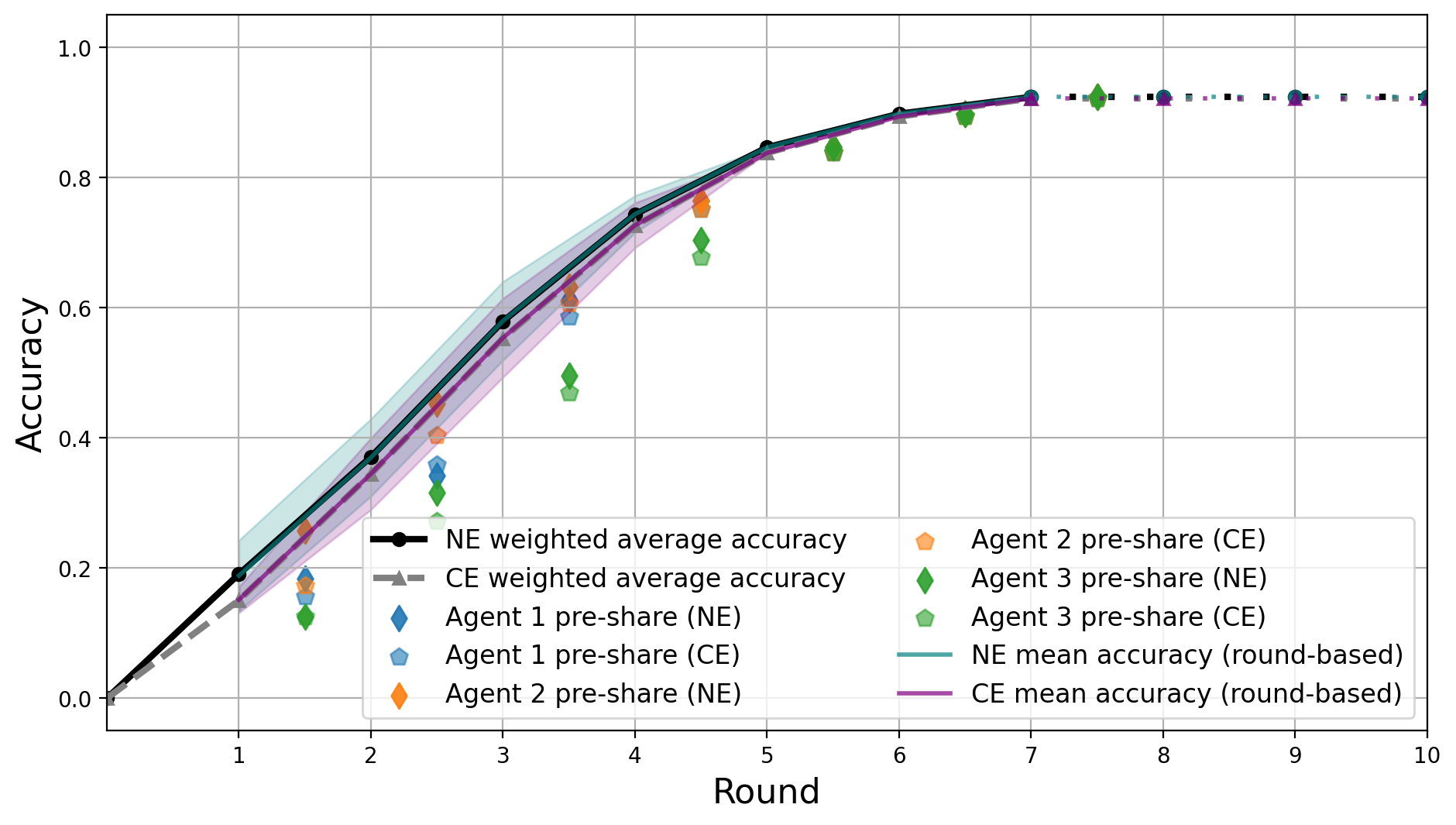}\hfill
    \includegraphics[width=.33\textwidth]{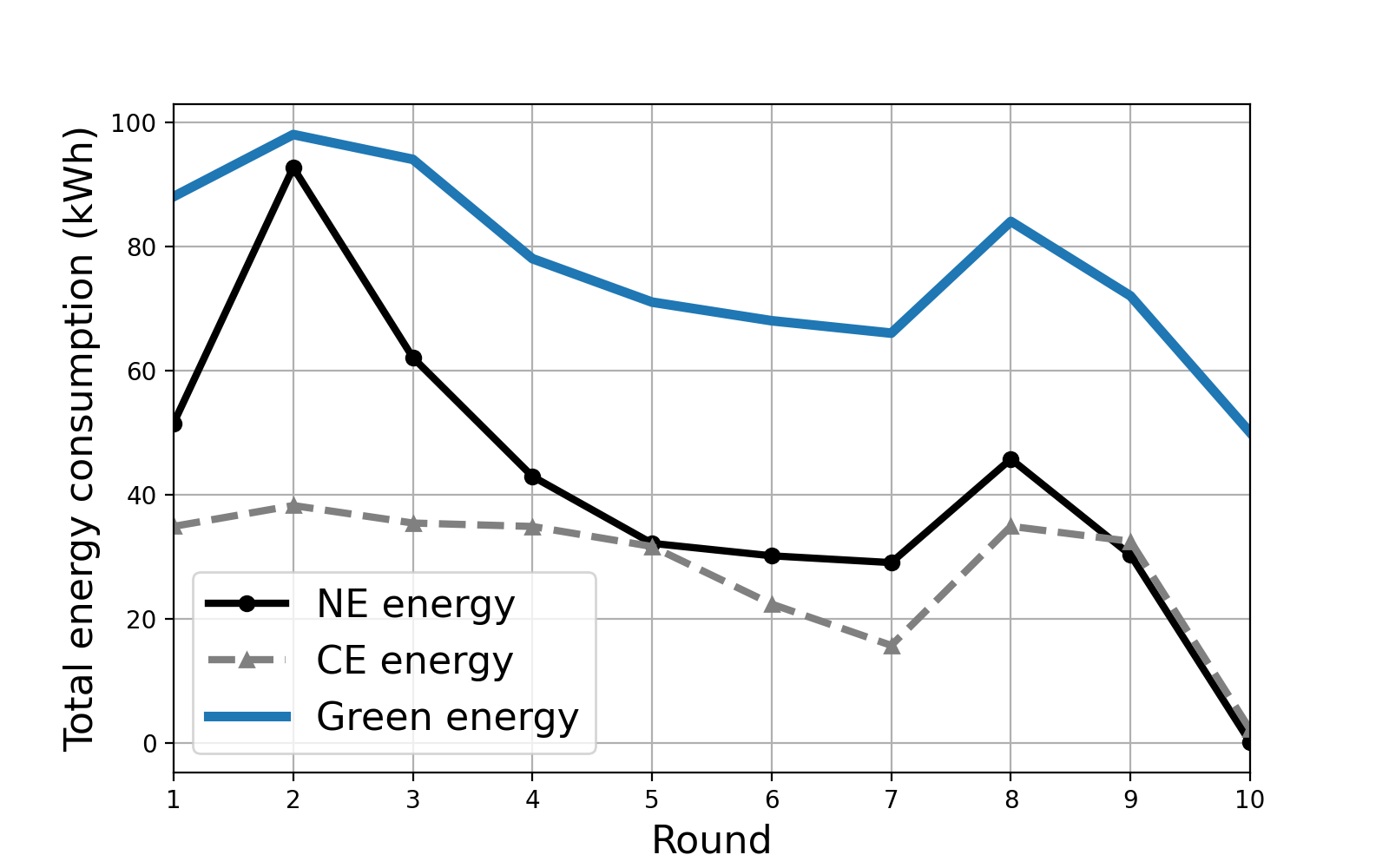}\hfill
    \includegraphics[width=.33\textwidth]{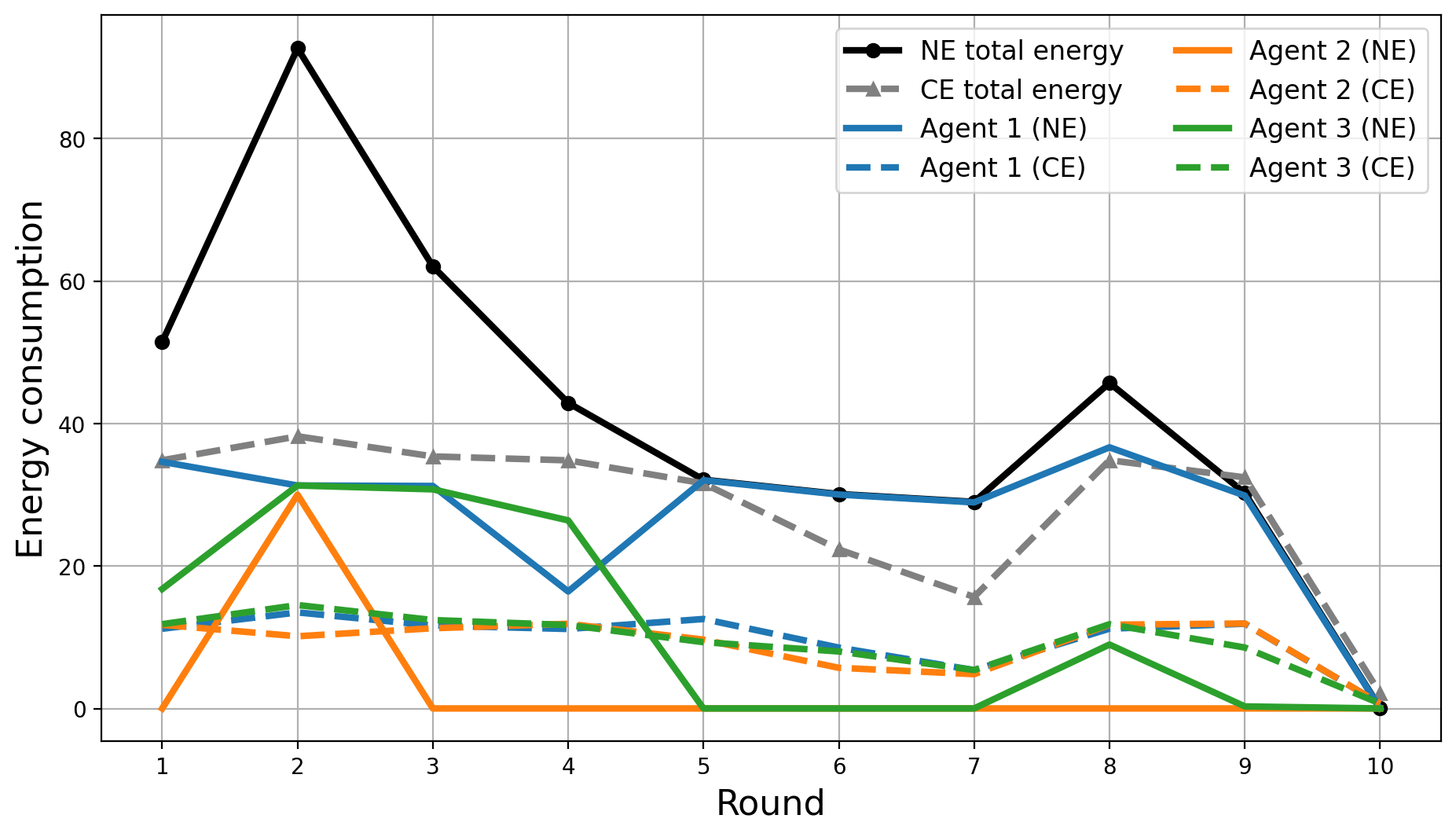}\hfill
    \includegraphics[width=.33\textwidth]{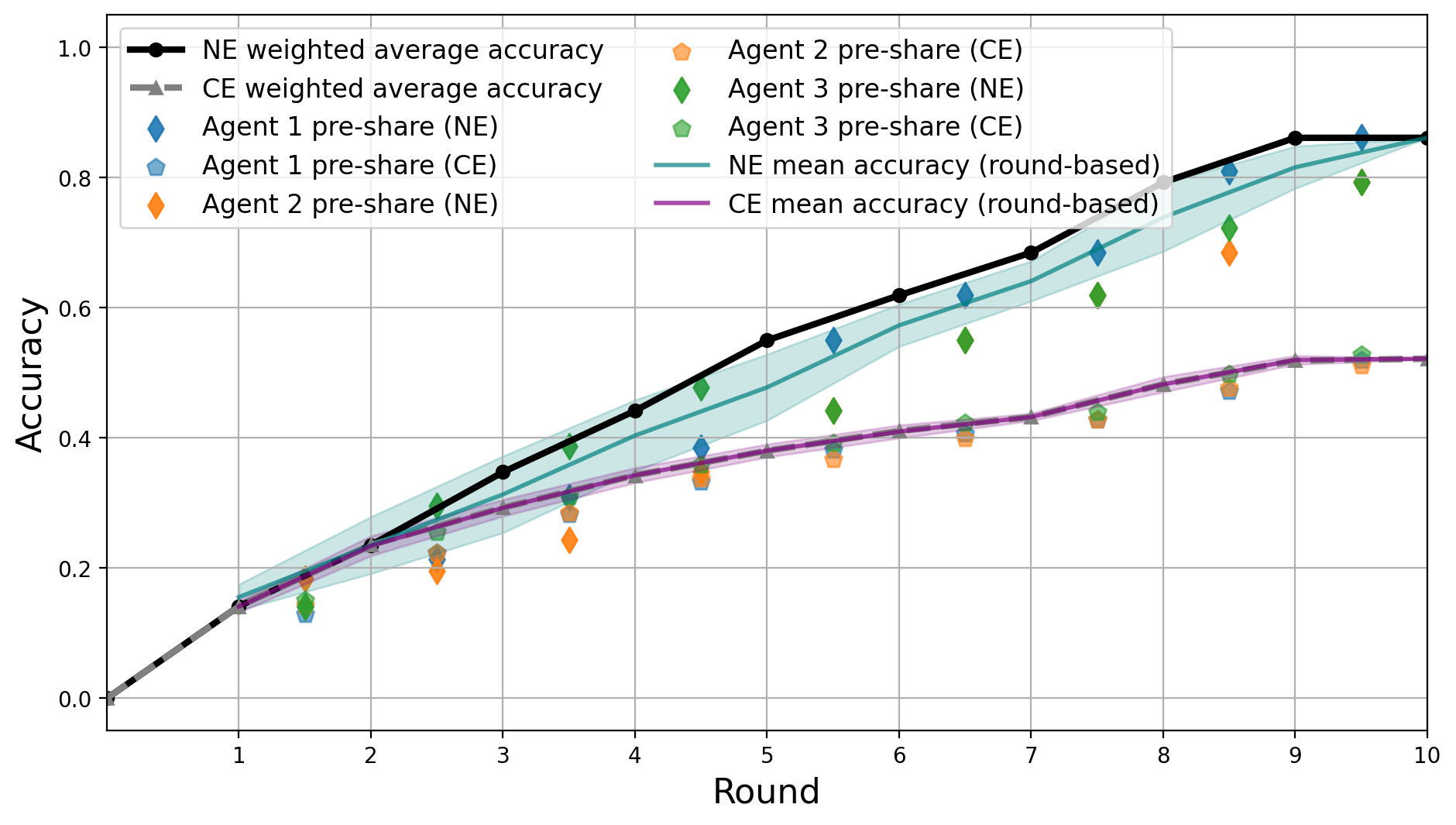}\hfill
    \includegraphics[width=.33\textwidth]{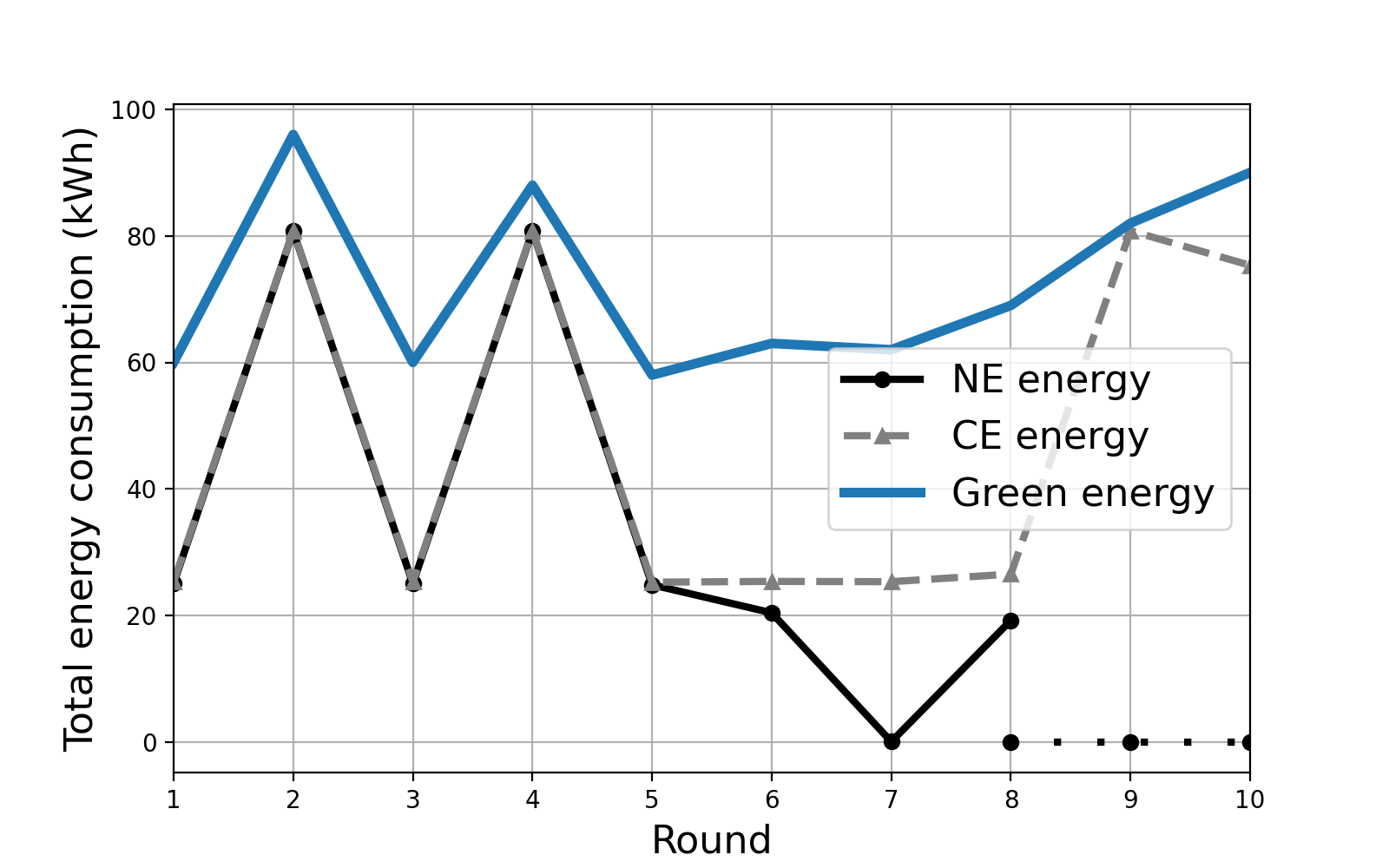}\hfill
    \includegraphics[width=.33\textwidth]{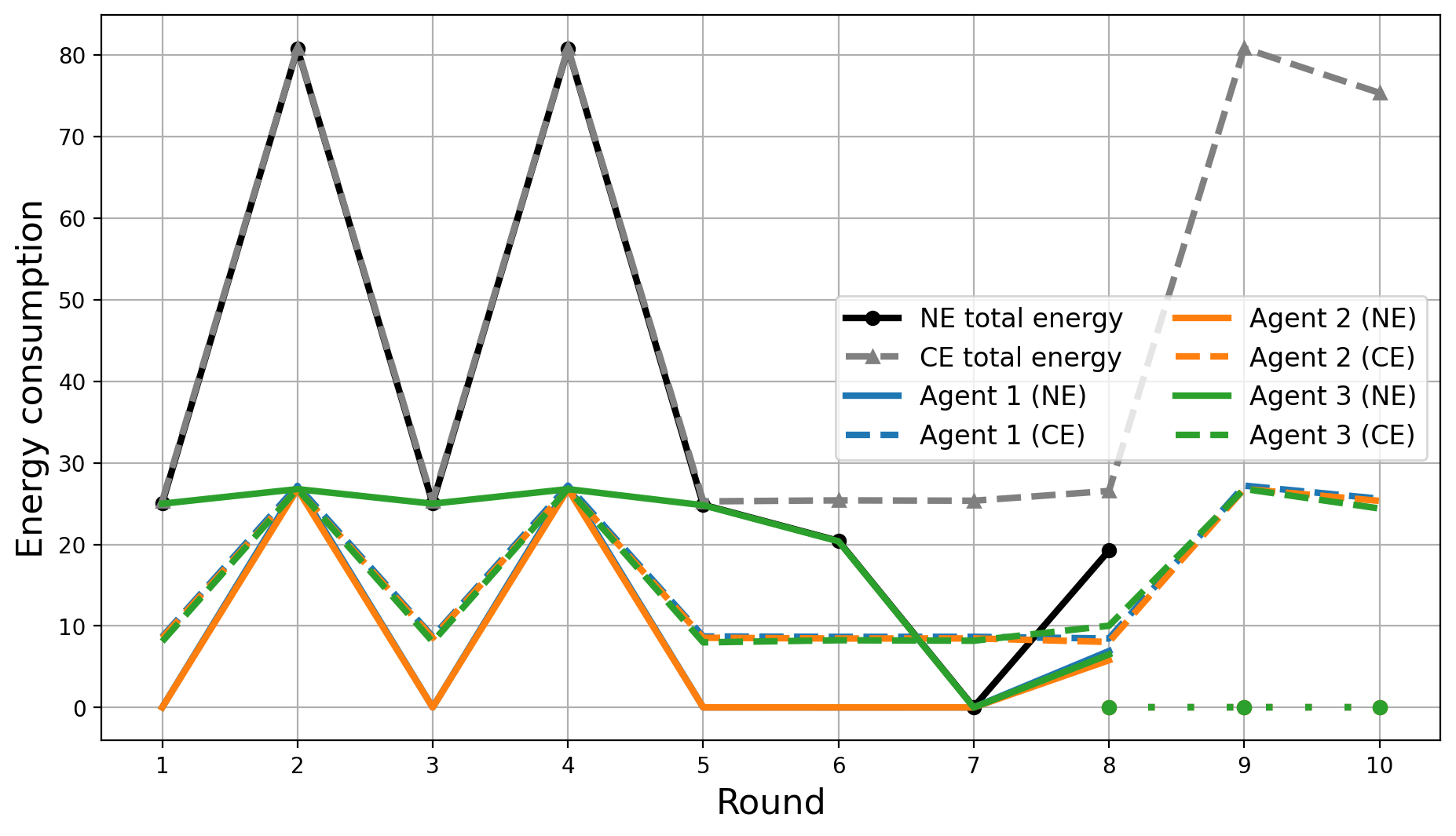}\hfill
    \includegraphics[width=.33\textwidth]{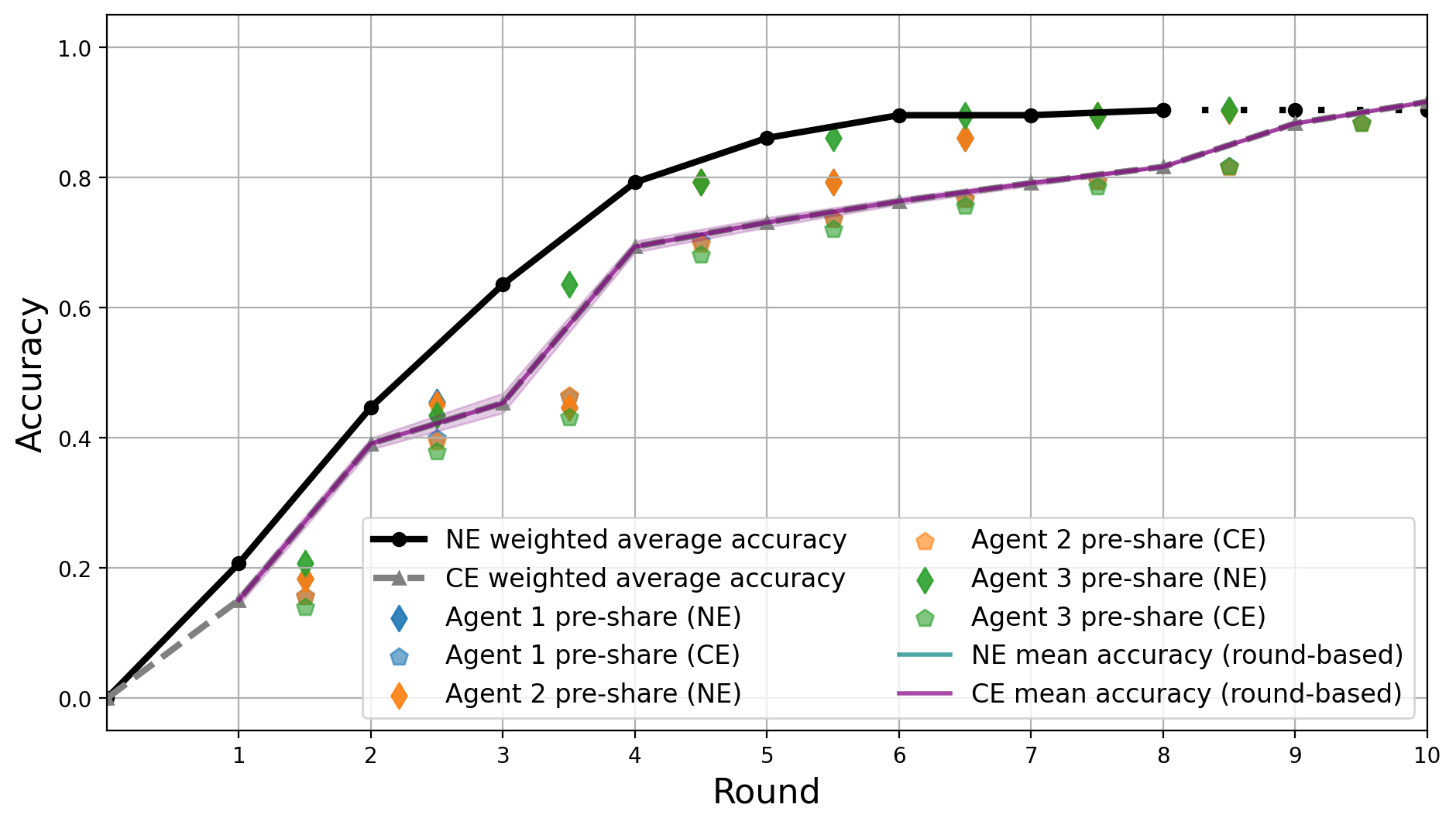}\hfill
    \caption{Quality of data $\theta_i \sim U[0, 0.5]$: $\gamma = 1$ (upper row), $\gamma = 10$ (middle row), and  $\gamma = 100$ (lower row).}
\label{fig:q_0_0_5}
\end{figure}

\begin{figure}[htp!]
    \centering
    \includegraphics[width=.33\textwidth]{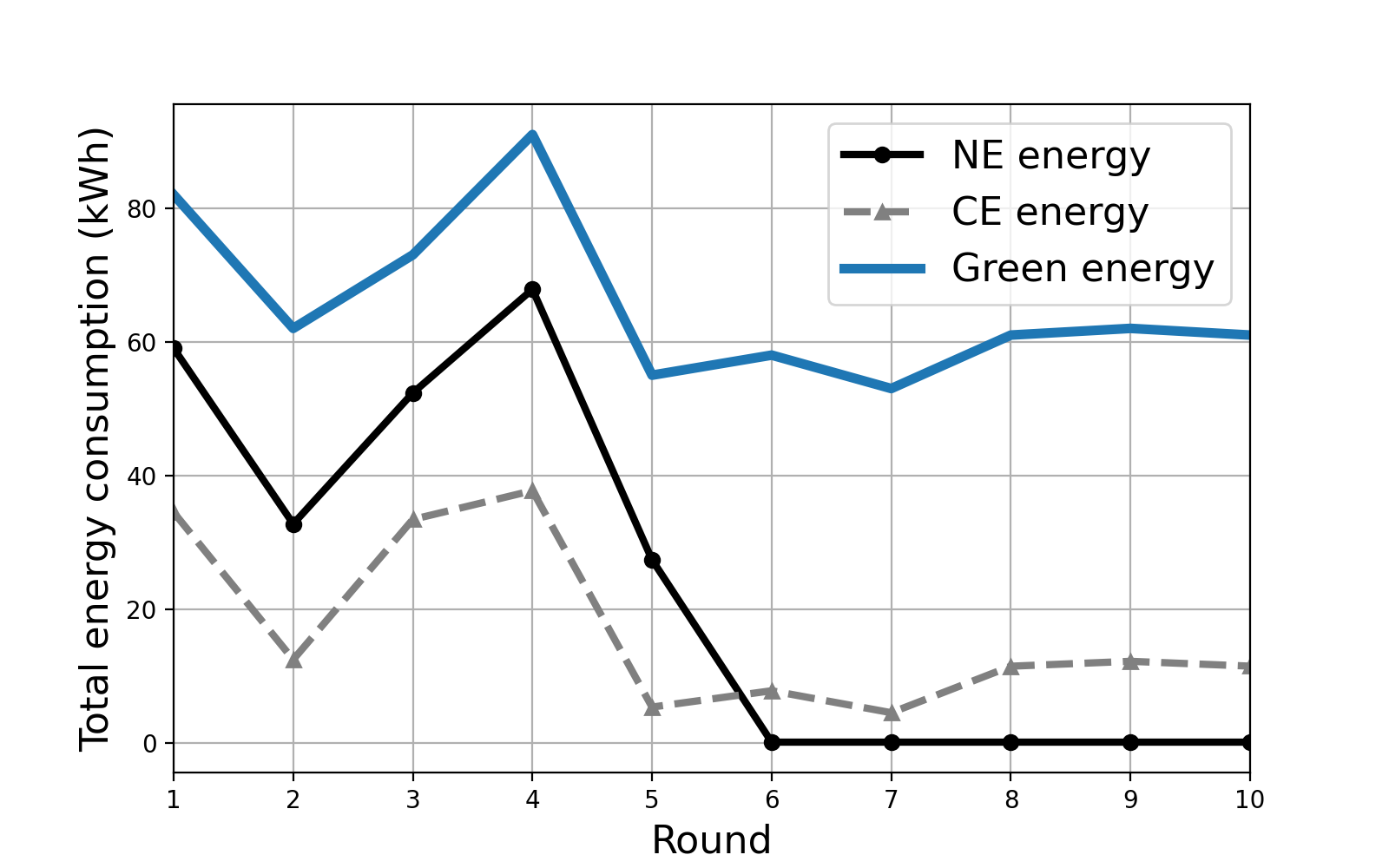}\hfill
    \includegraphics[width=.33\textwidth]{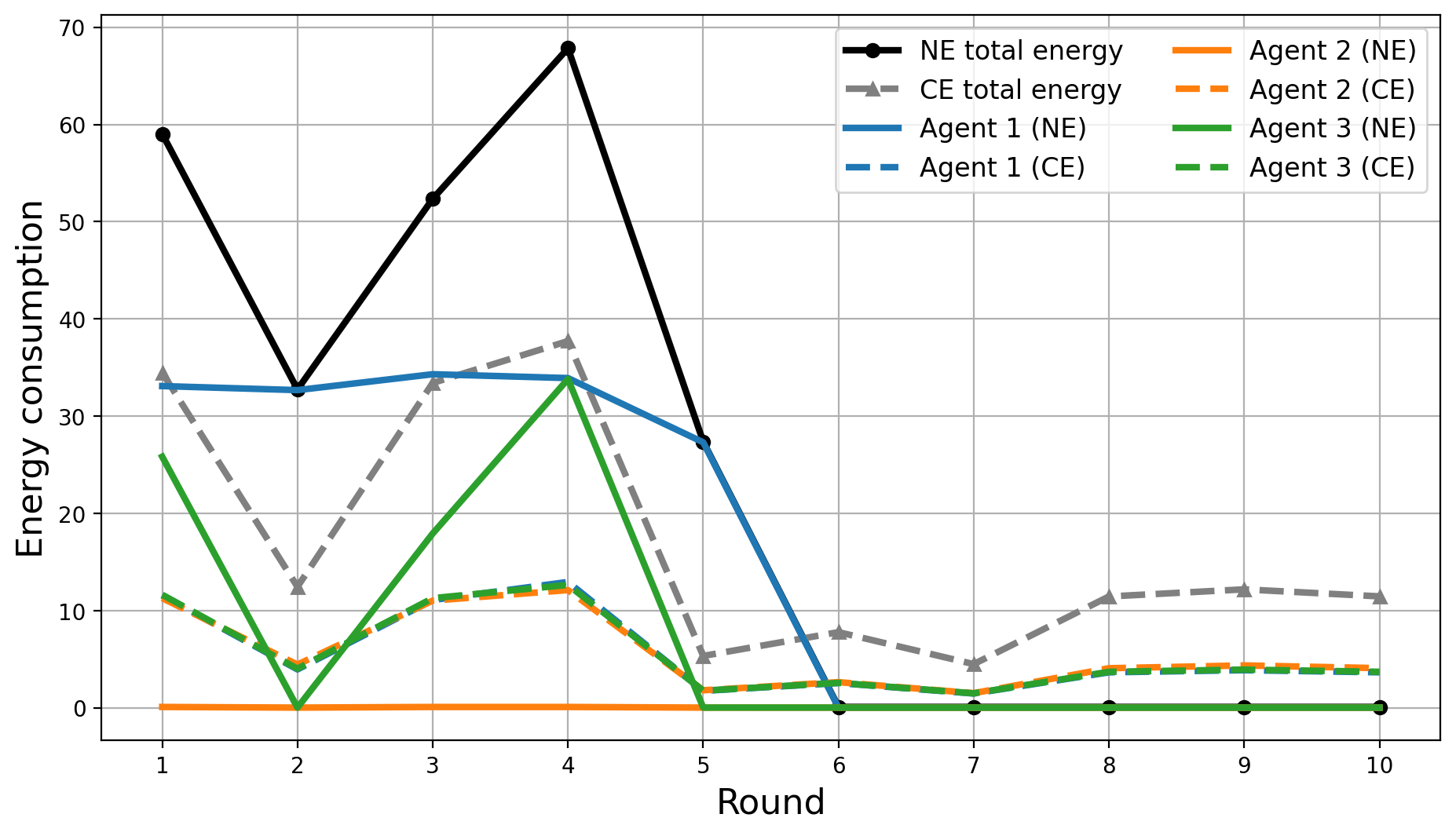}\hfill
    \includegraphics[width=.33\textwidth]{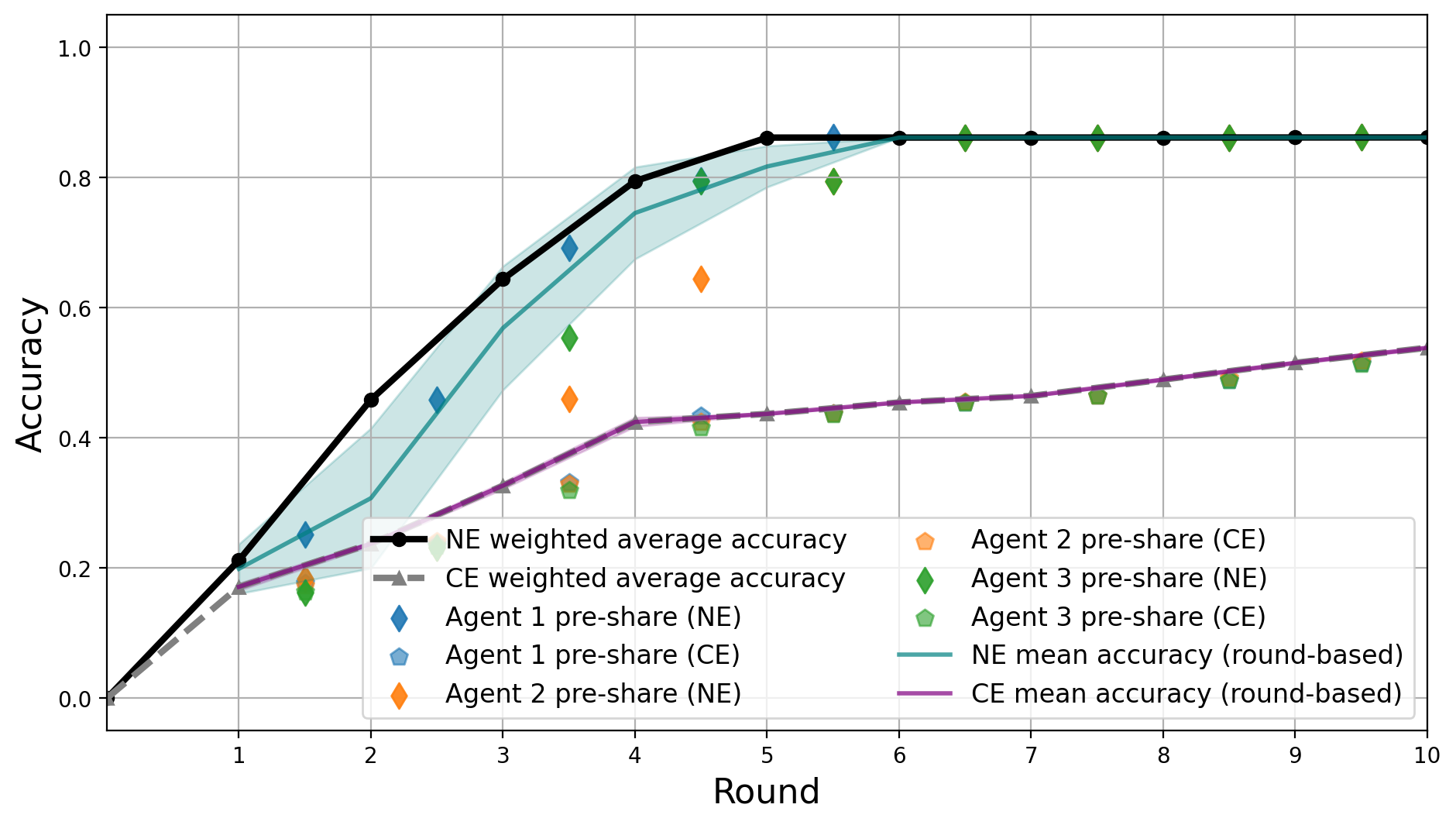}\hfill
    \includegraphics[width=.33\textwidth]{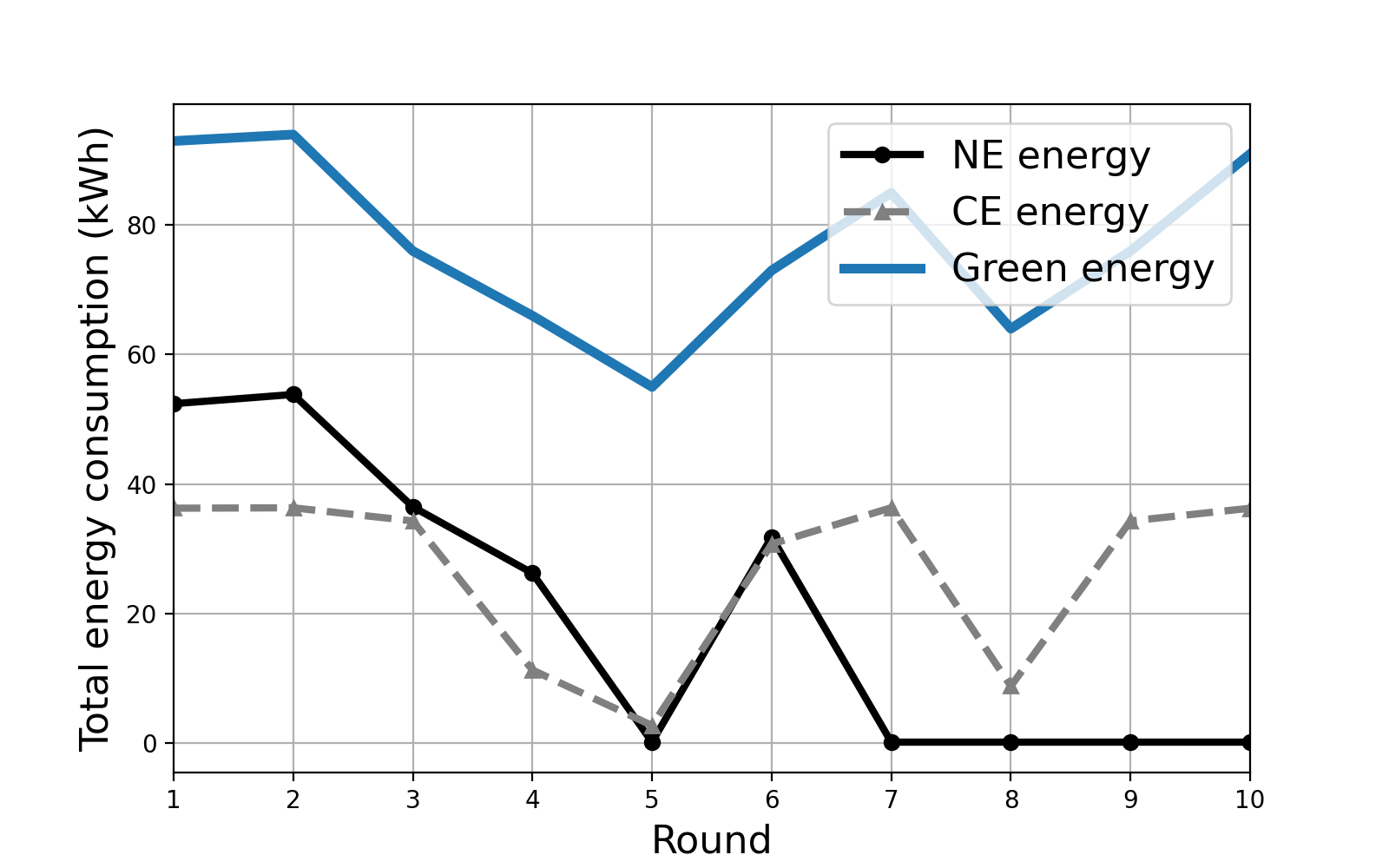}\hfill
    \includegraphics[width=.33\textwidth]{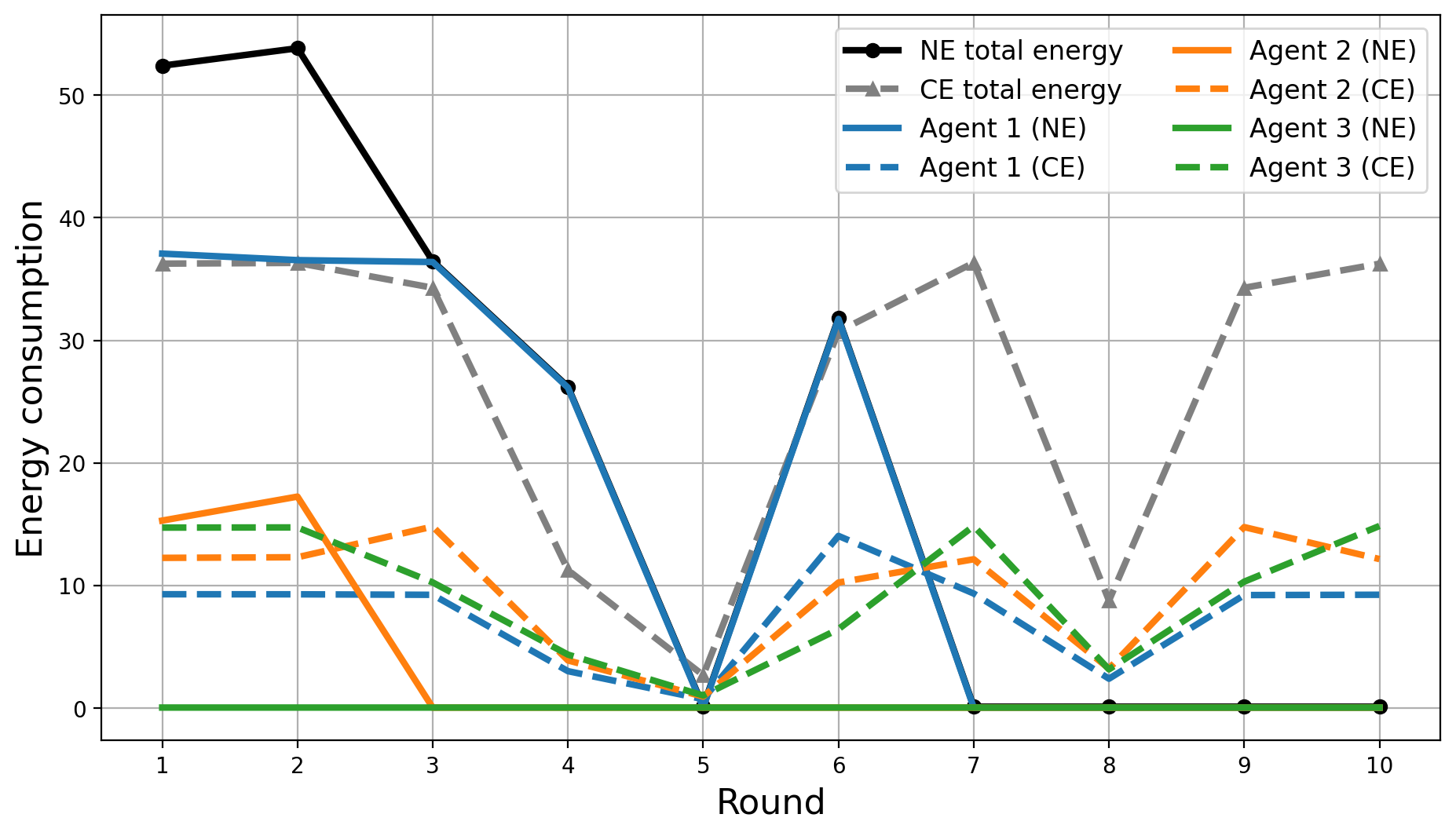}\hfill
    \includegraphics[width=.33\textwidth]{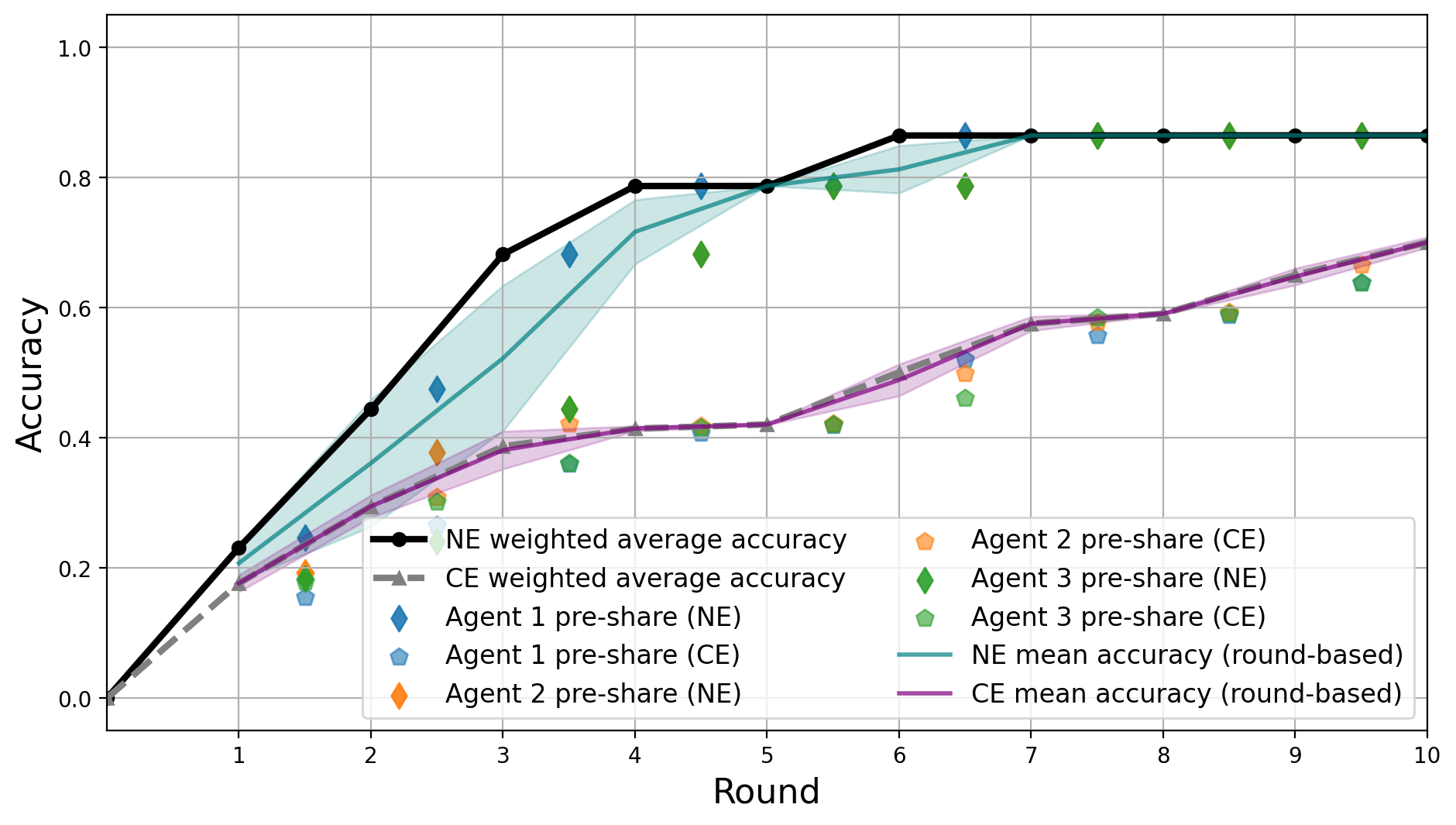}\hfill
    \includegraphics[width=.33\textwidth]{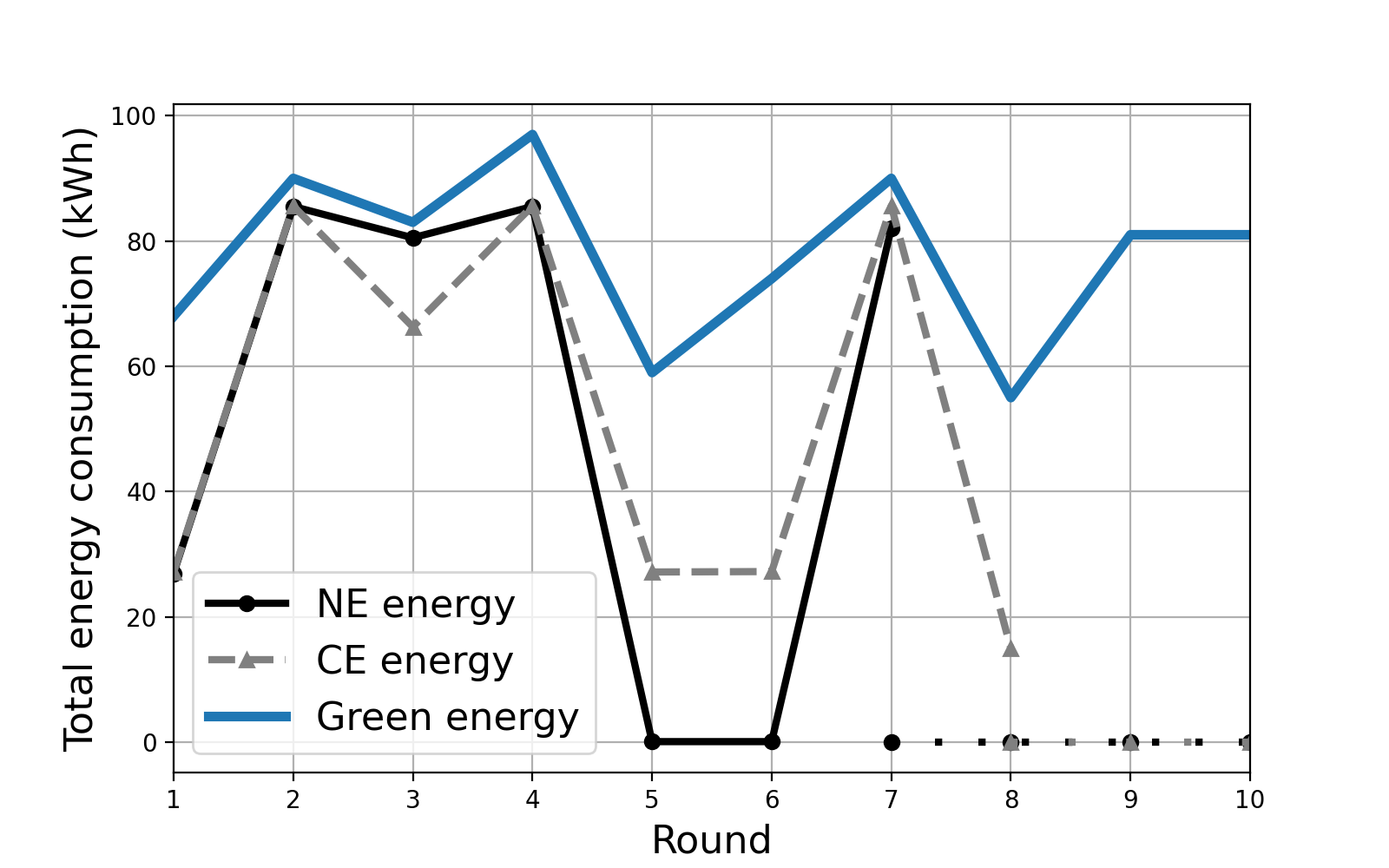}\hfill
    \includegraphics[width=.33\textwidth]{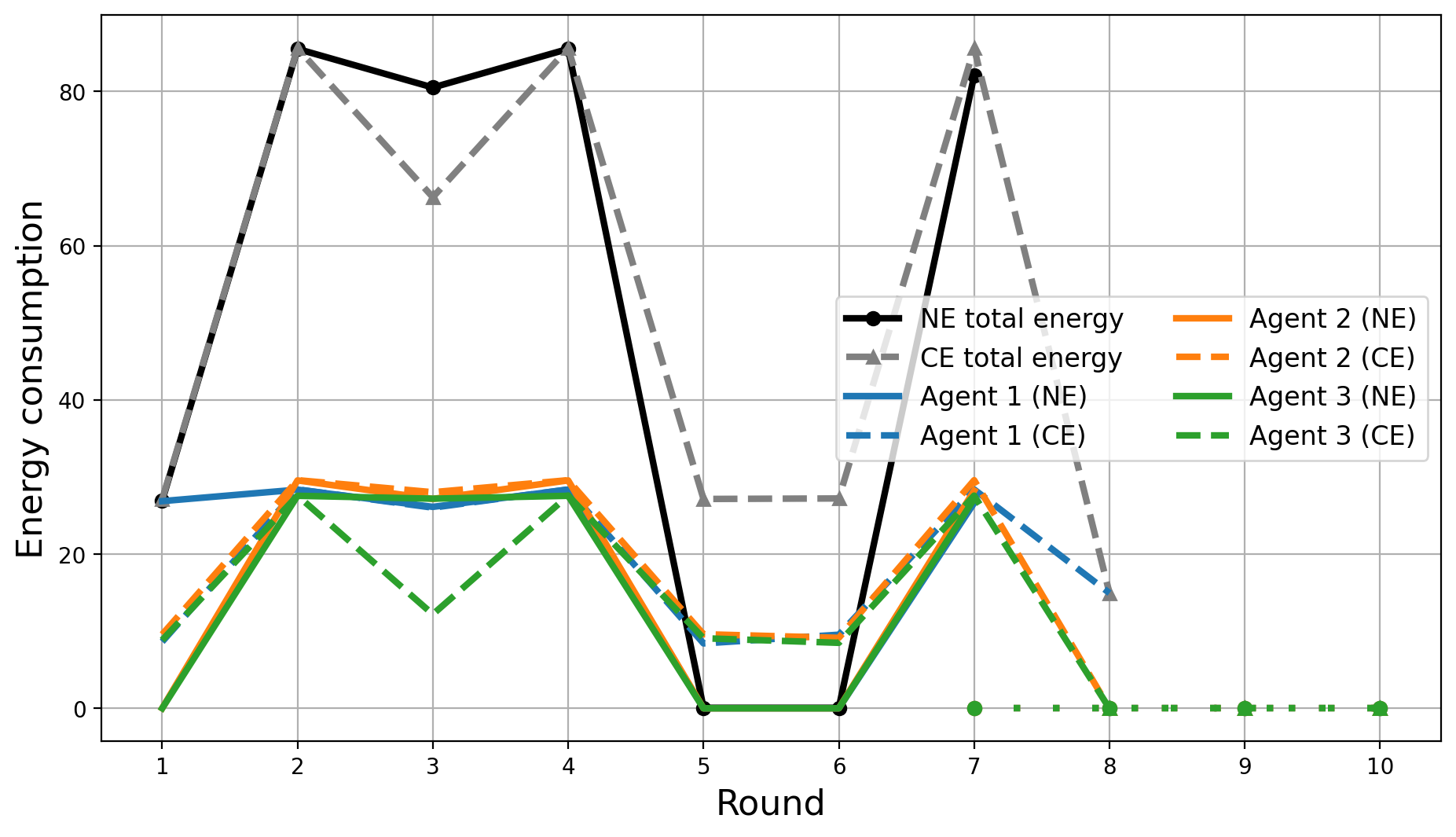}\hfill
    \includegraphics[width=.33\textwidth]{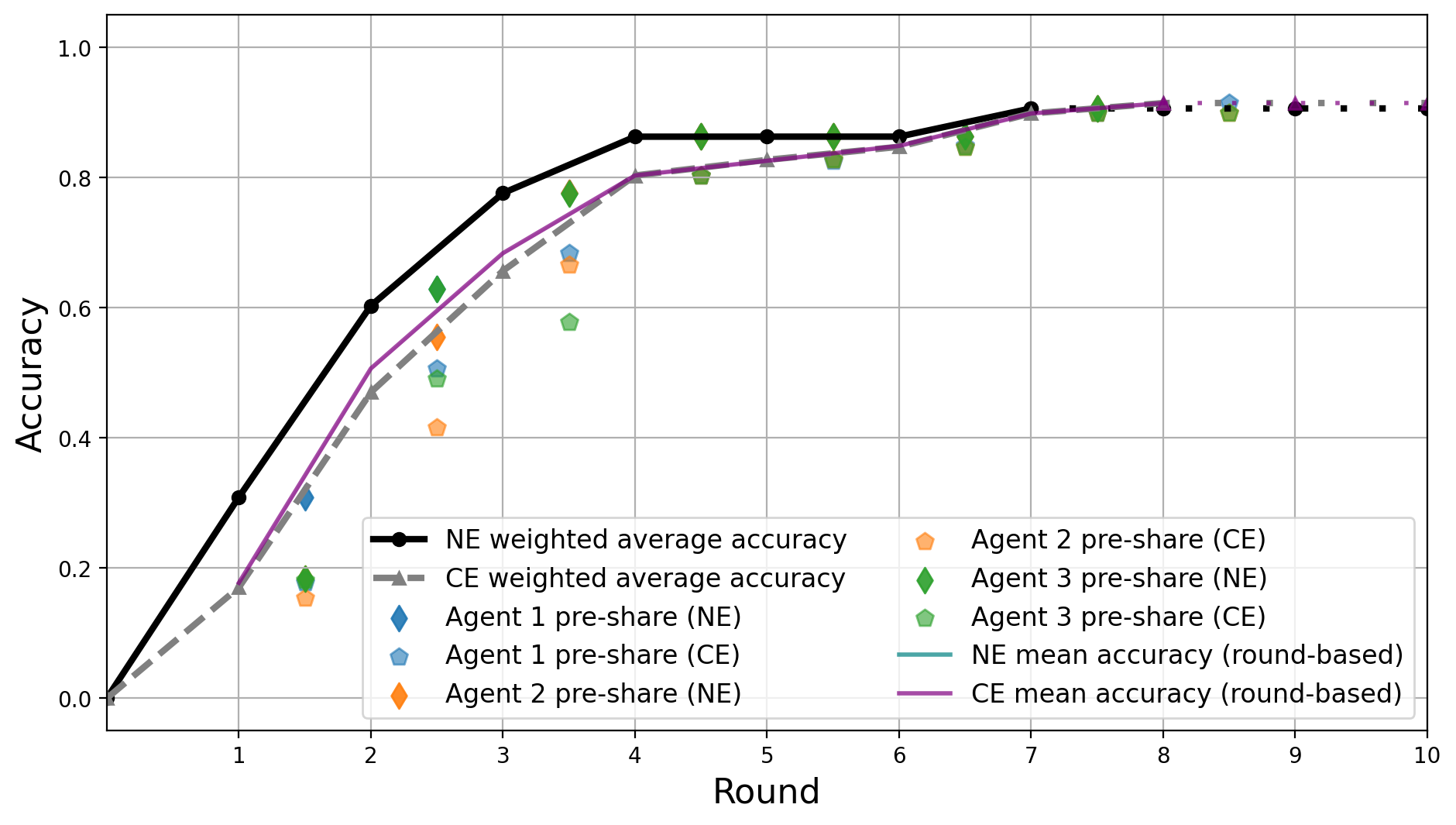}\hfill
    \caption{Quality of data $\theta_i \sim U[0.25, 0.75]$: $\gamma = 1$ (upper row), $\gamma = 10$ (middle row), and  $\gamma = 100$ (lower row).}
\label{fig:q_0_25_0_75}
\end{figure}

\begin{figure}[htp!]
    \centering
    \includegraphics[width=.33\textwidth]{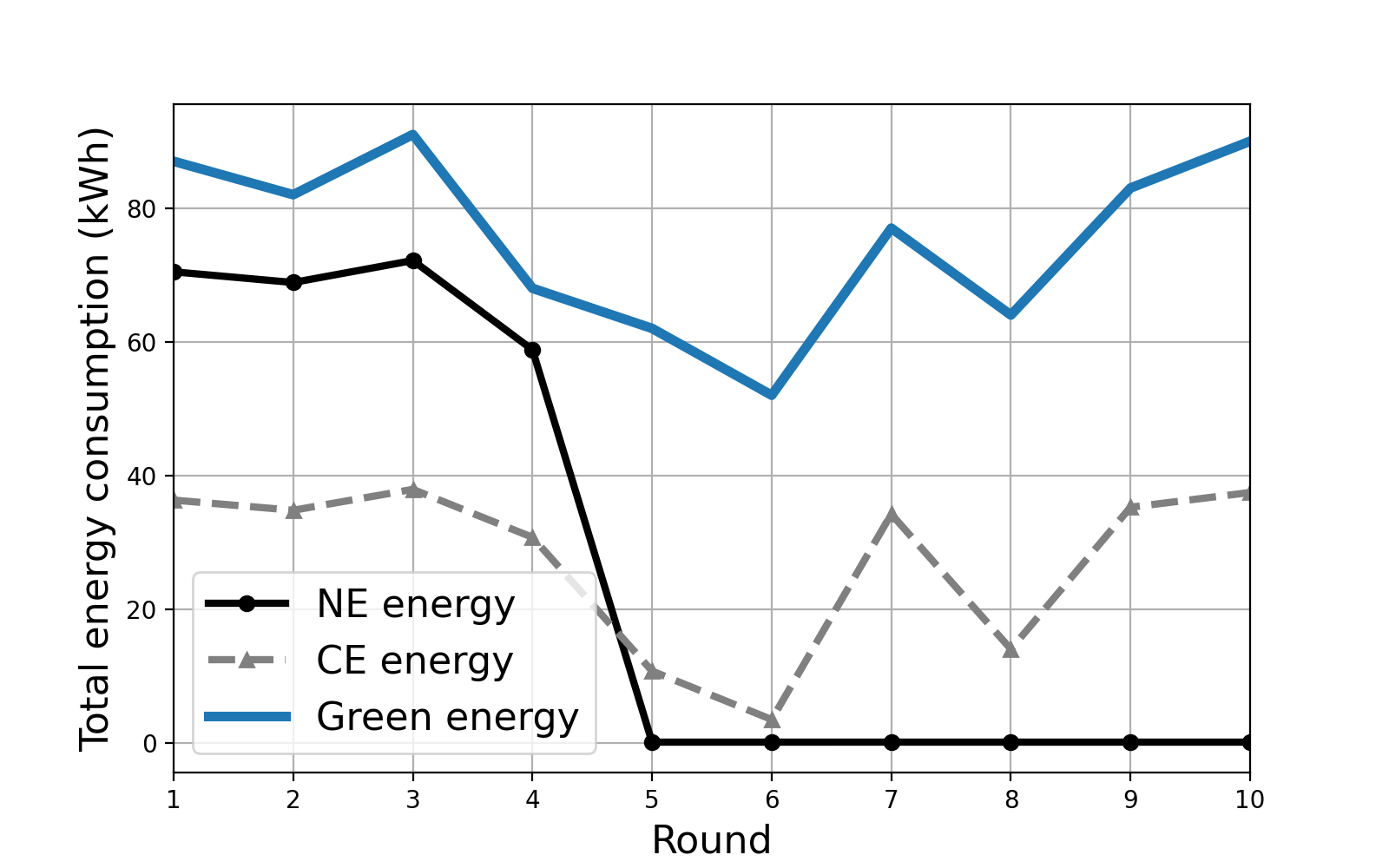}\hfill
    \includegraphics[width=.33\textwidth]{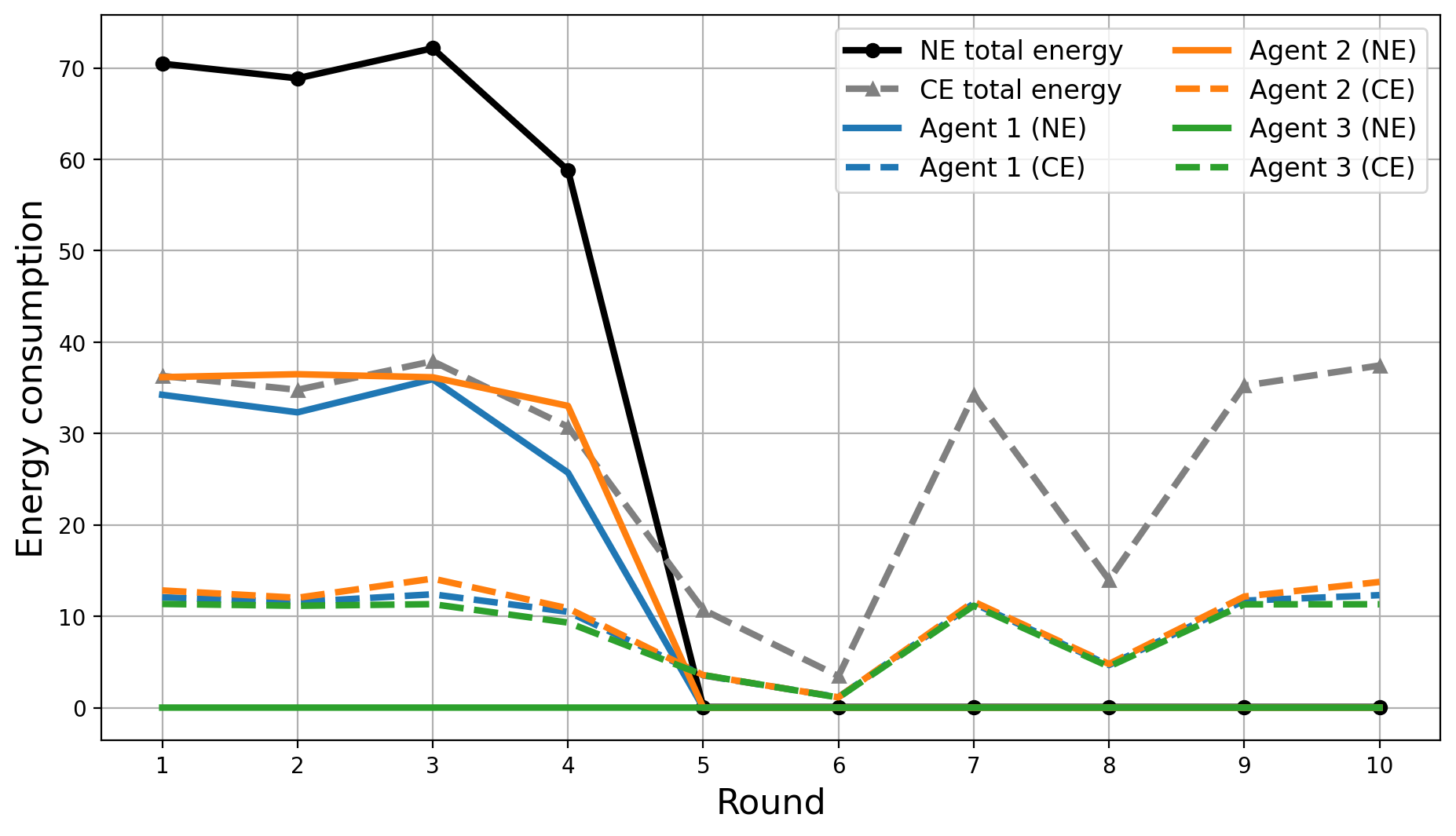}\hfill
    \includegraphics[width=.33\textwidth]{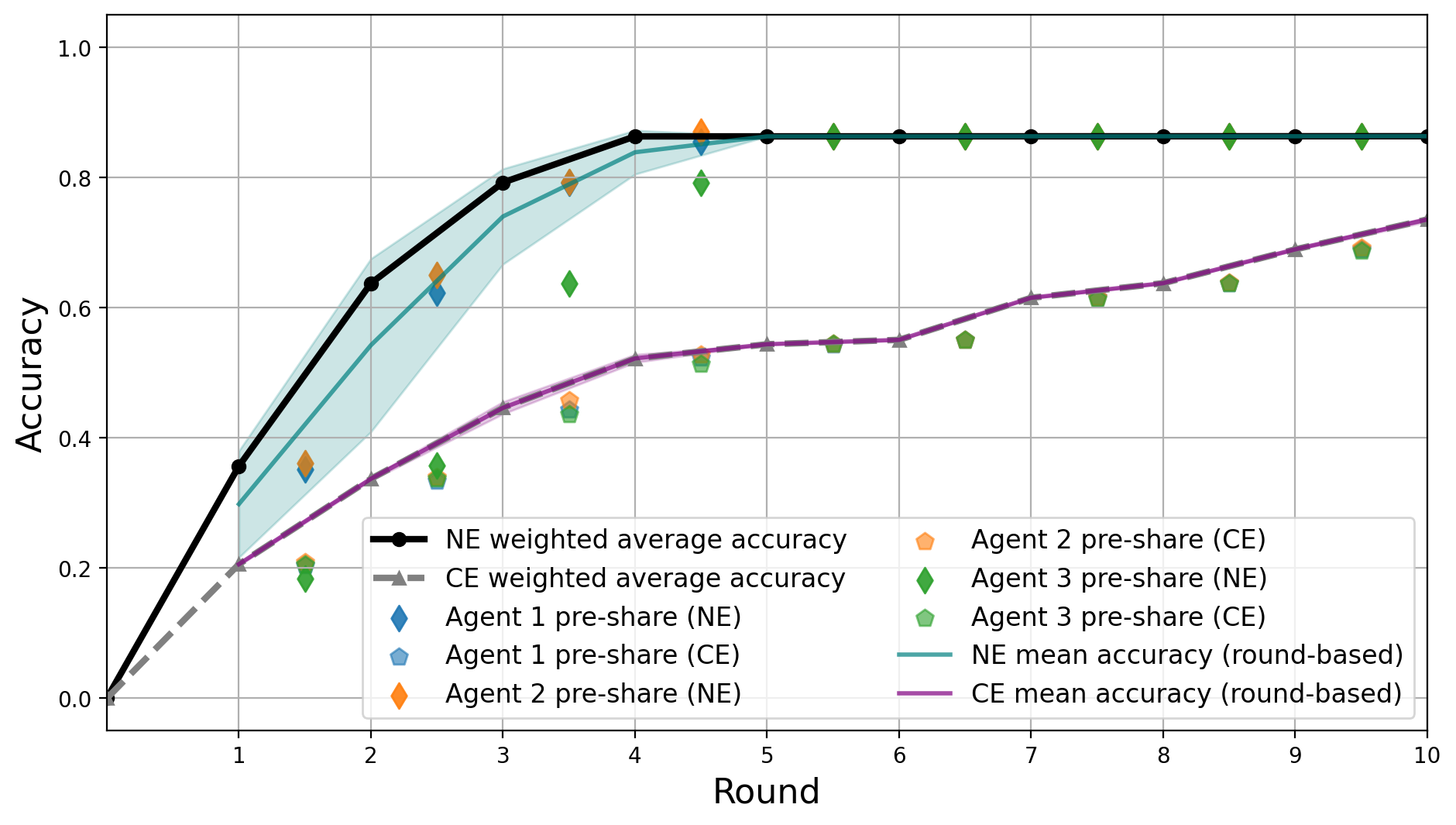}\hfill
    \includegraphics[width=.33\textwidth]{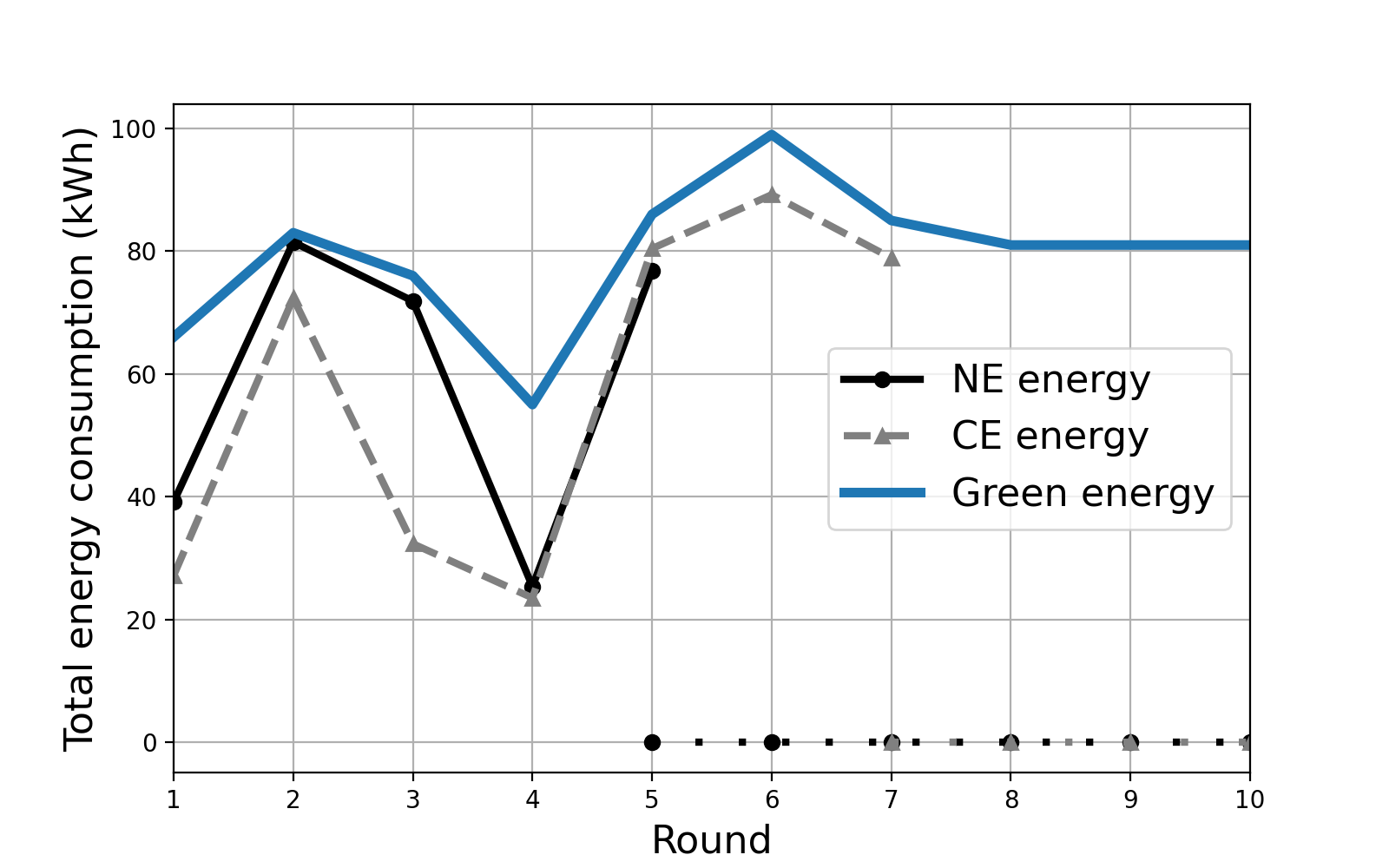}\hfill
    \includegraphics[width=.33\textwidth]{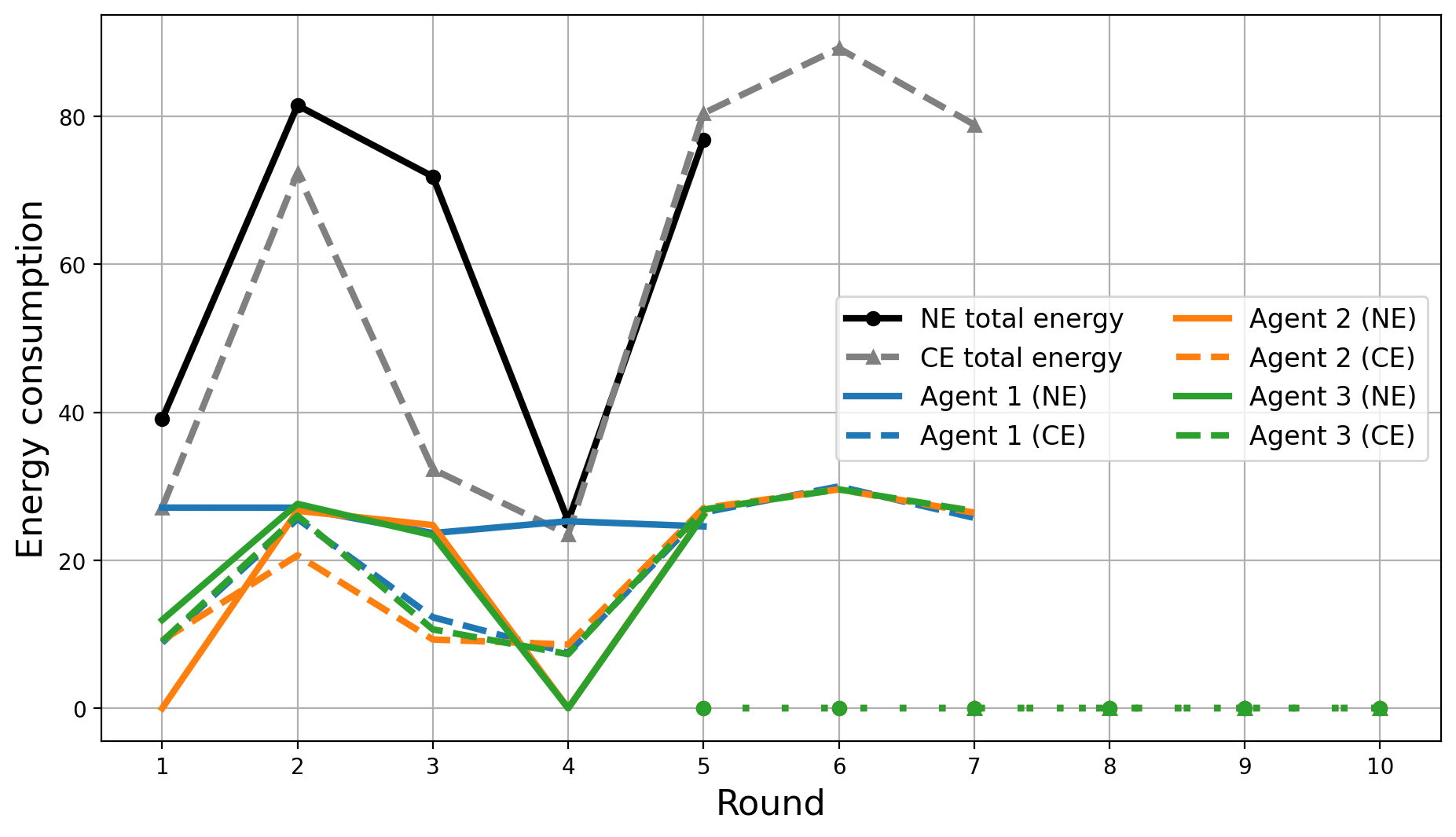}\hfill
    \includegraphics[width=.33\textwidth]{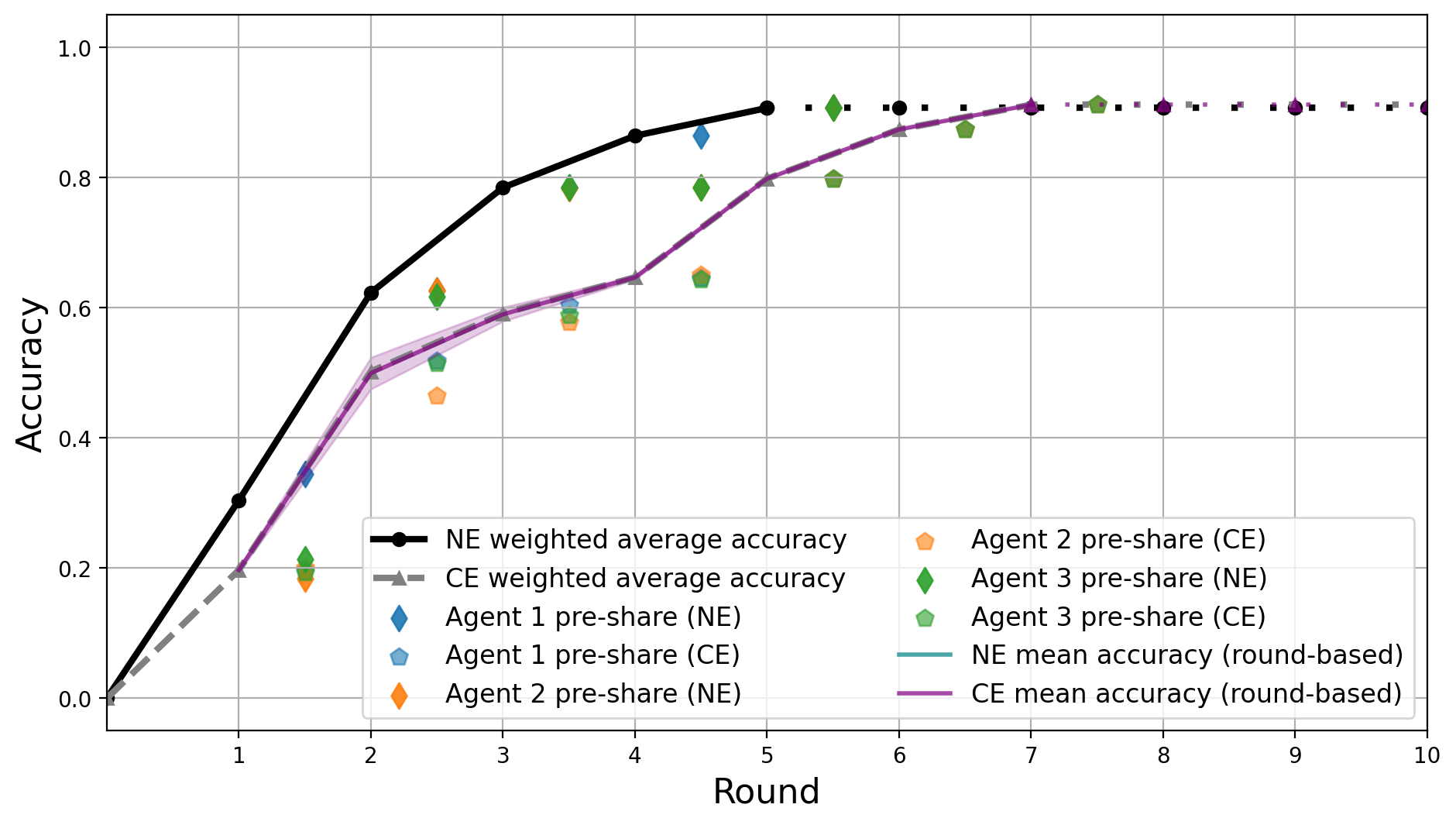}\hfill
    \includegraphics[width=.33\textwidth]{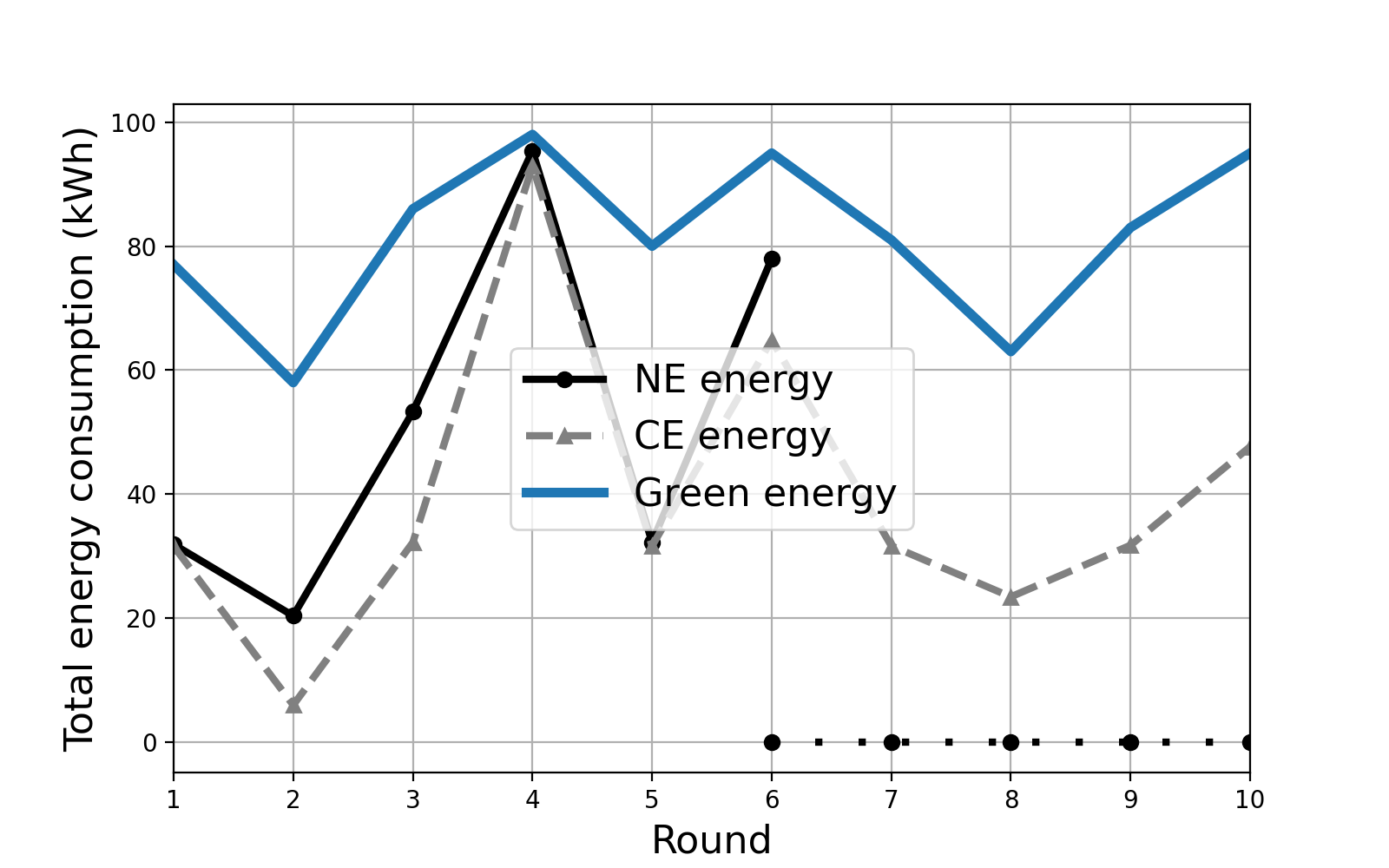}\hfill
    \includegraphics[width=.33\textwidth]{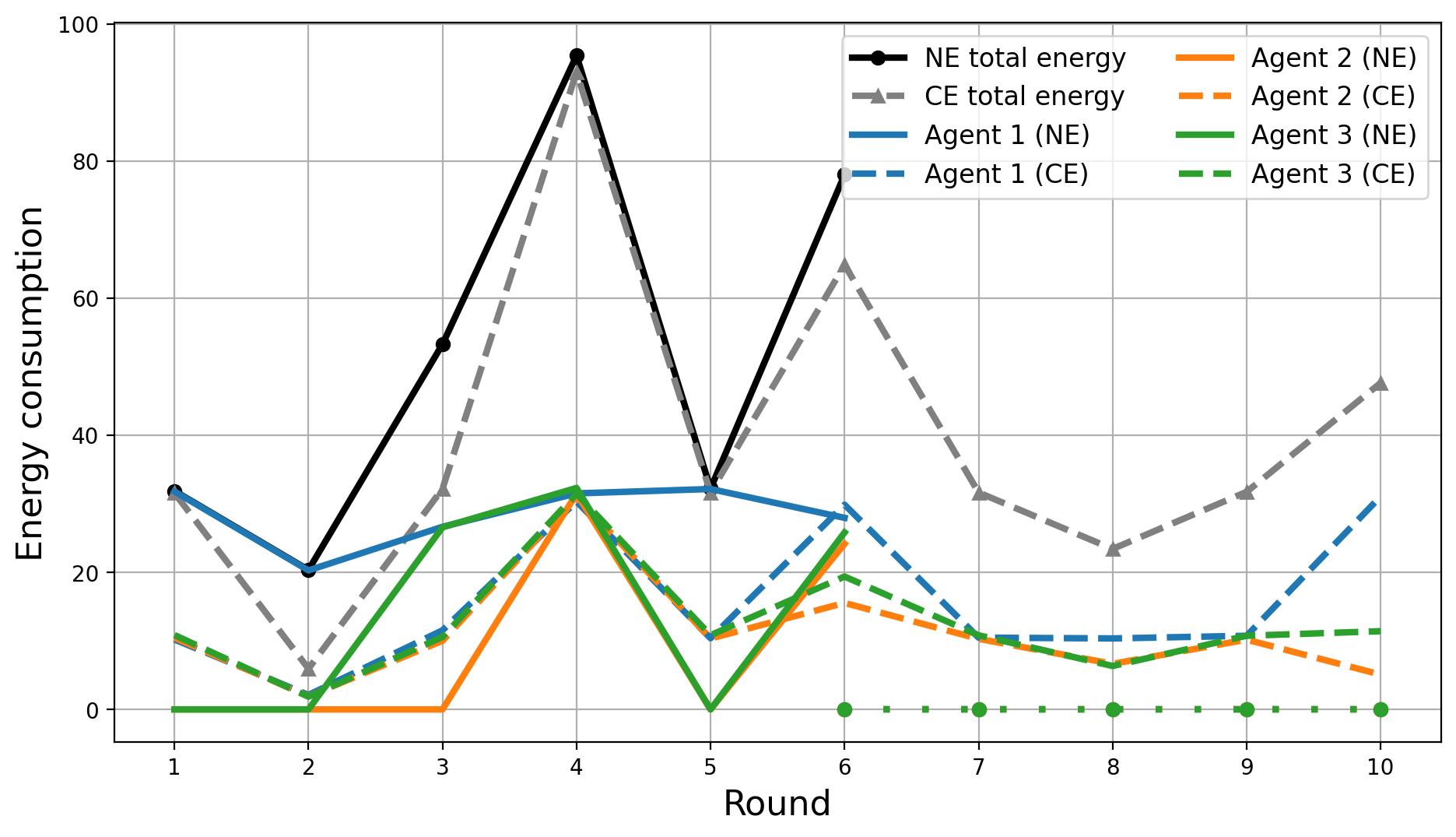}\hfill
    \includegraphics[width=.33\textwidth]{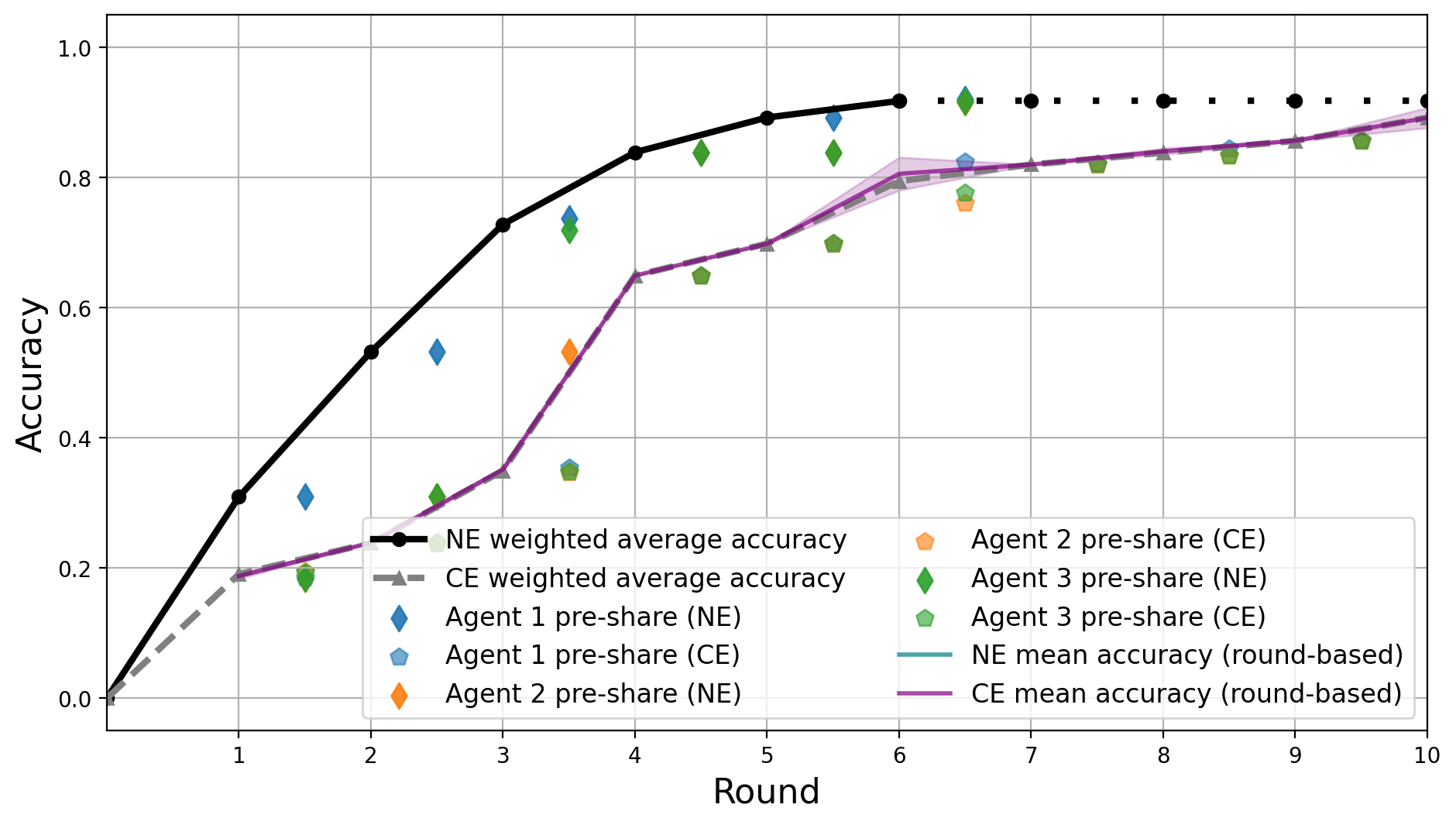}\hfill
    \caption{Quality of data $\theta_i \sim U[0.5, 1]$: $\gamma = 1$ (upper row), $\gamma = 10$ (middle row), and  $\gamma = 100$ (lower row).}
\label{fig:q_0_5_1}
\end{figure}

\paragraph*{Quality of data vs accuracy.} In the right columns in Figures \ref{fig:q_0_0_5}, \ref{fig:q_0_25_0_75}, and \ref{fig:q_0_5_1} we illustrate the training performance of each agent for the three different data-quality regimes and two penalty values, $\gamma \in \{10, 100\}$. We observe that the dispersion of agent performance is larger under the heuristic approach than under the correlation device. In the latter case, performances are more concentrated, which is consistent with the fact that agents with similar characteristics tend to follow more similar training behavior when coordination is imposed.

Figures \ref{fig:q_0_0_5} - \ref{fig:q_0_5_1} show that data quality primarily affects the speed of convergence, rather than the qualitative behavior of the two mechanisms. Under the heuristic approach, the agents generally reach higher final accuracy, with the improvement becoming faster and smoother as data quality increases. This is particularly evident in Figure \ref{fig:q_0_5_1}, where the higher-quality local data allow the target accuracy to be approached with fewer training rounds. This observation is consistent with the previous experiments: when participation decisions are made independently, agents can adapt their training activity to their individual conditions, which generally results in higher accuracy but also greater dispersion across agents.

The correlation device produces more synchronized accuracy trajectories, as expected from the coordination of participation decisions. However, its relative performance varies with data quality. In particular, Figure \ref{fig:q_0_25_0_75} shows that the correlation mechanism performs best in the medium-quality regime, whereas the heuristic approach benefits most clearly from the high-quality data in Figure \ref{fig:q_0_5_1}. We therefore do not observe a uniformly increasing performance advantage for the correlation mechanism as data quality improves. A possible explanation is that coordination imposes a more uniform participation pattern and may therefore prevent the mechanism from fully exploiting differences in the informativeness of individual updates. This interpretation is consistent with the earlier experiments, where coordination was found to produce more synchronized behavior, while decentralized decisions allowed agents to adapt more directly to their individual conditions. However, the experiments do not by themselves establish that this is the sole cause of the observed difference

\paragraph*{Quality of data vs energy.} The effect on energy consumption is considerably weaker. Across Figures \ref{fig:q_0_0_5}, \ref{fig:q_0_25_0_75}, and \ref{fig:q_0_5_1}, the heuristic approach generally consumes more energy than the correlation mechanism, while the overall trajectories remain qualitatively similar across the three data-quality regimes. This is consistent with the previous energy experiments, where energy consumption was primarily determined by the agents' participation decisions and the available renewable energy. Here, data quality affects the learning outcome more directly than the energy allocation, since energy is constrained through the equilibrium incentives rather than being explicitly optimized as a function of $\theta_i$. Thus, within the considered settings, improving data quality mainly translates into faster and more stable learning, while its effect on total energy consumption remains comparatively limited.

Overall, Figures \ref{fig:q_0_0_5} - \ref{fig:q_0_5_1} indicate that data quality and participation incentives interact primarily through the learning dynamics. Higher-quality data allow the heuristic mechanism to exploit individual contributions more effectively, whereas coordinated participation can produce different outcomes depending on the data-quality regime. This complements the earlier observations on energy heterogeneity and model drift: the quality of the available updates affects how valuable continued participation is, while the incentive mechanism determines how that participation is distributed across training rounds. The experiments therefore support an effect of data quality on convergence and accuracy, but they do not justify concluding that higher data quality necessarily reduces energy consumption or that one participation mechanism is universally preferable.

\end{document}